# THERMOELECTRIC AND HEAVY QUARK TRANSPORT COEFFICIENTS OF HOT QCD MATTER IN THE PRESENCE OF MAGNETIC FIELD

**Ph.D. THESIS**

*by*

**DEBARSHI DEY**

**DEPARTMENT OF PHYSICS**
**INDIAN INSTITUTE OF TECHNOLOGY ROORKEE**
**ROORKEE 247667 (INDIA)**

# THERMOELECTRIC AND HEAVY QUARK TRANSPORT COEFFICIENTS OF HOT QCD MATTER IN THE PRESENCE OF MAGNETIC FIELD

**A THESIS**

***Submitted in partial fulfilment of the requirements for the award of the degree***

*of*

**DOCTOR OF PHILOSOPHY**

*in*

**PHYSICS**

*by*

**DEBARSHI DEY**

**DEPARTMENT OF PHYSICS**
**INDIAN INSTITUTE OF TECHNOLOGY ROORKEE**
**ROORKEE 247667 (INDIA)**
**FEBRUARY, 2024**

# INDIAN INSTITUTE OF TECHNOLOGY ROORKEE

## STUDENT'S DECLARATION

I hereby certify that the work presented in this thesis entitled "**THERMOELECTRIC AND HEAVY QUARK TRANSPORT COEFFICIENTS OF HOT QCD MATTER IN THE PRESENCE OF MAGNETIC FIELD**" is my own work carried out during a period from July, 2018 to February, 2024 under the supervision of **Dr. Binoy Krishna Patra**, Professor, Department of Physics, Indian Institute of Technology Roorkee, India.

The matter presented in this thesis has not been submitted for the award of any other degree of this or any other Institute.

**Date:** 19 February, 2024 **(DEBARSHI DEY)**

## SUPERVISOR'S DECLARATION

This is to certify that the above mentioned work is carried out under my supervision.

**Date:** 19 February, 2024 **(BINOY KRISHNA PATRA)**

# ABSTRACT

The aim of this thesis is twofold: a) A comprehensive study of the thermoelectric response in a hot QCD medium [called quark gluon plasma (QGP)] in the absence and presence of a background magnetic field, b) Exploring the dynamics of heavy quarks traversing in QGP in the presence of a weak background magnetic field.

Ultra-relativistic heavy ion collisions (ULRHICs) create huge energy densities in the reaction zone causing nucleons to "melt", thereby leading to the creation of a medium of deconfined quarks and gluons. Fluctuations in the initial energy densities can lead to significant temperature differences between the central and peripheral regions of the expanding fireball. Coupled with a small but finite quark chemical potential, this sets the ground for the QGP to exhibit Seebeck effect. Also, non-central nucleus-nucleus collisions give rise to magnetic fields ($B$) which affect the evolution of QGP. A temperature gradient coupled with a finite magnetic field sets the conditions for the system to exhibit the other thermoelectric phenomenon-The Nernst effect. Thermoelectricity in QGP provides for a new source of electric current, hence, a new source of magnetic field and entropy.

We have evaluated the strength of the aforementioned response in QGP quantified by the Seebeck and Nernst coefficients, first in the absence of a background magnetic field, and then in the presence of a strong magnetic field ($eB \gg T^2$). This is followed by the evaluation of the coefficients in the presence of a weak magnetic field ($eB \ll T^2$). Each of the above-mentioned scenarios is investigated under the assumption that the QGP is isotropic. Realistically, however, the fireball expands anisotropically due to anisotropic pressure gradients caused by geometrical anisotropies in the overlap zone between the two nuclei. We try to incorporate this by using an anisotropic distribution function (Romatchke-Strickland form) for the quarks and evaluate the Seebeck and Nernst coefficients to investigate the effects of anisotropy. The formalism adopted in the calculation of these coefficients is that of kinetic theory. Specifically, the system is assumed to deviate slightly away from equilibrium due to the temperature gradient, and the deviation is evaluated using the relativistic Boltzmann transport equation (RBTE) within the relaxation time approximation (RTA). Interactions among partons are incorporated via the quasiparticle masses of thermal quarks and gluons. We find that the thermoelectric response is the strongest when the background magnetic field is strong, and weakest in the absence of magnetic field. We also observe that a finite anisotropy reduces the strength of the thermoelectric response.

The other part of this thesis deals with HQ dynamics in the QGP. HQs have been recognised as very good probes of the QGP owing to their large masses ($M_Q \gg T^2$). As a result, they do not fully thermalize with the medium, and hence, carry a memory of their interactions therein. We have calculated the HQ energy loss $dE/dx$, longitudinal and transverse *momentum* diffusion coefficients $\kappa_{L/T}$, and *spatial* diffusion coefficient $D_s$, to leading order in the strong coupling $\alpha$, for both Charm and Bottom quarks. We consider Coulomb scattering of the HQ with thermal quarks and gluons, and ignore Compton scatterings to evaluate the scattering rate $\Gamma$ from which, all the aforementioned coefficients are obtained. We find that the values of normalised momentum diffusion coefficients ($\kappa_{L/T}/T^3$) increase in the presence of a weak magnetic field (compared to the $B = 0$ case), and the anisotropy therein ($\kappa_L/T^3 - \kappa_T/T^3$) is also heightened in the presence of the magnetic field. $D_s$ is found to decrease in the presence of magnetic field, compared to its value at $B = 0$.

# LIST OF PUBLICATIONS

## REFEREED JOURNALS

1. Seebeck effect in a thermal QCD medium in the presence of strong magnetic field,

   **Debarshi Dey** and Binoy Krishna Patra,

   Phys. Rev. D **102**, 096011 (2020).

2. Thermoelectric response of a weakly magnetized thermal QCD medium,

   **Debarshi Dey** and Binoy Krishna Patra,

   Phys. Rev. D **104**, 076021 (2021).

3. Thermoelectric response in a thermal QCD medium with chiral quasiparticle masses,

   **Debarshi Dey** and Binoy Krishna Patra,

   Int. J. Mod. Phys. E **31**, 250097 (2022).

### Under review

1. Dynamics of open heavy flavour in a weakly magnetized thermal QCD medium,

   **Debarshi Dey** and Binoy Krishna Patra,

   arXiv:2307.00420.

2. Thermoelectric response of a hot and weakly magnetized anisotropic QCD medium,

   Salman Ahamad Khan, **Debarshi Dey** and Binoy Krishna Patra,

   arXiv:2309.08467 [hep-ph] (2023).

## CONFERENCE PROCEEDINGS

1. Thermoelectric effects in a weakly magnetized QGP,

   **Debarshi Dey** and Binoy Krishna Patra,

   Proceedings of the DAE Symposium on Nuclear Physics 65, 688 (2021).

2. 1-D and 2-D Seebeck coefficient of a thermal QCD medium with chiral-mode dependent quasiparticle massesn,

   **Debarshi Dey** and Binoy Krishna Patra,

   Proceedings of the DAE Symposium on Nuclear Physics 66, 928 (2022).

## WORKSHOPS/CONFERENCES/SYMPOSIA ATTENDED

1. 67$^{th}$ DAE Symposium on Nuclear Physics, IIT Indore, Indore, Madhya Pradesh, India, December 09-13, 2023.

2. 8$^{th}$ International Conference on Physics and Astrophysics of Quark Gluon Plasma (ICPAQGP 2023), Puri, Odisha, India, February 07-10, 2023.

3. 66$^{th}$ DAE Symposium on Nuclear Physics, Cotton University, Guwahati, Assam, India, December 01-05, 2022.

4. 18th International Conference on QCD in Extreme Conditions (XQCD 2022), The Norwegian University of Science and Technology, Trondheim, Norway, July 27-29, 2022.

5. Hot QCD matter 2022, School of Physical and Applied Sciences, Goa University, Goa, India, May 12-14, 2022.

6. 65$^{th}$ DAE Symposium on Nuclear Physics, Bhabha Atomic Research Centre, Mumbai, India, December 01-05, 2021 (Online).

7. ECT* Doctoral Training Program: High-Energy and Nuclear Physics within Quantum Technologies, June 28-July 23, 2021 (Online).

8. ECT* TALENT school on Machine Learning and Data Analysis for Nuclear Physics, June 22-July 03, 2020 (Online).

9. The 2020 JETSCAPE online summer school, July 12-22, 2020 (Online).

# ACKNOWLEDGMENTS

I never thought that one day I would be writing my thesis acknowledgement, when I got into the Physics Hons. B.Sc program of my local college (Karimganj College, Karimganj, Assam). God has been kind, and in addition, I have had numerous wonderful people along the way, whom I take this opportunity to acknowledge.

First and foremost, I take this opportunity to express my sincere and deepest gratitude to my Ph.D. advisor ***Prof. Binoy Krishna Patra***, for providing me with the wealth of knowledge, support, and encouragement throughout this endeavor. I am grateful for his constructive feedback, guidance, and instructions during my stay at IIT Roorkee (IITR). It has been possible to make this thesis in its current shape owing to his suggestions and assistance. I have learnt a lot about the subject and beyond from him, which will definitely guide me in future. I thank him for everything he has enabled me to do in the past five years.

I consider myself fortunate to have had a lovely batchmates here at IIT Roorkee. Life here would have been barren without the *Nakli* group, the members of which have added flavor, fervour and inspiration in my life. I have been enjoying Jaikhomba's delicious food since our time together at Tezpur University (M.Sc) and that has continued here. I shall always remember the bulk of horror movies we saw during our initial days at IITR. I will cherish the time I spent with Chandan, especially during the lockdown; studying, playing, eating together. Sachin and Alok have been inspirational in their own ways. I thank Rishabh for being the entertainer of the group, and a helpful friend. I also acknowledge the conversations I have enjoyed with Bhanu. The lovely ladies of the group Priyanka, Madhu, Malvika, Richa and Jinti deserve special mention for bearing with the madness of the male members, and in many instances, joining them. They are great examples of fun-loving, hard-working and confident women.

I have been blessed with a very good research group. My seniors Shubhalaxmi Rath, Mujeeb Hasan and Salman Ahamad Khan have always been approachable, supportive and encouraging, throughout my journey. Apart from learning the tricks of the trade, I have also found succour speaking with them during stressful academic situations. My most recent senior, Sarthak, has been an inspiration to me because of his energy, enthusisasm and leadership abilities. I thank him for being the friendly person he is, and for being supportive, and look forward to learning from him. I have a wonderful set of juniors as well: Pushpa, Sumit and Abhishek Tiwari, all of whom, I believe, are better than me, and bright prospects for the future. They are more friends than juniors, and I have spent some quality time with them attending conferences, visiting places, which I shall always cherish. Outside my research group, I would thank the members of Research scholar Hall 1 throughout my journey for maintaining a very friendly atmosphere in our cabin: Aloka, Nitish, Hemant, Ayushi, Shikha, Neelam Didi, Swati Didi, Dibyendu Da, Shubham bhaiyya. I have been enriched by the company of all of them. Additionally, I have spent quality time with the ever helpful Ashutosh and the cheerful Himanshu, which I shall always cherish.

The biggest superpower of a **good** teacher is the ability to inspire, and hence, shape the future of students. For me , that person is Dr. Sujit Tewari. Learning physics from him (in 12th standard and beyond) is what set me on this path. I thank him for igniting in me the sense of curiosity and wonder for Physics. I would also like to acknowledge the encouragement and support of other professors of Karimganj College, *viz.* Dr. Sabyasachi Roy, Dr. Suchismita Dutta Sarkar, and Dr. Nirmal Kumar Sarkar, for their lessons, guidance, and blessings. I shall always remember the skill and perfection of Mr. Vinod Yadav, my high school (Kendriya Vidyalaya, Karimganj) Mathematics teacher, who was an artist with mathematics, and someone whom I looked up to. Going further back, I would like to acknowledge my secondary school (Maharishi Vidya Mandir, Karimganj) teachers of Mathematics, English and Chemistry: Mr. Dibakar Dutta, Mr. Amritava Purkayastha, and Mrs. Mita Sinha Das for their dedication to teaching and their contribution towards improving me as a student.

One can never forget their school friends, and I would like to remember the many who still continue to be a source of happiness, companionship and support. I go back the longest with Shiuli (23 years and counting) who has been one of my best friends, along with Sunasree, whom I thank for all the madness and laughs she gave rise to. I also acknowledge the continouous support of Bikram, Arka, Abhishek and Debojyoti in my life. Bikram is that doctor friend that all the non doctor people dream of having (and not the other way round). I hope someday I become rich enough to repay all the free consultations he has provided to me, my relatives, my friends, my friends' relatives, etc. Additionally, he has always been supportive and available, and I thank him for that. These are people who have inspired me, whom I trust, and find comfort spending time with. I also thank Debarpita for being a sweet friend-cum-sister to me through all these years despite being away, and making me smile with her beautiful *Raakhis*.

At the end, I wish to acknowledge three special people who were involved the most with me in this journey of mine. Firstly, Anish, my friend from the M.Sc days, with whom I have had countless conversations regarding work, life, relationships, and mostly, cricket! I thank him for sharing countless of his happy and sad moments with me, and also for his words of comfort and encouragement during my bad times. Secondly, Pushpa, who has been a sister-like friend and a lovely junior. She has always been helpful and supportive, and one of the few persons I felt comfortable sharing a lot of things with. I will cherish the memories of the many places we visited together, both national and abroad, the many conversations on spirituality, religion, and life in general. She is a strong woman, better than me on many fronts, who continues to inspire me through her dedication and hard work.

I have saved the most important person for the last: Priyanka. Throughout my entire journey, she has been a constant and the most intense source of support. She has always believed in me more than I believe in myself. She has always been the first person for me to turn to, in moments of happiness and distress both. She has taught me alot of things about myself, and in the process, helped me become a better person. I look up to her as an inspiration when it comes to breaking glass ceilings and fighting one's fears. I shall never forget the many wonderful moments we have spent together; learning, fighting, laughing, crying. The places we visited, the food we cooked, the movies we saw, are memories that I shall always keep going back to. Thank you for everything.

**Date: November 30, 2023** **(DEBARSHI DEY)**

*Dedicated to my parents*

Mrs. Seema Dey

&

Mr. Dipak Ranjan Dey

# Contents

# List of Figures

# List of Tables

# Chapter 1

# Introduction

ॐ असतो मा सद्गमय । तमसो मा ज्योतिर्गमय । मृत्योर्मा अमृतं गमय । ॐ शान्तिः शान्तिः शान्तिः ॥

*Lead me from ignorance to truth. Lead me from darkness to light. Lead me from death to immortality. May peace be, may peace be, may peace be.*

– **Bṛhadāraṇyaka Upaniṣad**

Delving deeper, both literally and figuratively, into the mysteries of nature, has been a constant history of mankind at all points of time. In particular, the fundamental constituents that make up the observable universe has been a subject of perennial intrigue among scientists and philosophers alike. This quest led to Rutherford's discovery of the nucleus (proton) [1], Chadwick's discovery of the neutron [2], and consequently the discovery of internal structure of the proton via Deep Inelastic $e$-$p$ scatterings [3]. In addition to the most recent revelation of *quarks* and *gluons* constituting the nucleons, the quest has also led to the discovery of fundamental forces or interactions that govern the universe. As of today, four fundamental forces are proven to exist: Gravitaional, Electromagnetic, Strong, and Weak. The theory of Quantum Chromodynamics (QCD) describing the strong force, will be the focus of attention in this thesis.

In the cosmic scale, the origin and subsequent evolution of the universe has always been the centre of scientific investigation. The Big Bang theory proposed by G. Lemaitre [4] in 1927 is currently the most reliable theory to that end, which suggested that the universe was much hotter and smaller (denser) towards the beginning of time, and gradually cooled and diluted as it expanded. That the universe has been expanding since the Big Bang was ultimately confirmed by the red-shift measurements of E. Hubble [5] in 1929.

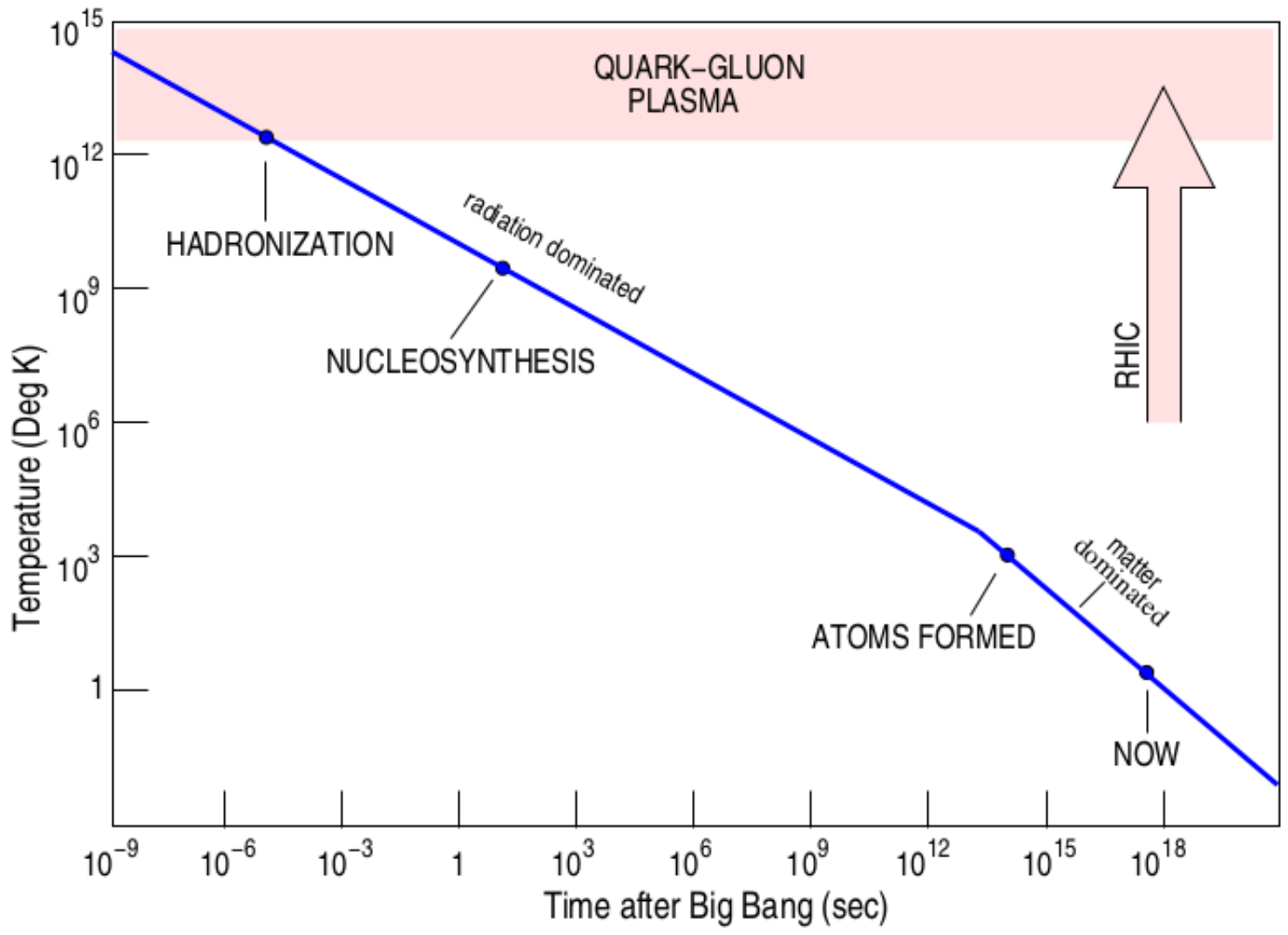

**Figure 1.1:** Conjectured temperature history of the universe

Fig.(1.1) shows the temperature history of the evolution of the universe from the Big Bang to the present day. In the due course of its expansion, the universe underwent many phase transitions, such as grand unification, electroweak, etc. The relevant one in our context is the QCD phase transition from a state called the Quark-Gluon Plasma (QGP) to hadronic matter (Hadronization point in Fig.(1.1)]), which takes place several $\mu$ sec after the Big Bang [6]. Subsequently, following the Big Bang nucleosynthesis, protons and neutrons combined to form atomic nuclei which forms the basis of most of the observed matter today. Also the QCD phase transition is the only phase transition that can be accessed directly via experiments in terrestrial laboratories today. This is achieved by colliding nuclei at ultra relativistic energies (aptly called The Little Bang) at state-of-the-art colliders, which leads to the creation of an extremely hot, dense and short lived QGP [7, 8], reminiscent of the nascent universe. This thesis deals primarily with the properties of this medium called the QGP.

## 1.1 The Quark-Gluon Plasma

Ultra relativistic nucleus-nucleus collisions create energy densities so high ($> 1\,\mathrm{GeV/fm^3}$) that matter no longer consists of separate hadrons (protons, neutrons, etc.), but of their fundamental constituents, quarks and gluons. The existence of such a 'partonic matter' can be motivated from the theory that describes strong interactions, *i.e.* QCD, which has two key aspects: *Asymptotic freedom* and *confinement*.

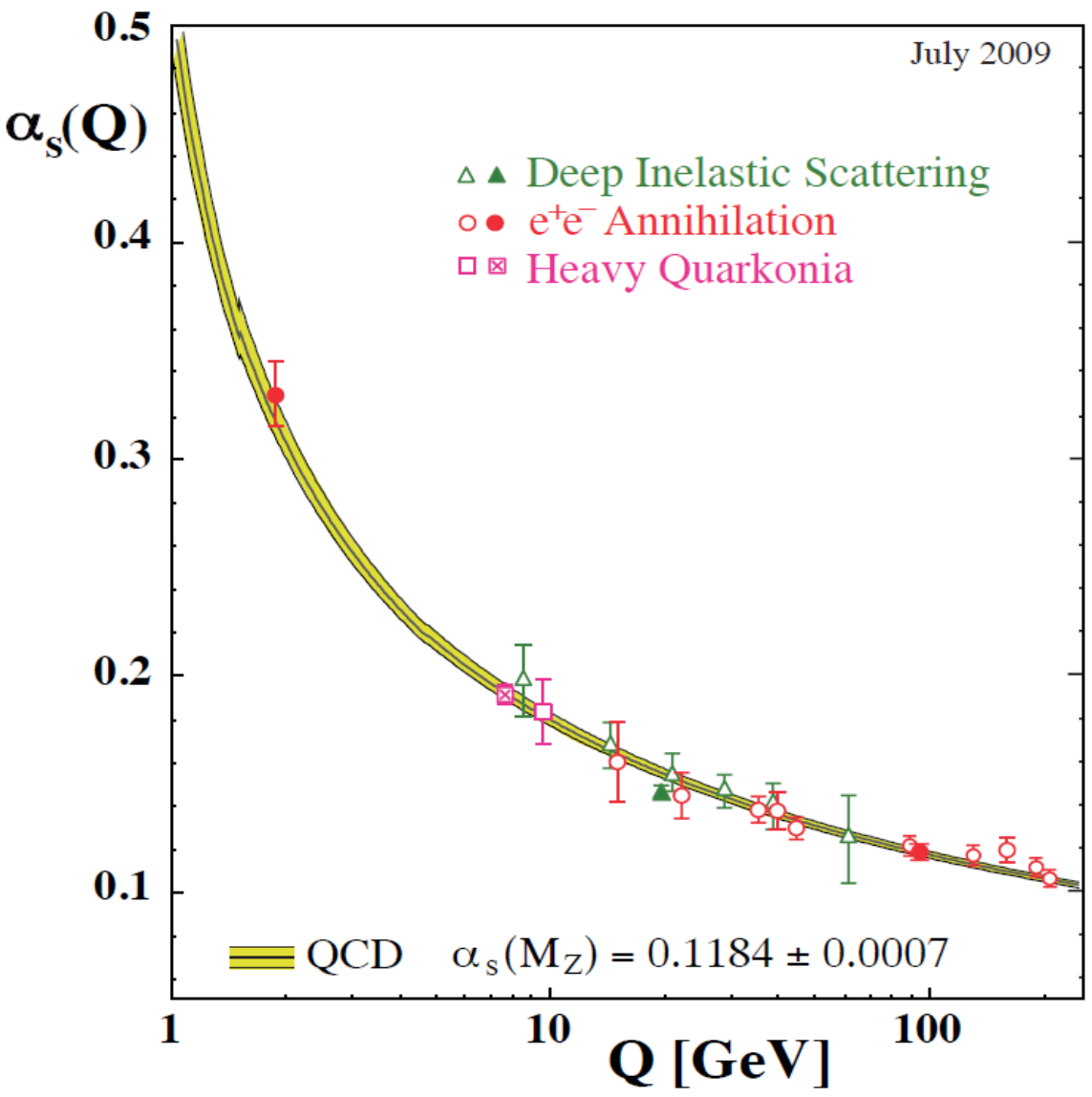

**Figure 1.2:** QCD coupling constant as a function of momentum transfer [9]

Fig.(1.2) shows the phenomenon of asymptotic freedom, which is a feature of non-abelian gauge theories involving color charges [10, 11]. The QCD coupling constant asymptotically approaches 0 in the limit of high momentum transfers (or equivalently at short distance scales). This suggests that at sufficiently high energies, quarks and gluons may not stay confined within the nucleons thereby giving rise to a deconfined collection of quarks and gluons. The QGP is thus aptly defined as a locally thermally equilibrated state of matter in which quarks and gluons are deconfined from hadrons, so that color degrees of freedom become manifest over nuclear, rather than merely nucleonic volumes [12].

## Recipes for QGP

Asymptotic freedom implies the existence of quark-gluon system that can come into being from a hadronic system, in two ways:

1. **High temperature QGP**: Hadrons are thermal excitations of the QCD vacuum. If a QCD vacuum in an enclosure is heated, the number of hadrons increases exponentially as a function of the resonance mass $m$ as $\rho(m) \sim \exp(m/T_c)$ [13, 14], where, $\rho(m)dm$ is the mass spectrum of hadrons, *i.e.* the number of hadrons with mass between $m$ and $m + dm$, and $T_c$ is the critical temperature. Eventually, the hadrons start overlapping with each other at $T_c$. Beyond $T_c$, the hadronic system ceases to exist and we are left with a quark-gluon plasma. This is shown in Fig.(1.3). The QGP produced in such a fashion has roughly equal number of quarks and antiquarks ($n_q \sim n_{\bar{q}}$). Monte Carlo lattice simulations yield $T_c \sim 155$ MeV [15].

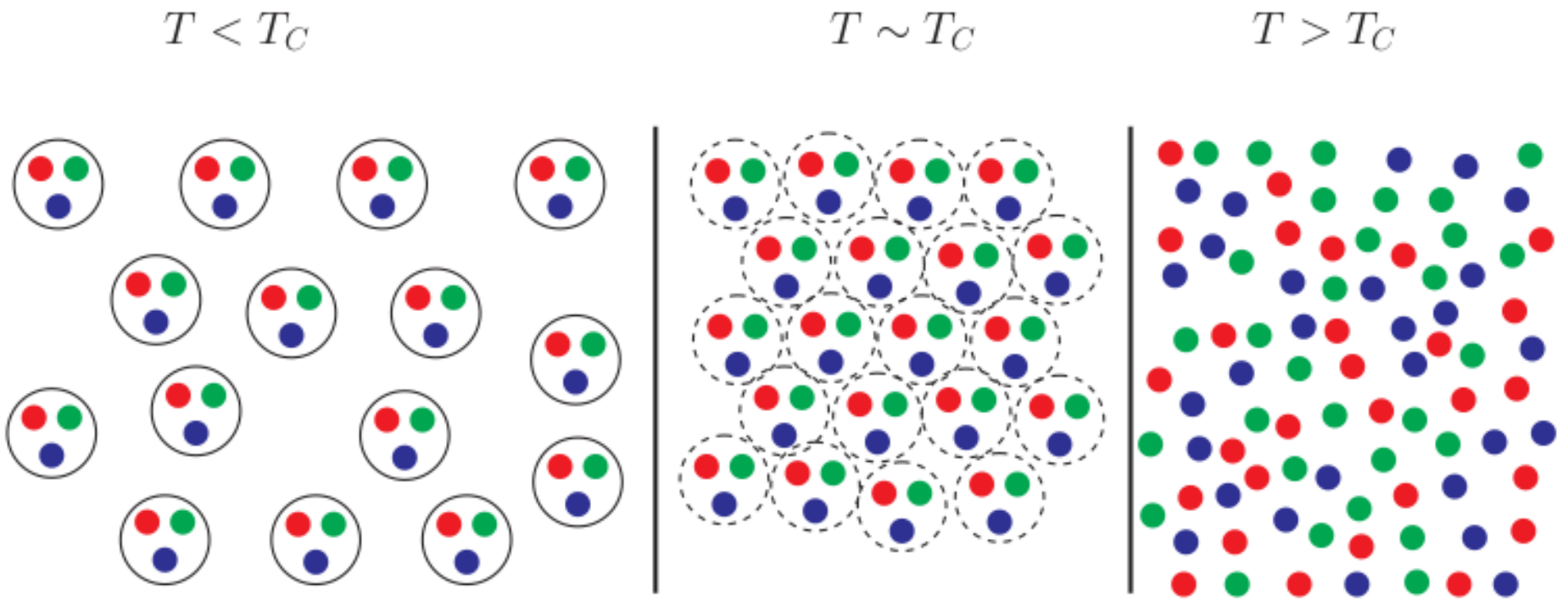

**Figure 1.3:** QGP formation by increasing the temperature of QCD vacuum

2. **High baryon-density QGP:** If a system consisting of a large number of baryons in an enclosure is compressed adiabatically (keeping $T \sim 0$), then, the baryons will start to overlap at a critical baryon density $\rho_c$. For $\rho > \rho_c$, the system becomes that of a degenerate quark matter. Such matter has a high surplus of quarks over antiquarks ($n_q \gg n_{\bar{q}}$). Model calculations show that $\rho_c$ is several times of $\rho_{\rm nm}$, where, $\rho_{\rm nm} = 0.16\,{\rm fm}^{-3}$ is the density of nuclear matter.

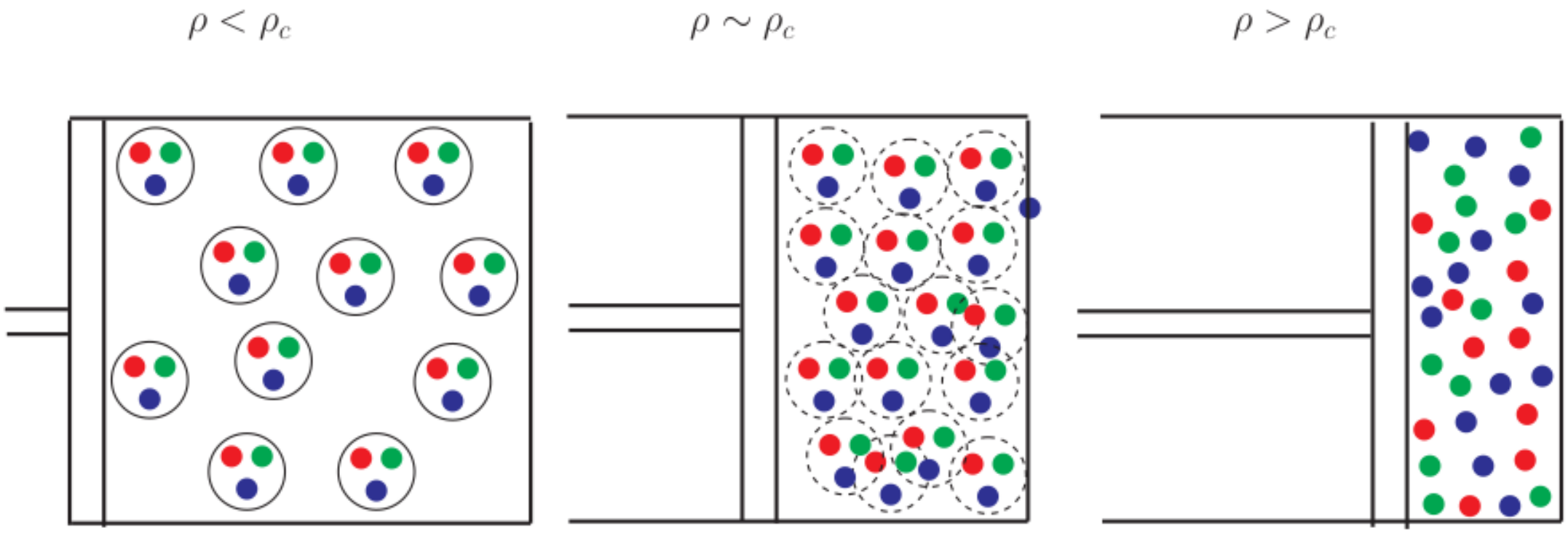

**Figure 1.4:** QGP formation at high baryon density.

Zero baryon density QGP is believed to have existed around 1-2 $\mu$ sec after the Big-Bang, whereas it is expected that the cores of super dense astrophysical objects such as neutron stars may contain baryon-rich QGP.

## 1.2 QGP in the laboratory: Heavy-ion collisions

Heavy nuclei are made to collide head-on in powerful colliders with the aim of depositing a large energy in a small volume (very large energy density) with the aim of creating QGP. The two scenarios of QGP formation mentioned in the earlier section are both realizable in laboratories. At relatively small centre of mass energies, $\sqrt{s}/A \sim 100$ GeV, the colliding Lorentz contracted nuclei stop each other, and a QGP with a finite chemical potential is created. At much higher energies though, the nuclei penetrate each other completely, leaving behind a highly excited vacuum in the space-time volume between the nuclei. This then leads to violent formation of quarks and gluons at zero baryon density ($n_q \sim n_{\bar{q}}$). Both the scenarios are schematically presented in Figs. (1.5a) and (1.5b), respectively.

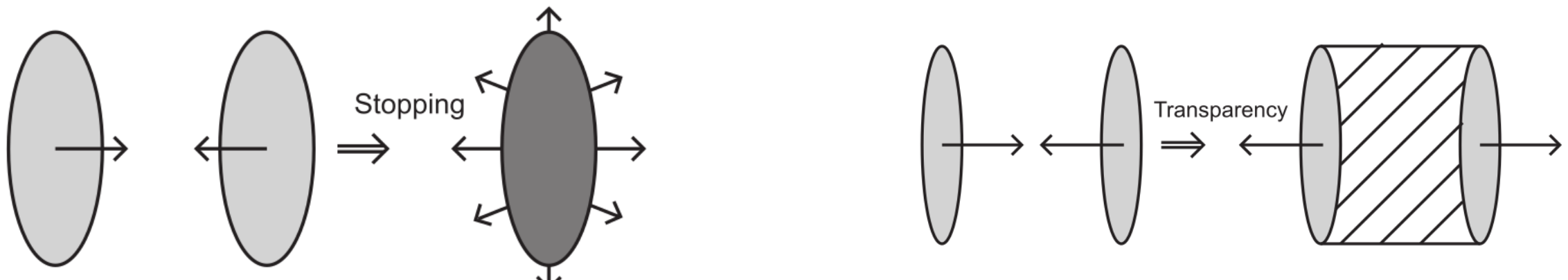

(a) $\sqrt{s}/A \sim 100$ GeV. (b) $\sqrt{s}/A > 100$ GeV.

**Figure 1.5:** Different collision energies leading to production of 'baryon-rich' and baryonless QGP [16].

Schematically, an ultrarelativistic heavy-ion collision (ULRHIC), and its evolution is shown in Fig.(1.6). The journey from nuclei to QGP to hadrons involves wide ranging, rich and inter-disciplinary physics. The various stages involved are discussed in what follows.

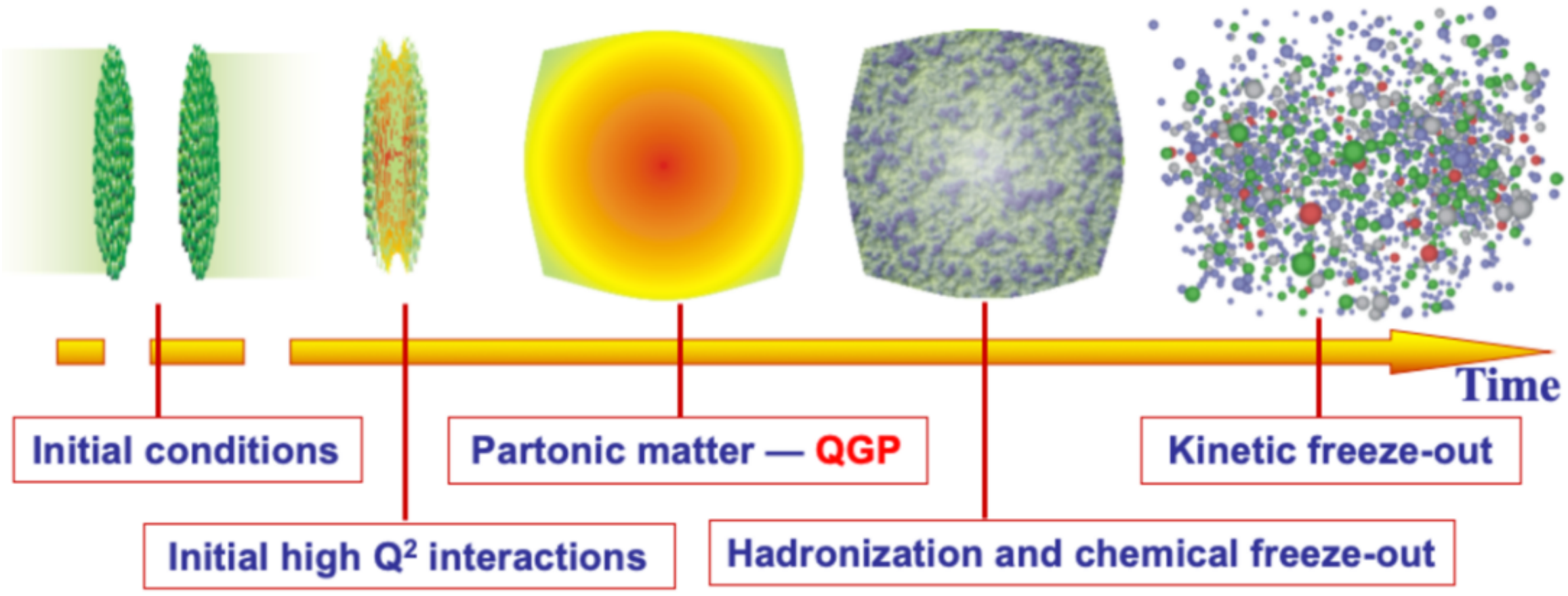

**Figure 1.6:** Evolution of heavy-ion collisions [17].

**The initial state:** The Color Glass Condensate (CGC) is used to describe the highly dense partonic (largely gluonic) matter in a high energy hadron, and is the initial state in relativistic hadron-hadron collisions. It is an effective theory which describes the high energy (or small Bjorken $x$) components of the hadronic wave function in QCD [18, 19]. The word color is because gluons are colored. It is a condensate because of the high density of gluons at high energies. The terminology 'glass' comes from the fact that these small $x$ partons are classical fields generated by random sources with lifetimes significantly greater than characteristic scattering timescales. This is analogous to glass: a system evolving much slower relative to natural time scales [20]. The CGC framework provides for an ab-initio approach to study thermalization in heavy-ion collisions.

**Glasma:** A large fraction of the incident energy of the incoming nuclei gets deposited in a small volume, called the reaction volume, leading to the formation of a high-energy-density non-equilibrated state called Glasma (fireball). The CGC dynamics leads naturally to Glasma field configurations at early stages of HICs [23]. It can be thought of as the precursor to the thermalized QGP. Glasma is a coherent state from which, partons are liberated after a finite amount of time (a fraction of a fm/c).

**QGP:** Intense collisions among the partons liberated from the glasma leads to a nearly thermalized (locally) state of matter called the quark-gluon plasma. QGP is estimated to form at a time of the order of 1 fm/c. Perturbation theory still works in this regime and has been used to calculate several quantities. The system thereafter undergoes expansion, cooling and dilution, and is best described by relativistic viscous hydrodynamics. Extensive reviews can be found here [21, 22].

**Hadronization:** As the plasma expands and the temperature drops, the QCD coupling grows, and the system eventually hadronizes. The degrees of freedom now are color neutral hadrons. Hadronization is a low-energy phenomenon that is essentially non-perturbative, and hence, is not amenable to perturbative investigation methods. The exact details of this transition are far from settled.

**Chemical and kinetic freezeout:** Hadrons interact among themselves both elastically and inelastically. While the former changes the energy-momenta, the latter alters the abundance of individual hadron species. Chemical freezeout refers to the cessation of inelastic processes. This is followed by kinetic freezeout when elastic interactions stop, as well. This happens in a timescale of around 10-15 fm/c.

Table (1.1) lists some details of the various collider experiments carried out in search of QGP. The first experimental evidence of deconfined partonic matter was found in the RHIC data [12, 24, 25]. The LHC at CERN led to further confirmation of QGP creation and added to the available data [26, 27]. It is the most powerful accelerator on earth today. Both RHIC and LHC are designed to create a high temperature, vanishing baryon number QGP. Accelerators to produce low temperature-high baryon number QGP have been proposed and are expected to come up shortly. FAIR is one such upcoming facility in GSI, Germany that will be devoted to the creation of compressed baryonic matter (CBM) with $\sqrt{s}$ = 2.7-8.3 GeV/A [28, 29]. Another facility devoted to the exploration of baryon-rich QGP is the NICA, slated to come up at JINR, Dubna, Russia [30, 31].

**Table 1.1:** Past and present collider experiments that have provided valuable data for the study of QGP.

| Accelerator | Location | Projectile-Target | $\sqrt{s}$ (A.GeV) | $T_{\rm max}$ (MeV) |
|---|---|---|---|---|
| AGS (1986-1996) | BNL, USA | Si-Au, Au-Au | 4-5 | 150 |
| SPS | CERN, Geneva | S-U, Pb-Pb | 17-20 | 190 |
| RHIC (2000-ongoing) | BNL, USA | Au-Au | 200 | 230 |
| LHC (2010-ongoing) | CERN, Geneva | Pb-Pb | 5500 | 260 |

## 1.3 The nature of transition: QCD phase diagram

There are several transition phenomena that can be exhibited in the ULRHIC experiments, as briefly described below:

- As described in the previous section, at high $T$ and/or $\mu$, a deconfining transition from hadronic matter to quark-gluon matter takes place. The degrees of freedom in the deconfined matter are the colored quarks and gluons and thus, the transition is the QCD analog of the insulator-conductor transition in atomic matter.

- In addition to a deconfinement transition, a transition related to the restoration of chiral symmetry is also expected. Inside hadrons, the quarks are dressed with gluons, leading to the bare quark mass $m_q \sim 0$ being replaced by the constituent quark mass $M_q \sim 300$ MeV. Since the QCD Lagrangian is chirally symmetric for $m_q \sim 0$, a non-zero $M_q$ corresponds to the breaking of chiral symmetry. At high $T$, this dressing melts ($M_q \to 0$), and this corresponds to the restoration of chiral symmetry for massless bare quarks. The transition temperatures related to deconfinement and chiral symmetry restoration may or may not coincide.

- Another type of transition that is expected at high $\mu$ and low $T$ is the formation of colored bosonic diquark pairs (analogous to Cooper pairs in superconductors), due to the attractive interaction between quarks in the QGP phase. This color superconducting phase would turn into a normal color conductor phase on dissociation of the QCD Cooper pairs due to heating.

All of the above mentioned transitions are schematically represented in Fig.(1.7).

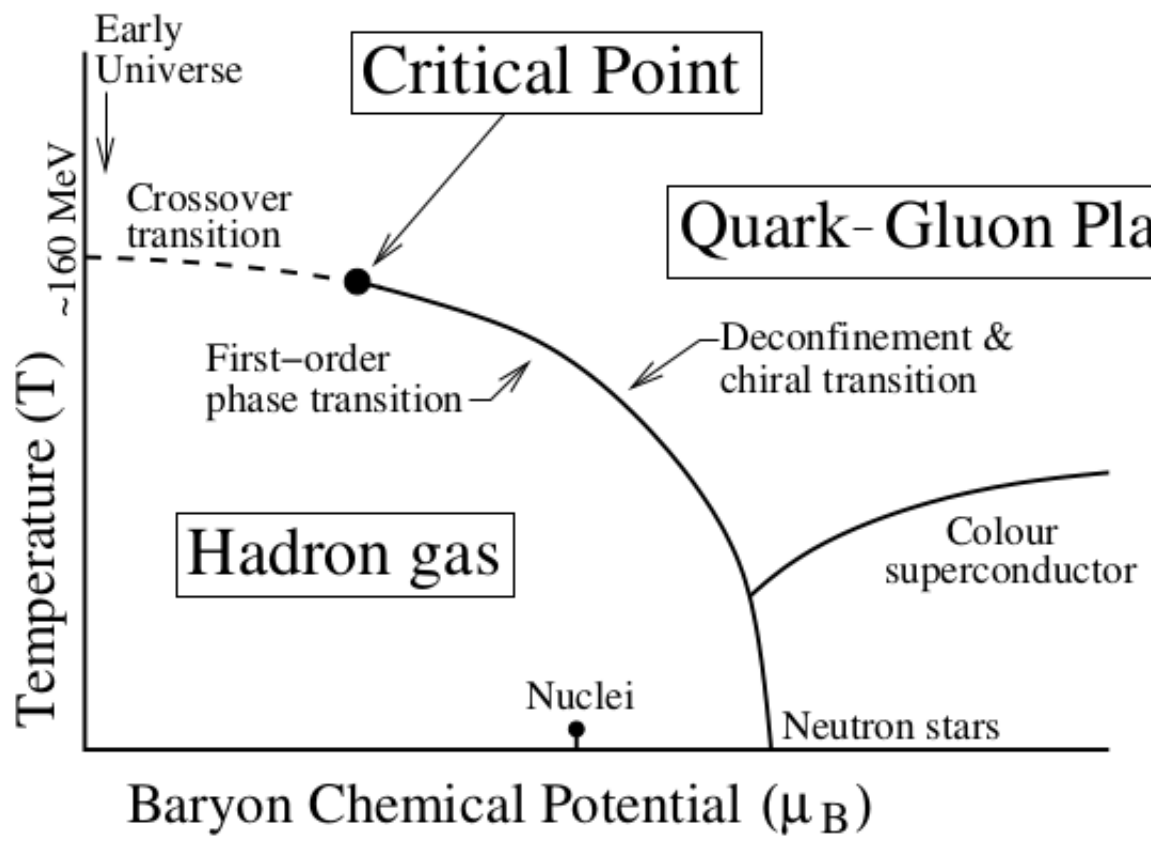

**Figure 1.7:** Schematic QCD phase diagram

Arguments based on a variety of phenomenological models and effective theories [32, 33] indicate that the deconfinement transition from hadronic to partonic system, as a function of $T$ is first order at large $\mu$. Since a line of first order phase transition must necessarily end at a critical point, the QCD phase diagram also features one [Fig.(1.7)]. The critical point is however, yet to be determined experimentally. At the critical point, the transition is believed to be second-order [34]. Along and near the $T$ axis ($\mu \approx 0$), lattice QCD is applicable, and its observations [35, 36] are summarised below:

- For $\mu \approx 0$, the color deconfinement transition is coincident with chiral symmetry restoration at $T_c \simeq 150 - 200$ MeV. In particular, for 3 massless quarks ($m_u, m_d, m_s \to 0$), the phase transition is first order, and corresponds to chiral symmetry restoration.

- In the limit $m_q \to \infty$ ($q \equiv u, d, s$), QCD approaches pure $SU(3)$ gauge theory which is invariant under a global $Z_3$ symmetry [37]. The deconfinement transition in such a scenario is first order, and is provided by spontaneous breaking of the $Z_3$ symmetry.

- For the realistic case of $0 < m_q < \infty$, the transition is not a true phase transition in the sense that there is no singular behaviour, but a rapid and analytic crossover. The chiral symmetry restoration in this case is only approximate.

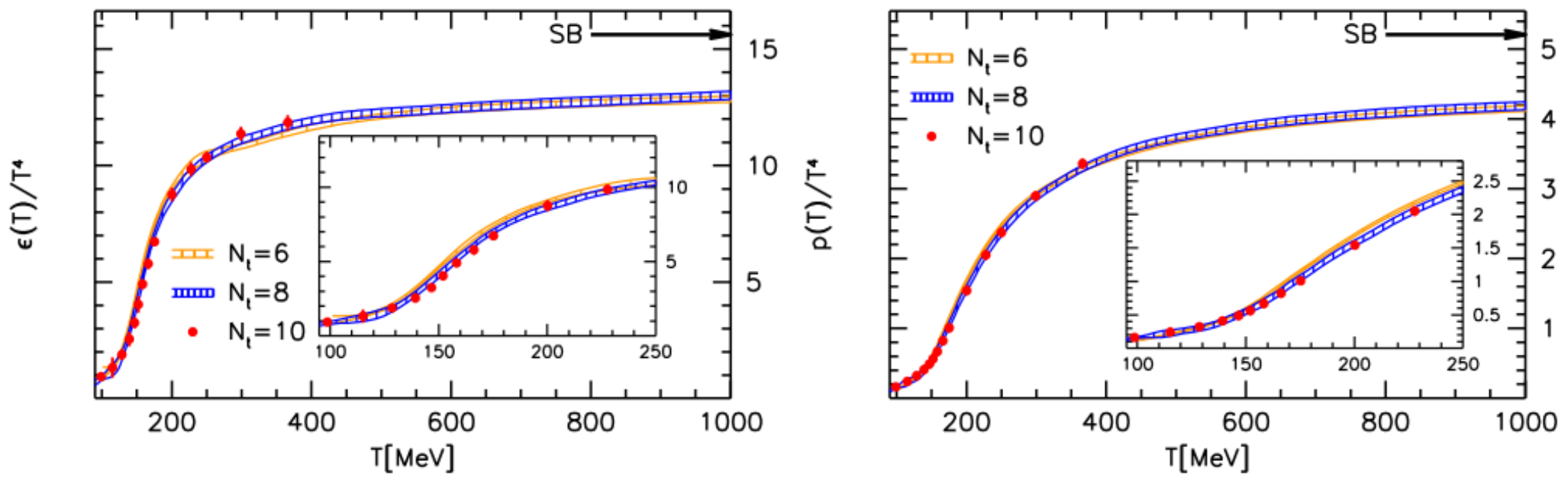

**Figure 1.8:** Energy density and pressure normalized by $T^4$ as a function of $T$ on $N_t = 6$, 8 and 10 lattices. Figure from [38].

Fig.(1.8) shows a rapid rise in $\epsilon$ and $P$ (scaled by $T^4$) around $T = 160$ MeV, which can be interpreted as increase in entropy or degrees of freedom due to deconfinement. They don't, however, reach the Stefan-Boltzmann limit at high $T$, suggesting that the partons interact significantly among themselves.

## Order parameters

Order parameters are variables that are used to identify phase transitions. Below, we briefly discuss the order parameters associated with the confinement-deconfinement transition and the chiral symmetry breaking-restoration transition.

- **Polyakov Loop:** In Lattice QCD studies, the variable corresponding to the deconfinement measure is the Polyakov loop [39, 40].

$$L(T) \sim \lim_{r \to \infty} \exp(-V(r)/T). \tag{1.1}$$

$V(r)$ is the static potential between an infinitely heavy ($m_q \to \infty$) quark-antiquark pair. For a pure gauge theory, $V(r) \sim \sigma r$, where, $\sigma$ is the string tension. So, $V(\infty) = \infty$, so that $L = 0$. This is the case for $0 < T < T_c$. When $T > T_c$, the string melts due to color screening among gluons. $V(r)$ is thus finite for large $r$, and $L \neq 0$. The temperature $T_c$ at which $L$ becomes finite, is thus identified as the deconfinement temperature. For finite quark mass, $V(r)$ remains finite as $r \to \infty$ so that in the confined phase, $L(T)$ is not exactly 0, but very small ($\sim 10^{-2}$). $L(T)$ and its corresponding susceptibility $\chi_L(T) \sim \langle L^2 \rangle - \langle L \rangle^2$ are shown in Fig.(1.9a). $L(T)$ exhibits a sudden increase as $T$ rises (onset of deconfinement). Also, $\chi_L(T)$ features a prominent peak, which defines the transition temperature quite well.

- **Chiral condensate:** The other variable that is used as an order parameter for chiral symmetry restoration is the chiral condensate, which is a measure of the effective quark mass, and is given by the expectation value $\langle \bar{\psi}\psi \rangle$, where, $\psi$ is the quark wave function in the QCD Lagrangian. In the limit $m_q \to 0$, the Lagrangian becomes chirally symmetric and $\langle \bar{\psi}\psi \rangle$ acts as an order parameter. The chiral symmetry is spontaneously broken in the confined phase (where effective quark mass $M_q \sim 0.3$ GeV), and restored in the deconfined phase (at high $T$). For the realistic situation of finite $m_q$, $\langle \bar{\psi}\psi \rangle$ does not vanish completely, so the symmetry restoration is only approximate. $\langle \bar{\psi}\psi \rangle$, and the corresponding susceptibility $\chi_m \sim \partial \langle \bar{\psi}\psi \rangle / \partial m_q$ are shown as functions of temperature in Fig.(1.9b). As expected, there is a steep drop in the effective quark mass, along with a sharp peak in the susceptibility, thereby defining well, the chiral symmetry restoration temperature, which coincides with the deconfinement temperature (for $\mu \approx 0$).

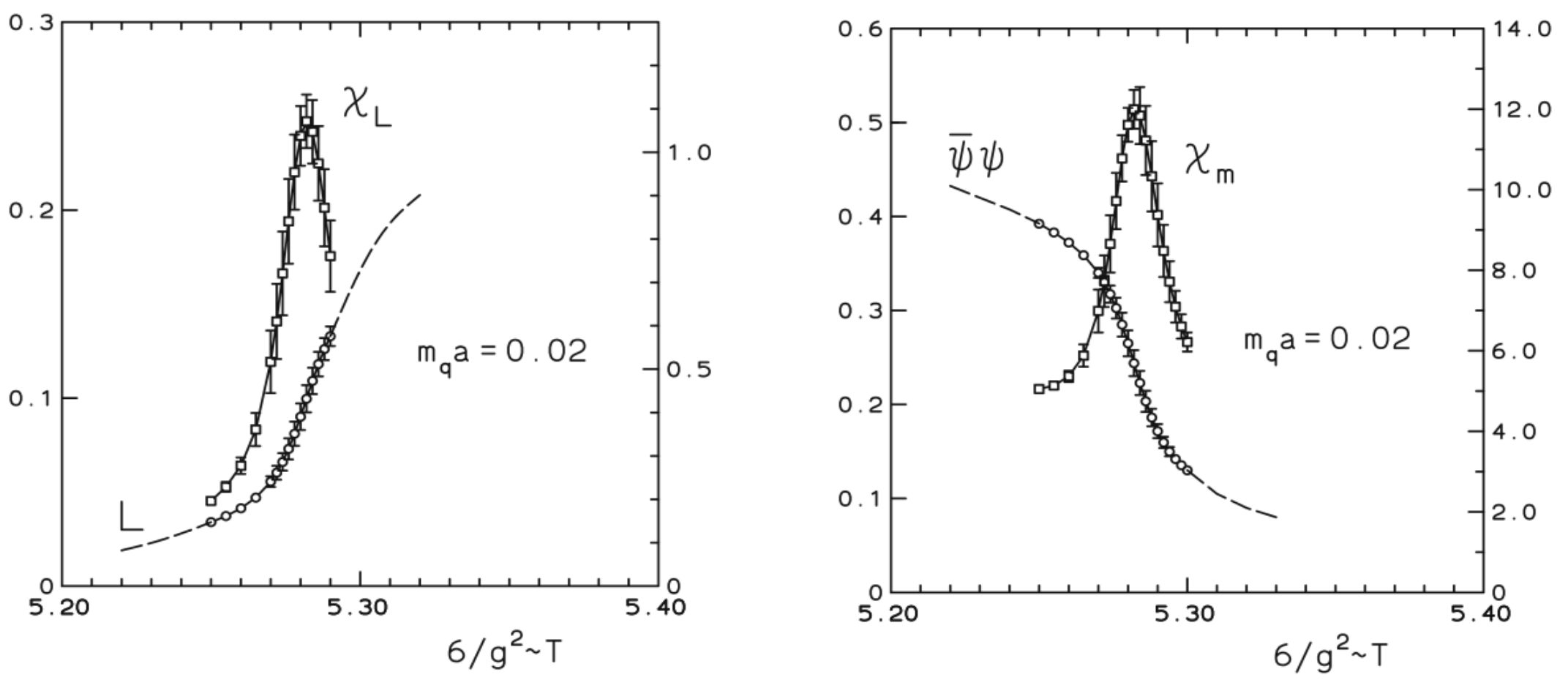

**(a)** **(b)**

**Figure 1.9:** Polyakov loop (a) and chiral condensate (b) in 2 flavor QCD, using using a current quark mass about four times larger than that needed for the physical pion mass.

# 1.4 Signatures of QGP

## 1.4.1 Anisotropic flow:

The invariant distribution of particles in the azimuthal plane emitted in the final state of an HIC is Fourier decomposed as [41]

$$E\frac{d^3N}{d^3p} = \frac{d^3N}{p_T\,dp_T dy\,d\phi} = \frac{d^2N}{p_T\,dp_T dy}\frac{1}{2\pi}\left[1+\sum_{n=1}^{\infty}2v_n\cos n(\phi-\Phi_R)\right],$$

where, $p_T$ is the transverse momentum, $y$ the rapidity, $\phi$ the azimuthal angle of the outgoing particle momentum, and $\Phi_R$ the reaction-plane angle. The beam axis is the $z$- axis, so that the $x$-$y$ plane in Fig.(1.10) corresponds to the azimuthal or transverse plane. The leading term in the square brackets corresponds to a flow profile that is azimuthally symmetric. The first two harmonic coefficients $v_1$ and $v_2$ are called *directed* and *elliptic* flows, respectively, and correspond to azimuthally asymmetric flow profiles. Non-zero values of the flow coefficients ($v_n$) is interpreted as being indicative of collective flow.

In a non-central HIC event (finite impact parameter), the overlap region of the colliding nuclei is spatially anisotropic in the azimuthal plane, as is shown in Fig.(1.10).

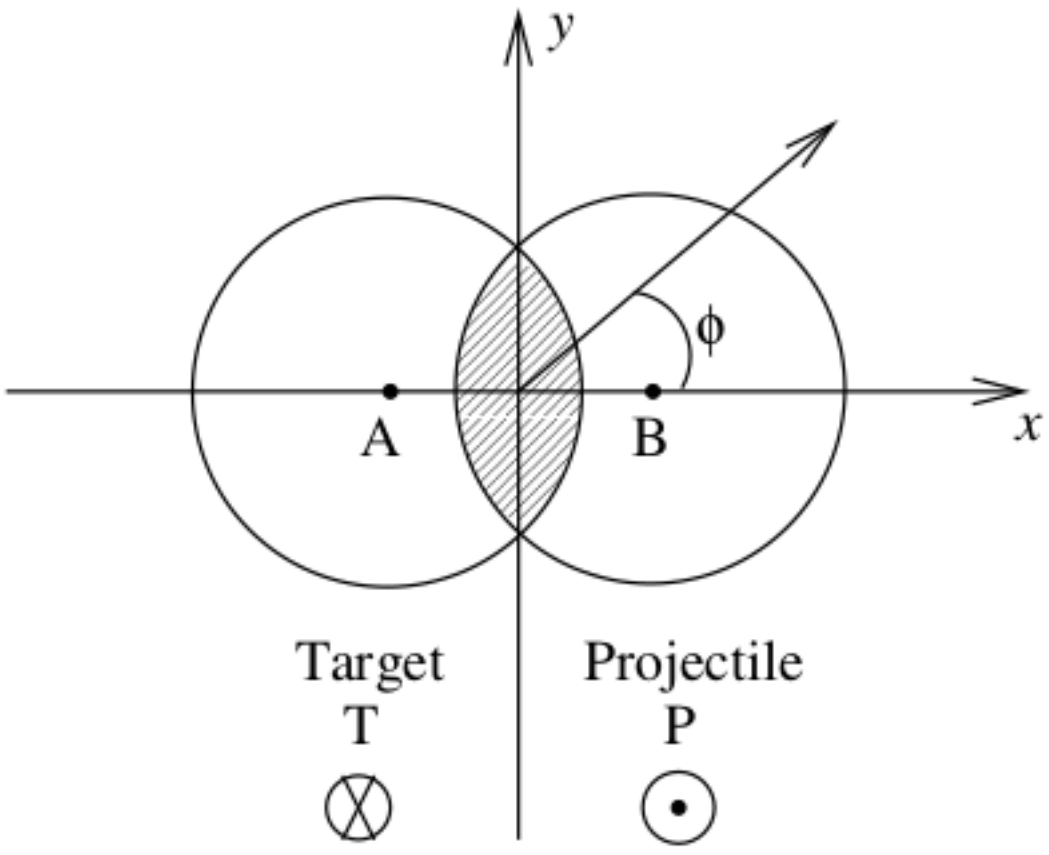

**Figure 1.10:** Spatial anisotropy of the overlap region in a non-central HIC event [42].

Just before collision, the initial momenta of the particles in the overlap zone are predominantly longitudinal, with the transverse momenta (if any) being distributed isotropically in the azimuthal plane. If the particles passed through without interacting, the final state momentum distributions would also be azimuthally isotropic. If however, they interact strongly enough leading to thermal equilibrium of the cre-ated fireball, the position space initial anisotropy in Fig.(1.10) will be converted to a final state momentum space anisotropy (reflected by non-zero $v_n$) of the emitted particles. Thus, anisotropic or collective flow is a consequence of the creation of a locally thermally equilibrated medium- a *sine qua non* for QGP discovery.

A large azimuthal asymmetry in soft ($p_T < 2\,\mathrm{GeV}$) hadron production was indeed observed by RHIC.

### 1.4.2 Jet quenching:

Jet quenching refers to the suppression of high $p_T$ hadron production in HICs. Hard partonic interactions lead to the formation of two back-to-back hard partons with a large $p_T$. They fragment and emerge as jets of high $p_T$ hadrons. If the partons encounter a dense colored medium in their path, they loose energy or get quenched. Partons can lose energy in the medium via two processes: Collisional energy loss via elastic scatterings, and radiative energy loss via inelastic scatterings, as shown in Fig.(1.11).

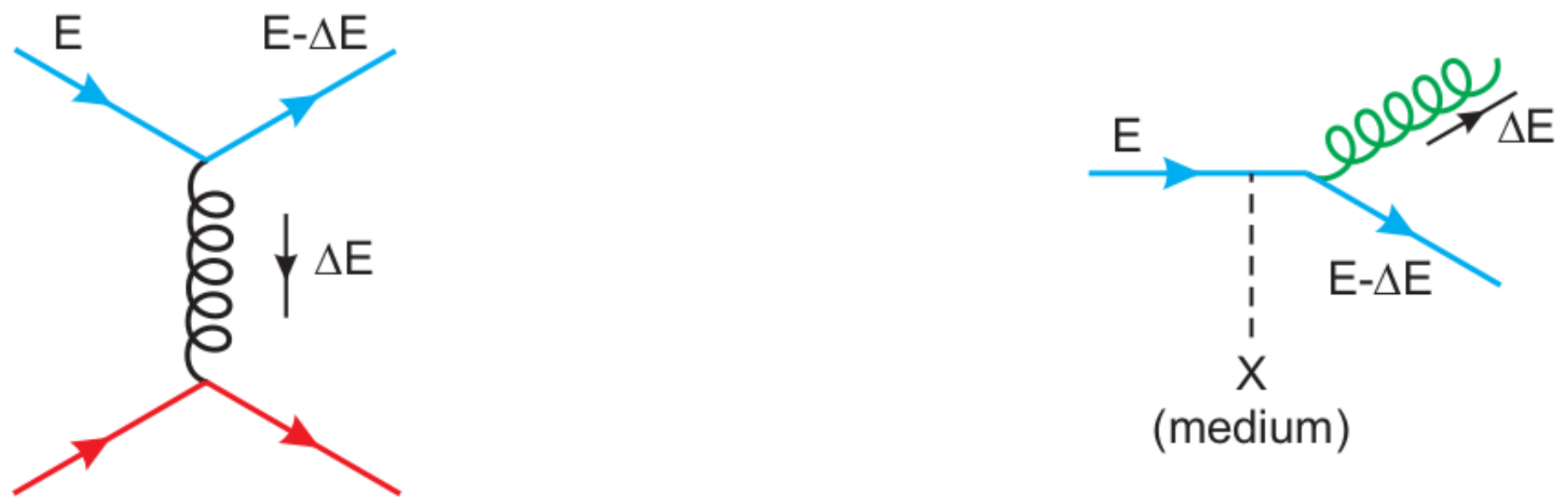

**Figure 1.11:** Energy loss mechanisms of a hard parton traversing through QGP.

An experimental variable of interest is the nuclear modification factor $R_{AA}$, defined schematically as

$$R_{AA}(p_T) = \text{Yield in } AA/\langle N_{\text{coll}}\rangle \text{Yield in } pp,$$

where, $\langle N_{\text{coll}}\rangle$ is the mean number of nucleon-nucleon collisions occurring in a single nucleus-nucleus (AA) collision. if nucleus-nucleus collision were simply a superposition of nucleon-nucleon collisions, the ratio $R_{AA}$ would be unity, thus indicating that no medium is formed. Suppression is characterized by an $R_{AA}$ of less than 1, which is a signal of medium formation.

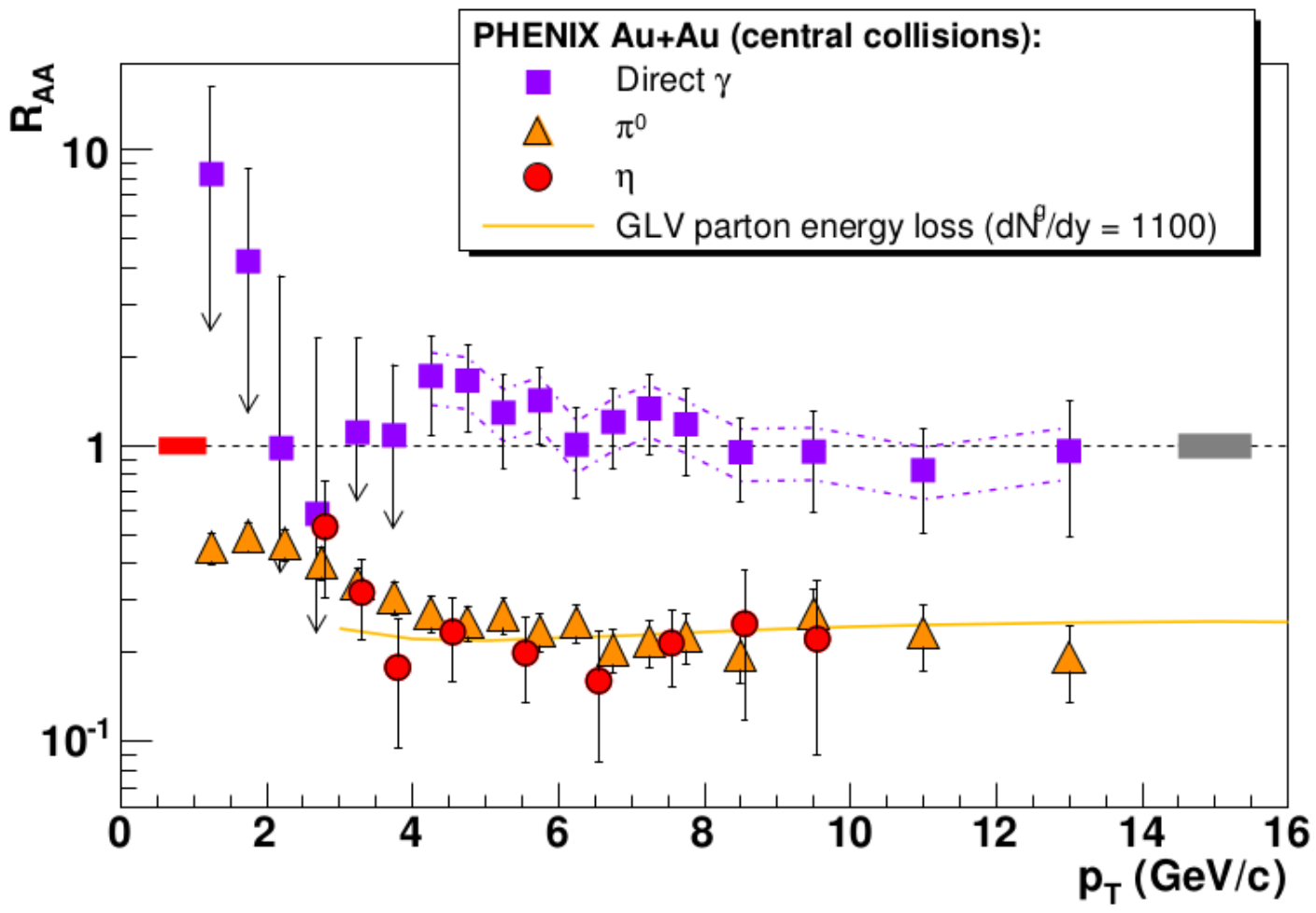

**Figure 1.12:** $R_{AA}$ vs $P_T$ data for central Au-Au collisions with $\sqrt{S_{NN}} = 200$ GeV. Figure from [43].

Fig.(1.12) shows $R_{AA}$ data as a function of $p_T$ in central Au-Au collisions. The solid orange line is the jet-quenching calculation for leading pions in a medium with initial effective gluon density $dN^g/dy = 1100$. As can be seen, there is no suppression of the direct photon production, which is consistent with NLO pQCD calculations. However, there is a significant suppression of pions and etas at high $p_T$, where the yields are suppressed by a factor of $\sim 5$.

### 1.4.3 Quarkonium suppression:

In the seminal work by Matsui and Satz [44], they argued in relation to $J/\psi$ ($c$-$\bar{c}$) that in a deconfined medium of quarks and gluons, the confining potential of the $c$-$\bar{c}$ pair becomes color screened, effectively limiting the range of the strong interaction, ultimately leading to the dissociation of the bound state. This manifests itself in the suppression of $J/\psi$ yield relative to that without QGP formation, which was considered to be a 'smoking gun' signal of QGP formation. It was later discovered that the contribution to quarkonium dissociation also comes from the static $q$-$\bar{q}$ potential governing the evolution of the quarkonium system.The potential contains an imaginary part [45], leading to a finite decay width which exceeds the quarkonium binding energy much before the latter vanishes.

### 1.4.4 Quark recombination and quark number scaling:

At intermediate $p_T$ (2-3 GeV), the yield of baryons (containing 3 valence quarks) is strongly enhanced compared to that of mesons (containing a valence quark and antiquark) in nucleus-nucleus collisions relative to $p$-$p$ collisions. his observation is well-described by models in which baryons and mesons are generated by the recombination of quarks drawn from a collectively flowing, thermally equilibrated partonic medium. The recombination model also suggests that the elliptic flow of mesons and baryons scale according to their constituent quark numbers [46] (assuming quarks and antiquarks exhibit the same collective flow)

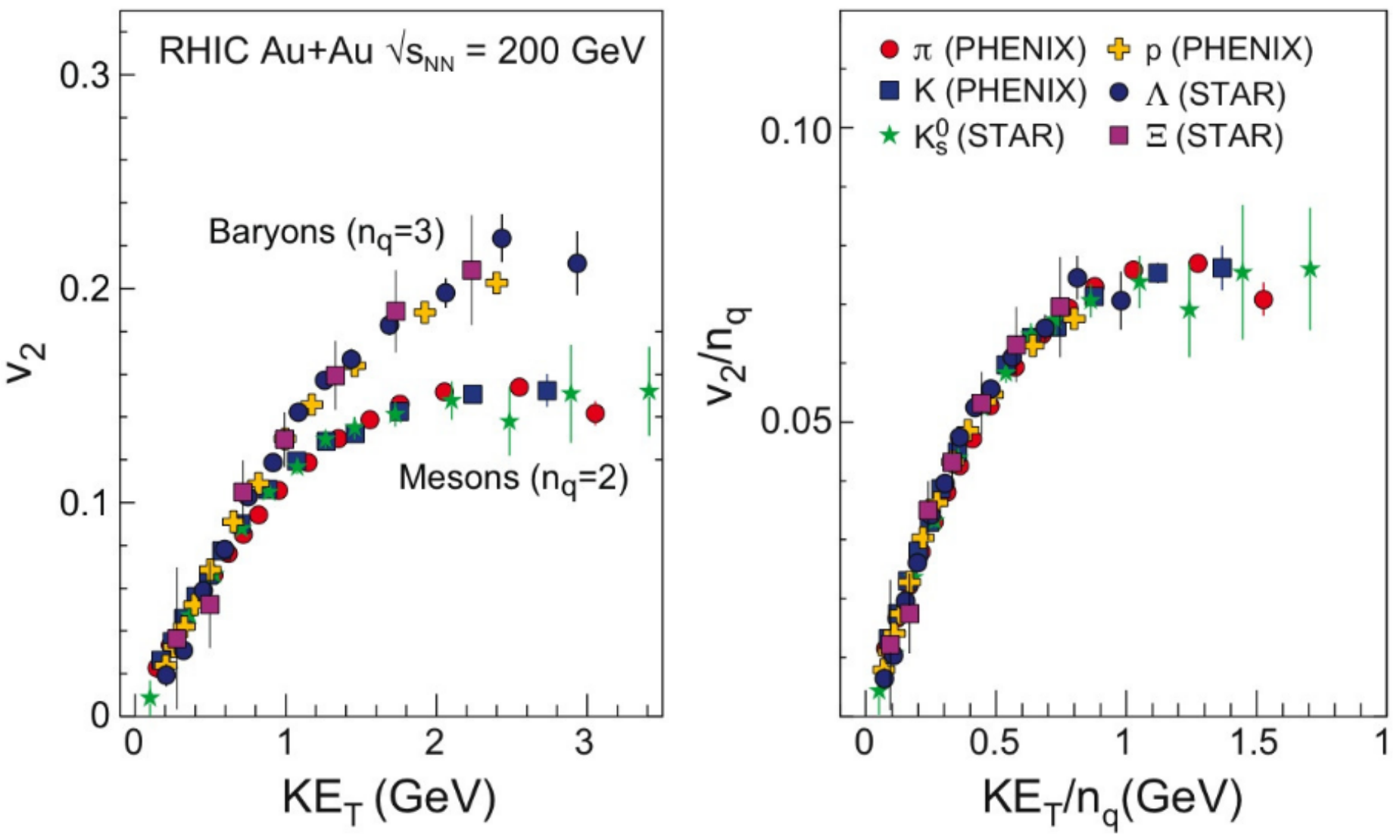

**Figure 1.13:** (Left) $KE_T$ dependence of $v_2$ of mesons and baryons. (Right) Both $v_2$ and $KE_T$ scaled by the number of constituent quarks $n_q$. Figure from [47]

Fig.(1.13) shows that when the flow parameters of mesons and baryons are scaled, the curves merge into one universal curve, suggesting that the flow develops in the partonic stage. This is considered to be one of the most direct evidences of deconfinement.

### 1.4.5 Strangeness enhancement:

Since the colliding nuclei are devoid of strange quarks, the initial strangeness is zero. This also means that a strange quark $s$ is always accompanied by its antiparticle $\bar{s}$ when they are produced via light quark and gluon scatterings post HICs, ensuring conservation of strangeness. The scatterings leading to strangeness production are:

$$u + \bar{u} \rightarrow s + \bar{s}$$
$$d + \bar{d} \rightarrow s + \bar{s}$$
$$g + g \rightarrow s + \bar{s}$$

Production of strange hadrons is expected to be enhanced in nucleus-nucleus collisions relative to scaled $pp$ collisions because of the following reasons:

- The temperature of an equilibrated QGP $T$ is such that $T \gg m_s$, where, $m_s$ is the mass of the strange quark.

- Gluon fusion $gg \longrightarrow s\bar{s}$ is expected to be an efficient production mechanism for strange quarks in QGP, where the gluon density is high. Thus, $gg \longrightarrow s\bar{s}$ processes are expected to dominate the $q\bar{q} \rightarrow s\bar{s}$ processes with respect to production of strange quarks.

- Compared to a pure hadronic medium, the energy threshold for strangeness production is much lower for a QGP medium.

Abundance of strange quarks and antiquarks in QGP should be manifested via enhanced number of strange and multi-strange hadrons detected in the final state, which is exactly what the experiments at SPS and RHIC bring out [48]. In addition, the ratio of number of strange hadrons to non-strange hadrons is also used as an observable to study strangeness enhancement. For instance, the ratios $K/\pi$, $\phi/\omega$, $\bar{\Lambda}/\Lambda$, $\bar{\Xi}/\Xi$ have been measured to study strangeness enhancement [49].

### 1.4.6 Dilepton and photon spectra

A quark annihilates its antiparticle counterpart to produce a photon (along with a gluon), which subsequently decays into lepton-antilepton pair (dilepton). Radiation of photons and dileptons are considered as efficient tools to characterize the deconfined QCD plasma. This is because the photons and the dileptons interact only electromagnetically, and hence, undergo minimum interactions in the QGP ($\alpha \ll \alpha_s$); their mean free paths being larger than the typical system size ($\sim$ 10 fm). The electromagnetic interaction is strong enough to lead to a detectable signal, and yet, weak enough to not lead to much reinteraction of the produced photons and dileptons with the medium constituents, thereby allowing them to carry largely undistorted information about the QGP to the detector.

Photons have been studied to measure the temperature of the QGP, its evolution by intensity interferometry [50], momentum anisotropy of the initial partons [51], as well as formation time of quark–gluon plasma [52]. The dominant processes responsible for production of photons from QGP are the annihilation process $q + \bar{q} \to g + \gamma$ and Compton processes $(q(\bar{q}) + g \to g(\bar{q}) + g)$. Dileptons are considered as the most reliable messengers of the medium modification of vector mesons [53]. Apart from being produced as decay products of photons, they are also produced from the decay of hadronic resonances such as $\rho$, $\omega$, $J/\psi$.

## 1.5 Creation of magnetic fields in HICs

If two nuclei collide with a finite impact parameter, the collision is said to be non-central. Fig.(1.14) represents a cross-sectional image of a non-central nucleus-nucleus collision. The distance between the centers of the nuclei defines the impact parameter ($b$ in the figure). The overlap region of the two nuclei contains the participating nucleons ($N_{\text{part}}$) which gives rise to the QGP fireball. The remaining particles, called spectators pass through almost without interacting at all, at relativistic velocities, and are responsible for creating large magnetic fields in direction perpendicular to the reaction plane. If two heavy ions are approaching each other at relativistic speed, the strength of the generated magnetic field using Biot-Savart can be written as [54]

$$B = \gamma Z e \frac{b}{R^3}, \tag{1.2}$$

where, $Ze$ and $R$ are the charge and radius of the colliding ions, respectively, $b$ is the impact parameter, and $\gamma = \sqrt{s_{NN}}/2m_N$ is the Lorentz factor with $m_N$ being the nuclear mass.

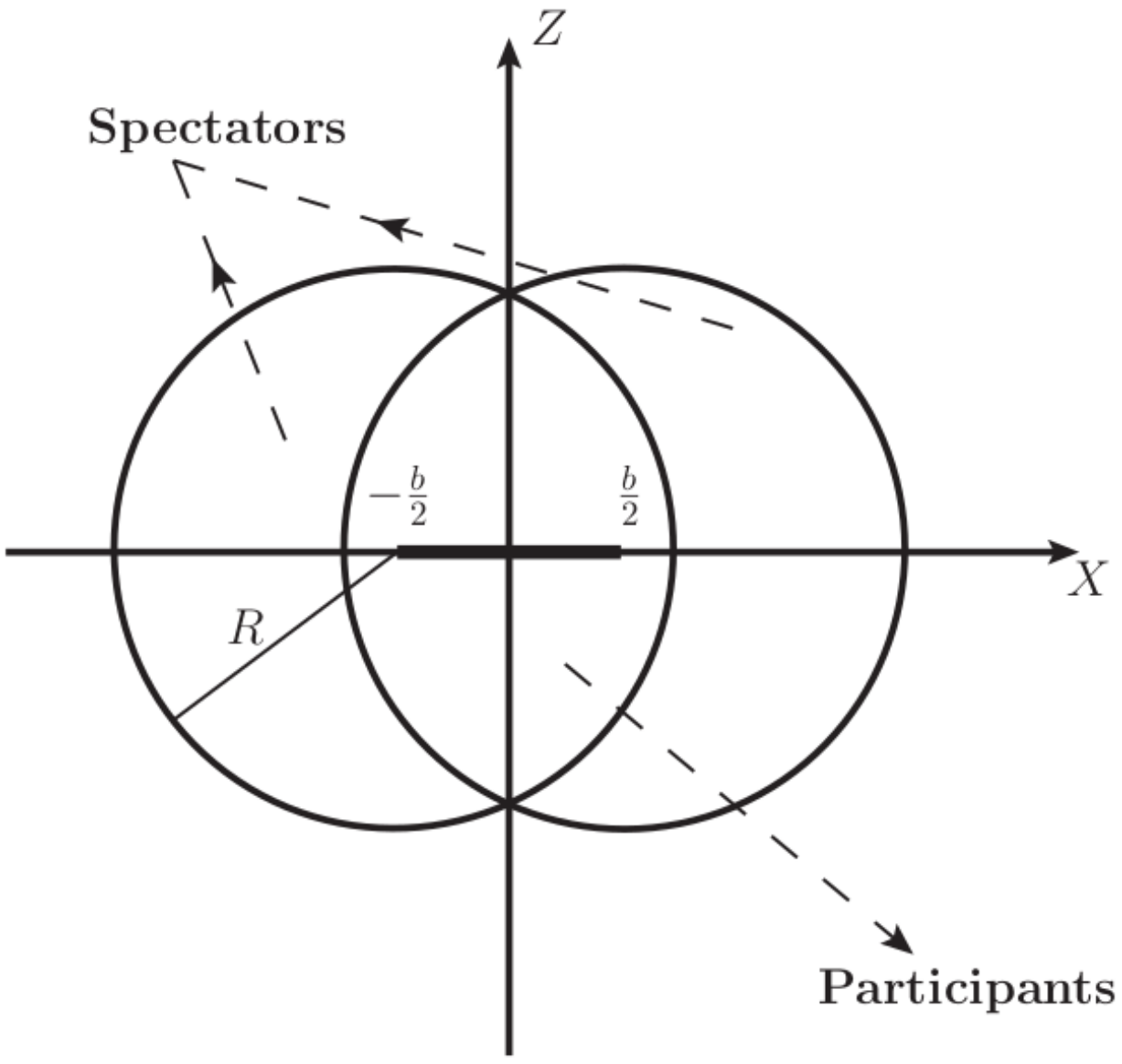

**Figure 1.14:** A schematic non-central nucleus-nucleus collision.

The first estimate followed by a numerical evaluation of strength of magnetic fields generated in HICs were carried out in [55, 56]. The strength of the field is estimated to be around $eB \approx 10^{-1} m_\pi^2$ ($m_\pi^2 = 10^{18}$ Gauss) at SPS energies, $eB \approx m_\pi^2$ at RHIC and $eB \approx 15 m_\pi^2$ at LHC. This has spurred the investigation of some unique strong magnetic driven phenomena in the context of QGP such as the chiral magnetic effect [55, 57], which come about due to quantum anomalies. The generation of magnetic fields in HICs is very relevant to this thesis, since it deals with the evaluation of certain transport coefficients of a magnetized QGP, both in the limit of strong and weak background magnetic fields.

## 1.6 Transport phenomena in QGP

Transport properties describe how a system which is slightly out of equilibrium, reverts to equilibrium via transport of energy, momentum, baryon number, etc. from one part of the system to another. The shear and bulk viscosities ($\eta$ and $\zeta$), respectively govern the momentum transport transverse to, and along the hydrodynamic flow. Electrical conductivity ($\sigma_{\rm el}$) and thermal conductivity ($\kappa$) are the responses of the medium to electric field and thermal gradients, respectively.

Transport coefficients of the hot QCD matter created post HICs, play an important role in determining its evolution and observation of many exotic phenomena therein.

For instance, as described in the previous section, large magnetic fields are generated in non-central HICs. Initially, it was believed that the magnetic field strength would decay extremely rapidly and thereby would be too short lived to have any observable effects in the evolution and properties of the QGP. However, it was later realized that a finite electrical conductivity of the QGP medium slows down this decay significantly [54, 58]. The magnetic field thus remains strong enough for a longer period of time which leads to some novel phenomena in the QGP such as chiral magnetic effect [55], chiral vortical effect [59], magnetic catalysis [60] and inverse magnetic catalysis [61] , etc. Likewise, $\kappa$ controls the attenuation of sound through the QGP medium, thus affecting the dynamics of the phase transition. Evaluation of QGP transport coefficients is a tough task. This is because at temperatures relevant to the QGP, the coupling constant $g \approx$ 1-2, hence, weak coupling ($g \ll 1$) pQCD calculations provide only an order-of-magnitude estimate of these coefficients [62]. Extraction of $\eta$ and $\zeta$ from Lattice QCD also is challenging because of the requirement of analytic continuation from imaginary to real times, and yield results with large error bands [63]. Exploiting the correspondence between strongly coupled quantum field theories to classical gravity in AdS space-time (AdS/CFT correspondence), the strong coupling limits of specific shear and bulk viscosities have been established: $\eta/s \simeq 1/(\pi)$, $\zeta/s \simeq 2(\frac{1}{3}-c_s^2)$ [64–66]. Calculations from several transport and hadronic interaction models reveal interesting features of transport coefficients near the transition temperature $T_c$: that $\eta/s$ has a minimum near $T_c$, whereas $\zeta/s$ peaks near $T_c$ [67, 68] .

A major part of this thesis deals with the evaluation of thermoelectric transport coefficients of the QGP medium, *viz.* Seebeck and Nernst coefficients. These effects are described in detail, in the next section.

## 1.7 Thermoelectric phenomena

The first of the thermoelectric effects was discovered by T.J. Seebeck in 1821 wherein he showed that an electromotive force was generated on heating the junction between two dissimilar metals [69]. This phenomenon of conversion of a temperature-gradient in a conducting medium into an electric current is termed as the Seebeck effect and depends on the bulk properties of the medium involved. Let us try to understand the Seebeck effect using a simple schematic.

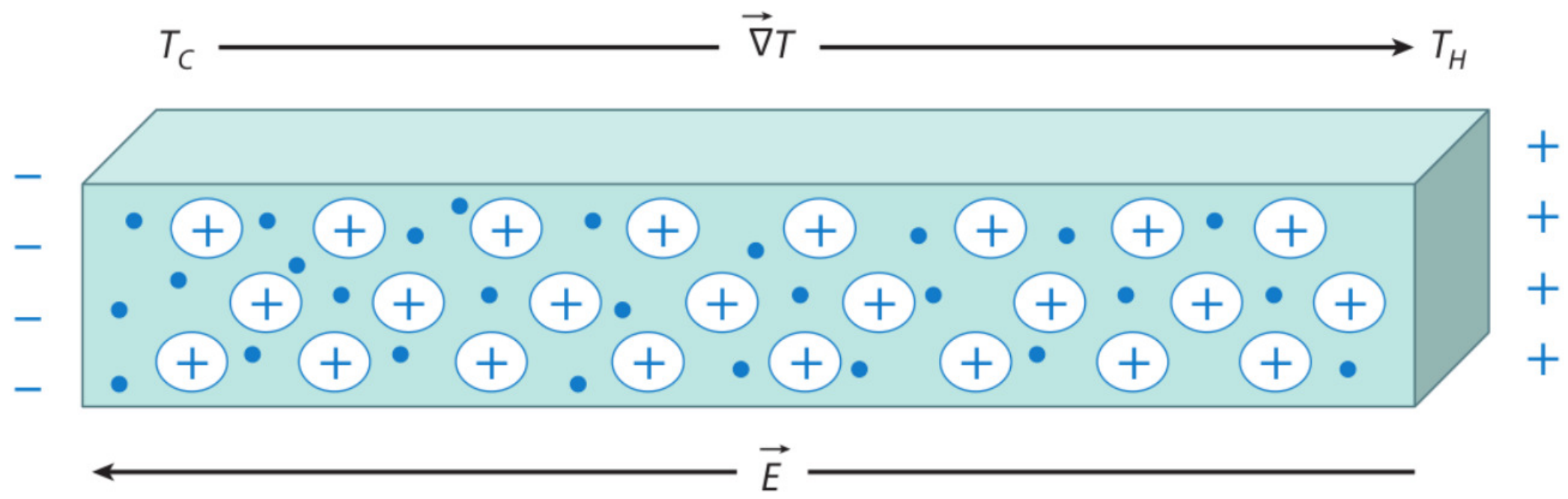

**Figure 1.15:** A schematic of a material consisting of immobile ions and mobile electrons under the influence of a temperature gradient. Figure from [70].

Consider, as in Fig.(1.15), a block consisting of immobile ions (positively charged) and mobile electrons. When a temperature-gradient is established, the more energetic charge carriers diffuse from the region of higher temperature to the region of lower temperature, thus constituting the thermoelectric current. The diffusion of mobile charges causes a relative abundance of electrons in the hotter side, leading to the creation of an induced electric field in the direction opposite to that of the temperature gradient[1]. The diffusion stops when the created electric field becomes strong enough to impede the further flow of charges. The Seebeck coefficient is defined as the electric field (magnitude) generated in a conducting medium per unit temperature-gradient when the electric current is set to zero [71, 72], *i.e.* $\boldsymbol{E} = S\boldsymbol{\nabla}T$ ($S$ is the Seebeck coefficient). In the convention described above, the Seebeck coefficient is positive for positive charge carriers and negative for negative charge carriers. This can be verified by considering positively charged mobile charge carriers in the setup of Fig.(1.15). In this case, the direction of the induced electric field is reversed; it is in the same direction as the temperature gradient. It is to be noted that while for condensed matter systems, a temperature gradient is sufficient to give rise to an induced current, this is not the case with a system such as the electron-positron plasma or the QGP. This is because in a condensed matter system, the ions are stationary and the majority charge carriers are responsible for conduction of electric current. However, in a medium consisting of mobile charged particles and antiparticles, a temperature gradient will cause them to diffuse in the same direction, giving rise to equal and opposite currents, which cancel. In such media, therefore, in addition to a temperature-gradient, a finite chemical potential is also required for a net induced current to exist.

The Seebeck coefficient magnitude is very low for metals (only a few micro volts per degree Kelvin temperature-gradient) whereas much higher for semiconductors (typically a few hundred micro volts per degree Kelvin temperature-gradient) [73]. Thermoelectric properties have been an extensive area of investigation in the field of condensed matter physics over the past three decades. Some of the notable works include the study of the Seebeck effect in superconductors [74–78], Seebeck effect in the graphene superconductor junction [79], electric and thermoelectric transport properties of correlated quantum dots coupled to superconducting electrode [80], transport coefficients of high temperature cuprates [81], thermoelectric properties of a ferromagnet- superconductor hybrid junction [82], Seebeck coefficient in low dimensional correlated organic metals [83], etc.

[1]The direction of temperature gradient is a convention. In this case, it is taken to be directed from the colder end to the hotter end.

The situation in the presence of a background magnetic field is schematically shown in Fig.(1.16)

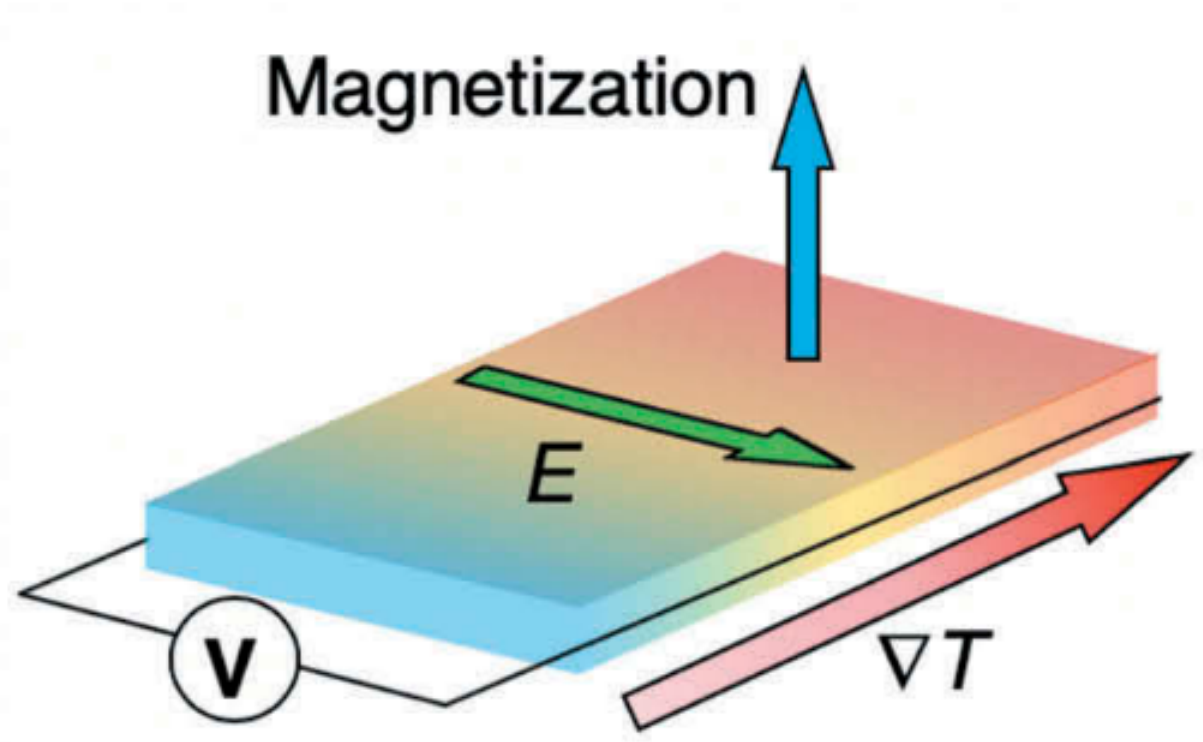

**Figure 1.16:** A schematic showing the thermoelectric phenomenon in the presence of a background magnetic field. Figure from [84].

Under the influence of a temperature gradient, the mobile charge carriers start diffusing as earlier. Due to the presence, however, of a transverse magnetic field, the charge carriers experience a Lorentz force which makes them drift in a direction perpendicular to both the temperature gradient ($\boldsymbol{\nabla} T$) and the external (constant) magnetic field $B$. The mutually transverse components of the induced electric field and the temperature gradient (as in Fig.1.16) are connected by the Nernst coefficient: $E_y = NB_z(\nabla T)_x$. A major difference between the Seebeck and Nernst effects pertains to the sign of the coefficients. As was discussed before, the Seebeck coefficient is positive (negative) for positively (negatively) charged particles, reflecting the fact that the direction of the induced electric field gets flipped as the electric charge of the carriers flips. In contrast, the Nernst coefficient is independent of the electric charge of the mobile charge carriers. To understand this, consider first, positively charged particles as majority charge carriers in Fig.(1.16). Let the directions of $B$ and $\nabla T$ be $\hat{z}$ and $\hat{x}$, respectively. Then, under the influence of $\nabla T$, the positive charges start diffusing in the $-\hat{x}$ direction. Due to the magnetic field, the charge carriers experience a Lorentz force and consequently pile up in the $-\hat{y}$, leading to the electric field configuration as depicted in Fig.(1.16). Now, if one replaces the positively charged particles with negatively charged ones, they now pile up in the $+\hat{x}$ direction, giving rise to the same electric field configuration as earlier. Thus, the direction of the induced electric field, and hence, the sign of the Nernst coefficient, does not change with the electric charge of the charge carriers.

Seebeck and Nernst coefficients can thus be thought of as the thermoelectric analogue of the electric and Hall conductivities (coefficients).

The deconfined hot QCD medium created post heavy ion collisions can possess a significant temperature-gradient between the central and peripheral regions of the collisions [85]. Majority of collisions in such experiments being non-central, a strong magnetic field perpendicular to the reaction plane is also expected to be created and could be sustained by a finite electrical conductivity of the medium[2]. As such, the quark gluon plasma medium (QGP) is expected to exhibit the thermoelectric effects described above. They provide for a new source of electric current, and hence, a new source of entropy. Study of thermoelectric properties in the context of heavy ion collisions is relatively recent, with the first work done in a hadron gas medium in Ref [86]. Thereafter, the same was studied in QGP in the absence and presence of magnetic fields in [87–92], using different models such as the quasiparticle model, the effective fugacity model, etc.

## 1.8 Overview of the thesis

The research work in this thesis has been presented in the form of seven chapters. A brief introduction of the quark-gluon plasma, its characteristic features and experimental signatures have been presented in **chapter 1**.

**In chapter 2**, we have introduced the theoretical techniques that form the basis of the subsequent chapters. Among other things, we give a brief description of the Imaginary Time Formalism (ITF) of finite temperature field theory and discuss frequency summation techniques. This is followed by a discussion of Hard Thermal Loop

[2]This will be discussed in detail in the next chapter

(HTL) resummation which is essential for obtaining consistent results via perturbation theory at finite temperature. This is followed by a discussion on the Langevin description of heavy quarks in a QCD plasma. We also present a brief section on the problem of fermions in a constant magnetic field.

**In chapter 3**, we study the thermoelectric response in a QGP in the absence of magnetic field as well as in the presence of a strong magnetic field, which is quantified by the Seebeck coefficient. We use the relativistic Boltzmann transport equation within the relaxation time approximation in our calculation. We study the Seebeck coefficient both using current quark masses an quasiparticle masses to explore how the interactions among partons described in perturbative thermal QCD in the quasiparticle framework affect thermoelectric response.

**In chapter 4**, we dial down the strength of the magnetic field and study the same response, first by using the $B = 0$ quasiparticle masses as in the previous section. In the presence of weak $B$, the thermoelectric response becomes a $2 \times 2$ matrix, instead of a scalar, with the diagonal elements representing the Seebeck coefficient and the off-diagonal elements representing the Nernst coefficient. We analyse the variation of these coefficients with temperature, chemical potential, $B$ field strength, and quark mass. Thereafter, we use the quasiparticle masses derived explicitly from the one-loop quark self energy in weak magnetic field to evaluate the same coefficients. We find that these quasiparticle masses can be used to define the range of values for $T$ and $B$ that can be used simultaneously in the weak field limit.

**In chapter 5**, we try to incorporate the effects of an anisotropically expanding weakly magnetized QGP into the the Seebeck and Nernst coefficients. Weak momentum anisotropy is incorporated via the spheroidal deformation of the quark equilibrium distribution functions. We find that a non-zero anisotropy decreases the thermoelectric response of the medium.

**In chapter 6**, we use individual heavy quarks to probe the QGP. We calculate the momentum diffusion coefficients and energy loss of a heavy quark (HQ) traversing through the quark-gluon plasma in the presence of a weak magnetic field, upto leading order in the strong coupling $\alpha_s$. Coulomb scatterings of the HQ with the thermal quarks and gluons are considered, whereas Compton scatterings and gluon radiation are neglected. The calculations are carried out in a perturbative framework where the interaction rate $\Gamma$ is calculated from the imaginary part of the HQ self energy.

**Chapter 7** contains the summary of the present thesis along with possible future directions of research.

Chapter 2

# Theoretical tools and formalisms

एकाद्येकोत्तर पद सङ्कलितं समं पदवर्गत्तिन्टे पकुति।

*The integral of a variable equals half of its square.*

**– Yuktibhāṣā by Jyeṣṭadeva (1500 CE-1575 CE)**

## 2.1 Quantum Chromodynamics (QCD)

As already mentioned, the relevant interaction in the case of QGP is the strong interaction, and the theory governing it is Quantum Chromodynamics (QCD). QCD consists of quarks as the fundamental matter particles that come in 3 ($= N_c$) color charges (instead of one electric charge in QED), along with 8 ($= N_c^2 - 1$) intermediate vector gauge bosons, called gluons. The gauge group of QCD is $SU(3)$, the elements of which, do not commute with each other. So, gluons, unlike photons are (color) charged and interact among themselves. This makes QCD a non-Abelian gauge theory, as is breifly described below.

The Lagrangian of QCD is

$$\mathcal{L}_{\text{QCD}} = \sum_f \bar{\psi}_f (i\not{D} - m_f)\psi_f - \frac{1}{4} F_c^{\mu\nu} F_{\mu\nu}^c. \tag{2.1}$$

Here, $\psi$ and $\bar{\psi}$ are the quark and antiquark fields, and $F_{\mu\nu}^c$ is the field strength tensor given by

$$F_{\mu\nu}^c = \partial_\mu A_\nu^c - \partial_\nu A_\mu^c + g f_{abc} A_\mu^a A_\nu^b, \tag{2.2}$$

where, $A_\mu^c$ is the gluon field carrying color index $c$ ($c = 1$ to 8), and $f$ is the quark flavor index. In the standard model, there are six quark flavours, so that $a \equiv u, d, s, c, b, t$.

$D_\mu = \partial_\mu - igT_cA^c_\mu$ is the covariant derivative, the first term of which, corresponds to the kinetic term in the Lagrangian [Eq.(2.1)], and the second term gives rise to the coupling between the quark and gluon fields via the minimal coupling scheme with $g$ as the gauge coupling constant. $T^c$ are the generators of the gauge group $SU(3)$ which follow the lie algebra given by

$$[T_a, T_b] = if_{abc}T_c, \tag{2.3}$$

where, $f_{abc}$ are the structure constants. Tha Lagrangian in Eq.(2.1) is invariant under the following simultaneous local transformations of the matter and gauge fields [93]

$$\psi \longrightarrow U\psi\,, \qquad U = \exp\big(-iT^c\theta^c(x)\big) \tag{2.4}$$

$$T^cA^c_\mu(x) \longrightarrow U\left(T^cA^c_\mu(x) - \frac{i}{g}U^{-1}\partial_\mu U\right)U^{-1}. \tag{2.5}$$

Here, $\theta$ is the spacetime dependent gauge parameter. The generators $T^c$ are $3\times 3$ matrices, in contrast to the scalar charge belonging to the $U(1)$ gauge group in electrodynamics. Since matrices are non commutative in general (as opposed to scalars which are commutative), QCD is a non-Abelian gauge theory as opposed to the Abelian nature of electrodynamics. The $F^{\mu\nu}_cF^c_{\mu\nu}$ term in Eq.(2.1) with $F^{\mu\nu}_c$ given by Eq.(2.2) shows that the Lagrangian contains terms that are cubic and quartic in the $A^i_\mu$'s, describing the 3-point and 4-point gluon vertex functions. Self interactions among the gauge fields is the main source of asymptotic freedom in QCD. In the 3-dimensional fundamental representation of $SU(3)$ the quark wave function can be represented as

$$\psi_f = \begin{pmatrix}\psi_{red}\\ \psi_{blue}\\ \psi_{green}\end{pmatrix}_f, \tag{2.6}$$

where, the 3 color degrees of freedom associated with each quark flavour $f$ are represented by the color names red, blue and green. The gauge fields $A^\mu_c$, however, belong to the adjoint representation of the gauge group $SU(3)$.

There are some subtleties involved with quantizing the QCD lagrangian. Just like in the abelian QED case, the gauge boson propagator is ill-defined as a consequence of the gauge freedom. To see this, we consider pure QCD (no matter fields). The non-interacting part of such a Lagrangian will be

$$\mathcal{L}_0 = \frac{1}{4}\left(\partial_\mu A^c_\nu - \partial_\nu A^c_\mu\right)^2 . \tag{2.7}$$

To obtain the gauge boson propagator, one needs to invert the matrix corresponding to the differential operator in Eq.(2.7). However, the matrix is singular, and hence, non-invertible. Thus, the propagator cannot be defined. The issue is remedied by fixing the gauge. In QCD, the mechanism of gauge fixing gives rise to terms that are functions of $A^\mu_c$, and hence, results in addition of new terms in the Lagrangian[1]. The new fields are called ghost fields since they follow the wrong spin-statistics (scalars following fermi statistics), so, are not physical. The new QCD Lagrangian thus becomes

$$\mathcal{L}_{\rm QCD} = \sum_f \bar{\psi}_f(i\not{D} - m_f)\psi_f - \frac{1}{4}F^{\mu\nu}_c F^c_{\mu\nu} - \frac{1}{2\xi}(\partial_\mu A^\mu_c)^2 + \partial_\mu \bar{\eta}_a(\partial_\mu + gf^{abc}A^\mu_c)\eta_b \tag{2.8}$$

Here, $\eta$ and $\bar{\eta}$ are the ghost and anti-ghost fields, and $\xi$ is the gauge parameter. $\xi = 0$ and $\xi = 1$ correspond to the Landau and Feynman gauges, respectively.

## 2.2 QCD at finite $T$: Imaginary time formalism

In the previous section, a brief introduction of QCD at $T = 0$ (vacuum) was presented. In the context of studying QGP created in the experiments, a consistent description of QCD at finite temperature is needed. The first such description is due to Matsubara, and is known as the Imaginary Time Formalism (ITF).

The key point in the formulation of the ITF is the observation that the evolution operator in quantum mechanics[2] $e^{-iHt}$ can be related to the Boltzmann factor $e^{-\beta H}$ in statistical thermodynamics, when the former is subjected to the transformation (analytic continuation)

$$t \longrightarrow -i\tau, \tag{2.9}$$

followed by the identification

$$\tau = \beta \equiv 1/T,$$

*i.e.* imaginary time is identified with the inverse temperature. Because of the trade-off

[1] In QED, the same gauge fixing procedure produces a term in the generating functional that is indepepndent of the gauge potential $A^\mu$, and hence, is treated as just another contribution to the normalisation factor.

[2] We work in natural units where $c = \hbar = k_B = 1$.

of the time variable for the temperature, the ITF is most suitable to study equilibrium phenomena and static quantities that do not evolve with time. Geometrically, the transformation Eq.(2.9) corresponds to a wick rotation

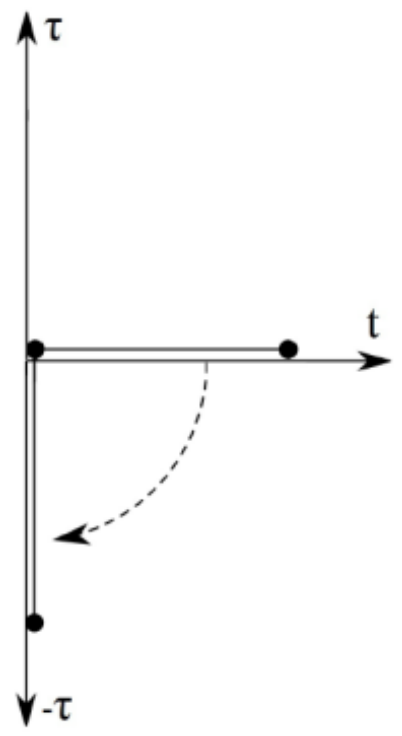

**Figure 2.1:** Analytical continuation of real to imaginary time represented as a rotation in the complex time plane.

The starting point is the path integral representation of the partition function ($\mathcal{Z}$), from which, all thermodynamical quantities can be obtained. The finite $T$ result for $\mathcal{Z}$ can be obtained from its $T = 0$ counterpart with the following recipe [94]

- Wick rotation of real times to imaginary times, $\tau = it$. This amounts to going from Minkowski metric to the Euclidean metric: $t^2 - \boldsymbol{x}^2 \longrightarrow -(\tau^2 + \boldsymbol{x}^2)$.

- A Euclidean Lagrangian is introduced. For a $1+1$ D system, we have

$$\mathcal{L}_E \equiv -\mathcal{L}_M(\tau = it) = \frac{m}{2}\left(\frac{dx}{d\tau}\right)^2 + V(x).$$

- Restriction of the imaginary time $\tau$ to the interval $(0, \beta)$: Because the time variable is restricted, it is no longer at an equal footing with the space variable. Hence, Lorentz invariance is lost.

- Periodicity of the path $x(\tau)$; *i.e.* $x(\beta) = x(0)$.

### 2.2.1 Matsubara frequencies and 2-point functions

Any operator in the Heisenberg picture is given by

$$A_H(t) = e^{iHt} A e^{-iHt}, \tag{2.10}$$

where, $A$ is the operator in the Schrödinger picture. Performing a wick rotation

$(t = -i\tau)$ on Eq.(2.10) yields

$$A_H(-i\tau) = e^{H\tau} A e^{-H\tau}. \tag{2.11}$$

Using the property of cyclicity of trace, the thermal correlation function between two Heisenberg operators $A_H(-i\tau)$ and $B_H(-i\tau')$ can be shown to be related as [95].

$$\langle A_H(-i\tau) B_H(-i\tau') \rangle_\beta = \langle B_H(-i\tau') A_H(-i(\tau - \beta)) \rangle_\beta. \tag{2.12}$$

The subscript $\beta$ implies a thermal average, *e.g.* $\langle A_H \rangle_\beta = Z^{-1} \mathrm{Tr}\rho(\beta) A_H$, where, $Z = \mathrm{Tr}\rho(\beta)$ is the partition function, and $\rho = e^{-\beta H}$ is the density matrix. Relations of the type of Eq.(2.12) are called *Kubo-Martin-Schwinger* (KMS) relations.

The two-point Green's function is defined as

$$\Delta(\tau, \tau') = \left\langle T\left(\phi_H(\tau)\phi_H^\dagger(\tau')\right)\right\rangle_\beta, \tag{2.13}$$

where, $\phi$ could be a bosonic or a fermionic field operator, and $T$ denotes time-ordering. The functions $\Delta$ depend only on the difference $\tau - \tau'$, and consequently have the range $-\beta \le \tau - \tau' \le \beta$. From the KMS relation, it follows that for $\tau > 0$,

$$\Delta(0, \tau) = \pm\Delta(\beta, \tau), \tag{2.14}$$

where, the + and − signs refer to bosons and ferions, respectively, i.e., the Euclidean bosonic and fermionic 2-point green's functions are respectively periodic, and anti-periodic. Since the domain of the Green's functions is restricted, its Fourier decomposition has to be in terms of discrete frequencies

$$\Delta(\tau) = \frac{1}{\beta} \sum_{n=-\infty}^{\infty} \Delta(\omega_n) e^{-i\omega_n \tau} \tag{2.15}$$

$$\Delta(\omega_n) = \int_0^\beta d\tau \Delta(\tau) e^{i\omega_n \tau}, \tag{2.16}$$

where,

$$\omega_n = \begin{cases} \frac{2n\pi}{\beta}, & \text{for bosons} \\ \frac{(2n+1)\pi}{\beta}, & \text{for fermions} \end{cases}$$

The discrete values $\omega_n$ are the *Matsubara frequencies*, and the corresponding coefficients $\mathcal{G}_\beta(\omega_n)$ are the *Matsubara modes.* The discrete Matsubara frequencies are a key feature of ITF which, as we have seen, arises due to the restricted range of the imaginary time variable.

### 2.2.2 Frequency sum

The familiar Feynman rules in zero temperature theory carry over to the finte $T$ case with the exception that since the time ($\tau$) domain is restricted, the energy values (fourier conjugate to $\tau$) are quantized, leading to intermediate energy integrals being replaced by sums:

$$\int \frac{d^4p}{(2\pi)^4} \longrightarrow iT \sum_{p_0} \int \frac{d^3p}{(2\pi)^3} \equiv T \sum_{n} \int \frac{d^3p}{(2\pi)^3}, \tag{2.17}$$

with $p_0 = i\omega_n = \frac{i2\pi n}{\beta}$ for bosons and $p_0 = i\omega_n = \frac{i\pi(2n+1)}{\beta}$ for fermions. There are different ways in which these frequebcy sums can be carried out. A convenient way is the Saclay method [96] which is described below briefly, and has also been used in this thesis.

For a bosonic propagator in momentum space $\Delta(P) = 1/P^2$, with $P^2 = (p^0)^2 + p^2$ and $p^0 = 2\pi nT$, fourier transformation of the $p^0$ variable leads to[3]

$$\Delta(\tau, p) = T \sum_{\substack{n=-\infty \\ p^0=2\pi nT}}^{\infty} e^{-ip^0\tau} \Delta(P). \tag{2.18}$$

By observing that $\Delta(P)$ has poles at $p^0 = \pm ip$ with residues $\mp i/2p$, the sum above can be evaluated by contour integration techniques [97] which yields

$$\Delta(\tau, p) = \frac{1}{2p} \left[ \{1 + n_B(p)\} e^{-p\tau} + n_B(p) e^{p\tau} \right], \tag{2.19}$$

where, $n_B(p) = \left(\exp(p/T) - 1\right)^{-1}$ is the Bose-Einstein distribution function. The inverse of Eq.(2.18) is

$$\Delta(P) = \int_0^\beta d\tau e^{ip_o\tau} \Delta(\tau, p). \tag{2.20}$$

The sum over $p^0$ produces a delta function in $\tau$, thus making the $\tau$ integration trivial.

[3]Equivalently, one can have $p_0 = i\omega_n = \frac{i\pi(2n+1)}{\beta}$, and $\Delta(\tau, p)$ defined as a Laplace transform: $\Delta(\tau, p) = T \sum_{\substack{n=-\infty \\ p^0=i2\pi nT}}^{\infty} e^{-p^0\tau} \Delta(P)$

This leaves an integral over the three-momentum $\boldsymbol{p}$. For instance, to evaluate the sum-integral $S = T\sum_n \int \frac{d^3p}{(2\pi)^3}\Delta(P)$, we use Eq.(2.20) for $\Delta(p)$, and Eq.(2.19) for $\Delta(\tau, p)$. The sum over $p^0$ produces $\delta(\tau)$, rendering the $\tau$ integration trivial. We are finally left with

$$S = \int \frac{d^3p}{(2\pi)^3}\Delta(\tau = 0, p) = \int \frac{d^3p}{(2\pi)^3}\frac{1}{2p}[1 + 2n_B(p)]. \tag{2.21}$$

For fermions, Eq.(2.19) becomes

$$\tilde{\Delta}(\tau, p) = \frac{1}{2p}\left[\{1 - n_f(p)\}e^{-p\tau} - n_f(p)e^{p\tau}\right], \tag{2.22}$$

where, $n_f(p) = \big(\exp(p/T) + 1\big)^{-1}$ is the Fermi-Dirac distribution function. The propagator is

$$\tilde{\Delta}(P) = 1/P^2\,; \quad p^0 = (2n + 1)\pi T. \tag{2.23}$$

## 2.3 Hard Thermal Loop (HTL) resummation

The familiar perturbation theory in vacuum ($T = 0$) breaks down at finite temperature. This was realized during attempts to calculate the gluon damping rate to leading order in the strong coupling $g$ in hot QCD. Naive perturbation theory yielded results that were gauge dependent, and which possessed instabilities [98–100]. The reason for this was that the calculations were 'incomplete'. At finite temperature, the usual connection between the the order of the loop expansion and powers of $g$ no longer exists: contribution of order $g$ arise not only from one-loop diagrams, but from every order of loop expansion. Therefore, it is needed to resum an infinite subset of graphs in order to get the complete result in any order in $g$. The systematic procedure of carrying out such a resummation was introduced by Braaten and Pisarski, using which, the results of gluon damping rate to leading order in $g$ came out to be gauge independent and devoid of instabilities [101].

The bare and effective propagators are related via the Schwinger Dyson equation as

$$\Delta_{\text{eff}}^{-1}(P) = \Delta^{-1}(P) + \Sigma, \tag{2.24}$$

For a hot[4] scalar theory with quartic self coupling $g^2$, the bare propagator is $\Delta(P) = 1/P^2$, and the effect of finite temperature is that the tadpole diagram (correction to

[4] Hot implies that the scalar is massless at tree level.

the bare inverse propagator) generates a $T$ dependent mass term $\Sigma \equiv m_T^2 \sim g^2T^2$. For hard momenta ($P \sim T$), $\Sigma \ll \Delta^{-1}(P)$ (provided $g \ll 1$), and it suffices to use the bare propagator in calculations. However, when $P$ is soft ($P \sim gT$), $\Sigma$ is as large as the bare inverse propagator and must be resummed to obtain an effective propagator. Resummation in scalar theory involves just replacing the bare propagator $1/P^2$ with the effective propagator $1/(P^2 + m_T^2)$. For gauge theories, matters are complicated by the fact that the self energy that enters into the effective propagator is no longer a simple mass term, but carries non-trivial $P$ dependence. Also, effective vertices need to be used instead of bare vertices. The diagrams that must be resummed are called **hard thermal loops**, since they originate from integration regions where the loop momentum is hard. These are loop corrections which are $g^2T^2/P^2$ times the corresponding tree amplitude, where $P$ is the momentum of an external line. An example of HTL is described below:

We consider the bosonic propagator $\Delta(K)$ of Sec.(2.2.2). For one loop diagrams with loop momentum $K^\mu$, we define the sum integral as

$$\mathrm{Tr} = T\sum_n \int \frac{d^3k}{(2\pi)^3} \tag{2.25}$$

Then, as was shown in Sec.(2.2.2),

$$\mathrm{Tr}\,\Delta(K) = \int \frac{d^3k}{(2\pi)^3}\frac{1}{2k}[1 + 2n_B(k)]. \tag{2.26}$$

The first term is $T$ independent and quadratically UV divergent, and is removed by renormalization at zero temperature. Using $\int\limits_0^\infty dk\, k\, n_B(k) = \pi^2T^2/6$, Eq.(2.26) becomes

$$\mathrm{Tr}\,\Delta(K) \approx \frac{T^2}{12}. \tag{2.27}$$

As salready stated, HTLs are $g^2T^2/P^2$ times the corresponding tree diagram. For one-loop diagrams, the $g^2$ automatically comes from the Feynman rules, so, the HTL diagrams are those which are proportional to $T^2$. Thus, $\mathrm{Tr}\,\Delta(K)$ [Eq.(2.27)] is an example of an HTL.

## 2.4 Quasiparticle model

Quasiparticle description is a phenomenological description of quarks and gluons in a thermal QCD medium, in which thermal masses of partons are generated, apart from their current masses in QCD Lagrangian. These masses are generated due to the interaction of a given parton with other partons in the medium, therefore, quasiparticle description describes the collective properties of the medium. It can be applied to study several thermal properties of QGP near the crossover temperature, $T_c$, where perturbation theory cannot be used directly. Such a model was initially proposed by Goloviznin and Satz [102]. Different versions of quasiparticle description exist in the literature based on different effective theories, such as Nambu-Jona-Lasinio (NJL) model and its extension Polyakov-loop extended Nambu Jona Lasinio model [103–105], Gribov-Zwanziger quantization [106, 107], thermodynamically consistent quasiparticle model [108], etc. The results arrived at using these models suggest that it is possible to describe the high temperature QGP phase by a thermodynamically consistent quasiparticle model. Our description relies on perturbative thermal QCD, where the medium generated masses for quarks and gluons are obtained from the poles of dressed propagators calculated by the respective self energies at finite temperature. In the quasiparticle description of quarks and gluons in a thermal medium, all quark flavors (with current/vacuum masses, $m_i \ll T$) acquire the same thermal mass

$$m_T^2 = \frac{g(T)^2 T^2}{6}. \tag{2.28}$$

which is, however, modified in the presence of a finite chemical potential [109] as

$$m_{T,\mu}^2 = \frac{g(T)^2 T^2}{6}\left(1 + \frac{\mu^2}{\pi^2 T^2}\right). \tag{2.29}$$

We take the quasiparticle mass (squared) of $i$th flavor to be [108, 110–112]:

$$m'^{\,2}_{iT} = m_i^2 + \sqrt{2}\, m_i\, m_T + m_T^2. \tag{2.30}$$

In the case of a strong background magnetic field, the quark quasparticle masses are different from Eq.(2.28). The quark self energy in the presence of a strong magnetic field is first evaluated, followed by the effective quark propagator. Taking the static limit of the denominator of the effective quark propagator yields the quasiquark masses

as:

$$m_{iT,B}^2 = \frac{g_s^2|q_iB|}{3\pi^2}\left[\frac{\pi T}{2m_i} - \ln(2)\right]. \tag{2.31}$$

The general procedure of evaluating the quasifermion masses involves expanding the fermion self energy and the fermion effective propagator in terms of chiral projection operators $P_{R/L} = (\mathbb{I} \pm \gamma_5)/2$. Consequently, the effective propagator has a left handed ($L$) term and a right-handed ($R$) term. Correspondingly, we obtain $L$ mode and $R$ mode quasifermion masses. In the case of strong magnetic field, the two masses come out to be the same, as given by Eq.(2.31). In the case of a weak magnetic field however, this degeneracy in the mass is lifted and we obtain:

$$m_{L/R}^2 = m_{th}^2 \pm 4g^2 C_F M^2, \tag{2.32}$$

where,

$$M^2 = \frac{|q_fB|}{16\pi^2}\left(\frac{\pi T}{2m_f} - \ln 2 + \frac{7\mu^2\zeta(3)}{8\pi^2T^2}\right), \tag{2.33}$$

$$m_{th}^2 = \frac{1}{8}g^2 C_F\left(T^2 + \frac{\mu^2}{\pi^2}\right). \tag{2.34}$$

This lifting of mass degeneracy has interesting consequences which will be discussed in Chapter(4).

## 2.5 Boltzmann transport equation

Each parton in the plasma is associated with a one-particle distribution function, $f(x,p) \equiv f(\boldsymbol{r},\boldsymbol{p},t)$ which is a Lorentz invariant density in phase space, so that $f(\boldsymbol{r},\boldsymbol{p},t)\, d^3r\, d^3p$ gives the number of partons in the spatial volume element $d^3r$ about $\boldsymbol{r}$ and with momenta in a range $d^3p$ about $\boldsymbol{p}$. The evolution of this distribution function towards equilibrium via collisions is described by the Boltzmann transport equation

$$\frac{df(r,p,t)}{dt} = \frac{\partial f}{\partial t} + \frac{\boldsymbol{p}}{m}\cdot\boldsymbol{\nabla} f + \boldsymbol{F}\cdot\frac{\partial f}{\partial \boldsymbol{p}} \tag{2.35}$$

$$= \left(\frac{\partial f}{\partial t}\right)_{\text{coll}} \equiv C(f), \tag{2.36}$$

where, $C(f)$ denotes the rate of change of the distribution function by virtue of collisions, $\boldsymbol{F}$ is the force field acting on the particles in the medium. If the collision term $C(f)$ is zero then the particles do not collide, and individual collisions get replaced by long-range aggregated (Coulomb) interactions, and the equation is then referred to as the collisionless Boltzmann equation or Vlasov equation.

Each collision undergone by a parton changes its momentum $p$, and thereby transfers it out of a particular range $dp$ in the momentum space. The total number of binary collisions $p, p_1 \rightarrow p', p'_1$ with all possible values of $p_1, p', p'_1$ for a given $p$, occurring in a volume $dV$ per unit time, is

$$dV\, dp \int w(p, p_1 : p', p'_1) f f_1\, dp_1 dp' dp'_1, \tag{2.37}$$

with $f \equiv f(r, p, t)$, $f_1 \equiv f(r, p_1, t)$. Such collisions are referred to as "losses", since they result in loss of partons from the concerned phase-space element. Similarly, there are collisions which lead to "gain" of particles. These are the collisions $p', p'_1 \rightarrow p, p_1$ with all possible $p_1, p', p'_1$ for a given $p$. The total number of such collisions in the volume $dV$ per unit time is given by

$$dV\, dp \int w(p', p'_1 : p, p_1) f f_1\, dp_1 dp' dp'_1 \tag{2.38}$$

Subtracting the loss term from the gain term yields the increase in the number of molecules in the concerned phase space per unit time. This is given by

$$dV\, dp \int (w' f' f'_1 - w f f_1)\, dp_1 dp' dp'_1, \tag{2.39}$$

where,

$$w \equiv w(p', p'_1 : p, p_1), \qquad w' \equiv w(p, p_1 : p', p'_1) \tag{2.40}$$

The expression of the collision integral is thus

$$C(f) = \int (w' f' f'_1 - w f f_1)\, dp_1 dp' dp'_1 \tag{2.41}$$

By the principle of detailed balance, one can write [113]

$$\int w(p', p'_1 : p, p_1) dp'\, dp'_1 = \int w(p, p_1 : p', p'_1) dp'\, dp'_1 \tag{2.42}$$

In the second term of Eq.(2.41), $f$ and $f$' do not depend on $p'$ and $p'_1$, and so, the integration over $p'$ and $p'_1$ relates only to $w$. Using Eq.(2.42) and Eq.(2.41), the Boltzman transport equation can be written as

$$\frac{\partial f}{\partial t} + \frac{\boldsymbol{p}}{m} \cdot \boldsymbol{\nabla} f + \boldsymbol{F} \cdot \frac{\partial f}{\partial \boldsymbol{p}} = \int w'(f'f'_1 - ff_1)\, dp_1 dp' dp'_1. \tag{2.43}$$

Since the QGP is a relativistic plasma, the relativistic version of the Boltzmann transport equation is relevant. Further, for the study of thermoelectric response, only the quark distribution is relevant. For a quark carrying flavor index $i$, the transport equation reads [114]

$$p^\mu \frac{\partial f_i(x,p)}{\partial x^\mu} + q_i F^{\mu\nu} p_\nu \frac{\partial f_i(x,p)}{\partial p^\rho} = C[f_i(x,p)], \tag{2.44}$$

where, $q_i$ is the electric charge of flavour $i$, $F^{\mu\nu}$ is the electromagnetic field strength tensor and $C[f_i(x,p)]$ is the collision term already discussed earlier. $x$ and $p$ in Eq.(2.44) are now 4-vectors. The transport equation [Eq.s(2.43) and (2.44)] is an integro-differential equation, non linear in $f$, and is thus very difficult to solve. To simplify this, the equation needs to be linearized. A way of linearizing the Boltzman equation is via the relaxation time approximation (RTA) in the form

$$C[f_i(x,p)] = -\frac{p^\mu u_\mu}{\tau_i}(f_i - f_i^0) = -\frac{p^\mu u_\mu}{\tau_i}\, \delta f_i. \tag{2.45}$$

where, $\tau$ is the relaxation time, $u^\mu$ is the fluid 4-velocity, and $f^0$ the equilibrium distribution function given by (suppressing the functional dependence)

$$f_i^0 = \frac{1}{\exp(\frac{u_\nu p^\nu - \mu_i}{T}) - 1}. \tag{2.46}$$

In order to study dissipative processes, it is required to allow a slight deviation of the local distribution function $f$ away from equilibrium, *i.e.*, $f = f_0 + \delta f$, with $\delta f \ll f_0$. The correction $\delta f$ must in principle be determined by solving the transport equation linearized with respect to the correction [115]. The macroscopic properties of the gas such as its velocity $v$, temperature $T$, pressure $P$ are assumed to vary appreciably only over very large distances. Since the gradients of these quantities are assumed small, it is sufficient (in this approximation) to replace $f$ by $f_0$ on the left of the transport equation (Eq. 2.44). We shall be working out the thermoelectric transport coefficients of the QGP in this approximation.

## 2.6 Fermions in constant magnetic field

The problem of fermions in a magnetic field is relevant in the case of QGP due to the creation of large magnetic fields in off-central nucleus-nucleus collisions. The non-relativistic Hamiltonian corresponding to a free fermion in the presence of a constant magnetic field is (ignoring spin) [116]:

$$\hat{H} = \frac{1}{2m}\left(\hat{\boldsymbol{p}} - q\boldsymbol{A}/c\right)^2 + q\phi, \tag{2.47}$$

where, $\hat{\boldsymbol{p}} = -i\hbar\boldsymbol{\nabla}$ is the momentum operator, $q$ is the electric charge, and $\boldsymbol{A}$, $\phi$ are the vector and scalar potentials, respectively. In the Landau gauge, the vector potential corresponding to a uniform magnetic field $B$ in the $\hat{z}$ direction is taken to be

$$\boldsymbol{A} = (-By, 0, 0) \tag{2.48}$$

With this, the Schrödinger equation becomes

$$\frac{1}{2m}\left[\left(\hat{p_x} + \frac{qBy}{c}\right)^2 + \hat{p_y}^2 + \hat{p_z}^2\right]\psi = E\psi \tag{2.49}$$

Since the Hamiltonian does not contain the coordinates $x$ and $z$, we have, $[\hat{H}, \hat{p_x}] = [\hat{H}, \hat{p_z}] = 0$. This leads to the ansatz

$$\psi = e^{i/\hbar(p_x x + p_z z)}\chi(y) \tag{2.50}$$

The eigenvalues $p_x$ and $p_y$ in Eq.(2.50) are such that $0 < p_x, p_y < \infty$. Thus, motion along the field is not quantized. Using Eq.(2.50) in Eq.(2.49), one obtains

$$\chi''(y) + \frac{2m}{\hbar^2}\left[\left(E - \frac{p_z^2}{2m}\right) - \frac{1}{2}m\omega_c^2(y - y_0)^2\right]\chi(y) = 0, \tag{2.51}$$

with $y_0 = -cp_x/qB$, and $\omega_c = qB/mc$. Eq.(2.51) is identical to the Schrödinger equation for a 1-D harmonic oscillator with centre at $y = y_0$. One can thus immediately

write

$$E_n = (n + 1/2)\hbar\omega_c + p_z^2/2m, \tag{2.52}$$

where, $n$ are the Landau levels.

For the relativistic case, one can start with the Dirac equation for a free fermion in the presence of an external electromagnetic field

$$\hat{H}\psi \equiv [\boldsymbol{\alpha} \cdot \boldsymbol{\pi} + \beta m]\psi = E\psi, \tag{2.53}$$

where,

$$\boldsymbol{\alpha} = \begin{pmatrix} 0 & \boldsymbol{\sigma} \\ \boldsymbol{\sigma} & 0 \end{pmatrix}, \quad \boldsymbol{\pi} = \boldsymbol{p} - e \cdot A \tag{2.54}$$

Remembering that the motion along the field is free, the above equation can be written as

$$\hat{H}\psi \equiv [\alpha_x \pi_x + \alpha_y \pi_y + \hbar k_x + \beta m]\psi = E\psi \tag{2.55}$$

Applying the Hamiltonian operator once more from the left, we get

$$\hat{H}^2\psi = E^2\psi. \tag{2.56}$$

Using the ladder operators

$$a = \frac{\pi_x + i\pi_y}{\sqrt{2eB\hbar}}, \quad a^\dagger = \frac{\pi_x - i\pi_y}{\sqrt{2eB\hbar}}, \tag{2.57}$$

and with the identification $a^\dagger a = n$, one can show that the energy eigenvalues are

$$E_n = \sqrt{p_z^2 + m_f^2 + 2nq_f B}, \tag{2.58}$$

where, $n$ denotes the Landau levels. Thus, the motion in the plane perpendicular to the magnetic field gets quantized. More will be discussed about the Landau levels in Chapter(3).

## 2.7 Langevin dynamics for heavy quark

Heavy quarks (charm, bottom) are considered to be excellent probes of the QGP primarily because of their large masses compared to the temperature of the plasma ($M_Q \gg T$), which confers them several advantages when it comes to probing the

QGP.

- Owing to their very large masses, heavy quarks (HQs) are expected to be produced almost exclusively at the primordial stages of the HICs, and not via recombination in the plasma [117].

- The thermal relaxation time for HQs is expected to be larger than lifetime of QGP, and hence, ought not to thermalise completely with the medium. The HQs therefore carry a 'memory' of their interactions in the evolving plasma.

- Theoretically, the description of HQs in a plasma of light quarks and gluons is amenable to a diffusion treatment; more specifically, Brownian motion of a heavy test particle in a medium of light particles.

The justification of using Langevin dynamics to describe HQs in QGP, and its formulation is described below:

The HQ thermal momentum $p \sim \sqrt{M_Q T} \gg T$ translates to a thermal velocity $v \sim \sqrt{T/M_Q} \ll 1$. Even if one considers hard scatterings of the HQ with the light medium particles (characterised by a momentum transfer of $\mathcal{O}(T)$), it takes a large number of collisions ($\sim M_Q/T$) to change the HQ momentum by $\mathcal{O}(1)$, since $p \gg T$. This implies that the momentum changes accumulate over time from uncorrelated "kicks", and the HQ momentum therefore evolves according to Langevin dynamics:

$$\frac{dp_i}{dt} = \xi_i(t) - \eta_D p_i\,, \qquad \langle \xi_i(t)\xi_j(t')\rangle = \kappa\,\delta_{ij}\delta(t-t'), \tag{2.59}$$

where, $(i,j) = (x,y,z)$. The interaction of the HQ with all the other degrees of freedom in the QGP is hopelessly difficult to explain in detail. According to Langevin dynamics, the influence of the medium on the motion of HQ can be split into two parts: first, a part representing the dynamical friction experienced by the HQ [$-\eta_D p$ in Eq.(2.59)] that drives the HQ towards equilibrium, and second, a rapidly fluctuating force $\xi(t)$ that is characteristic of Brownian motion. Although the functional form of $\xi(t)$ cannot be specified, certain assumptions are made [118]:

1. $\xi(t)$ is independent of $v$, the HQ velocity.

2. $\xi(t)$ varies much more rapidly than $v$.

The second assumption implies that in a time interval $\Delta t$ in which $v$ varies by a negligible amount, $\xi(t)$ would have undergone many fluctuations, so that no correlation exists between $\xi(t)$ and $\xi(t+\Delta t)$. This is reflected by the $\delta$ function in Eq.(2.59), and such random forces are referred to as white noises.

$\eta_D$ is called the momentum drag coefficient and $\kappa$ is the momentum diffusion coefficient. The solution of Eq.(2.59) under the assumption $\eta_D^{-1} \ll t$ is given as

$$p_i(t) = \int_{-\infty}^{t} dt'\, e^{\eta_D(t'-t)} \xi_i(t'). \tag{2.60}$$

Dynamical quantities such as the momentum diffusion coefficient $\kappa$, HQ energy loss $dE/dx$, spatial diffusion coefficient $D_s$ can be calculated from the Langevin's equation, and the same is described in Chapter(6).

The key object needed in the quantum field theoretical evaluation of HQ dynamics in QGP is the scattering rate $\Gamma$. In particular, $\frac{d\Gamma(\boldsymbol{q})}{d^3\boldsymbol{q}}$ is the scattering rate of the HQ with thermal particles (light quarks and gluons)via gluon exchange per unit volume with momentum transfer $q$. $\Gamma$ can be evaluated from the imaginary part of the HQ self energy, as was shown in [119]

$$\Gamma(P \equiv E, \mathbf{v}) = -\frac{1}{2E}[1 - n_F(E)]\, \mathrm{Tr}\left[(\not{P} + M_Q)\, \mathrm{Im}\, \Sigma\, (p_0 + i\epsilon, \boldsymbol{p})\right]. \tag{2.61}$$

The formula above can be motivated by realizing that the imaginary part of the self energy is related to the amplitude (squared) for Coulomb scattering processes, via the cutting rules. For the purpose of explanantion ,let us consider the following QED diagrams. For the case of HQ in QGP, one simply has to replace the photon lines with gluon lines.

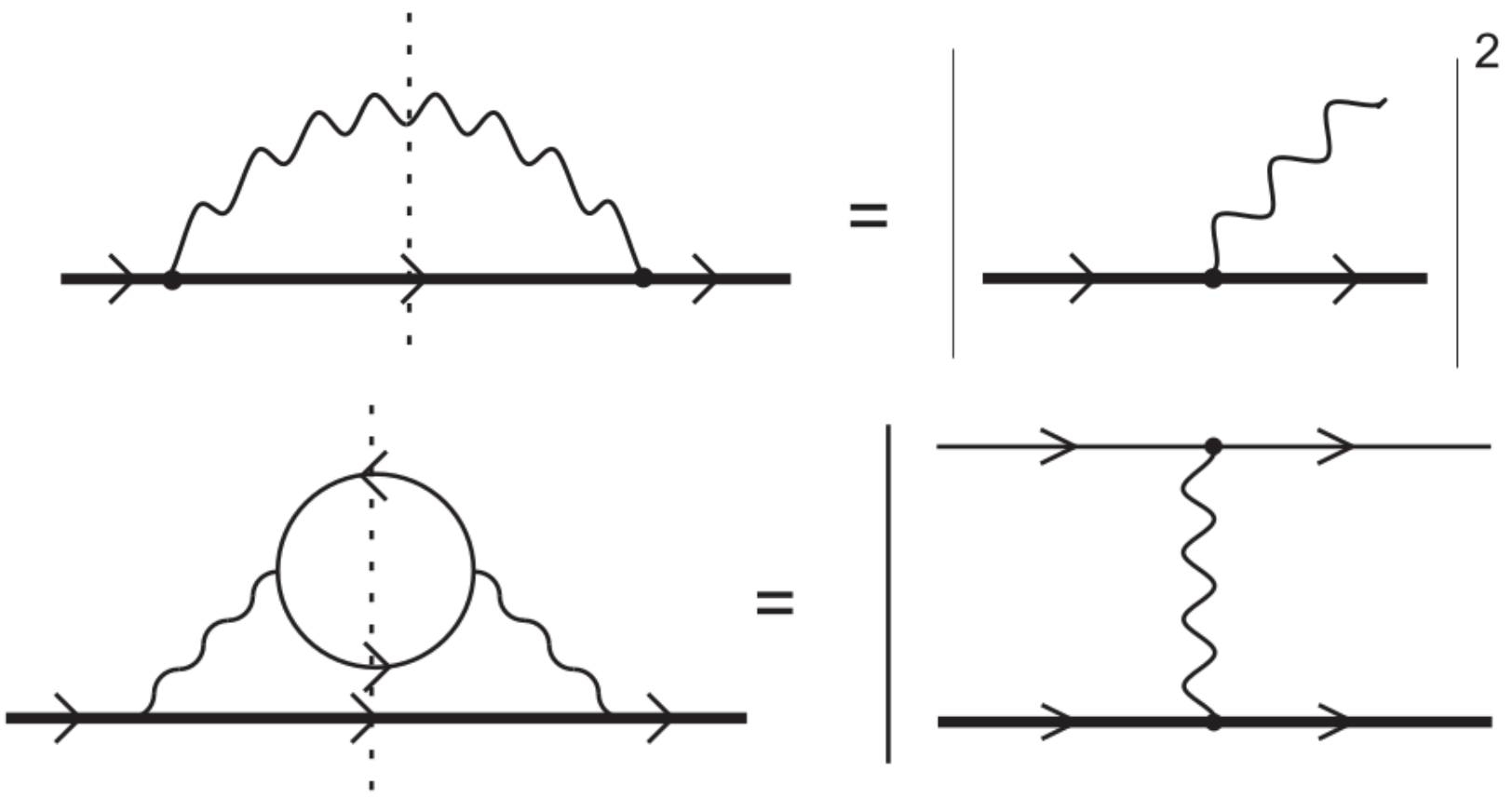

**Figure 2.2:** 1 and 2 loop heavy fermion self energy diagrams and the scattering processes they correspond to.

The cut part of 1-loop self-energy diagram in Fig.(2.2) corresponds to a process that is unphysical, since it violates energy-momentum conservation. Thus, its imaginary part is 0. The cut part of the 2-loop self-energy diagram corresponds to Coulomb scattering of the heavy fermion with light fermions and has a finite imaginary part. However, $\Gamma$ evaluated from this diagram comes out to be quadratically infrared divergent due to the exchange of a bare gauge boson. This is remedied by using an effective gauge boson propagator instead of a bare one. Thus, in our work, we calculate $\Gamma$ from the one-loop HQ self-energy with an effective gluon propagator which takes care of the scatterings of HQ with both quarks and gluons of the medium.

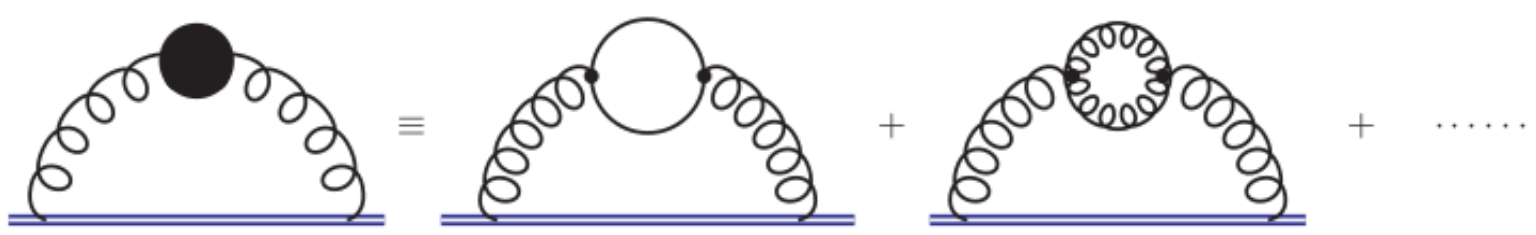

**Figure 2.3:** Heavy quark self energy with effective gluon propagator.

Once $\Gamma$ is evaluated, we use it to calculate quantities such as $\kappa$ and $dE/dx$ as

$$3\kappa = \int d^3q \frac{d\Gamma(q)}{d^3q} q_L^2 \tag{2.62}$$

$$\frac{dE}{dx} = \frac{1}{v} \int d^3q \frac{d\Gamma(q)}{d^3q} q_0 \tag{2.63}$$

where, $q$ is the magnitude of the 3-momentum of the exchanged gluon, and $q_0$ is its energy.

Chapter 3

# Seebeck effect in a thermal QCD medium in the presence of strong magnetic field

दीर्घचतुरश्रस्याक्ष्णया रज्जुः पार्श्वमानी तिर्यग् मानी च यत् पृथग् भूते कुरूतस्तदुभयं करोति ॥

*The areas produced separately by the length and the breadth of a rectangle together equal the areas produced by the diagonal.*

(The Pythagorean theorem)

– **Baudhāyana Śulbasûtra (800 BC-740 BC)**

In this chapter, we calculate the Seebeck coefficient of a thermal medium of quarks and gluons, which quantifies the strength of the thermoelectric response. Interactions among partons are accomodated by using thermally generated masses of quarks and gluons instead of bare masses. The coefficient has been calculated in the absence ($B = 0$) as well as in the presence of a strong ($eB \gg T^2$) magnetic field. This chapter is based on the paper *Seebeck effect in a thermal QCD medium in the presence of strong magnetic field* by Debarshi Dey and Binoy Krishna Patra, **Phys. Rev. D, 104, 076021 (2021).**

In the present work, we wish to calculate the Seebeck coefficient of a color deconfined medium of quarks and gluons. In addition, we also explore the effects of a strong magnetic field and quasiparticle description of the medium constituents, on the Seebeck effect, where the quasiparticle/effective masses of the partons (*mainly* quarks) are evaluated from perturbative thermal QCD up to one-loop. The noteworthy differences from a hadronic medium are two-fold: i) The degrees of freedom are more fundamental, *i.e.* the elementary quarks and gluons instead of mesons and baryons. ii) The system is relativistic, *i.e.* $m_f \ll T$ ($m_f$ refers to the mass of $f^{\rm th}$ flavour).

The chapter is organised as follows: In Section (3.1.1) and Section (3.1.2), we quantify the effect by the Seebeck coefficients with the current quark masses of the partons in the absence and presence of a strong magnetic field, respectively. In Section (3.2), we calculate the same, taking into account the interactions in the medium by perturbative thermal QCD in a strong magnetic field, which in turn generates (quasiparticle) masses for the partons. In Section (3.3), the conclusions are drawn and the results are summarized.

## 3.1 Seebeck effect in hot partonic medium with current quark masses

In this section, we construct a general framework for studying the thermoelectric effect for a hot partonic medium and then use the framework to estimate the Seebeck coefficient for the individual species as well as for the composite medium. Then we study the effects of strong magnetic field on the aforesaid study of the thermoelectric effect.

The spacetime evolution of the single particle distribution function with a finite quark chemical potential $\mu_i$, immersed in a thermal medium of quarks and gluons at temperature $T$ is given by the relativistic Boltzmann transport equation RBTE

$$p^\mu \frac{\partial f_i(x,p)}{\partial x^\mu} + q_i F^{\rho\sigma} p_\sigma \frac{\partial f_i(x,p)}{\partial p^\rho} = C[f_i(x,p)], \tag{3.1}$$

where, $i$ is the flavor index, $C[f_i(x,p)]$ is the collision term and $F^{\rho\sigma}$ is the electromagnetic field strength tensor. As has already been discussed, we consider an infitesimal deviation of $f$ from its equilibrium value, *i.e.* $f + f^0 + \delta f$, and employ the relaxation time approximation (RTA) to linearise the Boltzmann equation. In this approximation, $C[f_i(x,p)]$ reduces to

$$C[f_i(x,p)] \simeq -\frac{p_\nu u^\nu}{\tau_i} \delta f_i, \tag{3.2}$$

where $u^\nu$, is the fluid 4-velocity, and $\tau_i$ is the relaxation time given by [120]

$$\tau(T) = \frac{1}{5.1 T \alpha_s^2 \log\left(\frac{1}{\alpha_s}\right)[1 + 0.12(2N_f + 1)]}, \tag{3.3}$$

We choose to work in the rest frame of the medium so that $u^\nu = (1, 0, 0, 0)$. Here, the strong coupling runs with the temperature and chemical potential as [121]

$$\alpha_s(T, \mu_i) = \frac{g^2(T, \mu_i)}{4\pi} \tag{3.4}$$

$$= \frac{6\pi}{(33 - 2N_f) \ln\left(\frac{2\pi\sqrt{T^2 + \mu_i^2/\pi^2}}{\Lambda_{QCD}}\right)}, \tag{3.5}$$

where, $\Lambda_{QCD} = 176$ MeV, $N_f$ is the number of quark flavors in the medium, which, in our case is 2 ($u$, $d$).

### 3.1.1 Seebeck coefficient in the absence of magnetic filed

For the thermoelectric effect, we are interested only in the quark distribution for which the equilibrium distribution for $i^{\text{th}}$ flavour, $f_i^0$, is given by

$$f_i^0 = \frac{1}{\exp\left(\frac{\omega_i - \mu_i}{T(\boldsymbol{r})}\right) + 1}. \tag{3.6}$$

The disturbance ($\delta f$) due to an eectric field $\boldsymbol{E}$ is obtained by considering the $\rho = i$ and $\sigma = 0$ components of the RBTE (Eq. 3.1), and replacing $f$ by $f^0$ in the L.H.S. of Eq.(3.1).

$$\delta f_i = \frac{\tau_i}{\omega_i} f_i^0(1 - f_i^0)\left(-\frac{1}{T^2}\right) \boldsymbol{p} \cdot \nabla_{\boldsymbol{r}} T(\boldsymbol{r}) + 2q_i f_i^0(1 - f_i^0)(\boldsymbol{E}.\boldsymbol{p}) \frac{\tau_i}{\omega_i T}, \tag{3.7}$$

which, in turn, produces the induced four-current through the relation

$$J_\mu = \sum_i q_i g_i \int \frac{d^3\,\mathrm{p}}{(2\pi)^3\,\omega_i} p_\mu \left[\delta f_i^q(x,p) - \delta f_i^{\bar{q}}(x,p)\right], \tag{3.8}$$

where $q_i$ and $g_i$ are respectively the charge and degeneracy factors of quark flavour $i$.

We thus obtain the spatial-part of the induced four-current, *i.e.* the induced

current density

$$\boldsymbol{J} = \frac{g_i q_i \tau_i}{2\pi^2}\Bigg[\int_0^\infty \mathrm{dp}\, \frac{p_i^4}{\omega_i^2(p)} \left\{f_i^0(1-f_i^0)(\omega_i(p)-\mu_i) + \bar{f}_i^0(1-\bar{f}_i^0)(\omega_i(p)+\mu_i)\right\}\left(-\frac{1}{T^2}\right)\nabla_{\boldsymbol{r}} T(\boldsymbol{r}) \\ + 2q_i\int_0^\infty \mathrm{dp}\, \frac{p_i^4}{\omega_i^2(p)}\left\{f_i^0(1-f_i^0) + \bar{f^0}_i(1-\bar{f^0}_i)\right\}\frac{\boldsymbol{E}}{T(\boldsymbol{r})}\Bigg]. \tag{3.9}$$

The above current density is set equal to zero to obtain the relation between the temperature-gradient and the induced electric field in the coordinate space [122]:

$$\boldsymbol{E} = \frac{1}{2Tq_i}\frac{\int_0^\infty \mathrm{dp}\, \frac{p_i^4}{\omega_i^2(p)}\left\{f_i^0(1-f_i^0)(\omega_i(p)-\mu_i) - \bar{f}_i^0(1-\bar{f}_i^0)(\omega_i(p)+\mu_i)\right\}}{\int_0^\infty \mathrm{dp}\, \frac{p_i^4}{\omega_I^2(p)}\left\{f_i^0(1-f_i^0) + \bar{f}_i^0(1-\bar{f}_i^0)\right\}}\nabla_{\boldsymbol{r}}\, T(\boldsymbol{r}), \tag{3.10}$$

For a hypothetical medium consisting of a single quark flavor, the degeneracy factor and the relaxation time cancel out from the numerator and denominator. Thus, the induced electric field can be recast in the form

$$\boldsymbol{E} = \frac{1}{2Tq}\frac{I_2}{I_1}\nabla_{\boldsymbol{r}} T(\boldsymbol{r}), \tag{3.11}$$

where the integrals $I_1$ and $I_2$ are defined by

$$I_1 \equiv \int_0^\infty \mathrm{dp}\, \frac{p^4}{\omega^2(p)}\left\{f^0(1-f^0) + \bar{f}^0(1-\bar{f}^0)\right\} \tag{3.12}$$

$$I_2 \equiv \int_0^\infty \mathrm{dp}\, \frac{p^4}{\omega^2(p)}\left\{f^0(1-f^0)(\omega(p)-\mu) - \bar{f}^0(1-\bar{f}^0)(\omega(p)+\mu)\right\}, \tag{3.13}$$

where the chemical potential ($\mu_i$) for all flavours are taken the same, *i.e.* $\mu_i = \mu$.

Therefore, the coefficient of the temperature-gradient in Eq.(3.11) gives the Seebeck coefficient ($S$) for a single species

$$S = \frac{1}{2Tq}\frac{I_2}{I_1}. \tag{3.14}$$

We compute the coefficient, $S$ as a function of temperature at fixed chemical potentials, $\mu_q$ =30 MeV, 40 MeV and 50 MeV to observe the Seebeck effect in a hot and dense medium. As per recent studies, the transition temperature for the transition from hadron phase to QGP phase for $2+1$ flavours is $154 \pm 9$ MeV [15]. The temperature range considered here, is $T$ = 165 MeV- 450 MeV.

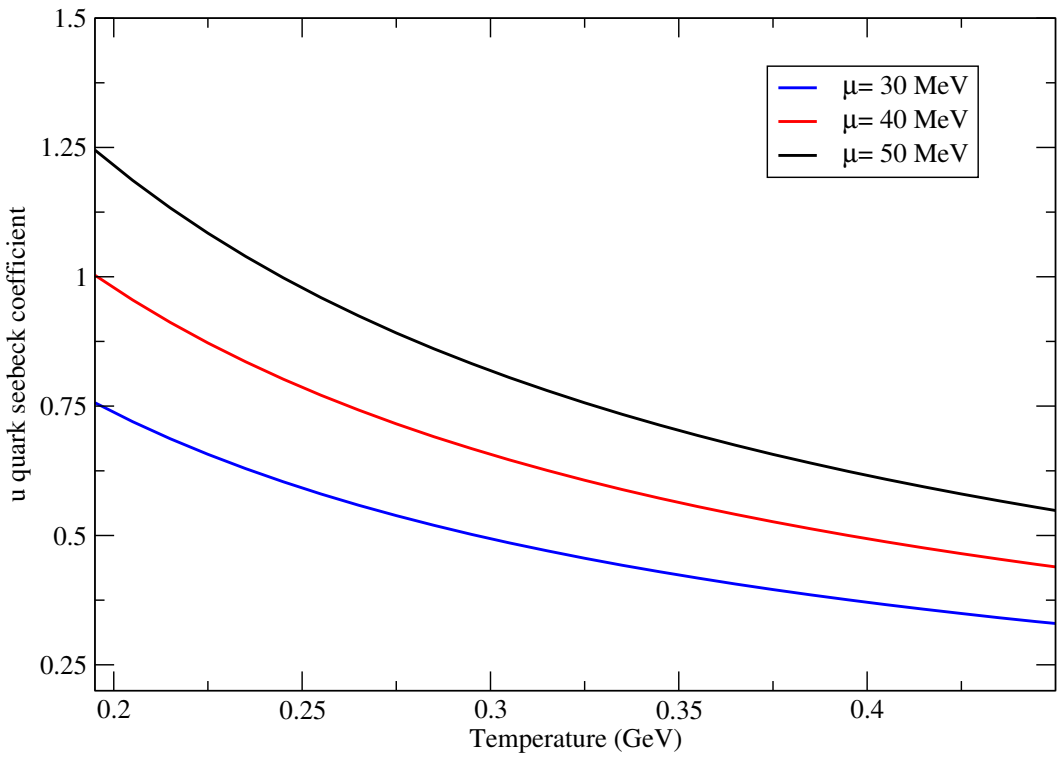

**(a)**

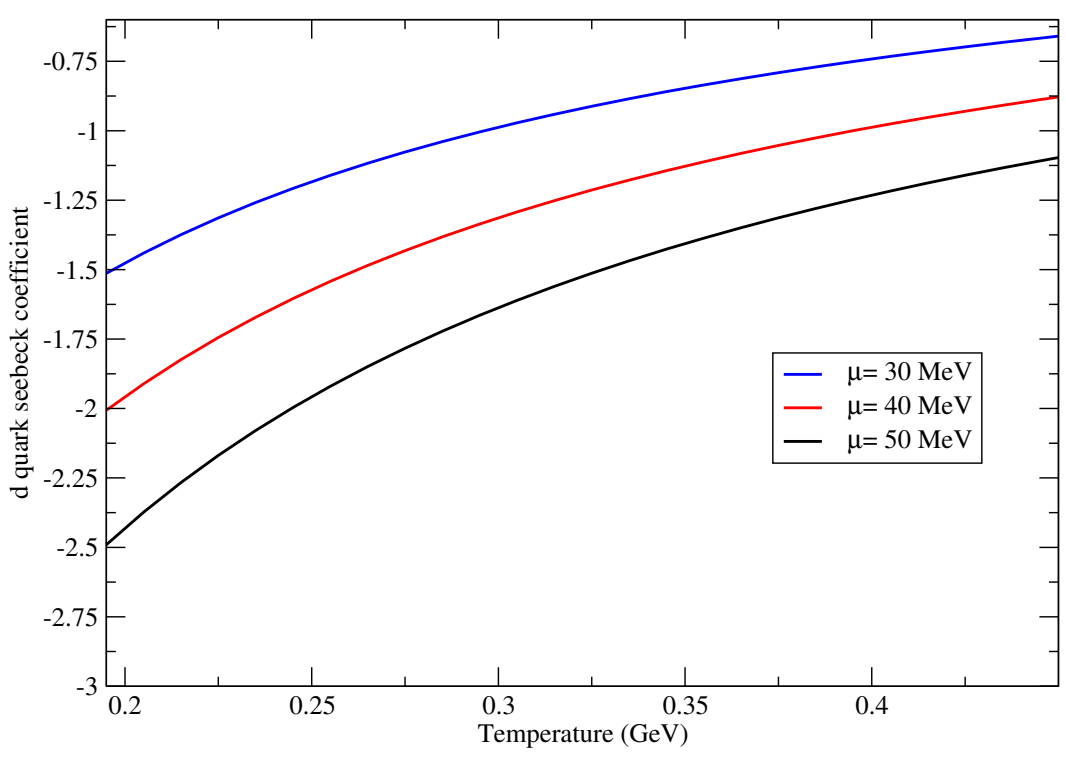

**(b)**

**Figure 3.1:** Variation of Seebeck coefficient of $u$ (a) and $d$ (b) quarks with temperature for different fixed values of chemical potentials.

We observe that the Seebeck coefficients (magnitudes) for $u$ and $d$ quark (Figs. 3.1a and 3.1b) decrease with the temperature for a fixed chemical potential, which is due to the fact that the net number density, $(n - \bar{n})$ (which is proportional to the net charge) decreases with the temperature for a fixed $\mu$. However, the coefficient is found to increase with chemical potential at a given temperature. This is because a larger $\mu$ is indicative of a larger surplus of particles over anti-particles, which, in the case of $u$ quark implies a larger abundance of positive over negative charges, leading to a larger thermoelectric current and hence a larger $S$. However, for the $d$ quark, a larger $\mu_q$ would mean a larger abundance of negative charges (particles) over positive charges (anti-particles), leading to a more negative value of $S$. Here, the sign of the Seebeck coefficient is solely determined by the sign of the electric charge the particle carries, because the other factors in the coefficient- the integrals $I_1$ and $I_2$, for both the quarks, are positive as can be seen from Fig(3.2) and Fig(3.3). The current quark masses of $u$ and $d$ quarks being very close to each other leads to almost identical values of the $I_1$ and $I_2$ integrals for both the quarks. As such, the magnitude of the electric charge of $u$ quark being twice that of $d$ quark is directly reflected in the magnitude of Seebeck coefficient of $d$ quark being almost twice that of the $u$ quark.

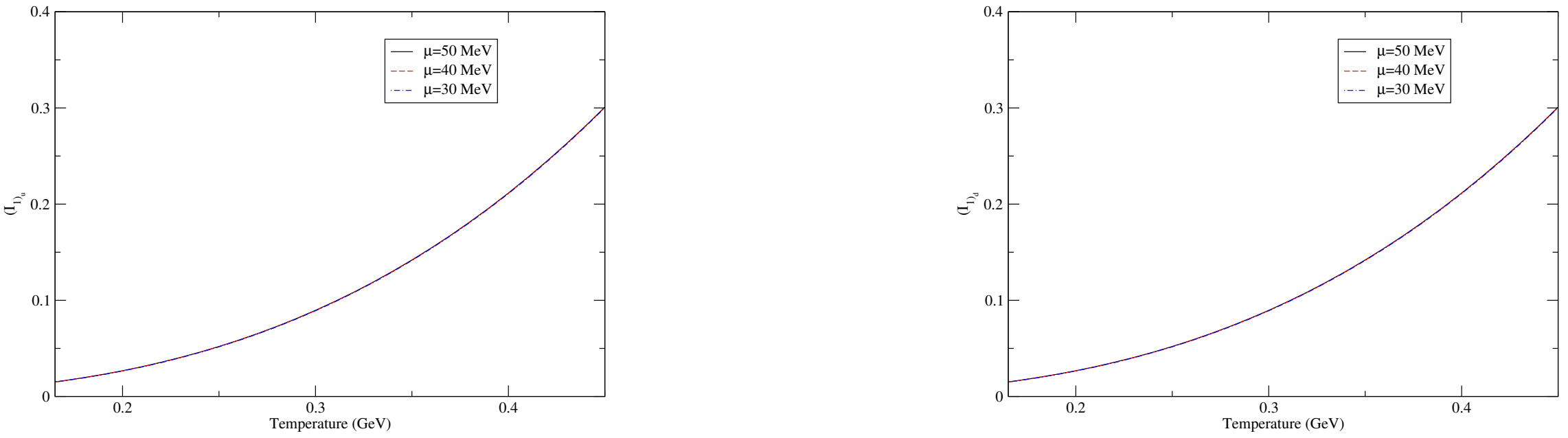

**(a)** **(b)**

**Figure 3.2:** Variation of $I_1$ integrals for $u$ (left) and $d$ (right) quarks with temperature for different fixed values of chemical potential.

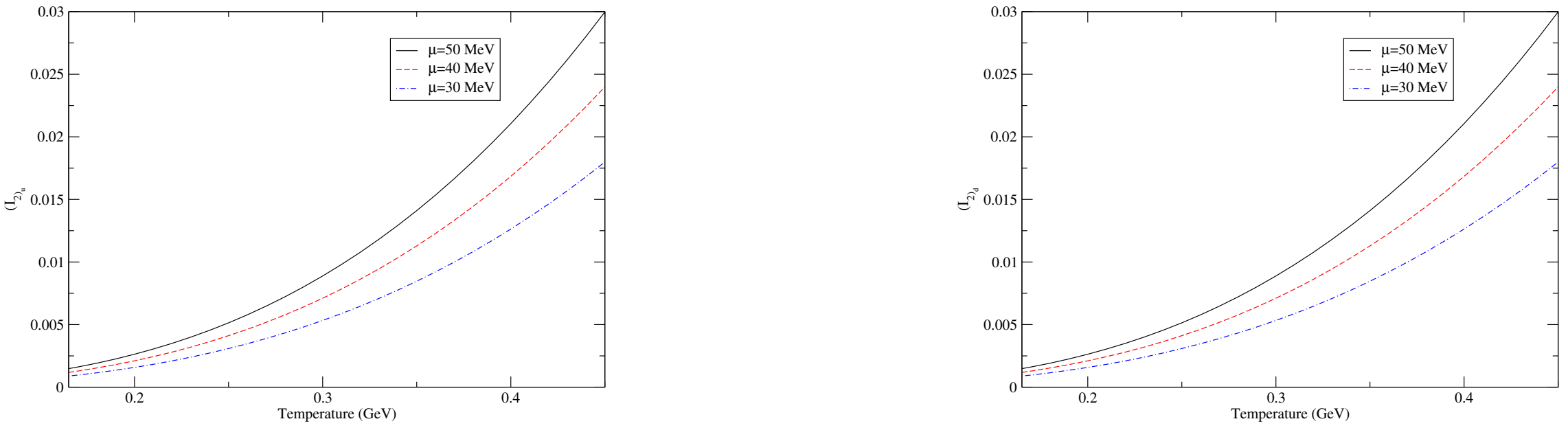

**(a)** **(b)**

**Figure 3.3:** Variation of $I_2$ integrals for $u$ (left) and $d$ (right) quarks with temperature for different fixed values of chemical potential.

After having calculated the Seebeck coefficient for a thermal medium consisting of a single species, we move on to the more realistic case of a multi-component system, which in our case corresponds to multiple flavours of quarks in the QGP. However, gluons being electrically neutral, do not contribute to the thermoelectric current, therefore, the total electric current in the medium is the vector sum of currents due to individual species:

$$\boldsymbol{J} = \boldsymbol{J_{(1)}} + \boldsymbol{J_{(2)}} + \boldsymbol{J_{(3)}} + \cdots \tag{3.15}$$

$$= \left( \frac{q_1^2 g_1 \tau_1}{T\pi^2} (I_1)_1 + \frac{q_2^2 g_2 \tau_1}{T\pi^2} (I_1)_2 + ... \right) \boldsymbol{E} - \left( \frac{q_1 g_1 \tau_1}{2T^2\pi^2} (I_2)_1 + \frac{q_2 g_2 \tau_2}{2T^2\pi^2} (I_2)_2 + ... \right) \nabla_{\boldsymbol{r}} T(\boldsymbol{r}). \tag{3.16}$$

Setting the total current, $\boldsymbol{J} = 0$ as earlier, we get the induced electric field,

$$\boldsymbol{E} = \frac{\sum_i \frac{q_i g_i \tau_i (I_2)_i}{2T}}{\sum_i q_i^2 g_i \tau_i (I_1)_i} \nabla_{\boldsymbol{r}} T(\boldsymbol{r}), \tag{3.17}$$

which yields the Seebeck coefficient for the multi-component medium:

$$S = \frac{1}{2T} \frac{\sum_i q_i g_i \tau_i (I_2)_i}{\sum_i q_i^2 g_i \tau_i (I_1)_i}. \tag{3.18}$$

All quarks have the same degeneracy factor and their relaxation times (seen from Eq.(3.3)) are also identical for each flavour. Hence, the total Seebeck coefficient for the multi-component systems can be rewritten as

$$S = \frac{\sum_i S_i q_i^2 (I_1)_i}{\sum_i q_i^2 (I_1)_i}, \tag{3.19}$$

which could be viewed as a weighted average of the Seebeck coefficients of individual species ($S_i$) present in the medium. In our calculation, we have considered only two flavours of quarks, viz: $u$, and $d$ , thus, the explicit expression comes out to be:

$$S = \frac{4S_u (I_1)_u + S_d (I_1)_d}{4(I_1)_u + (I_1)_d}, \tag{3.20}$$

where $S_u$, $S_d$ denote individual Seebeck coefficients for the $u$ and $d$ quarks respectively. Likewise, the $I_1$ integrals for different flavours are denoted by the respective flavour indices.

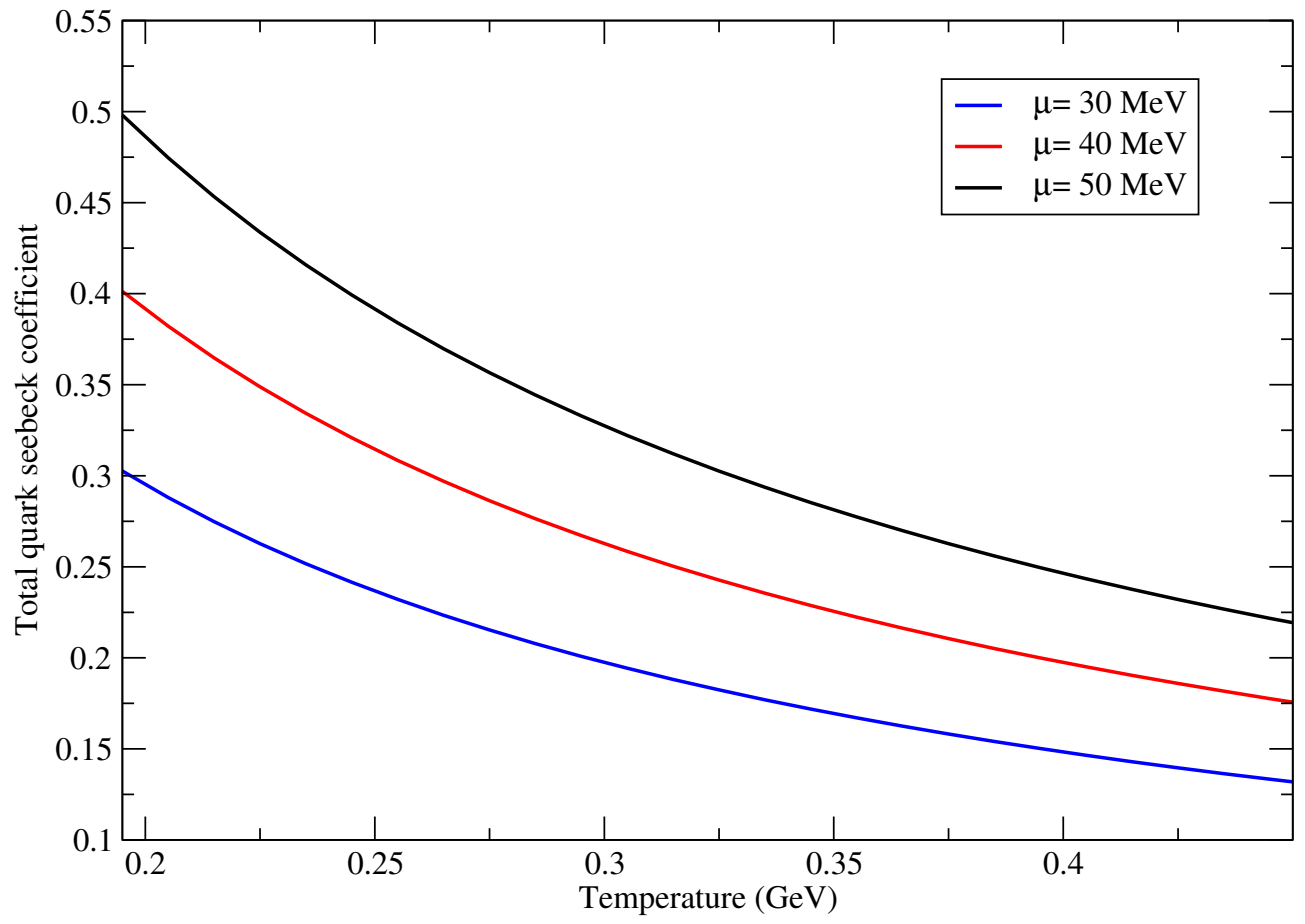

**Figure 3.4:** Variation of total Seebeck coefficient with temperature for different fixed values of chemical potentials.

As can be seen from Fig(3.4), the Seebeck coefficient of the medium is positive and decreases with the temperature. Like earlier for single species, it increases with the chemical potential. Although $S_d$ is negative, the relative magnitudes of $S_u$ and $S_d$ are such that eq.(3.20) renders the Seebeck coefficient of the medium positive.

The magnitude of the Seebeck coefficient is the magnitude of electric field produced in the medium for a unit temperature-gradient. Qualitatively, it is a measure of how efficiently a medium can convert a temperature-gradient into electricity. The sign of the Seebeck coefficient expresses the direction of the induced field with respect to the direction of temperature gradient, which is conventionally taken to point towards the direction of increasing temperature. A positive value of the Seebeck coefficient means that the induced field is in the direction of the temperature-gradient. In the convention mentioned above, this will happen when the majority charge carriers are positively charged. As expected, individual Seebeck coefficient is positive for a positively charged species ($u$ quark) and negative for a negatively charged species ($d$ quark).

### 3.1.2 Seebeck coefficient in the presence of a strong magnetic field

In the presence of magnetic field, we decompose the quark momentum into components longitudinal ($p_L$) and transverse ($p_T$) to the direction of the magnetic field. Quantum mechanically the energy levels of the $i^{\rm th}$ quark flavour get discretized into Landau levels, so the dispersion relation becomes

$$\omega_{(i,n)}(p_L) = \sqrt{p_L^2 + m_i^2 + 2n|q_i B|}, \tag{3.21}$$

where $n = 0, 1, 2, \cdots$ are quantum numbers specifying the Landau levels. It is well known that in the strong magnetic field (SMF) limit (characterised by $|q_f B| \gg T^2$, where $B$ is the magnetic field and $q_f$ is the electric charge of the $f^{\rm th}$ flavour), quarks are rarely excited thermally to higher Landau levels owing to the large energy gap between the levels, which is of the order of $\sqrt{|eB|}$ [116]. Therefore, they are constrained to be populated exclusively in the lowest Landau level (n=0), implying that the quark momentum in the presence of a strong magnetic field is purely longitudinal [123–125]. Taking the magnetic field to be in the $z$-direction, we identify $p_L$ with $p_z$, so the above dispersion relation is simplified into a relation for a one-dimensional free particle :

$$\omega_i(p_z) = \sqrt{p_z^2 + m_i^2}. \tag{3.22}$$

Thus, the equilibrium quark distribution function in SMF limit becomes:

$$f^0_{i,B} = \frac{1}{e^{\beta(\omega_i - \mu_i)} + 1}. \tag{3.23}$$

Owing to the quark momentum being purely longitudinal in the presence of a strong magnetic field, the electromagnetic current generated in response to the electric field ($J_z$) is also purely longitudinal.

$$J_z = \sum_i q_i g_i \int \frac{d^3\,\mathrm{p}}{(2\pi)^3\,\omega_i} p_z \left[\delta f_i^q(\tilde{x},\tilde{p}) - \delta f_i^{\bar{q}}(\tilde{x},\tilde{p})\right], \tag{3.24}$$

where, $\tilde{x} = (x^0, 0, 0, x^3 \equiv z)$ and $\tilde{p} = (p^0, 0, 0, p_z)$. In addition, as an artifact of strong magnetic field, the density of states in two spatial directions perpendicular to the direction of magnetic field becomes $|q_i B|$ [126, 127], *i.e.*

$$\int \frac{d^3p}{(2\pi)^3} \rightarrow \frac{q_i B}{2\pi} \int \frac{dp_z}{2\pi}. \tag{3.25}$$

The infinitesimal change in the distribution function in the strong magnetic field is thus obtained from the RBTE in the relaxation-time approximation

$$p^0 \frac{\partial f_{i,B}}{\partial t} + p_z \frac{\partial f_{i,B}}{\partial x^3} - q_i F^{03} p^z \frac{\partial f_{i,B}}{\partial p^0} + q_i F^{30} p_0 \frac{\partial f_{i,B}}{\partial p_3} = -\frac{p^0}{\tau^B} \delta f_{i,B}, \tag{3.26}$$

where $\tau_i$ denotes the relaxation-time for quarks in the presence of strong magnetic field, which, in the Lowest Landau Level (LLL) approximation is given by [128]:

$$\tau_i(T,B) = \frac{w_i\left(e^{\beta\omega_i} - 1\right)}{\alpha_s\left(\Lambda^2, eB\right) C_2 m_i^2 \left(e^{\beta\omega_i} + 1\right)} \left[\frac{1}{\int dp'^3 \frac{1}{w_i'\left(e^{\beta\omega'_i} + 1\right)}}\right], \tag{3.27}$$

which has been evaluated for massless quarks. However, it has been shown in Ref. [129] that the effect of finite quark mass in the evaluation of scattering cross sections is very small, and hence, the relaxation-time is largely unaffected. $C_2 = 4/3$ is the Casimir factor. We use a one loop running coupling constant $\alpha_s(\Lambda^2, eB)$, which runs with both the magnetic field and temperature. In the strong magnetic field (SMF) regime, its form is given by: [130]

$$\alpha_s(\Lambda^2, |eB|) = \frac{\alpha_s(\Lambda^2)}{1 + b_1\alpha_s(\Lambda^2)\ln\left(\frac{\Lambda^2}{\Lambda^2+|eB|}\right)}. \tag{3.28}$$

where, $\alpha_s(\Lambda^2)$ is the one- loop running coupling in the absence of a magnetic field.

$$\alpha_s(\Lambda^2) = \frac{1}{b_1 \ln\left(\frac{\Lambda^2}{\Lambda^2_{QCD}}\right)},$$

where $b_1 = (11N_c - 2N_f)/12\pi$ and $\Lambda_{QCD} = 0.176$ GeV. The renormalisation scale is chosen to be $\Lambda = 2\pi\sqrt{T^2 + \frac{\mu^2}{\pi^2 T^2}}$. Thus, via the strong coupling, the relaxation time acquires an implicit dependence on the chemical potential.

Now, the infinitesimal change for quark and anti-quark distribution functions can be obtained in SMF regime from the RBTE [Eq.(3.26)], again by replacing $f_{i,B}$ with $f^0_{i,B}$ in the L.H.S. This yields

$$\delta f_{i,B} = -\frac{\tau_i^B}{p^0}\frac{p_z f^0_{i,B}(1-f^0_{i,B})}{T}\left[\frac{\omega_i - \mu_i}{T}(\boldsymbol{\nabla} T)_z - 2q_iE_z\right] \tag{3.29}$$

$$\delta \bar{f}_{i,B} = -\frac{\tau_i^B}{p^0}\frac{p_z \bar{f}^0{}_{i,B}(1-\bar{f}^0{}_{i,B})}{T}\left[\frac{\omega_i + \mu_i}{T}(\boldsymbol{\nabla} T)_z + 2q_iE_z\right], \tag{3.30}$$

which gives the induced current density from (3.24) for a single species,

$$\begin{aligned} J_z = (\boldsymbol{\nabla} T)_z &\left[\frac{qg|qB|}{T(2\pi)^2}\int \frac{dp_z}{\omega^2}p_z^2(\tau^B)\left\{\bar{f}^0(1-\bar{f}^0)\frac{\omega+\mu}{T} - f^0(1-f^0)\frac{\omega-\mu}{T}\right\}\right] \\ &- 2qE_z\left[\frac{qg|qB|}{T(2\pi)^2}\int \frac{dp_z}{\omega^2}p_z^2(\tau^B)\left\{\bar{f}^0(1-\bar{f}^0) + f^0(1-f^0)\right\}\right], \end{aligned} \tag{3.31}$$

where, $g$ is the degeneracy factor. Defining the following integrals

$$H_1 = \int \frac{dp_z}{w_i^2}\,\tau_B^i\, p_z^2\left\{-\bar{f}^0(1-\bar{f}^0)(\omega+\mu) + f^0(1-f^0)(\omega-\mu)\right\}, \tag{3.32}$$

$$H_2 = \int \frac{dp_z}{w_i^2}\,\tau_B^i\;, p_z^2\left\{\bar{f}^0(1-\bar{f}^0) + f^0(1-f^0)\right\}, \tag{3.33}$$

the current density (3rd-component) from eq.(3.31) can be recast in the form

$$J_z = (\boldsymbol{\nabla} T)_z \frac{qg|qB|}{T^2(2\pi)^2} H_1 - E_z \frac{qg|qB|}{T(2\pi)^2} 2qH_2 \tag{3.34}$$

As earlier, the induced electric field due to the temperature-gradient is obtained by setting $J_z = 0$,

$$E_z = \frac{1}{2Tq}\frac{H_1}{H_2}(\boldsymbol{\nabla} T)_z. \tag{3.35}$$

The proportionally constant gives the Seebeck coefficient:

$$S = \frac{1}{2Tq}\frac{H_1}{H_2} \tag{3.36}$$

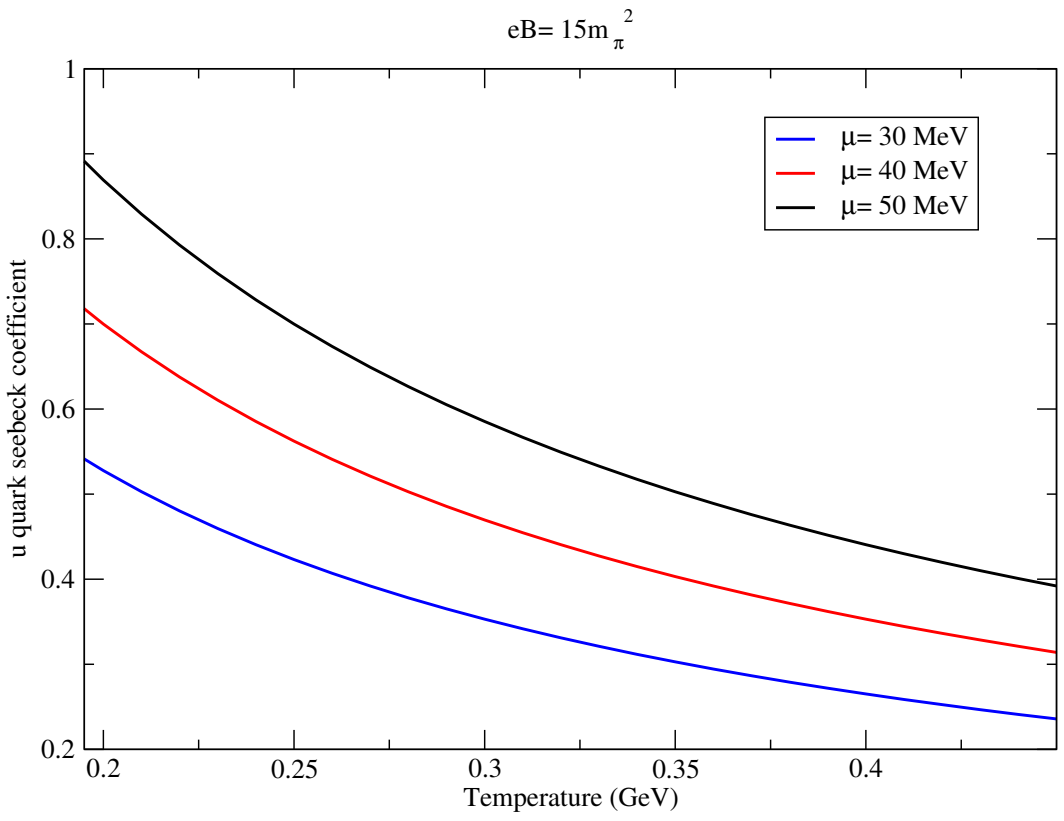

(a)

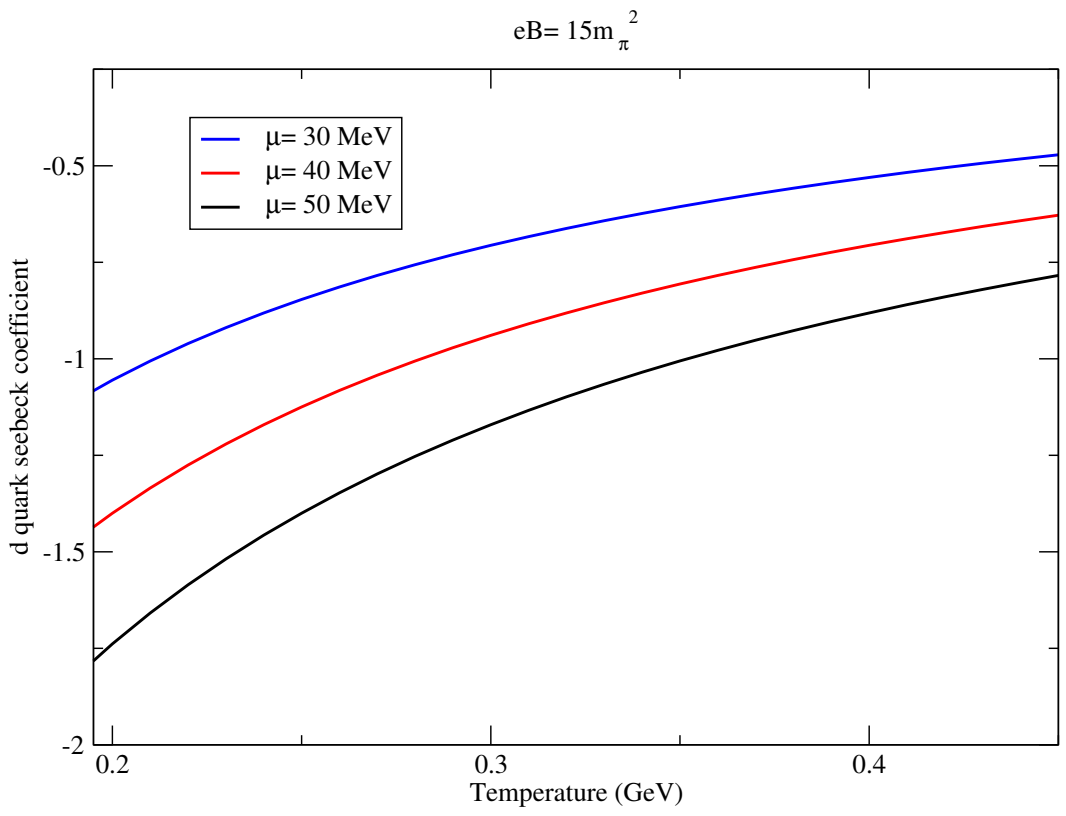

(b)

**Figure 3.5:** Variation of Seebeck coefficient of $u$ (left) and $d$ (right) quarks with temperature for a fixed chemical potential and magnetic field.

As can be seen from Fig.(3.5), the variation of the individual Seebeck coefficients (magnitudes) of the $u$ and $d$ quarks with temperature and chemical potential shows the same trend as in the earlier case. The Seebeck coefficients for the $u$ and $d$ quarks in this case turn out to be smaller ($\sim 25\%$), compared to their $B = 0$ counterparts. The $H_1$ and $H_2$ integrals are shown as a function of $T$ in Fig.(3.6) and Fig.(3.7). It should be noted that contrary to the case of pure thermal medium, the relaxation time here is momentum dependent. As such, it cannot be taken out of the momentum integrations $(H_1, H_2)$ and hence does not cancel out in $S$.

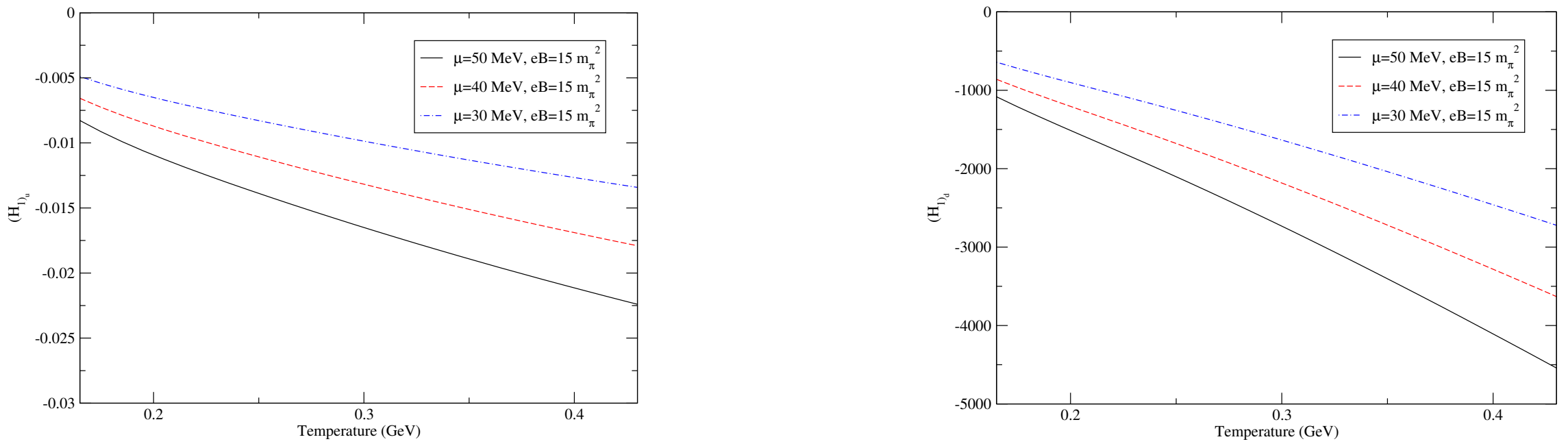

**(a)** **(b)**

**Figure 3.6:** Variation of $H_1$ integrals for $u$ (left) and $d$ (rght) quarks with temperature

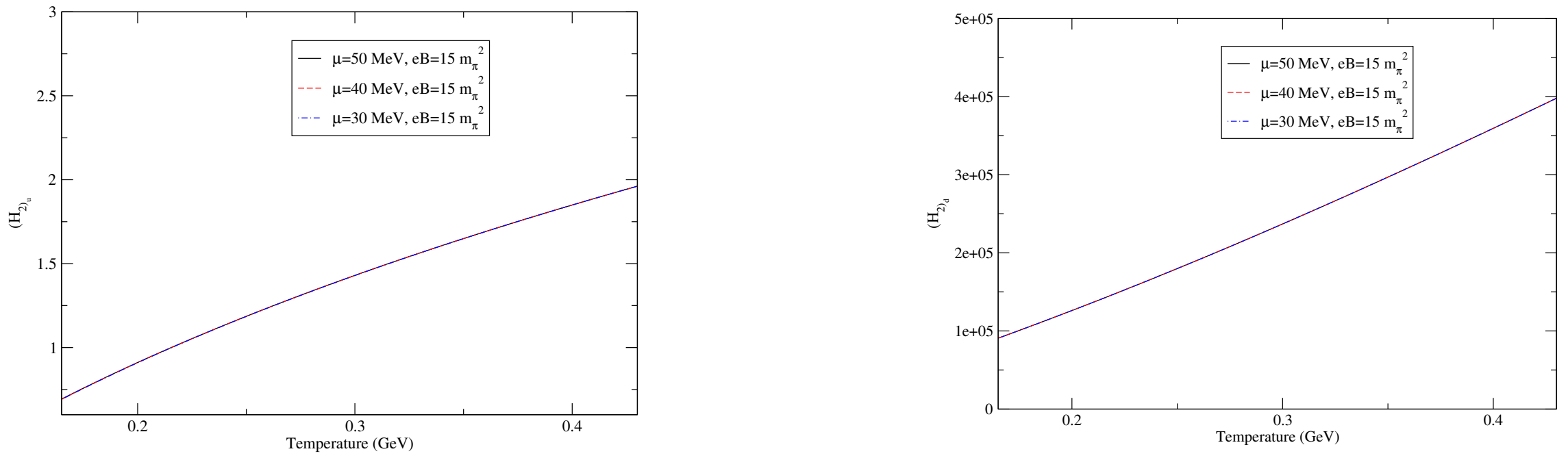

**(a)** **(b)**

**Figure 3.7:** Variation of $H_2$ integrals for $u$ (left) and $d$ (right) quarks with temperature.

Now we generalize our formalism to a medium consisting of multiple species, therefore, the total current is given by the sum of currents due to individual species:

$$\begin{aligned} J_z &= J_z^1 + J_z^2 + J_z^3 + \cdots \\ &= (\boldsymbol{\nabla} T)_z \left\{ \frac{q_1 g_1 |q_1 B|}{T^2 (2\pi)^2} (H_1)_1 + \frac{q_2 g_2 |q_2 B|}{T^2 (2\pi)^2} (H_1)_2 + \frac{q_3 g_3 |q_3 B|}{T^2 (2\pi)^2} (H_1)_3 + \cdots \right\} \\ &- E_z \left\{ \frac{q_1 g_1 |q_1 B|}{T (2\pi)^2} 2q_1 (H_2)_1 + \frac{q_2 g_2 |q_2 B|}{T (2\pi)^2} 2q_2 (H_2)_2 + \frac{q_3 g_3 |q_3 B|}{T (2\pi)^2} 2q_3 (H_2)_3 + \cdots \right\} \end{aligned} \tag{3.37}$$

Again, the Seebeck coefficient of the medium in a strong magnetic field is obtained by setting $J_z = 0$,

$$S = \frac{1}{2T} \frac{\sum_i q_i |q_i B| (H_1)_i}{\sum_i q_i^2 |q_i B| (H_2)_i}, \tag{3.38}$$

which could be further expressed in terms of the weighted average of individual Seebeck coefficients.

$$S = \frac{\sum_i S_i |q_i|^3 (H_2)_i}{\sum_i |q_i|^3 (H_2)_i}. \tag{3.39}$$

Thus, unlike the Seebeck coefficient in the absence of magnetic field [Eq.(3.19)],

both the individual as well as total Seebeck coefficient of the medium depend on the relaxation-time in the presence of a strong magnetic field.

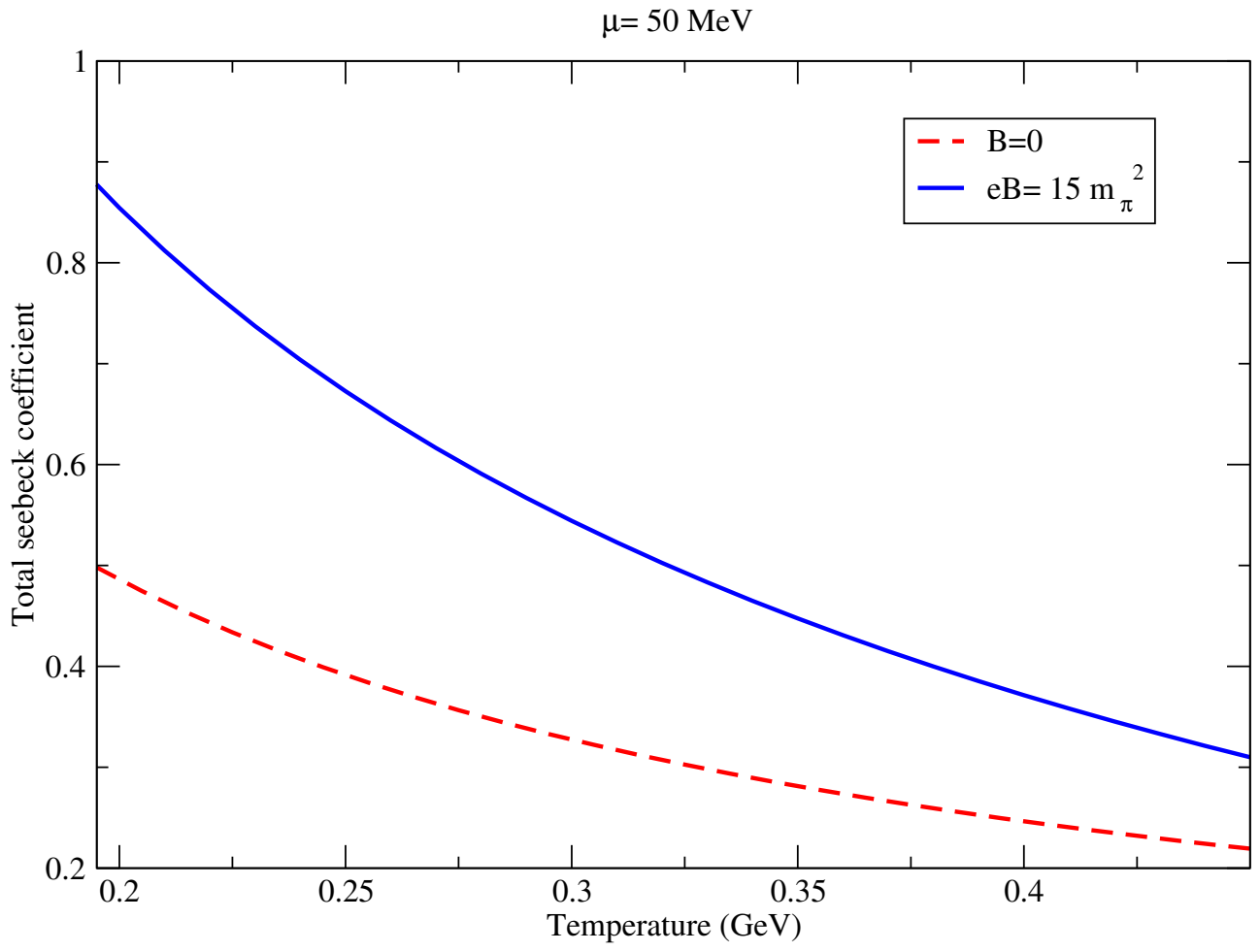

**Figure 3.8:** Variation of Seebeck coefficient of the medium with temperature for a fixed chemical potential in the absence and presence of a magnetic field.

We can now visualize the sole effect of strong magnetic field on the (total) Seebeck coefficient from the comparison of $B \neq 0$ and $B = 0$ results in Fig.(3.8). Like the $B = 0$ case, the total Seebeck coefficient in strong $B$ is positive, and is a decreasing function of temperature. However, the magnitude of $S$ is significantly enhanced (by $\sim$2 times), compared to the $B = 0$ case.

## 3.2 Seebeck effect of hot partonic medium in a quasiparticle model

Quasiparticle description of quarks and gluons in a thermal QCD medium in general, introduces a thermal mass, apart from their current masses in QCD Lagrangian. These masses are generated due to the interaction of a given parton with other partons in the medium, therefore, quasiparticle description in turn describes the collective properties of the medium. However, in the presence of strong magnetic field in the thermal QCD medium, different flavors acquire masses differently due to their different electric charges. Different versions of quasiparticle description exist in the literature based on different effective theories, as has been elaborated in Chapter(2) Our description relies on perturbative thermal QCD, where the medium generated masses for quarks and gluons are obtained from the poles of dressed propagators calculated by the respective self-energies at finite temperature and/or strong magnetic field.

### 3.2.1 Seebeck coefficient in the absence of magnetic field

In the quasiparticle description of quarks and gluons in a thermal medium with 3 flavours, all flavours (with current/vacuum masses, $m_i << T$) acquire the same thermal mass [131, 132]

$$m_T^2 = \frac{g^2(T)T^2}{6}, \tag{3.40}$$

which is, however, modified in the presence of a finite chemical potential [109]

$$m_{T,\mu}^2 = \frac{g^2(T)T^2}{6}\left(1 + \frac{\mu^2}{\pi^2 T^2}\right), \tag{3.41}$$

where $g$ is the running coupling constant already mentioned in Eq.(3.28).

We take the quasiparticle mass (squared) of $i^{\rm th}$ flavor in a pure thermal medium to be [108]:

$$m_{iT}'^{\,2} = m_i^2 + \sqrt{2}\, m_i\, m_T + m_T^2, \tag{3.42}$$

where $m_i$ is the current quark mass of the $i^{\rm th}$ flavour. So the dispersion relation for the $i^{\rm th}$ flavour takes the form

$$\omega_i^2(p) = \boldsymbol{p_i}^{\,2} + m_i^2 + \sqrt{2}\, m_i\, m_T + m_T^2. \tag{3.43}$$

Using this expression of $\omega_i(p)$ in the quark distribution functions as well as in the integrals $I_1$ and $I_2$ [Eqs.(3.12), (3.13)], we proceed in a similar fashion and evaluate the individual Seebeck coefficients for the $u$ and $d$ quarks in quasiparticle description from Eq.(3.14).

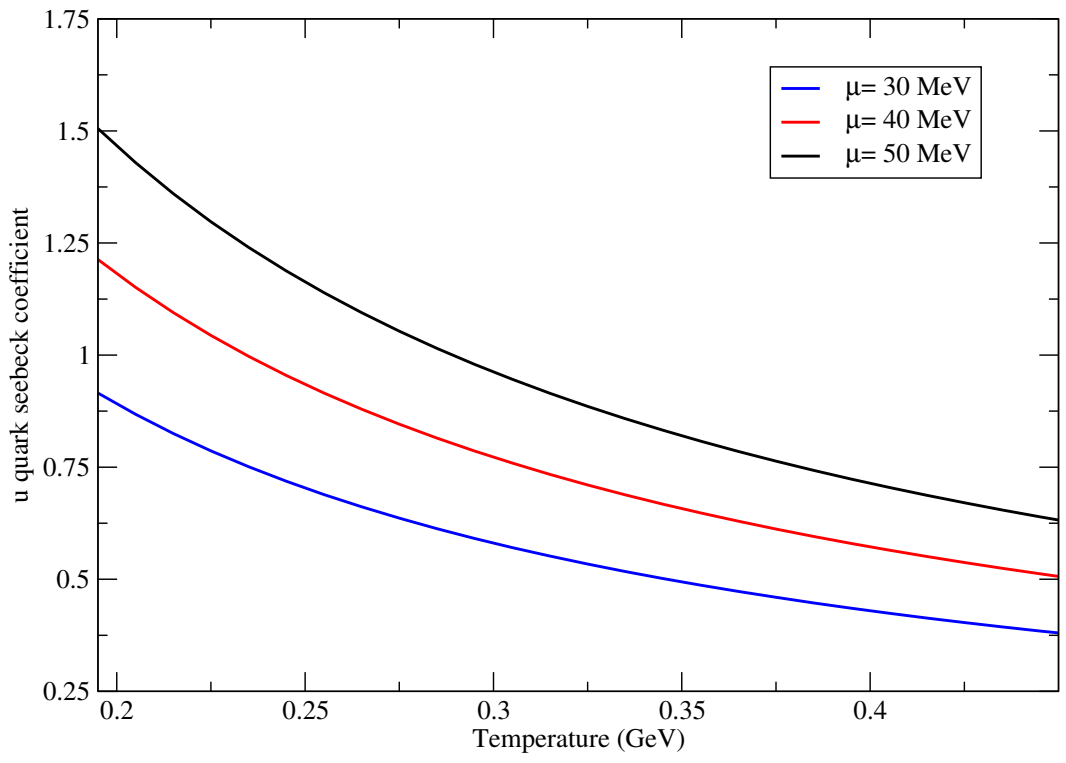

(a)

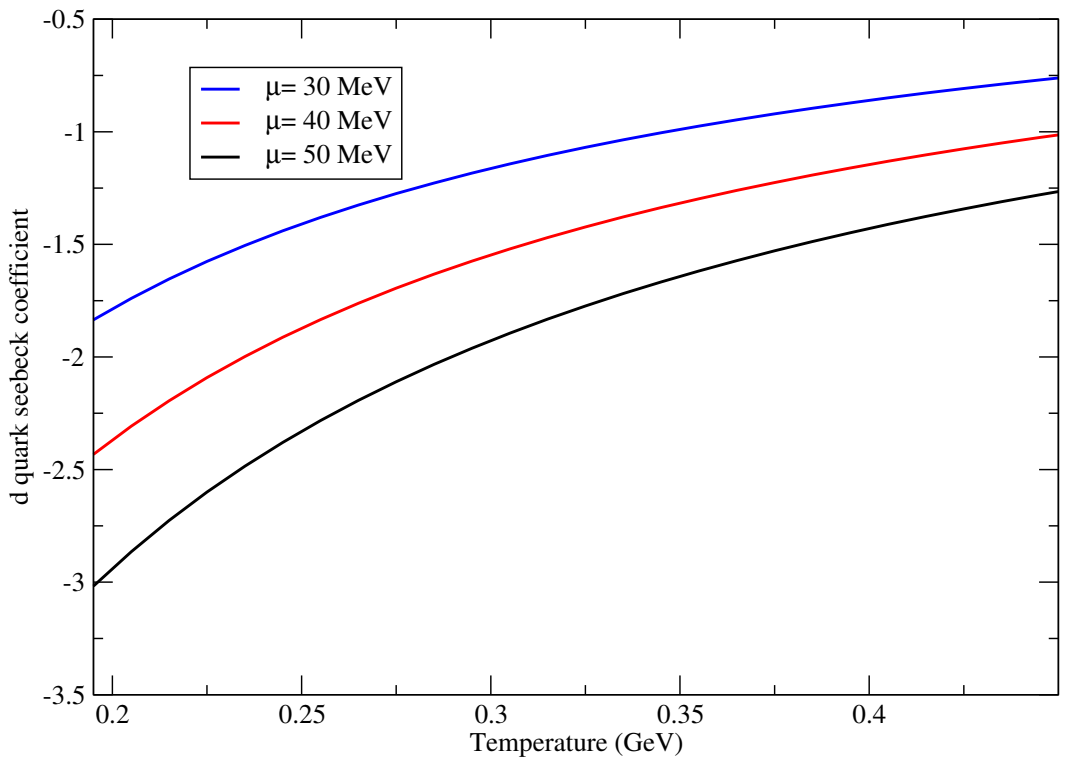

(b)

**Figure 3.9:** Variation of Seebeck coefficient of $u$ (left) and $d$ (right) quarks with temperature for different fixed values of chemical potentials.

As can be seen from Figs. (3.9a) and (3.9b), the Seebeck coefficients of $u$ and $d$ quarks show a trend similar to their current quark mass counterparts in the absence of magnetic field (Figs. 3.1a & 3.1b) and their magnitudes decrease with temperature and increase with chemical potential. The $I_1$ and $I_2$ integrals for both quarks are also found to be positive and as such, the sign of the coefficient is again determined by the electric charge of the particle. The change due to the quasiparticle description adopted here, is a slight increase in the magnitudes of the Seebeck coefficients for both quarks. Numerically the average percentage increase for the $u$ and $d$ quarks are around 17.25% and 14.81%, respectively.

Similarly, the total Seebeck coefficient of the medium in quasiparticle description

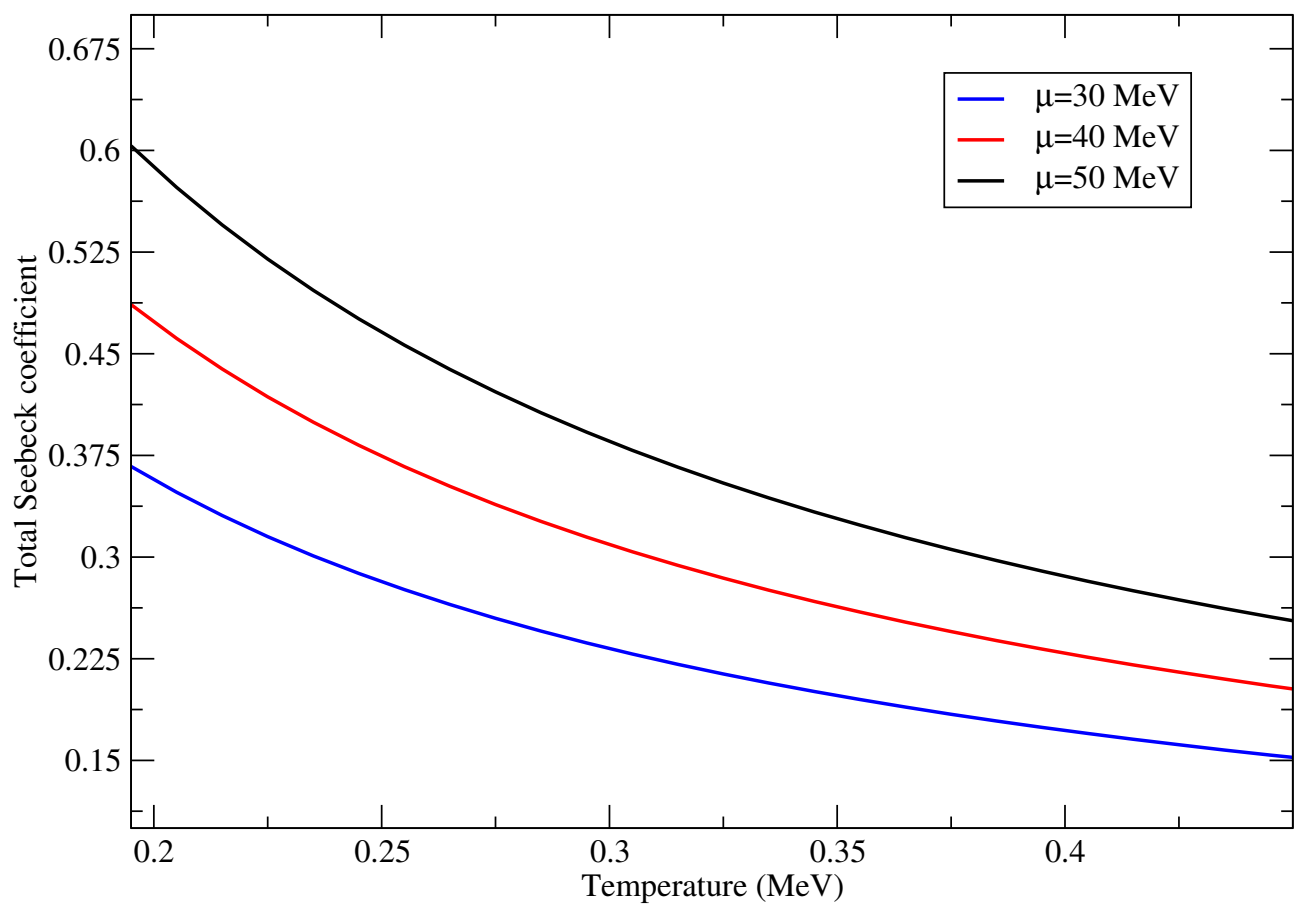

**Figure 3.10:** Variation of Seebeck coefficient of the medium with temperature for different fixed values of chemical potential.

(in Fig. 3.10) is found to have a positive value, which decreases with the temperature and increases with the chemical potential as earlier, but with a slightly elevated magnitude in comparison with the current quark mass case (in Fig 3.4).

### 3.2.2 Seebeck coefficient in the presence of strong magnetic field

In the presence of magnetic field, only quarks are affected while the gluons are not directly influenced. As a result, only the quark-loop of the gluon self-energy will be affected and the gluon-loops remains altered. Furthermore, only quarks contribute to the thermoelectric effect, and hence, we proceed to calculate the thermal quark mass in the presence of a strong magnetic field, which can be obtained from the pole ($p_0 = 0, \mathbf{p} \to 0$ limit) of the full propagator.

As we know, the full quark propagator can be obtained self-consistently from the Schwinger-Dyson equation (assuming massless flavours, which is assumed to be true at least for light flavours),

$$S^{-1}(p_\parallel) = \gamma^\mu p_{\parallel\mu} - \Sigma(p_\parallel) \ , \tag{3.44}$$

where $\Sigma(p_\parallel)$ is the quark self-energy at finite temperature in the presence of strong magnetic field. We can evaluate it up to one-loop

$$\Sigma(p) = -\frac{4}{3} g_s^2 i \int \frac{d^4k}{(2\pi)^4} \left[\gamma_\mu S(k) \gamma_\nu D^{\mu\nu}(p-k)\right], \tag{3.45}$$

where $4/3$ denotes the Casimir factor and $g_s = \sqrt{4\pi\alpha_s}$ represents the running coupling with the $\alpha_s$ already defined in Eq.(3.28). $D^{\mu\nu}(p-k)$ is the gluon propagator, which is not affected by the magnetic field, so its form is given by

$$D^{\mu\nu}(p-k) = \frac{ig^{\mu\nu}}{(p-k)^2} \ . \tag{3.46}$$

However, the quark propagator, $S(K)$ in the strong magnetic field limit, is affected and is obtained by the Schwinger proper-time method at the lowest Landau level (n=0) in momentum space,

$$S(k) = ie^{-\frac{k_\perp^2}{|q_i B|}} \frac{\left(\gamma^0 k_0 - \gamma^3 k_z + m_i\right)}{k_\parallel^2 - m_i^2} \left(1 - \gamma^0\gamma^3\gamma^5\right), \tag{3.47}$$

where the 4-vectors are defined as $k_\perp \equiv (0, k_x, k_y, 0)$, $\quad k_\parallel \equiv (k_0, 0, 0, k_z)$.

Next we obtain the form of quark and gluon propagators at finite temperature in the imaginary-time formalism and subsequently replace the energy integral ($\int \frac{dp_0}{2\pi}$) by Matsubara frequency sum. However, in a strong magnetic field along $z$-direction, the transverse component of the momentum becomes vanishingly small ($k_\perp \approx 0$), so the exponential factor in eq.(3.47) becomes unity and the integration over the transverse component of the momentum becomes $|q_f B|$. Thus, the quark self-energy in eq.(3.45) at finite temperature in the SMF limit will be of the form

$$\begin{aligned} \Sigma(p_\parallel) &= \frac{2g_s^2}{3\pi^2}|q_i B| T \sum_n \int dk_z \frac{\left[\left(1+\gamma^0\gamma^3\gamma^5\right)\left(\gamma^0 k_0 - \gamma^3 k_z\right) - 2m_i\right]}{\left[k_0^2 - \omega_k^2\right]\left[(p_0-k_0)^2 - \omega_{pk}^2\right]} \\ &= \frac{2g_s^2|q_i B|}{3\pi^2}\int dk_z \left[(\gamma^0 + \gamma^3\gamma^5)L^1 - (\gamma^3 + \gamma^0\gamma^5)k_z L^2\right], \end{aligned} \tag{3.48}$$

where $\omega_k^2 = k_z^2 + m_i^2$, $\omega_{pk}^2 = (p_z - k_z)^2$ and $L^1$ and $L^2$ are the two frequency sums, which are given by

$$L^1 = T\sum_n k_0 \frac{1}{\left[k_0^2 - \omega_k^2\right]} \frac{1}{\left[(p_0-k_0)^2 - \omega_{pk}^2\right]}, \tag{3.49}$$

$$L^2 = T\sum_n \frac{1}{\left[k_0^2 - \omega_k^2\right]} \frac{1}{\left[(p_0-k_0)^2 - \omega_{pk}^2\right]}. \tag{3.50}$$

We first do the frequency sums [133, 134] and then integrate the momentum $k_z$ to obtain the simplified form of quark self-energy eq.(3.48) [135] as

$$\Sigma(p_\parallel) = \frac{g_s^2|q_i B|}{3\pi^2}\left[\frac{\pi T}{2m_i} - \ln(2)\right]\left[\frac{\gamma^0 p_0}{p_\parallel^2} + \frac{\gamma^3 p_z}{p_\parallel^2} + \frac{\gamma^0\gamma^5 p_z}{p_\parallel^2} + \frac{\gamma^3\gamma^5 p_0}{p_\parallel^2}\right]. \tag{3.51}$$

To solve the Schwinger-Dyson equation eq.(3.44), one needs to first express the self-energy at finite temperature in magnetic field in a covariant form [136, 137],

$$\Sigma(p_\parallel) = A(\mathrm{p}_0, \mathbf{p})\gamma^\mu u_\mu + B(\mathrm{p}_0, \mathbf{p})\gamma^\mu b_\mu + C(\mathrm{p}_0, \mathbf{p})\gamma^5\gamma^\mu u_\mu + D(\mathrm{p}_0, \mathbf{p})\gamma^5\gamma^\mu b_\mu, \tag{3.52}$$

where $u^\mu$ (1,0,0,0) and $b^\mu$ (0,0,0,-1) denote the preferred directions of heat bath and magnetic field, respectively and these vectors mimic the breaking of Lorentz and rotational invariances, respectively. The form factors, $A$, $B$, $C$ and $D$ are computed in strong $B$ with LLL approximation as

$$A = \frac{g_s^2|q_iB|}{3\pi^2}\left[\frac{\pi T}{2m_i} - \ln(2)\right]\frac{p_0}{p_\parallel^2} \ , \tag{3.53}$$

$$B = \frac{g_s^2|q_iB|}{3\pi^2}\left[\frac{\pi T}{2m_i} - \ln(2)\right]\frac{p_z}{p_\parallel^2} \ , \tag{3.54}$$

$$C = -\frac{g_s^2|q_iB|}{3\pi^2}\left[\frac{\pi T}{2m_i} - \ln(2)\right]\frac{p_z}{p_\parallel^2} \ , \tag{3.55}$$

$$D = -\frac{g_s^2|q_iB|}{3\pi^2}\left[\frac{\pi T}{2m_i} - \ln(2)\right]\frac{p_0}{p_\parallel^2} \ . \tag{3.56}$$

Then the self-energy (3.52) can be expressed in terms of chiral projection operators ($P_R$ and $P_L$) as

$$\Sigma(p_\parallel) = P_R\left[(A-B)\gamma^\mu u_\mu + (B-A)\gamma^\mu b_\mu\right]P_L + P_L\left[(A+B)\gamma^\mu u_\mu + (B+A)\gamma^\mu b_\mu\right]P_R \, . \tag{3.57}$$

Hence, the Schwinger-Dyson equation is able to express the inverse of the full propagator in terms of $P_L$ and $P_R$,

$$S^{-1}(p_\parallel) = P_R\gamma^\mu X_\mu P_L + P_L\gamma^\mu Y_\mu P_R \ , \tag{3.58}$$

where

$$\gamma^\mu X_\mu = \gamma^\mu p_{\parallel\mu} - (A-B)\gamma^\mu u_\mu - (B-A)\gamma^\mu b_\mu \ , \tag{3.59}$$

$$\gamma^\mu Y_\mu = \gamma^\mu p_{\parallel\mu} - (A+B)\gamma^\mu u_\mu - (B+A)\gamma^\mu b_\mu \ . \tag{3.60}$$

Thus, the effective propagator is finally obtained by inverting eq. (3.58)

$$S(p_\parallel) = \frac{1}{2}\left[P_R\frac{\gamma^\mu Y_\mu}{Y^2/2}P_L + P_L\frac{\gamma^\mu X_\mu}{X^2/2}P_R\right], \tag{3.61}$$

where

$$\frac{X^2}{2} = X_1^2 = \frac{1}{2}\left[p_0 - (A-B)\right]^2 - \frac{1}{2}\left[p_z + (B-A)\right]^2 \ , \tag{3.62}$$

$$\frac{Y^2}{2} = Y_1^2 = \frac{1}{2}\left[p_0 - (A+B)\right]^2 - \frac{1}{2}\left[p_z + (B+A)\right]^2 \ . \tag{3.63}$$

Thus, the thermal mass (squared) for $i^{\text{th}}$ flavor at finite temperature and strong magnetic field is finally obtained by taking the $p_0 = 0, p_z \to 0$ limit in either $X_1^2$ or $Y_1^2$ (because both of them are equal),

$$m^2_{iT,B} = \frac{g_s^2|q_iB|}{3\pi^2}\left[\frac{\pi T}{2m_i} - \ln(2)\right], \tag{3.64}$$

which depends both on temperature and magnetic field. The quark distribution functions with medium generated masses in the absence and presence of magnetic field therefore manifest the interactions present in the respective medium in terms of modified occupation probabilities in the phase space and thus affect the Seebeck coefficients. The quasiparticle (or effective) mass of $i^{\text{th}}$ quark flavor is generalized in finite temperature and strong magnetic field into

$$m_i'{}^2_{T,B} = m_i^2 + \sqrt{2}\, m_i\, m_{iT,B} + m^2_{iT,B}. \tag{3.65}$$

Now, using the above quasiparticle mass in the distribution function and proceeding identically, we compute the individual Seebeck coefficients from Eq.(3.36) as a function of temperature, which is shown in Fig. 3.11.

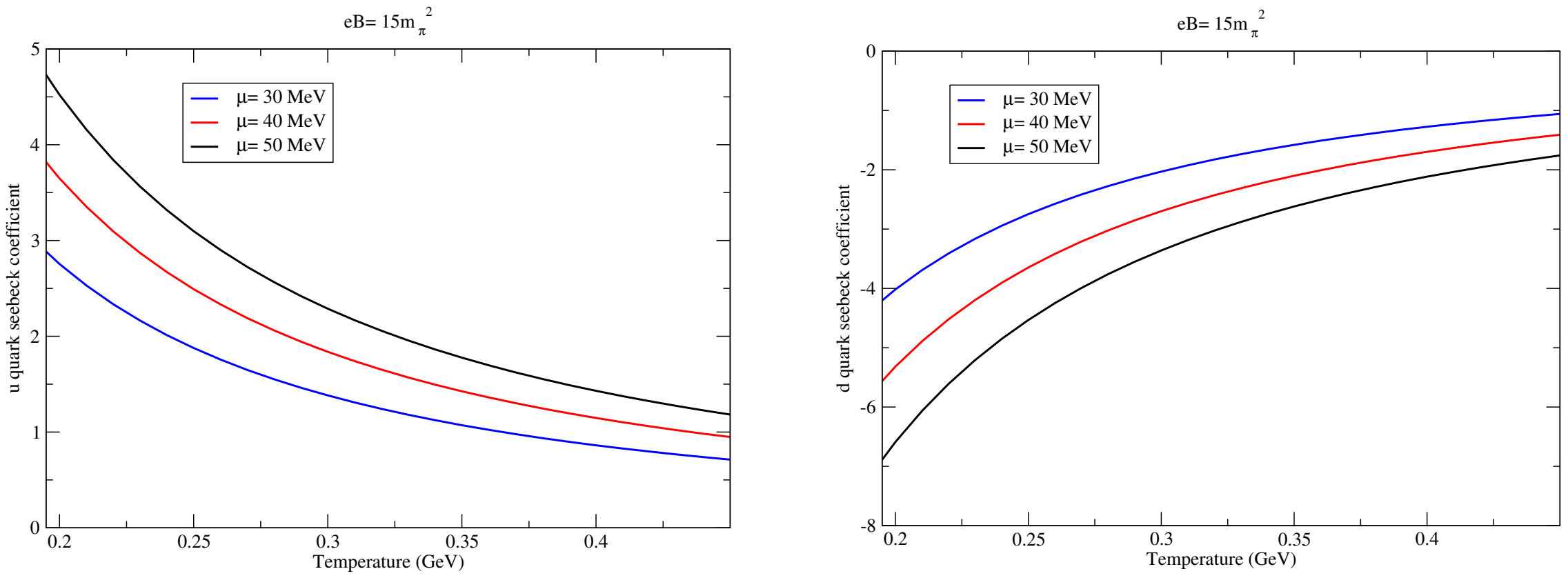

(a) (b)

**Figure 3.11:** Variation of Seebeck coefficient of $u$ (left) and $d$ (right) quarks with temperature for a fixed chemical potential and magnetic field.

We find that the magnitudes of both $u$ and $d$ quark Seebeck coefficients decrease with the temperature and increase with the chemical potential. The sign of the individual Seebeck coefficient is decided only by the electric charge of quarks, similar to the $B = 0$ case. The magnitudes of the Seebeck coefficients are found to increase by $\sim 3$ times over their current quark mass case counterparts (seen in Fig. 3.5), which could thus be attributed to the quasiparticle description.

Comparison between the $B = 0$ and $B \neq 0$ results within the quasiparticle description reveals summarily that the percentage increase is more pronounced at lower temperatures. The average percentage increase over the entire temperature range is 137.34% and 69.47% for $u$ and $d$ quarks, respectively. Hence, in the presence of strong $B$, the Seebeck effect depends strongly on the interactions among the constituents in the medium, encoded by the appropriate quasiparticle description. Once the individual Seebeck coefficients of $u$ and $d$ quarks have been evaluated in quasiparticle description, we compute the weighted average of the above individual coefficients to obtain the (total) Seebeck coefficient of the medium as a function of temperature from Eq.(3.39). This is shown in Fig. 3.12. To see the effects of magnetic field in quasiparticle description, we have also displayed the same in the absence of magnetic field in the same figure for better visual effects.

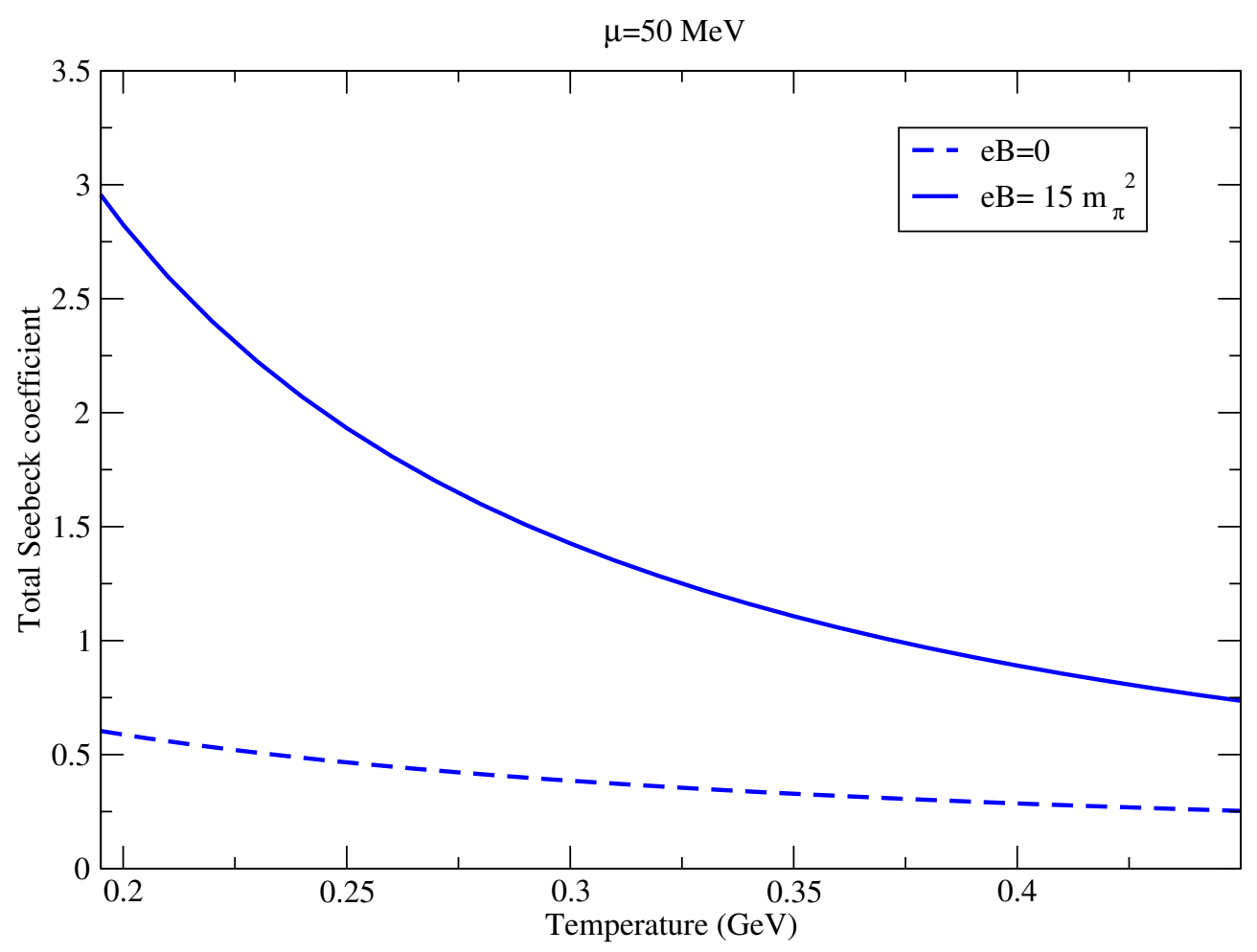

**Figure 3.12:** Variation of total Seebeck coefficient of the medium with temperature for a fixed chemical potential in the absence and presence of a magnetic field.

The seebeck coefficient starts out negative and one order of magnitude larger than its immediate counterpart in the absence of magnetic field. The magnitude decreases rapidly with increasing temperature, eventually crosses the zero mark and continues to higher positive values thereafter. Physically, the direction of the induced field is opposite to that of the temperature gradient to start with. As the temperature rises, the strength of this field gets weaker. At a particular value of the temperature, the individual seebeck coefficients have values so as to make the weighted average zero. For still higher values of temperature, the weighted average becomes positive, indicating that the induced electric field is now along the temperature gradient.

## 3.3 Summary and conclusions

In this paper, we have investigated the thermoelectric phenomenon of Seebeck effect in a hot QCD medium in two descriptions: i) when the quarks are treated in QCD with their current masses and ii) when the quarks are treated in quasiparticle model. The emergence of a strong magnetic field in the non-central events of the ultra-relativistic heavy ion collisions provides a further impetus to carry out the aforesaid investigations both in the absence and presence of a strong magnetic field, in order to isolate the effects of strong magnetic fields and interactions present among partons. For this purpose, the Seebeck coefficients are calculated individually for the $u$ and $d$ quarks, which, in turn, give the Seebeck coefficient of the medium via a weighted average of the individual coefficients. Thus, effectively, four different scenarios have been analysed:

1. Current mass description with $B = 0$.
2. Current mass description with $B \neq 0$.
3. Quasiparticle description with $B = 0$.
4. Quasiparticle description with $B \neq 0$.

Comparison between the cases 1 and 3 is able to decipher the effect of the intereactions among the partons through the quasiparticle description on the Seebeck effect in the absence of strong magnetic field, where the magnitudes of Seebeck coefficient of individual species as well as that of the medium get(s) sightly enhanced with respect to the current mass description. The sign of the individual Seebeck coefficients is positive for positively charged particles ($u$ quark) and negative for negatively charged particles ($d$ quark) for all the cases. The total Seebeck coefficient is positive for all cases. Comparison between cases 2 and 4 brings out the sole effect of the quasiparticle description in the presence of a strong magnetic field, where it is seen that the magnitude of the coefficients are amplified. Lastly, the comparison between cases 3 and 4 brings forth the sole effect of strong constant magnetic field on the Seebeck effect in the quasiparticle description. The variation of individual and total Seebeck coefficients with temperature and chemical potential are found to show similar trends in both the cases but with enhanced magnitudes in the latter case.

The trend of overall decrease (increase) of the magnitude of Seebeck coefficient with the increase in temperature (chemical potential) is seen for all cases. However, in the quasiparticle description, the magnitude of the coefficient gets enhanced in the presence of strong magnetic field, so the inclusion of interactions among partons plays a crucial role in thermoelectric phenomenon in thermal QCD.

Chapter 4

# Thermoelectric response of a weakly magnetized QGP

व्यस्तचतुर्थाद् बहुसः पृथक् स्थत त्रिपञ्चाष्टद्यायुगऋतनि
व्यस्ते चतुर्थे क्रमास्स्सत्वत्रिनां स्वाम चुर्युक्ततदास्यत परिधिस्सुच्ष्मः।।

*Divide the diameter multiplied by 4 severally and continually by the odd numbers 3, 5, 7, 9, 11, etc., and the quotients thus obtained, alternately subtract from and add to, the diameter multiplied by 4. The result is the precise circumference.*

$$\pi = 4\left(1 - 1/3 + 1/5 - 1/7 + \cdots\right)$$

**– Karanapaddhati by Puthumana Sómayaji (1465 CE-1545 CE)**

After having studied the thermoelectric response in the absence of magnetic field, as well as in the presence of a strong magnetic field, we study in this chapter, the effect of a weak constant background magnetic field on the thermoelectric response. This chapter is divided into two parts. In the first part, the quasiparticle masses are taken to be the same as in the $B = 0$ case, with magnetic field dependence appearing implicitly via the $B$ dependent coupling. In the second part, we make use of quasiparticle masses evaluated explicitly in the presence of a weak magnetic field, upto $\mathcal{O}(qB)$. This chapter is based on the papers *Seebeck effect in a thermal QCD medium in the presence of strong magnetic field* by Debarshi Dey and Binoy Krishna Patra, **Phys. Rev. D, 104, 076021 (2021)**, and *Thermoelectric response in a thermal QCD medium with chiral quasiparticle masses* by Debarshi Dey and Binoy Krishna Patra, **Int. J. Mod. Phys. E 31, 250097 (2022).**

## 4.1 Introduction

In the presence of a magnetic field, there will exist a Lorentz force on the moving charges, causing them to drift perpendicular to their original direction of motion. This transverse thermocurrent in response to a temperature gradient is called the Nernst effect. Like the Seebeck coefficient, the Nernst coefficient is also calculated at the condition of zero electric current, that is, under equilibrium conditions. The Nernst coefficient can be defined as the electric field induced in the $\hat{x}$ ($\hat{y}$) direction per unit temperature gradient in the $\hat{y}$ ($\hat{x}$) direction. In the context of heavy ion collisions, Nernst effect has been investigated in a few studies [87–89]. A comparison of the approach and results of our work with that of the other studies is also carried out in Sec. IV.B.

In this work, we have investigated the Seebeck effect and Nernst effect in a QGP medium in the presence of a weak magnetic field wherein the medium interactions are encoded in the quasiparticle masses of the partons derived from one loop perturbative thermal QCD. Since temperature is the largest scale in the case of a weak background magnetic field, the quasiparticle masses have been taken to be the thermal ($B = 0$) masses with magnetic field dependence appearing implicitly via the coupling constant. We make use of kinetic theory via the relativistic Boltzmann transport equation within the framework of relaxation time approximation for our study, using the electromagnetic Lorentz force field as the external force term in the L.H.S. of the Boltzmann equation. It may be noted that the thermalization of the matter created post heavy ion collisions is governed by QCD and as such, gluons play a dominant role in the thermalization process since the initial density of gluons is significantly larger than that of quarks or antiquarks. Magnetic field does not affect the gluons on account of their electrical neutrality and hence it is a reasonable assumption to consider the effect of magnetic field to be subdominant. Consequently, it can be argued that the distribution function of the particles in the medium never deviates significantly from equilibrium, which makes the relaxation time approximation of the Boltzmann transport equation, a suitable approach to calculate the Seebeck and Nernst coefficients (as well as other transport coefficients). Thus, in the present work, we assume that the phase space and dispersion relation of the particles are not affected by magnetic field via Landau quantization [138–140]. The magnetic field is taken to be homogeneous and time-independent. The baryon chemical potential is also considered to be homogeneous.

The chapter is organized as follows: In Sec. (4.2), we discuss the relativistic Boltzmann transport equation (RBTE) in the relaxation-time approximation and set the framework for deriving the Seebeck and Nernst coefficients of the medium considering the background magnetic field to be weak. In Sec. (4.3), we use quasiparticle masses of the $B = 0$ case, where $B$ dependence enters via the coupling constant $g(T, B)$, to evaluate the Seebeck and Nernst coefficients.In Sec. (4.4), evaluation of quasiquark masses in the presence of a weak magnetic field is discussed and subsequently used to evaluate the same coefficients. We finally conclude in Sec. (4.5).

## 4.2 The Boltzmann equation

We start with the relativistic Boltzmann transport equation which reads for the $i$-th parton:

$$p^\mu \frac{\partial f_i(x,p)}{\partial x^\mu} + q_i F^{\mu\nu} p_\nu \frac{\partial f_i(x,p)}{\partial p^\rho} = C[f_i(x,p)], \tag{4.1}$$

where, $f_i(x,p)$ and $q_i$ are the distribution function and electric charge, respectively, of the $i$th quark flavour, $F^{\mu\nu}$ is the electromagnetic field strength tensor and $C[f_i(x,p)]$ is the collision term. As earlier, we use the relaxation time approximation to write

$$C[f_i(x,p)] \simeq -\frac{p^\mu u_\mu}{\tau_i} \delta f_i. \tag{4.2}$$

where, $u^\mu$ is the heat bath velocity, $f_0$ is the equilibrium distribution function, and $\delta f$ is a small deviation such that $f = f_0 + \delta f$, with $\delta f \ll f_0$. To study the Seebeck effect, we need consider only the quarks, as gluons do not contribute to the electric current. Thus, the equilibrium one-particle distribution function for a plasma moving with a macroscopic four-velocity $u^\nu$ is the Fermi-Dirac distribution given by:

$$f_i^0(\boldsymbol{r}, \boldsymbol{p}) \equiv f_i^0 = \frac{1}{\exp(\frac{u_\nu p^\nu - \mu_i}{T}) - 1}, \tag{4.3}$$

where, $\mu$ refers to the quark chemical potential. In the local rest frame of the plasma, $u^\nu = (1, 0, 0, 0)$ and the distribution function reduces to:

$$f_i^0 = \frac{1}{\exp(\frac{\epsilon_i - \mu_i}{T}) - 1}, \tag{4.4}$$

with $\epsilon(\boldsymbol{p}) = \sqrt{\boldsymbol{p}^2 + m^2}$ and $\beta(\boldsymbol{r}) = 1/T(\boldsymbol{r})$. The collision term under RTA is simplified to Eq.(4.1) can then be written as

$$p^\mu \frac{\partial f_i(x,p)}{\partial x^\mu} + F'^\mu \frac{\partial f_i(x,p)}{\partial p^\mu} = -\frac{p^\mu u_\mu}{\tau_i}\,\delta f_i. \tag{4.5}$$

Here, $F'^\mu = (p^0 \boldsymbol{v}.\boldsymbol{F}, p^0 \boldsymbol{F})$ is a 4-vector that can be thought of as the relativistic counterpart of the classical electromagnetic force with $\boldsymbol{F} = q(\boldsymbol{E} + \boldsymbol{v} \times \boldsymbol{B})$ being the background classical electromagnetic force field. By using $F^{0i} = -E^i$ and $2F_{ij} = \epsilon_{ijk}B^k$, we can show:

$$F'^\mu = qF^{\mu\nu}p_\nu, \tag{4.6}$$

where, $\epsilon_{ijk}$ is the completely antisymmetric Levi-Civita tensor. Writing RBTE in 3-notation, we have (dropping the particle label $i$):

$$\left(p^0 \frac{\partial}{\partial x^0} + p^j \frac{\partial}{\partial x^j} + F^0 \frac{\partial}{\partial p^0} + F^j \frac{\partial}{\partial p^j}\right) f = -\frac{p^0}{\tau}(f - f_0). \tag{4.7}$$

*i.e.*

$$\frac{\partial f}{\partial t} + \boldsymbol{v}.\frac{\partial f}{\partial \boldsymbol{r}} + \frac{\boldsymbol{F}.\boldsymbol{p}}{p^0}\frac{\partial f}{\partial p^0} + \boldsymbol{F}.\frac{\partial f}{\partial \boldsymbol{p}} = -\frac{(f - f_0)}{\tau}. \tag{4.8}$$

Under steady state assumption, $\frac{\partial f}{\partial t} = 0$. Thus, we get:

$$\left(\boldsymbol{v}.\frac{\partial}{\partial \boldsymbol{r}} + \frac{\boldsymbol{F}.\boldsymbol{p}}{p^0}\frac{\partial}{\partial p^0} + \boldsymbol{F}.\frac{\partial}{\partial \boldsymbol{p}}\right) f = -\frac{(f - f_0)}{\tau}. \tag{4.9}$$

Considering $p^0$ to be an independent variable, we make use of the chain rule [114]

$$\frac{\partial}{\partial \boldsymbol{p}} \rightarrow \frac{\partial p^0}{\partial \boldsymbol{p}}\frac{\partial}{\partial p^0} + \frac{\partial}{\partial \boldsymbol{p}} = \frac{\boldsymbol{p}}{p^0}\frac{\partial}{\partial p^0} + \frac{\partial}{\partial \boldsymbol{p}}. \tag{4.10}$$

Thus, Eq.(4.9) becomes

$$\boldsymbol{v}.\frac{\partial f}{\partial \boldsymbol{r}} + \boldsymbol{F}.\frac{\partial f}{\partial \boldsymbol{p}} = -\frac{(f - f_0)}{\tau}, \tag{4.11}$$

with $\boldsymbol{F} = q(\boldsymbol{E} + \boldsymbol{v} \times \boldsymbol{B})$. Eq.(4.11) is the Boltzmann equation in 3-vector form. The relevant derivatives of the equilibrium distribution function, which will be used in the calculation are given below

$$\frac{\partial^2 f_0}{\partial p_y \partial p_x} = \frac{\beta f_0 p_x p_y}{\epsilon^2}\left(\beta + \frac{1}{\epsilon}\right), \tag{4.12}$$

$$\frac{\partial^2 f_0}{\partial p_y \partial p_z} = \frac{\beta f_0 p_y p_z}{\epsilon^2}\left(\beta + \frac{1}{\epsilon}\right), \tag{4.13}$$

$$\frac{\partial^2 f_0}{\partial p_y^2} = -\frac{\beta f_0}{\epsilon}\left(1 - \frac{p_y^2}{\epsilon^2} - \frac{\beta p_y^2}{\epsilon}\right), \tag{4.14}$$

$$\frac{\partial^2 f_0}{\partial p_x^2} = -\frac{\beta f_0}{\epsilon}\left(1 - \frac{p_x^2}{\epsilon^2} - \frac{\beta p_x^2}{\epsilon}\right), \tag{4.15}$$

where we have neglected $f_0^2$ at high $T$.

## 4.3 Part I: $B$ independent quasiparticle masses

### 4.3.1 Seebeck and Nernst coefficients

A conducting medium subjected to mutually perpendicular magnetic field and temperature gradient develops a thermocurrent perpendicular to both the magnetic field and temperature gradient. This phenomenon is called the Nernst effect. While the Seebeck coefficient determined from the 'open circuit' condition relates the electric field component in a particular direction to the temperature gradient component in the same direction, the Nernst coefficient can be thought of as a Hall type thermoelectric coefficient that relates the electric field and the temperature gradient in mutually transverse directions. Thus, evaluating the Nernst coefficient requires a 2-dimensional formulation of the problem. Here, we consider the electric field and temperature gradient to exist in the $x$-$y$ plane with the magnetic field pointing exclusively in the $z$ direction. Also, we consider a two flavour quark gluon plasma medium with $u$ and $d$ quarks (and their antiquarks). We first evaluate the Seebeck and Nernst coefficients for a QGP medium composed of a single quark species.

With $\boldsymbol{E} = E_x\,\hat{\boldsymbol{x}} + E_y\,\hat{\boldsymbol{y}}$, the Boltzmann equation [Eq.(4.11)] reads:

$$f - qB\tau\left(v_x\frac{\partial f}{\partial p_y} - v_y\frac{\partial f}{\partial p_x}\right) = f_0 - \tau\boldsymbol{v}\cdot\frac{\partial f}{\partial \boldsymbol{r}} - \tau q\boldsymbol{E}\cdot\frac{\partial f}{\partial \boldsymbol{p}}, \tag{4.16}$$

where, $f_0$ is the equilibrium quark distribution function given by Eq.(4.4) and $f$ is the total distribution function satisfying $f = f_0 + \delta f$. We use an ansatz for $\delta f$ given by:

$$f = f_0 + \delta f = f_0 - \tau q \boldsymbol{E} \cdot \frac{\partial f_0}{\partial \boldsymbol{p}} - \boldsymbol{\chi} \cdot \frac{\partial f_0}{\partial \boldsymbol{p}}. \tag{4.17}$$

This is the generalisation of the Ansatz used in [138]. Using Eq.(4.17), Eq.(4.16) becomes:

$$\boldsymbol{\chi} \cdot \frac{\partial f_0}{\partial \boldsymbol{p}} - qB\tau \left( v_y \frac{\partial f}{\partial p_x} - v_x \frac{\partial f}{\partial p_y} \right) = \tau \boldsymbol{v} \cdot \frac{\partial f_0}{\partial \boldsymbol{r}}. \tag{4.18}$$

The terms in the parenthesis, after using the ansatz and retaining only linear velocity terms, simplify to

$$v_y \frac{\partial f}{\partial p_x} - v_x \frac{\partial f}{\partial p_y} = v_y \chi_x + v_y \tau q E_x - v_x \chi_y - v_x \tau q E_y. \tag{4.19}$$

This finally leads to

$$v_x \left[ \frac{\chi_x}{\tau} - \omega_c \tau q E_y - \omega_c \chi_y + \frac{\epsilon - \mu}{T} \frac{\partial T}{\partial x} \right] + v_y \left[ \frac{\chi_y}{\tau} + \omega_c \tau q E_x + \omega_c \chi_x + \frac{\epsilon - \mu}{T} \frac{\partial T}{\partial y} \right] = 0, \tag{4.20}$$

where, $\omega_c = qB/\epsilon$ is the cyclotron frequency. Equating coefficients of $v_x$ and $v_y$, we get

$$\frac{\chi_x}{\tau} - \omega_c \tau q E_y - \omega_c \chi_y + \frac{\epsilon - \mu}{T} \frac{\partial T}{\partial x} = 0. \tag{4.21}$$

$$\frac{\chi_y}{\tau} + \omega_c \tau q E_x + \omega_c \chi_x + \frac{\epsilon - \mu}{T} \frac{\partial T}{\partial y} = 0. \tag{4.22}$$

Solving for $\chi_x$ and $\chi_y$ yields:

$$\chi_x = \frac{-\omega_c^2 \tau^3}{1 + \omega_c^2 \tau^2} q E_x - \frac{\tau}{1 + \omega_c^2 \tau^2} \left( \frac{\epsilon - \mu}{T} \right) \frac{\partial T}{\partial x} + \frac{\omega_c \tau^2}{1 + \omega_c^2 \tau^2} q E_y - \left( \frac{\epsilon - \mu}{T} \right) \frac{\omega_c \tau^2}{1 + \omega_c^2 \tau^2} \frac{\partial T}{\partial y}. \tag{4.23}$$

$$\chi_y = \frac{-\omega_c \tau^2}{1 + \omega_c^2 \tau^2} q E_x + \frac{\omega_c \tau^2}{1 + \omega_c^2 \tau^2} \left( \frac{\epsilon - \mu}{T} \right) \frac{\partial T}{\partial x} - \frac{\omega_c^2 \tau^3}{1 + \omega_c^2 \tau^2} q E_y - \frac{\tau}{1 + \omega_c^2 \tau^2} \left( \frac{\epsilon - \mu}{T} \right) \frac{\partial T}{\partial y}. \tag{4.24}$$

Substituting in Eq.(4.17), we obtain:

$$\delta f = \frac{\partial f_o}{\partial \epsilon}\left[-\tau q v_x + \frac{\omega_c^2\tau^3}{1+\omega_c^2\tau^2}q v_x + \frac{\omega_c\tau^2}{1+\omega_c^2\tau^2}q v_y\right]E_x + \frac{\partial f_o}{\partial \epsilon}\Big[-\tau q v_y + \frac{\omega_c^2\tau^3}{1+\omega_c^2\tau^2}q v_y \tag{4.25}$$

$$-\frac{\omega_c\tau^2}{1+\omega_c^2\tau^2}q v_x\Big]E_y + \frac{\partial f_o}{\partial \epsilon}\left[\frac{\tau}{1+\omega_c^2\tau^2}\left(\frac{\epsilon-\mu}{T}\right)v_x - \frac{\omega_c\tau^2}{1+\omega_c^2\tau^2}\left(\frac{\epsilon-\mu}{T}\right)v_y\right]\frac{\partial T}{\partial x} \tag{4.26}$$

$$+\frac{\partial f_o}{\partial \epsilon}\left[\frac{\tau}{1+\omega_c^2\tau^2}\left(\frac{\epsilon-\mu}{T}\right)v_y + \frac{\omega_c\tau^2}{1+\omega_c^2\tau^2}\left(\frac{\epsilon-\mu}{T}\right)v_x\right]\frac{\partial T}{\partial y}. \tag{4.27}$$

$\overline{\delta f}$ is obtained by replacing $q$ by $-q$ in Eq.(4.58). The induced 4-current as earlier is given by:

$$J^\mu = qg\int \frac{d^3\mathrm{p}}{(2\pi)^3\epsilon}p^\mu\left[\delta f - \overline{\delta f}\right]. \tag{4.28}$$

Substituting the expressions for $\delta f$ and $\overline{\delta f}$ above, we obtain:

$$J_x = \frac{qg}{6\pi^2}\left[(q\beta I_1)E_x + (q\beta I_2)E_y + (\beta^2 I_3)\frac{\partial T}{\partial x} + (\beta^2 I_4)\frac{\partial T}{\partial y}\right], \tag{4.29}$$

$$J_y = \frac{qg}{6\pi^2}\left[(q\beta I_1)E_y + (-q\beta I_2)E_x + (\beta^2 I_3)\frac{\partial T}{\partial y} + (-\beta^2 I_4)\frac{\partial T}{\partial x}\right], \tag{4.30}$$

where,

$$I_1 = \int \mathrm{dp}\, p^4\frac{\tau}{\epsilon^2(1+\omega_c^2\tau^2)}\left\{f_0(1-f_0) + \bar{f}_0(1-\bar{f}_0)\right\} \tag{4.31}$$

$$I_2 = \int \mathrm{dp}\, p^4\frac{\omega_c\tau^2}{\epsilon^2(1+\omega_c^2\tau^2)}\left\{f_0(1-f_0) - \bar{f}_0(1-\bar{f}_0)\right\} \tag{4.32}$$

$$I_3 = \int \mathrm{dp}\, p^4\frac{\tau}{\epsilon^2(1+\omega_c^2\tau^2)}\left\{(\epsilon+\mu)\bar{f}_0(1-\bar{f}_0) - (\epsilon-\mu)f_0(1-f_0)\right\} \tag{4.33}$$

$$I_4 = \int \mathrm{dp}\, p^4\frac{\omega_c\tau^2}{\epsilon^2(1+\omega_c^2\tau^2)}\left\{-(\epsilon+\mu)\bar{f}_0(1-\bar{f}_0) - (\epsilon-\mu)f_0(1-f_0)\right\} \tag{4.34}$$

In equilibrium, we have, $J_x = 0 = J_y$. This leads to

$$C_1 E_x + C_2 E_y + C_3 \frac{\partial T}{\partial x} + C_4 \frac{\partial T}{\partial y} = 0, \tag{4.35}$$

$$-C_2 E_x + C_1 E_y - C_4 \frac{\partial T}{\partial x} + C_3 \frac{\partial T}{\partial y} = 0, \tag{4.36}$$

where, $C_1 = qI_1$, $C_2 = qI_2$, $C_3 = \beta I_3$ and $C_4 = \beta I_4$.
The electric field components are related to the components of the temperature gradients via the Seebeck and Nernst coefficients via a matrix equation

$$\begin{pmatrix} E_x \\ E_y \end{pmatrix} = \begin{pmatrix} S & N|\boldsymbol{B}| \\ -N|\boldsymbol{B}| & S \end{pmatrix} \begin{pmatrix} \frac{\partial T}{\partial x} \\ \frac{\partial T}{\partial y} \end{pmatrix}. \tag{4.37}$$

Here, $S$ and $N$ refer to the Seebeck and Nernst coefficients respectively. The relative minus sign among the Nernst coefficients is necessitated by the Onsager reciprocity theorem [72]. Using Eq.(4.35) and Eq.(4.36), we finally obtain:

$$E_x = \left[ -\frac{C_1 C_3 + C_2 C_4}{C_1^2 + C_2^2} \right] \frac{\partial T}{\partial x} + \left[ \frac{C_2 C_3 - C_1 C_4}{C_1^2 + C_2^2} \right] \frac{\partial T}{\partial y} \tag{4.38}$$

$$E_y = \left[ -\frac{C_1 C_3 + C_2 C_4}{C_1^2 + C_2^2} \right] \frac{\partial T}{\partial y} - \left[ \frac{C_2 C_3 - C_1 C_4}{C_1^2 + C_2^2} \right] \frac{\partial T}{\partial y}. \tag{4.39}$$

Thus,

$$S = -\frac{C_1 C_3 + C_2 C_4}{C_1^2 + C_2^2}, \tag{4.40}$$

$$N|\boldsymbol{B}| = \frac{C_2 C_3 - C_1 C_4}{C_1^2 + C_2^2}. \tag{4.41}$$

For the physical medium consisting of $u$ and $d$ quarks, the total currents are given as:

$$J_x = \sum_{a=u,d} \left[ q_a (I_1)_a E_x + q_a (I_2)_a E_y + \beta (I_3)_a \frac{\partial T}{\partial x} + \beta (I_4)_a \frac{\partial T}{\partial y} \right] \tag{4.42}$$

$$J_y = \sum_{a=u,d} \left[ -q_a (I_2)_a E_x + q_a (I_1)_a E_y - \beta (I_4)_a \frac{\partial T}{\partial x} + \beta (I_3)_a \frac{\partial T}{\partial y} \right]. \tag{4.43}$$

Setting the currents equal to 0 as earlier, we arrive at the Seebeck and Nernst coeffi-cients of the composite medium:

$$S = -\frac{K_1K_3 + K_2K_4}{K_1^2 + K_2^2}\,, \tag{4.44}$$

$$N|\boldsymbol{B}| = \frac{K_2K_3 - K_1K_4}{K_1^2 + K_2^2}. \tag{4.45}$$

where,

$$K_1 = \sum_{a=u,d} q_a(I_1)_a\,, \qquad K_2 = \sum_{a=u,d} q_a(I_2)_a\,, \tag{4.46}$$

$$K_3 = \sum_{a=u,d} \beta(I_3)_a\,, \qquad K_4 = \sum_{a=u,d} \beta(I_4)_a. \tag{4.47}$$

In the quasiparticle description of quarks and gluons in a thermal medium, all quark flavours (with current/vacuum masses, $m_i << T$) acquire the same thermal mass [131, 132]

$$m_T^2 = \frac{g^2(T)T^2}{6}, \tag{4.48}$$

which is, however, modified in the presence of a finite chemical potential [109]

$$m_{T,\mu}^2 = \frac{g^2(T)T^2}{6}\left(1 + \frac{\mu^2}{\pi^2T^2}\right). \tag{4.49}$$

We take the pure thermal ($B = 0$) expressions of quasiparticle masses with magnetic field dependence coming in implicitly via the coupling constant. This is justified since we are working in a regime where $eB \ll T^2$. We use a one loop running coupling constant $\alpha_s(\Lambda^2, eB)$, which runs with both the magnetic field and temperature: [130]

$$\alpha_s(\Lambda^2, |eB|) = \frac{\alpha_s(\Lambda^2)}{1 + b_1\alpha_s(\Lambda^2)\ln\left(\frac{\Lambda^2}{\Lambda^2+|eB|}\right)}, \tag{4.50}$$

where, $\alpha_s(\Lambda^2)$ is the one-loop running coupling in the absence of a magnetic field

$$\alpha_s(\Lambda^2) = \frac{1}{b_1\ln\left(\frac{\Lambda^2}{\Lambda_{QCD}^2}\right)},$$

with $b_1 = (11N_c - 2N_f)/12\pi$ and $\Lambda_{QCD} \sim 0.2$ GeV. The renormalisation scale is chosen to be $\Lambda = 2\pi\sqrt{T^2 + \frac{\mu^2}{\pi^2}}$. $\alpha_s(\Lambda^2, |eB|)$ determines $g(T, |eB|)$ via the relation

$$\alpha_s(\Lambda^2, |eB|) = \frac{g^2(T, |eB|)}{4\pi}. \tag{4.51}$$

Thus, the thermally generated mass takes on an implicit dependence on the magnetic field via the coupling constant. We take the quasiparticle mass (squared) of $i$th flavor to be [108, 110–112]:

$$m'^{\,2}_{iT} = m_i^2 + \sqrt{2}\, m_i\, m_T + m_T^2. \tag{4.52}$$

### 4.3.2 Results

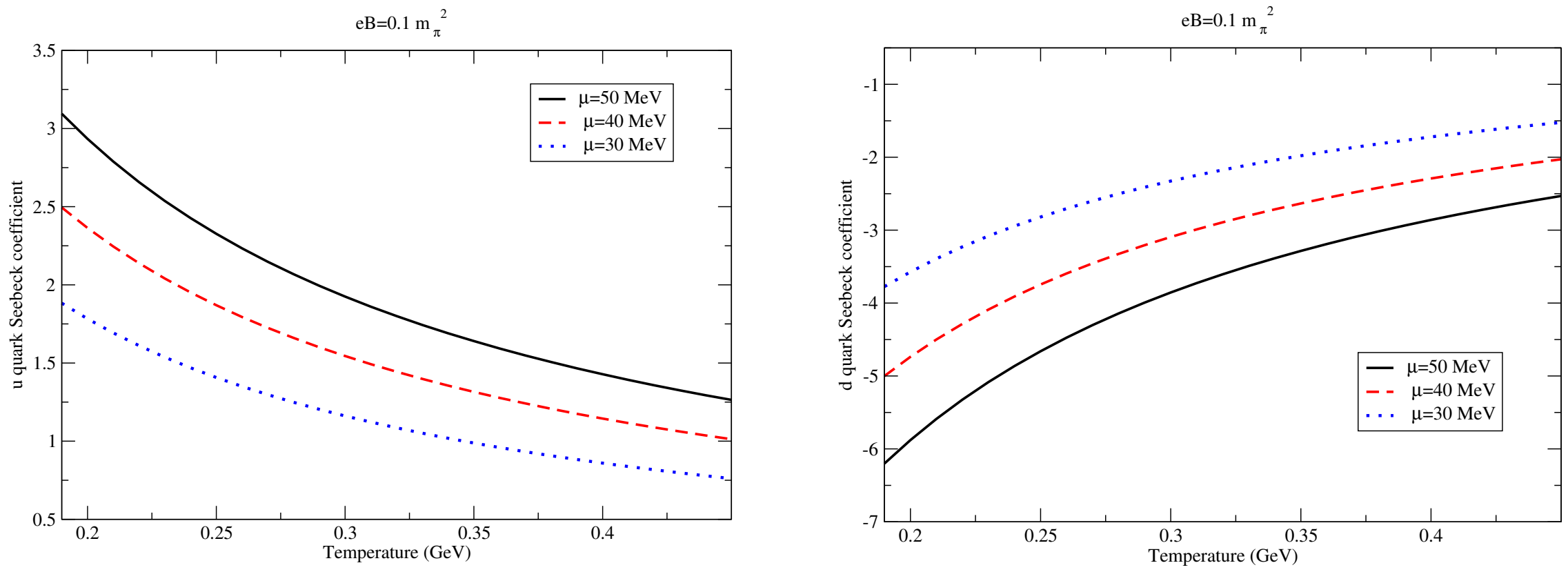

**(a)** **(b)**

**Figure 4.1:** Variation of Seebeck coefficient of $u$ (a) and $d$ (b) quarks with temperature for different fixed values of quark chemical potential.

As can be seen in Fig.(4.1a) and (4.1b), the magnitude of the individual coefficients decreases with temperature and increases with chemical potential, similar to the observations in the 1-D result. The individual Seebeck coefficients are ratios of two integrals [Eq.(4.40)]. The numerator and denominator of Eq.(4.40) for both $u$ and $d$ quarks are all increasing functions of temperature as far as the absolute values are concerned. However, the ratios are monotonically decreasing functions of temperature for the temperature range considered here. As expected, the coefficient is positive for the positively charged $u$ quark and negative for the negatively charged $d$ quark. The magnitudes of numerators and denominators in Eq.(4.40) for both $u$ and $d$ quarks are increasing functions of chemical potential as well. However, the rate of increase is more pronounced for the numerator than the denominator. This explains the overall increase of the individual coefficients with chemical potential.

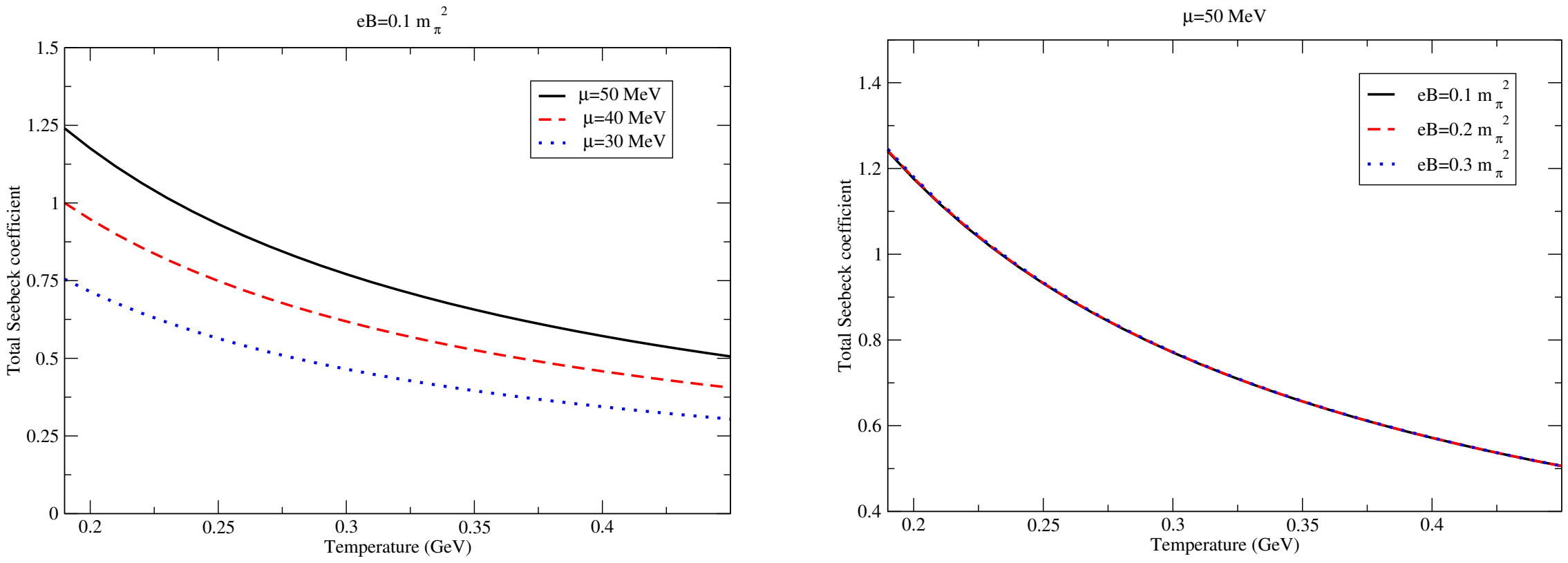

(a) (b)

**Figure 4.2: Left panel:** Variation of total Seebeck coefficient of the composite medium with temperature at $eB = 0.1m_\pi^2$ for different fixed values of chemical potential. **Right panel:** Variation of total Seebeck coefficient with temperature at $\mu = 50$ MeV for different fixed values of magnetic field.

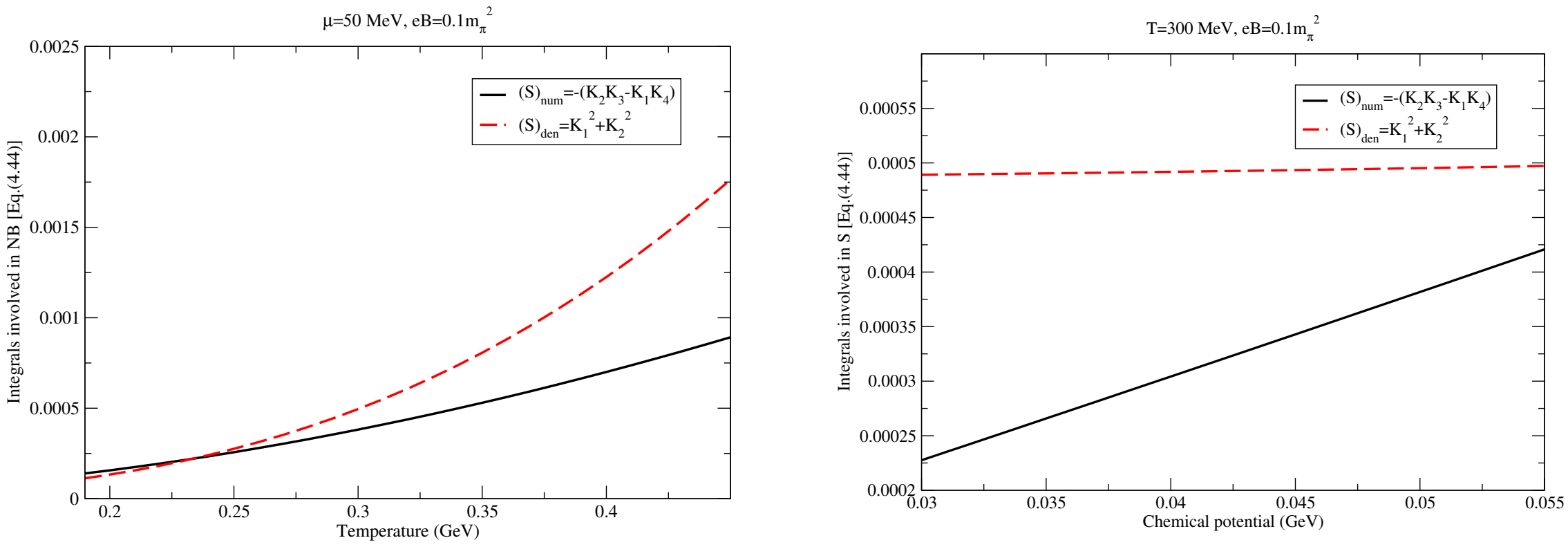

(a) (b)

**Figure 4.3: Left panel:** Variation of numerator and denominator of total Seebeck coefficient [Eq.(4.44)] with temperature for $eB = 0.1m_\pi^2$, $\mu = 50$ MeV. **Right panel:** Variation of numerator and denominator of total Seebeck coefficient [Eq.(4.44)] with chemical potential for $eB = 0.1m_\pi^2$, $T = 300$ MeV.

Fig.(4.2a) shows the variation of Seebeck coefficient of the composite medium composed of $u$ and $d$ quarks with temperature. We see the earlier trend of decrease of coefficient magnitude with temperature. Also, the coefficient is positive and increases with increasing chemical potential. These trends can be understood from analysing the integrals in the numerator and denominator of Eq.(4.44). Fig.(4.3a) shows the variation of the numerator and denominator integrals in Eq.(4.44) with temperature. It can be seen that both the numerator and the denominator are increasing functions of temperature. The comparative increase is such that the ratio is rendered a decreasing function of temperature. The variation with chemical potential $\mu$ is shown in Fig.(4.3b). As $\mu$ increases, the net thermoelectric current increases which leads to a larger value of the induced electric field, reflected by a larger $S$. Fig.(4.2b) shows the effect of magnetic field on the temperature dependence of the total Seebeck coefficient. As can be seen, the total Seebeck coefficient increases with increasing background magnetic field and records the same decreasing trend with temperature for each value of the magnetic field.

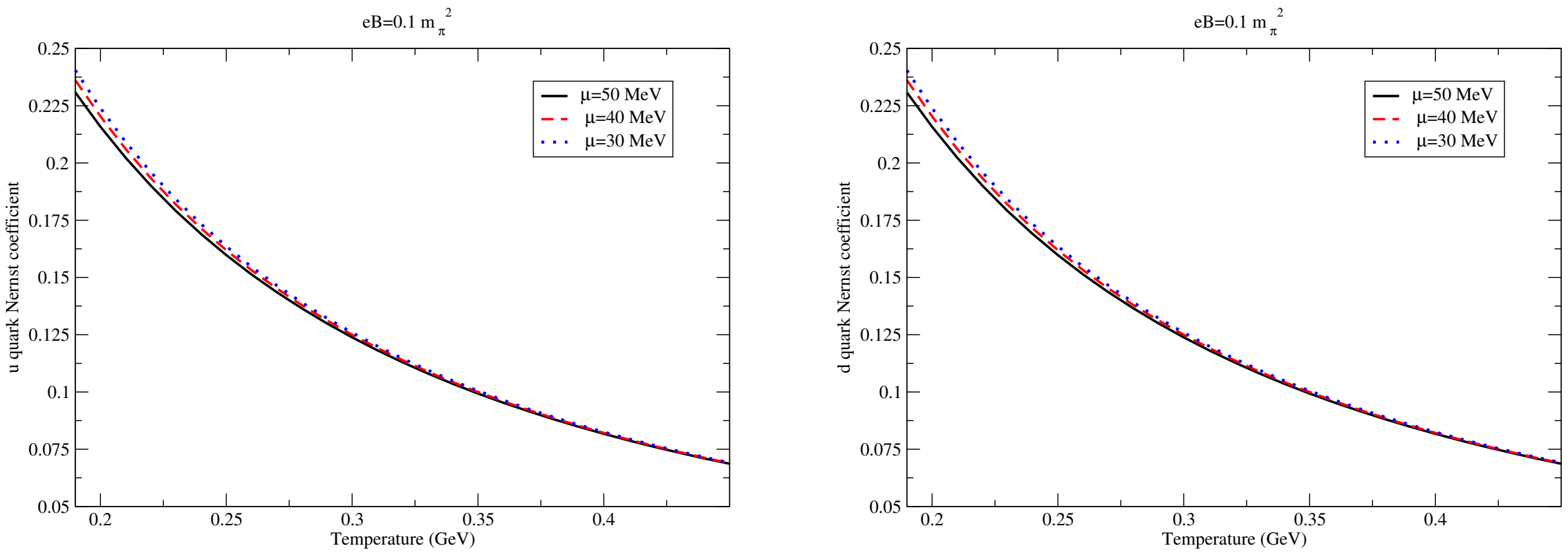

**(a)** **(b)**

**Figure 4.4:** Variation of Nernst coefficient of $u$ (a) and $d$ (b) quarks with temperature for different fixed values of quark chemical potential.

Figures (4.4a) and (4.4b) show the individual Nernst coefficients for the medium composed exclusively of $u$ quarks and $d$ quarks respectively. The magnitude decreases almost monotonically with temperature for the entire temperature range. The impact of chemical potential is overall feeble, with it being discernible only near the transition temperature. From $2 - 2.5\ T_c$, there is negligible impact of chemical potential on the value of the coefficients. The almost identical values of Nernst coefficient for both the $u$ quark medium and the $d$ quark medium suggests that the value is fairly independent of the quantum of charge carried by the individual charge carrier. This is unlike the case of Seebeck coefficient where the difference in the quantum of charge carried by the $u$ and $d$ quarks is directly reflected in their respective Seebeck coefficients.

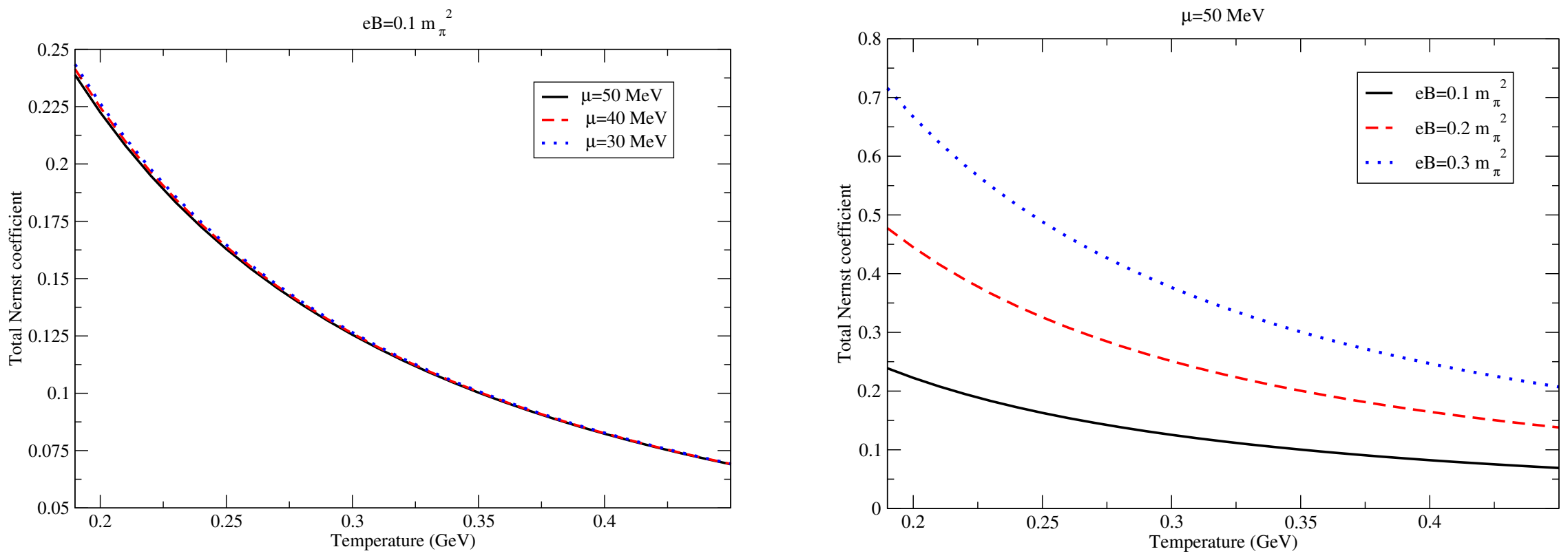

**(a)** **(b)**

**Figure 4.5: Left panel:** Variation of total Nernst coefficient with temperature at $eB = 0.1m_\pi^2$ for different fixed values of chemical potential. **Right panel:** Variation of total Nernst coefficient with temperature at $\mu = 30$ MeV for different fixed values of magnetic field.

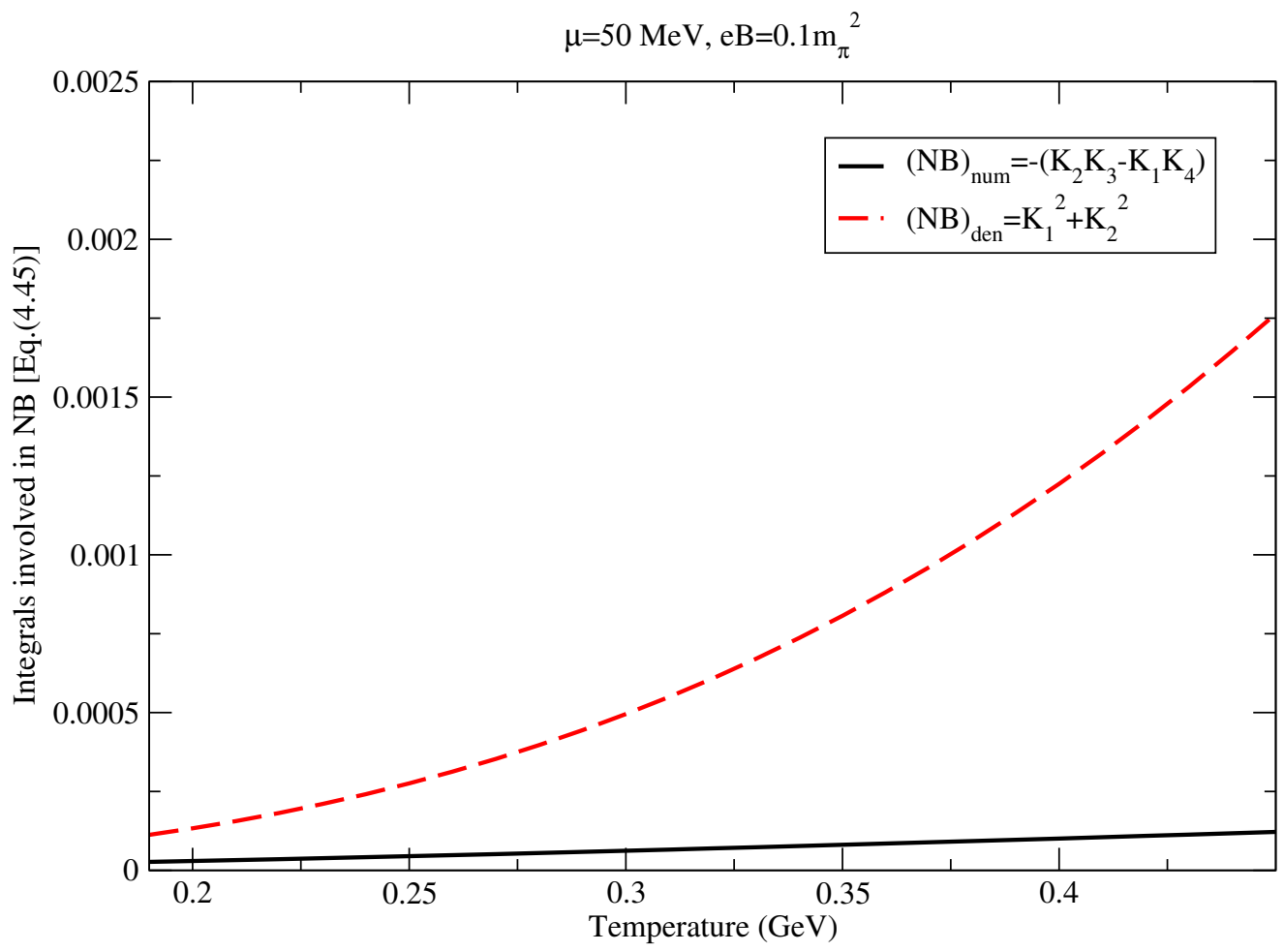

**Figure 4.6:** Variation of numerator and denominator of total Nernst coefficient [Eq.(4.45)] with temperature for $eB = 0.1m_\pi^2$, $\mu = 50$ MeV.

Fig.(4.5a) and (4.5b) show the variation of total Nernst coefficients with temperature for different values of chemical potential and magnetic fields respectively. As can be seen in Fig.(4.5a), the coefficient records an increase with temperature near the crossover temperature after which the slope of the curve changes sign and the coefficient records a decreasing trend with temperature. Both the numerator and the denominator of the Nernst coefficient [Eq.(4.45)] are increasing functions of temperature for a fixed chemical potential. This is shown in Fig.(4.6). The rate of increase of the denominator vis-a-vis the numerator is however, much greater. This explains the decreasing trends in Figs. (4.5a) and (4.5b). The Seebeck coefficient does not have a direct $B$ dependence; its $B$ dependence comes in via the quasiparticle mass, and the cyclotron frequency. The Nernst coefficient additionally carries an explicit $B$ dependence. As such, its sensitivity to changes in magnetic field strength is comparatively more pronounced, as can be seen by comparing Figs. (4.3a) and (4.5b).

Seebeck and Nernst coefficients have been calculated for both the hadron gas medium and the quark-gluon plasma medium Thermoelectric response in a hadron gas medium modelled by the Hadron Resonance Gas (HRG) model has been investigated in [89], the obvious difference with our work therefore being the choice of the medium. They record a negative total Seebeck coefficient with an increase in the absolute value with temperature along with a decrease in the absolute values with increasing magnetic field. The behaviour is thus different (both the sign and the trend) from that obtained by us for the QGP medium. This shows that the nature of the medium and the interactions therein affect both the direction of the induced electric field as well as the behaviour with temperature and magnetic field. The variation of Nernst coefficient with temperature and chemical potential is roughly similar to the results obtained by us, with the difference being the initial increase in the coefficient magnitude (near the transition temperature) in our work. In [87], a similar investigation has been carried out for the quark-gluon plasma medium. The medium interactions have been incorporated using the Effective fugacity Quasiparticle model (EQPM) with zero current quark masses for the $u$ and $d$ quarks, whereas in our work, a finite current quark mass has been considered for both the quarks and interactions are manifest via the quasiparticle masses obtained from perturbative QCD calculations. The other major difference is the ansatz [Eq.(4.17)] used for the calculation. The ansatz used by us is a natural generalization of the 1-D ansatz that has been used in multiple works before. The magnitudes of the total Seebeck coefficient obtained in the temperature range is comparable to the ones obtained in our study. However, the sign of the coefficient is negative. This could be because of the difference in ansatz or modelling of in-medium interactions. The variation of Nernst coefficient with temperature and chemical potential bears rough resemblances with our study. In [88], a similar study has been carried out, however, for an anisotropic QGP medium wherein the anisotropy of the medium has been incorporated via an anisotropy parameter $\xi$ in the distribution function. Further, another point of difference with our work is the expressions for total Seebeck and Nernst coefficients obtained from the individual ones. They have reported a negative total Seebeck coefficient with the magnitude decreasing with temperature and increasing with magnetic field. The Nernst coefficient reported by them bears a close resemblance with our study, which could be because of the similar structure of the ansatz used.

## 4.4 Part II: $B$ dependent quasiparticle masses

The difference in this part is that the quasiparticle masses are evaluated explicitly in the presence of a weak background magnetic field, and subsequently the Seebeck and Nernst coefficients are evaluated using those masses. As we shall see, the quasiparticle masses for the left ($L$) and right ($R$) handed chiral modes of quarks come out to be different. This is unlike the case of $B = 0$ as well as the strong $B$ limit ($eB \gg m_\pi^2$), where the $L$ and $R$ mode quasiquark masses are degenerate.

### 4.4.1 Dispersion Relations in weak $B$: Chiral modes

The dispersion relation of quarks is obtained from the zeros of the inverse resummed quark propagator. The Dyson-Schwinger equation relates the full/resummed/effective propagator ($S$), bare propagator ($S_0$) and the particle self energy ($\Sigma$) [141]

$$S^{-1}(K) = S_0^{-1}(K) - \Sigma(K) = \not{K} - \Sigma(K), \tag{4.53}$$

where the quark self-energy, $\Sigma(K)$ is to be calculated up to one-loop from thermal QCD in weak $B$ and the bare quark propagator, $S_0(K)$ in weak $B$, up to power $q_f B$, is given by [142]

$$iS_0(K) = \frac{i\left(\not{K} + m_f\right)}{K^2 - m_f^2} - \frac{\gamma_1\gamma_2\left(\gamma.K_{||} + m_f\right)}{\left(K^2 - m_f^2\right)^2}(q_f B), \tag{4.54}$$

where, the first term in Eq.(4.54) is the free fermion propagator and the second term is the $\mathcal{O}(q_f B)$ correction to it. The current quark mass $m_0$, the magnetic field $B$ and the temperature $T$ define the different scales present in the system. The scale hierarchy adopted in this work is $m_0^2 < qB < T^2$, which allows us to neglect $m_0$ in the numerator of Eq.(4.54), leading to:

$$iS_0(K) = \frac{i\left(\not{K}\right)}{K^2 - m_f^2} - \frac{\gamma_1\gamma_2\left(\gamma.K_{||}\right)}{\left(K^2 - m_f^2\right)^2}(q_f B) \tag{4.55}$$

The two 4-vectors that characterize the system are 1) the fluid four-velocity $u^\mu$, and 2) the direction of the external magnetic field $b_\mu$. In the rest frame of the heat bath, $u^\mu = (1, 0, 0, 0)$ and the magnetic field direction can be expressed as a projection of the electromagnetic field strength tensor $F^{\mu\nu}$ along $u^\mu$, $b^\mu = \frac{1}{B}\epsilon^{\mu\nu\rho\lambda}u^\nu F^{\rho\lambda} = (0, 0, 0, 1)$, which is in the $z$ direction. We express Eq.(4.54) in terms of $u^\mu$ and $b^\mu$ as

$$iS_0(K) = \frac{i\,(\not{K})}{K^2 - m_f^2} - \frac{i\gamma_5[(K.b)\not{u} - (K.u)\not{b}]}{(K^2 - m_f^2)^2}\big(q_f B\big). \tag{4.56}$$

Eq.(4.56) is derived from Eq.(4.55) with the help of the following relations:

$$K_\parallel^\mu = (k^0, 0, 0, k^3); \quad \not{K} = \gamma^\mu K_\mu$$
$$\not{K}_\parallel = \gamma^0 k_0 + \gamma^3 k_3; \quad \gamma_5 = i\gamma_0\gamma_1\gamma_2\gamma_3,$$

and by noting that

$$\gamma_5\not{u} = -i\gamma_1\gamma_2\gamma_3, \quad \gamma_5\not{b} = -i\gamma_0\gamma_1\gamma_2.$$

With the aforementioned definitions of $u^\mu$ and $b^\mu$, we can write $k^0 = (K \cdot u)$, $k^3 = -(K \cdot b)$. Using these relations in Eq.(4.55) ultimately leads to Eq.(4.56).

Diagrammatically, the one loop quark self energy is given by

$$\begin{aligned}\Sigma(P) =& g^2 C_F T \sum_n \int \frac{d^3k}{(2\pi)^3}\gamma_\mu\Bigg(\frac{\not{K}}{(K^2 - m_f^2)} - \\ & \frac{\gamma_5[(K.b)\not{u} - (K.u)\not{b}]}{(K^2 - m_f^2)^2}(|q_f B|)\Bigg)\gamma^\mu \frac{1}{(P-K)^2}\end{aligned} \tag{4.57}$$

In a covariant tensor basis, the above self energy can also be expressed generally in a tensor basis consisting of 4-vectors present in the system. The available 4-vectors are the fermion external 4-momentum $P$, heat bath velocity $u$ and the magnetic field direction $b$. In addition, the self enrgy $\Sigma(P)$ is itself a $4 \times 4$ matrix and thus can be expanded in terms of 16 basis matrices: $\{\mathbb{1},\ \gamma_5,\ \gamma_\mu,\ \gamma_\mu\gamma_5\}$. Using chiral symmetry arguments, the expansion can be reduced to 5 terms given below [143].

$$\Sigma(P) = -\mathcal{A}(p^0, p)\not{P} - \mathcal{B}(p^0, p)\not{u} - \mathcal{C}(p^0, p)\gamma_5\not{u} - \mathcal{D}(p^0, p)\gamma_5\not{b}, \tag{4.58}$$

where, $\mathcal{A}, \mathcal{B}, \mathcal{C}, \mathcal{D}$ are the structure functions. It should be noted that this tensor expansion is not unique. A different set of basis tensors has been used, for example, in [144]. Using Eqs.(6.14) and (4.58), the structure functions come out to be

$$\mathcal{A} = \frac{1}{4}\frac{\mathrm{Tr}(\Sigma(P)\not{P}) - (P\cdot u)\mathrm{Tr}(\Sigma(P)\not{u})}{(P\cdot u)^2 - P^2} \tag{4.59}$$

$$\mathcal{B} = \frac{1}{4}\frac{-(P\cdot u)\mathrm{Tr}(\Sigma(P)\not{P}) + P^2\mathrm{Tr}(\Sigma(P)\not{u})}{(P\cdot u)^2 - P^2} \tag{4.60}$$

$$\mathcal{C} = -\frac{1}{4}\mathrm{Tr}(\gamma_5\Sigma(P)\not{u}) \tag{4.61}$$

$$\mathcal{D} = \frac{1}{4}\mathrm{Tr}(\gamma_5\Sigma(P)\not{b}) \tag{4.62}$$

Next, we rewrite the self energy in terms of left and right handed chiral projection operators $P_L = (\mathbb{1} - \gamma_5)/2$ and $P_R = (\mathbb{1} + \gamma_5)/2$ as

$$\Sigma(P) = -P_R\not{A}'P_L - P_L\not{B}'P_R, \tag{4.63}$$

where,

$$\not{A}' = \mathcal{A}\not{P} + (\mathcal{B}+\mathcal{C})\not{u} + \mathcal{D}\not{b}, \tag{4.64}$$

$$\not{B}' = \mathcal{A}\not{P} + (\mathcal{B}-\mathcal{C})\not{u} - \mathcal{D}\not{b} \tag{4.65}$$

Using Eqs.(4.63) and (A.19), we can write the inverse fermion propagator as

$$S^{-1}(P) = \not{P} + P_R\left[\mathcal{A}\not{P} + (\mathcal{B}+\mathcal{C})\not{u} + \mathcal{D}\not{b}\right]P_L + P_L\left[\mathcal{A}\not{P} + (\mathcal{B}-\mathcal{C})\not{u} - \mathcal{D}\not{b}\right]P_R$$

Using the relations

$$P_{L,R}\gamma^\mu = \gamma^\mu P_{R,L}; \quad P_L\not{P}P_L = P_R\not{P}P_R = P_LP_R\not{P} = 0,$$

the inverse effective propagator simplifies to

$$S^{-1}(P) = P_R\not{L}P_L + P_L\not{R}P_R, \tag{4.66}$$

where,

$$\not{L} = (1+\mathcal{A})\not{P} + (\mathcal{B}+\mathcal{C})\not{u} + \mathcal{D}\not{b}, \tag{4.67}$$

$$\not{L} = (1+\mathcal{A})\not{P} + (\mathcal{B}-\mathcal{C})\not{u} - \mathcal{D}\not{b}. \tag{4.68}$$

Inverting Eq.(4.66) using Eqs.(4.67) and (4.68), we get,

$$S(P) = \frac{1}{2}\Big[P_L\frac{\not{L}}{L^2/2}P_R + \frac{1}{2}P_R\frac{\not{R}}{R^2/2}P_L\Big], \tag{4.69}$$

where,

$$L^2 = (1+\mathcal{A})^2P^2 + 2(1+\mathcal{A})(\mathcal{B}+\mathcal{C})p_0 \tag{4.70}$$
$$- 2\mathcal{D}(1+\mathcal{A})p_z + (\mathcal{B}+\mathcal{C})^2 - \mathcal{D}^2, \tag{4.71}$$

$$R^2 = (1+\mathcal{A})^2P^2 + 2(1+\mathcal{A})(\mathcal{B}-\mathcal{C})p_0 \tag{4.72}$$
$$+ 2\mathcal{D}(1+\mathcal{A})p_z + (\mathcal{B}-\mathcal{C})^2 - \mathcal{D}^2. \tag{4.73}$$

The $p_0 = 0$, $\mathbf{p} \to 0$ limit of the denominator of the effective propagator yields the quasiparticle masses as [145, 146]

$$m_L^2 = \frac{L^2}{2}|_{p_0=0,|\mathbf{p}|\to 0} = m_{th}^2 + 4g^2C_FM^2, \tag{4.74}$$
$$m_R^2 = \frac{R^2}{2}|_{p_0=0,|\mathbf{p}|\to 0} = m_{th}^2 - 4g^2C_FM^2, \tag{4.75}$$

thus, lifting the degeneracy. Here,

$$M^2 = \frac{|q_fB|}{16\pi^2}\left(\frac{\pi T}{2m_f} - \ln 2 + \frac{7\mu^2\zeta(3)}{8\pi^2T^2}\right), \tag{4.76}$$
$$m_{th}^2 = \frac{1}{8}g^2C_F\left(T^2 + \frac{\mu^2}{\pi^2}\right). \tag{4.77}$$

It should be noted that the same procedure is adopted for calculating the quasiparticle masses of fermions both in the absence of a magnetic field, and also in the strong magnetic field limit. In both the cases, the static limit of the $L^2$ and $R^2$ terms come out to be the same [58, 134]. It is only in the case of a weak magnetic field that these masses are different. The coupling constant, $g$ is used as in [130]. In the ultra relativistic limit, the chirality of a particle is the same as it's helicity, so that the right and left chiral modes can be thought of as the up/down spin projections in the direction of momentum ($s_z = \pm 1/2$) of the particle, suggesting that the medium generated mass of a collective fermion excitation at very high temperatures ($T \gg m_0$) is $s_z$ (helicity)-dependent. Such $s_z$ dependent quasiparticle masses are already known in condensed matter systems [147, 148].

### 4.4.2 Results and discussion

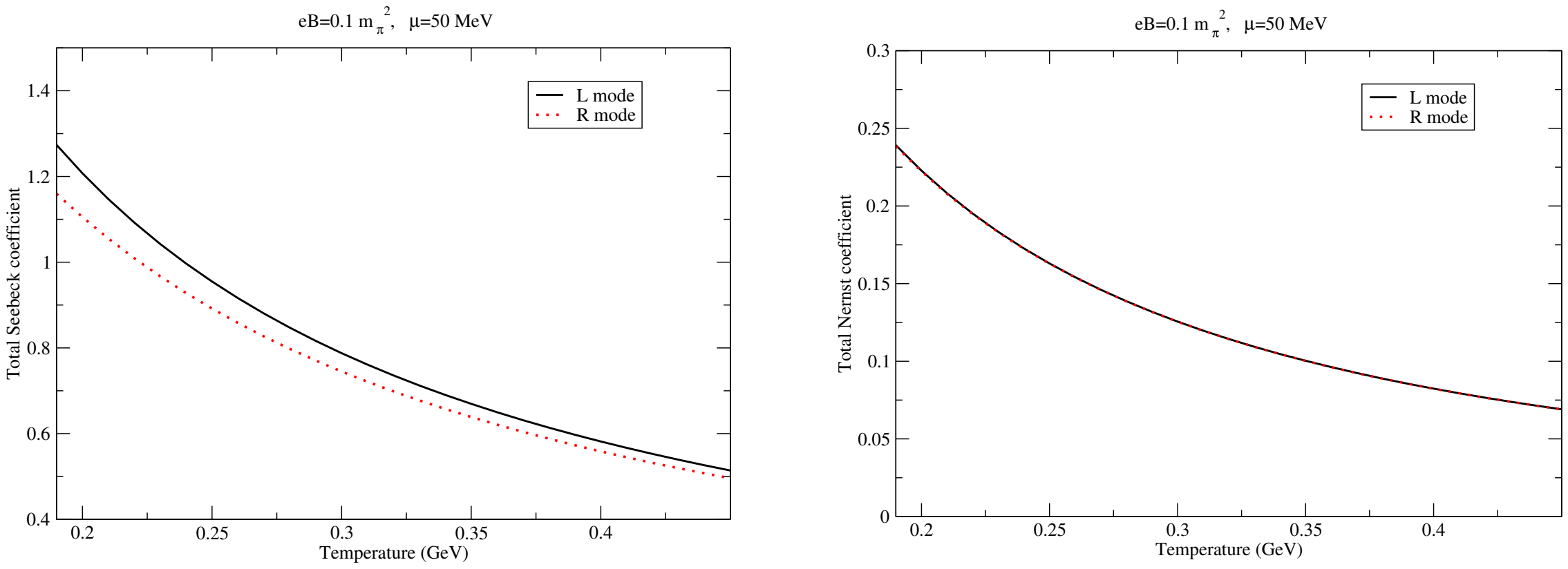

(a) (b)

**Figure 4.7: Left panel:** $L$ and $R$ mode medium Seebeck coefficients as a function of $T$, at $eB = 0.1m_\pi^2$ and $\mu = 50$ MeV. **Right panel:** $L$ and $R$ mode medium Nernst coefficients as a function of $T$ at $eB = 0.1m_\pi^2$ and $\mu = 50$ MeV.

Figures (4.7a) and (4.7b) show the variation of Seebeck and Nernst coefficients of the medium with temperature. It can be seen that the magnitude of the induced electric field in the longitudinal (along $\boldsymbol{\nabla T}$) and transverse (perpendicular to $\boldsymbol{\nabla T}$) directions decreases with temperature, for both the modes. Also, for both the coefficients, the $L$ mode elicits a larger comparative response[1]. This can be understood from a numerical perspective. Each of the integrals $C_1$, $C_2$, $C_3$, $C_4$ [Eqs. (4.35),(4.36)] are decreasing functions of the effective mass. However, the extent of decrease follows the hierarchy $\Delta C_2 > \Delta C_1 > \Delta C_4 > \Delta C_3$, where $\Delta C$ denotes change in the value of the integral due to a given change in mass. The mathematical expressions of $S$ and $N|\boldsymbol{B}|$ then imply that whereas both the numerator and denominator of the expressions decrease with increasing mass, the denominator decreases by a larger amount (because of the presence of $C_1$ and $C_2$) than the numerator. The value of the fraction, therefore, increases with increasing mass. For comparison, the $B = 0$ case is also shown for the Seebeck coefficient. The Nernst coefficient is, however zero for $B = 0$, as should be the case.

[1]The change is very feeble for the Nernst coefficient, which is reflected in both the curves overlapping in Fig.(4.7b)

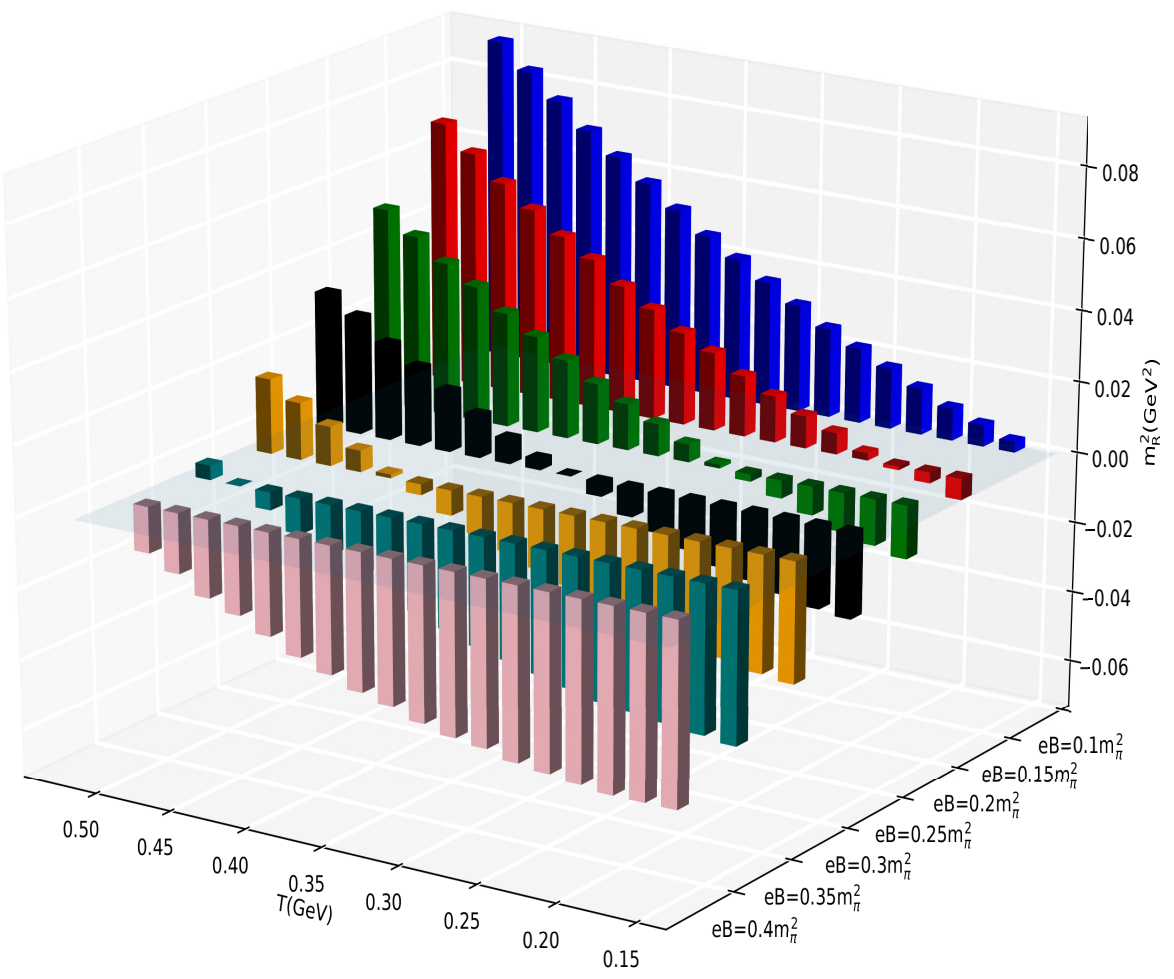

**Figure 4.8:** Medium generated mass (squared) of $R$ mode $u$ quark as a function of $T$ and $B$.

We found that the mass (squared) of the $R$ mode fermion, (Eq.(2.32)) comes out to be negative for a certain range of combinations of $T$ and $B$ values, which is brought out by Fig.(4.8). As the magnetic field is increased, the temperature (above $T_c \sim 155$ MeV) upto which $m_R^2$ is negative, also increases. This suggests that the perturbative framework used by us to study the chirality dependence of the thermoelectric response is valid only at regions sufficiently far (∼200 MeV) from the crossover region in the QCD phase diagram for $|eB| > 0.2m_\pi^2$. Another way to look at it is that the condition $eB \ll T^2$ is strictly enforced. For $T > T_c$, we find from Fig.(4.8) that $\left|\frac{eB}{T^2}\right|_{\rm max} \sim 0.07$ for $eB = 0.1m_\pi^2$ and $\sim 0.03$ for $eB = 0.35m_\pi^2$. For values of $T$ and $B$ leading to higher values of $\left|\frac{eB}{T^2}\right|$, $m_R^2$ is negative and thus unphysical.

## 4.5 Summary and Conclusions

In this chapter, we have investigated the thermoelectric phenomena of Seebeck effect and Nernst effect in a deconfined plasma of quarks and gluons in the presence of a weak magnetic field. In the first part of the chapter, we make use of the $B = 0$ results of quasiparticle masses with their $B$ dependence being implicit in the coupling constant. In the second part, we describe the evaluation of the quasiquark masses in the presence of a weak magnetic field upto $\mathcal{O}(qB)$. We have carried out the aforesaid analyses in the kinetic theory framework by applying the Boltzmann transport equation for a relativistic system in the relaxation time approximation wherein we assume that the phase space and dispersion relations of quarks are not affected by magnetic field via Landau quantization. The main takeaway is that the response becomes a $2 \times 2$ matrix [Eq.(4.37)] with the diagonal elements being the Seebeck coefficients and the off-diagonal elements the Nernst coefficients.

We observe that the magnitudes of the coefficients are all decreasing functions of $T$. The dependence on chemical potential is strong for the Seebeck coefficient and feeble for the Nernst coefficient. However, Seebeck coefficient depends weakly on the magnetic field strength, whereas Nernst coefficient is much more sensitive to changes in $B$ field strength. The physical medium consisting of different species of quarks, in the presence of a weak magnetic field has a feeble thermoelectric response in the entire temperature range considered, much less than that in the case of a strong magnetic field, as can be seen in Fig.(3.12). For the Nernst coefficients ($NB$), we find that the sign of the individual Nernst coefficient is independent of the charge of the majority charge carrier of the medium, unlike in the case of the individual Seebeck coefficients, as is explained in Chapter 1.

The quasiquark masses evaluated using perturbation theory in the presence of a weak magnetic field, turn out to be different for the $L$ and $R$ chiral modes. These masses are however degenerate ($m_L = m_R$) in the case of $B = 0$ and in the limit $eB \gg m_\pi^2$. As a result, there are 2 modes of the Seebeck and Nernst coeficients, both of which show similar trends with respect to variation wth $T$ and $eB$. The coefficient magnitudes are larger for the $L$ mode compared to the $R$ mode. Because of the negative sign in Eq.(4.75), the value of $m_R^2$ comes out to be negative for certain combination of values of $eB$ and $T$. The range of values of $eB$ and $T$ obtained by constraining $m_R^2$ to positive values serves as a consistency check of the weak magnetic field limit ($eB \ll T^2$).

# Chapter 5

# Thermoelectric response of a hot and weakly magnetized anisotropic QCD medium

वृत्तस्य चतुर्थ भागस्य दृष्टिरपि रेखा। सर्वस्य दृष्टिर्भूगोलस्य च।

*A quarter of the circle is seen as a line. The entire view of the Earth is a circle.*

**– Siddhānta Śiromaṇi by Bhāskara II (1114 CE-1185 CE)**

In the previous chapters, we investigated the thermoelectric response of a QGP medium ignoring the fact that the QGP fireball actually expands anisotropically. One way to take this anisotropy into account is by stretching or squeezing the equilibrium distribution functions. In this chapter, we have evaluated the thermoelectric transport coefficents for an anisotropically expanding QGP medium. This is based on *Thermoelectric response of a hot and weakly magnetized anisotropic QCD medium* by Salman Ahamad Khan, Debarshi Dey and Binoy Krishna Patra, **arXiv:2309.08467 [hep-ph](2023).**

## 5.1 Introduction

Heavy ion collisions at ultra-relativistic energies give rise to a state of matter comprising of asymptotically free quarks and gluons-the Quark Gluon Plasma (QGP). Several observables are considered as signatures of creation of such a medium; this includes photon and dilepton spectra [149, 150], quarkonium suppression [151–153], elliptic flow [154, 155], jet quenching [156–158], etc. Experimentally, significant evidence now exists of the observation of these signals, and thus, of the creation of QGP matter in Ultrarelativistic Heavy Ion collisions (URHICs) at experimental facilities such as the Brookhaven National Laboratory Relativistic Heavy Ion Collider (RHIC) [12, 24, 25] and Large Hadron Collider (LHC) [159, 160]. Right after the formation of QGP, it expands and cools and transitions into a mildly interacting collection of hadrons. At very small baryon chemical potentials ($\mu_B \sim 0$), with massive quarks, the results of lattice QCD indicate that the transition is actually an analytic crossover rather than a true phase transition [161–164]. The sign problem of lattice QCD at finite $\mu_B$ makes it an unreliable tool to explore large parts of the QCD phase diagram [165, 166].

The QGP lives for a very short period of time and does not expand at the same rate in all the directions. The colliding nuclei are Lorentz contracted due to their relativistic speeds. The overlap volume of such nuclei in a non-central collision is anisotropic in the plane perpendicular to the beam (transverse plane). This spatial asymmetry in coordinate space gets converted into an opposite asymmetry in the momentum space which is ultimately reflected in the hadron $p_T$ spectra. A convenient way of taking into account the anisotropy of the medium was introduced by Romatschke and Strickland wherein the anisotropic distribution function is obtained from an arbitrary isotropic one by the rescaling of only one direction in momentum space, *i.e.* by stretching or squeezing the isotropic distribution function of the medium constituents (partons) [167]. This parametrisation of the anisotropy involving a single direction and a single anisotropy parameter, $\xi$ (spheroidal momentum anisotropy), was used to calculate the collective modes of finite temperature QCD and study their impact in thermalization of the QGP medium [167–170]. Hard loop effective theories have also been used to study anisotropy; its equivalence with the kinetic theory approach was shown by Mrówczyński and Thoma, wherein they calculated the self-energies and dispersion relations for QGP partons [171]. Existence of instabilities associated with gluon collective modes in an anisotropic QGP has been observed and their growth rates have also been calculated [172, 173]. This description has also been used to study photon and dilepton production from the QGP [174, 175], the QGP heavy quark potential [176], and bottomonia suppression [177]. In addition to the case of spheroidal anisotropy, ellipsoidal momentum anisotropy has also been considered, which is characterised by two or more independent anisotropic parameters. In particular, parton self energies have been calculated in an ellipsoidally anisotropic QGP [178]. Anisotropic momentum distributions have also been used in relativistic hydrodynamic models to study the evolution of QGP to hadrons [179–183]. In fact, anisotropic hydrodynamics (aHydro) has been more accurate than its isotropic counterpart in the description of non-equilibrium dynamics [184–189]. As such, transport coefficients like electrical conductivity [58], heavy quark drag and diffusion coefficients [190] have also been evaluated in an anisotropic plasma.

Apart from causing an anisotropic expansion of the created matter, non-central heavy ion collisions also lead to creation of large magnetic fields [191]. These magnetic fields, produced mainly by the spectator protons moving away from each other at relativistic speeds, reach magnitudes upto $eB \sim 10^{-1}m_\pi^2\,(\simeq 10^{17}$ Gauss) for SPS energies, $eB \sim m_\pi^2$ for RHIC energies and $eB \sim 15m_\pi^2$ for LHC energies [56]. The decay rate of the magnetic field depends strongly on the electrical conductivity of the medium which is exposed to the field [192–201]. Depending on the strength of the background magnetic field, several interesting phenomena of the created matter can be probed. A strong background magnetic field causes separation of charges in a chiral QGP medium leading to a magnetic field dependent current, which is non-Maxwellian and has no analog in classical physics. This is the Chiral Magnetic effect [55, 57, 202]. Other phenomena induced by strong magnetic fields include magnetic catalysis [60], chiral magnetic wave [203], axial magnetic effect [204, 205], etc. A small electrical conductivity, however, would lead to only a small fraction of the initial magnetic field surviving when the created matter thermalizes. This has motivated several studies where the background magnetic field is considered to be weak. Further, such a weak field can give rise to novel phenomenological results like the lifting of mass degeneracy between the left and right handed quark effective masses [143]. Transport phenomena in the presence of weak magnetic field has been under investigation in the recent past [138–140, 146, 206, 207]

In this chapter, we study the thermoelectric response of an anisotropic QGP medium in the presence of a weak magnetic field, by incorporating the non-degenerate left and right chiral quasiparticle masses of quarks. Fluctuations in the initial energy density of heavy-ion collisions can create large temperature differences between the central and peripheral regions of the fireball [85]. This, coupled with a finite chemical potential can potentially give rise to thermoelectric phenomena: Seebeck and Nernst effects. The ability to convert a temperature gradient into an electric field is quantified by the Seebeck and Nernst coefficients.

In what follows, we present the calculation of the thermoelectric coefficients of a QGP medium, taking into account the expansion induced anisotropy of the momentum distribution of the partons, in the presence of a weak background magnetic field. The chapter is organized as follows: In section (5.2), the quasiparticle model used in this work is described. In section (5.3), the calculation of Seebeck and Nernst coefficients are outlined both for single component and multi component mediums. In section (5.4), the results are plotted and their interpretations discussed. Finally, we conclude in section (5.5).

## 5.2 Quasiparticle model

As has already been explained in chapter (1), the central feature of quasi particle models is that a strongly interacting system of massless quarks and gluons can be described in terms of massive, weakly interacting quasi-particles originating due to the collective excitations. In this study, we have used quasi-particle model [108] which has only a single adjustable parameter and the medium effects enter through the dispersion relations of the quark and gluon quasi-particles. The temperature and magnetic field-dependent masses of the quarks and gluons have been computed from the poles of their resummed propagators obtained from Dyson-Schwinger equations. The respective self-energies have been calculated using perturbative thermal QCD in a strong magnetic field background. The quasi particle mass of the $i^{th}$ flavor is written phenomenologically as [108]

$$m_i^2 = m_{i,0}^2 + \sqrt{2} m_{i,0} m_{i,T} + m_{i,T}^2, \tag{5.1}$$

where $m_{i,0}$ and $m_{i,T}$ are the current quark mass and medium generated quark mass. $m_{i,T}$ has been calculated using the HTL perturbation theory as [131, 132]

$$m_{iT}^2 = \frac{g'^2 T^2}{6} \left( 1 + \frac{\mu^2}{\pi^2 T^2} \right), \tag{5.2}$$

where $g' = \sqrt{4\pi\alpha_s}$ refers to the coupling constant which depends on the temperature as

$$\alpha_s(T) = \frac{g'^2}{4\pi} = \frac{6\pi}{(33 - 2N_f) \ln\left(\frac{Q}{\Lambda_{QCD}}\right)}, \tag{5.3}$$

where, $Q$ is set at $2\pi\sqrt{T^2+\frac{\mu^2}{\pi^2}}$. In the presence of a strong magnetic field, the coupling constant $g=\sqrt{4\pi\alpha_s}$ runs with the temperature, chemical potential and magnetic field as [130]

$$\alpha_s(\Lambda^2, eB) = \frac{g^2}{4\pi} = \frac{\alpha_s(\Lambda^2)}{1+b_1\alpha_s(\Lambda^2)\ln\left(\frac{\Lambda^2}{\Lambda^2+eB}\right)}, \tag{5.4}$$

with

$$\alpha_s(\Lambda^2) = \frac{1}{b_1\ln\left(\frac{\Lambda^2}{\Lambda^2_{\overline{MS}}}\right)}, \tag{5.5}$$

where $\Lambda$ is set at $2\pi\sqrt{T^2+\frac{\mu^2}{\pi^2}}$ for quarks, $b_1=\frac{11N_c-2N_f}{12\pi}$ and $\Lambda_{\overline{MS}}=0.176$ GeV. Similarly the effective quark mass for $i^{th}$ flavor in the case of a weak magnetic field can be parameterized like in the earlier cases as

$$m^2_{i,w} = m^2_{i0}+\sqrt{2}m_{i0}m_{i,L/R}+m^2_{i,L/R}, \tag{5.6}$$

where, $m_{i,L/R}$ refers to the thermal mass for the left- or right-handed chiral mode of $i^{th}$ flavor, the evaluation of which has already been elucidated in the previous chapter:

$$m_L^2 = m_T^2+4g^2C_FM^2, \tag{5.7}$$
$$m_R^2 = m_T^2-4g^2C_FM^2. \tag{5.8}$$

We will use these thermally generated masses in the dispersion relation of the quarks to calculate the Seebeck and Nernst coefficients in the forthcoming sections.

## 5.3 Thermoelectric response of an anisotropic QCD medium

QGP produced in relativistic heavy ion collisions can possess a significant temperature gradient between its central and peripheral regions. A temperature-gradient and a finite chemical potential in a conducting medium create the necessary conditions for the Seebeck effect. Charge carriers diffuse from regions of higher temperature to regions of lower temperature. This diffusion of charge carriers constitutes the Seebeck current, which leads to the generation of an electric field. The diffusion ceases when the strength of the created electric field balances the thermodynamic gradient. The magnitude of electric field thus generated per unit temperature gradient in the medium is termed as the Seebeck coefficient and is evaluated in the limit of zero electric current [71, 72]. The Seebeck coefficient is a quantitative estimate of the efficiency of conversion of a temperature gradient into electric field by a conducting medium. The sign of the Seebeck coefficient can be used to determine the sign of majority charge carriers in condensed matter systems, as it is positive for positive charge carriers and negative for negative charge carriers. Upcoming experimental programs such as the Facility for Antiproton and Ion Research (FAIR) in Germany and the Nuclotron-based Ion Collider fAcility (NICA) in Russia, where low-energy heavy ion collisions are expected to create a baryon-rich plasma, could be the perfect environment for the aforementioned thermoelectric phenomenon to manifest.

In the presence of a magnetic field, the charged particles drift perpendicular to their original direction of motion due to the Lorentz force acting on them. This leads to a thermocurrent that is transverse to both the direction of temperature gradient and the external magnetic field. This is called the Nernst effect. Like the Seebeck coefficient, the Nernst coefficient is also calculated at the limit of zero electric current, that is, by enforcing the equilibrium condition. The Nernst coefficient can be defined as the electric field induced in the $\hat{x}$ ($\hat{y}$) direction per unit temperature gradient in the $\hat{y}$ ($\hat{x}$) direction, in the presence of a magnetic field pointing in the $\hat{z}$ direction.

Due to a larger expansion rate of the medium along the longitudinal direction compared to the radial direction, one develops a local momentum anisotropy. This anisotropy can be taken into account by introducing an anisotropy parameter $\xi$ in the isotropic distribution function. For the case of weak momentum anisotropy ($\xi < 1$) the distribution function of quarks in an anisotropic medium in the presence of a finite quark chemical potential $\mu$ can be approximated as (suppressing the flavor index) [167].

$$f_a^0(\mathbf{p};T) = \frac{1}{e^{\beta\left(\sqrt{\mathrm{p}^2+\xi(\mathbf{p}\cdot\mathbf{n})^2+\mathrm{m}^2}\,-\mu\right)}+1}\,, \tag{5.9}$$

which can be expanded in a Taylor series about $\xi = 0$ (isotropic case) in the following

way:

$$f_a^0 = f^0 - \xi\beta\frac{(\mathbf{p}.\mathbf{n})^2}{2\epsilon}f^0(1-f^0), \tag{5.10}$$

with

$$f^0 = f_a^0(\xi=0) = \frac{1}{e^{\beta\left(\sqrt{\mathrm{p}^2+\mathrm{m}^2}-\mu\right)}+1}, \tag{5.11}$$

where, $\beta = 1/T$. The anisotropy parameter $\xi$ is defined as

$$\xi = \frac{< p_T^2 >}{2 < p_L^2 >} - 1, \tag{5.12}$$

where, $p_L$ and $\boldsymbol{p_T}$ refer to the longitudinal and transverse components of $\boldsymbol{p}$. The 2 in the denominator denotes the fact that there are two transverse directions with respect to any given vector. The condition of isotropy is when $< p_T^2 >= 2 < p_L^2 >$. For $p_T > 2p_L$, $\xi$ is positive. The aforementioned components are defined with respect to an arbitrary anisotropy direction, denoted by the vector $\boldsymbol{n} = (\sin\alpha, 0, \cos\alpha)$, with $\alpha$ being the angle between the direction of anisotropy and the $z$-axis. This parameter is arbitrary and hence physical quantities should be independent of it. The longitudinal and transverse momentum components are then defined as $p_L = \mathbf{p}.\mathbf{n}, \mathbf{p_T} = \mathbf{p} - \mathbf{n}.(\mathbf{p}.\mathbf{n})$. In spherical polar coordinates, $\mathbf{p} = (p\sin\theta\cos\phi, p\sin\theta\sin\phi, p\cos\theta)$, where, $\theta$ and $\phi$ are the polar and azimuthal angles, respectively. $(\mathbf{p}.\mathbf{n})^{\mathbf{2}} = p^2c(\theta,\alpha,\phi) = p^2(\sin^2\alpha\sin^2\theta\cos^2\phi + \cos^2\alpha\cos^2\theta + \sin 2\alpha\sin\theta\cos\theta\cos\phi)$.

## 5.3.1 In the absence of magnetic field

In this section, we evaluate the thermoelectric response of the anisotropic QGP medium in the absence of a background magnetic field within the kinetic theory framework. Our starting point is the time evolution of a single particle distribution function, which is given by the relativistic Boltzmann transport equation, which, in the relaxation time approximation reads

$$p.\frac{\partial f_a}{\partial \boldsymbol{r}} + q\mathbf{E}.\mathbf{p}\frac{\partial f_a}{\partial p^0} + qp_0\mathbf{E}.\frac{\partial f_a}{\partial \mathbf{p}} = -\frac{p^\mu u_\mu}{\tau}\left(f_a - f_a^0\right), \tag{5.13}$$

where $f_a = f_a^0 + \delta f_a$, $\delta f$ being the deviation in the equilibrium distribution function due to the electric field produced as a result of thermal gradients in the medium, and $f_a^0$ is as defined in Eq.(5.9) with $m$ as defined in Eq.(5.1). $\tau$ is the relaxation time. Once the system is infinitesimally disturbed from equilibrium, it takes on the average an amount of time $\tau$ to revert to equilibrium. The relaxation time has been calculated for the quarks considering $2 \to 2$ scatterings using perturbative QCD as [120]

$$\tau(T) = \frac{1}{5.1T\alpha_s^2 \log\left(\frac{1}{\alpha_s}\right)[1 + 0.12(2N_f + 1)]}, \tag{5.14}$$

where $\alpha_s$ is the running coupling constant (5.3). In order to calculate the deviation $\delta f$, we assume that the system deviates only infinitesimally away from equilibrium, *i.e.* $\delta f \ll f^0$. We then compute the relevant derivatives required to evaluate the left hand side of Eq. (5.13):

$$\frac{\partial f_a^0}{\partial p} = \frac{\partial f^0}{\partial p}\left[1 - \xi\beta\frac{c}{2}\left\{\frac{p^2}{2\beta\epsilon^2} - \frac{2}{\beta} + \frac{p^2}{\epsilon} - \frac{2p^2}{\epsilon}f^0\right\}\right] = \frac{\partial f^0}{\partial p}L_1(p,\xi) \tag{5.15}$$

$$\frac{\partial f_a^0}{\partial \epsilon} = \frac{\partial f^0}{\partial \epsilon}\left[1 - \xi\beta\frac{c}{2}\left\{\frac{p^2}{\epsilon} - \frac{2f^0p^2}{\epsilon} + \frac{p^2}{\beta\epsilon^2}\right\}\right] = \frac{\partial f^0}{\partial \epsilon}L_2(p,\xi) \tag{5.16}$$

$$\frac{\partial f_a^0}{\partial r} = \frac{\partial f^0}{\partial r}\left[1 - \xi\beta\frac{c}{2}\left\{\frac{p^2}{\epsilon} - \frac{2f^0p^2}{\epsilon} - \frac{p^2}{\epsilon(\epsilon - \mu)}\right\}\right] = \frac{\partial f^0}{\partial r}L_3(p,\xi), \tag{5.17}$$

where, $\epsilon = \sqrt{p^2 + m^2}$. As expected, the above derivatives reduce to their isotropic expressions on putting $\xi = 0$. We substitute these derivatives in Eq. (5.13) to get $\delta f$

$$\delta f_a = -\beta^2\tau(\epsilon - \mu)f_0(1 - f_0)\ \frac{\mathbf{p}}{\epsilon}.\nabla\mathbf{T}L_3(p,\xi) + \tau q\frac{\mathbf{E.p}}{\epsilon}\beta f_0(1 - f_0)L_2(p,\xi) \tag{5.18}$$

$$+ \tau q\frac{\mathbf{E.p}}{\epsilon}\beta f_0(1 - f_0)L_1(p,\xi). \tag{5.19}$$

Similarly, we calculate the deviation $\delta\bar{f}$ for the anti quarks. This is done by replacing the quark distribution function $f$ by the anti-quark distribution function $\bar{f}$ and changing the sign of the chemical potential $\mu$.

$$\delta\bar{f}_a = -\beta^2\tau(\epsilon + \mu)\bar{f}_0(1 - \bar{f}_0)L_3(p,\xi)\ \frac{\mathbf{p}}{\epsilon}.\nabla\mathbf{T} + \tau q\frac{\mathbf{E.p}}{\epsilon}L_2(p,\xi)\beta\bar{f}_0(1 - \bar{f}_0) \tag{5.20}$$

$$+ \tau\bar{q}\frac{\mathbf{E.p}}{\epsilon}L_1(p,\xi)\beta\bar{f}_0(1 - \bar{f}_0), \tag{5.21}$$

where, the anti-quark isotropic distribution function is

$$\bar{f}^0 = \frac{1}{e^{\beta(\epsilon+\mu)+1}}. \tag{5.22}$$

We consider the temperature gradient to exist in the $x$-$y$ plane, *i.e.*

$$\boldsymbol{\nabla} T = \frac{\partial T}{\partial x}\hat{\boldsymbol{x}} + \frac{\partial T}{\partial x}\hat{\boldsymbol{y}}. \tag{5.23}$$

Consequently, the induced electric field is also considered to be planar, *i.e.*, $\boldsymbol{E} = E_x\hat{\boldsymbol{x}} + E_y\hat{\boldsymbol{y}}$. The induced four current due to a single quark flavour can be written as

$$J^\mu = g\int \frac{d^3p}{(2\pi)^3}\frac{p^\mu}{\epsilon}[q\,\delta f_a + \bar{q}\,\delta\bar{f}_a], \tag{5.24}$$

where, $g$ is the quark degeneracy factor.

Now substituting $\delta f$ and $\delta\bar{f}$ in Eq.(5.24) and putting the induced current to zero in the steady state we get

$$\boldsymbol{E} = S\boldsymbol{\nabla} T. \tag{5.25}$$

Here, $S$ is the *individual* Seebeck coefficient, *i.e.* the Seebeck coefficient of a hypothetical medium consisting of a single quark flavor.

$$S = -\frac{H_1}{H_2}, \tag{5.26}$$

where,

$$H_1 = \frac{qg\beta^2}{3}\int \frac{d^3p}{(2\pi)^3}\frac{p^2\tau}{\epsilon^2}\Big\{(\epsilon+\mu)\bar{f}_0(1-\bar{f}_0)L_3(p,\xi) - (\epsilon-\mu)f_0(1-f_0)L_3(p,\xi)\Big\} \tag{5.27}$$

$$H_2 = \frac{q^2g\beta}{3}\int \frac{d^3p}{(2\pi)^3}\frac{p^2\tau}{\epsilon^2}\Big\{f_0(1-f_0)L_1(p,\xi) + \bar{f}_0(1-\bar{f}_0)L_1(p,\xi)\Big\}. \tag{5.28}$$

To evaluate the Seebeck coefficient of the composite medium, we need to take into account the total current due to multiple quark species. The spatial part of the total 4-current is given by

$$\boldsymbol{J} = \sum_i q_i g_i \int \frac{d^3\mathrm{p}}{(2\pi)^3\epsilon}\boldsymbol{p}\left[\delta f_a^i - \overline{\delta f_a^i}\right] \tag{5.29}$$

Setting the $x$ and $y$ components of the above equation to zero yields the following equations:

$$\sum_{i=u,d}\left[(H_2)_i E_x + (H_1)_i \frac{\partial T}{\partial x}\right] = 0, \tag{5.30}$$

$$\sum_{i=u,d}\left[(H_2)_i E_y + (H_1)_i \frac{\partial T}{\partial y}\right] = 0, \tag{5.31}$$

Solving the above equations and comparing with Eq.(5.25), we get

$$\begin{pmatrix} E_x \\ E_y \end{pmatrix} = \begin{pmatrix} S_{\rm tot} & 0 \\ 0 & S_{\rm tot} \end{pmatrix} \begin{pmatrix} \frac{\partial T}{\partial x} \\ \frac{\partial T}{\partial y}, \end{pmatrix} \tag{5.32}$$

where, the total Seebeck coefficient $S_{tot}$ is given by

$$S_{tot} = -\frac{C_1 C_2}{C_2^2}, \tag{5.33}$$

with

$$C_1 = \sum_{i=u,d}(H_1)_i, \qquad C_2 = \sum_{i=u,d}(H_2)_i \tag{5.34}$$

In the next subsection, we will see how the weak magnetic field in the background modulates the thermoelectric response of the hot QCD medium.

### 5.3.2 In the presence of weak magnetic field

The RBTE [Eq.(5.13)] in 3-vector notation, in the presence of the Lorentz force can be written as

$$\frac{\partial f_a}{\partial t} + \boldsymbol{v}.\frac{\partial f_a}{\partial \boldsymbol{r}} + q\left(\boldsymbol{E} + \boldsymbol{v}\times\boldsymbol{B}\right)\cdot\frac{\partial f_a}{\partial \boldsymbol{p}} = -\frac{\delta f_a}{\tau}, \tag{5.35}$$

In the first approximation, we replace $f_a$ by $f_0$ in the L.H.S. above, similar to the $B = 0$ case. However, we note that the equilibrium distribution function $f_a^0$ does not contribute to the Lorentz force as $\frac{\partial f_a^0}{\partial \boldsymbol{p}} \propto \boldsymbol{v}$, hence, $(\boldsymbol{v}\times\boldsymbol{B})\cdot\frac{\partial f_a^0}{\partial \boldsymbol{p}} = 0$. The contribution to the Lorentz force thus comes solely from $\delta f_a$. Additionally, we work in the static approximation where, both $f_a$ and $f_a^0$ do not depend on time, so that the first term in Eq.(5.35) drops out. The RBTE thus gets reduced to

$$\boldsymbol{v}\cdot\frac{\partial f_a^0}{\partial \boldsymbol{r}}+q\ \boldsymbol{E}\cdot\frac{\partial f_a^0}{\partial \boldsymbol{p}}+q(\boldsymbol{v}\times\boldsymbol{B})\frac{\partial(\delta f_a)}{\partial \boldsymbol{p}}=-\frac{\delta f_a}{\tau} \tag{5.36}$$

As earlier, $f_a=f_a^0+\delta f_a$ with $\delta f_a \ll f_a^0$. To solve for $\delta f_a$ we take an ansatz, similar to a trial solution for solving any differential equation [89]:

$$\delta f_a=(\mathbf{p}.\ )\ \frac{\partial f_a^0}{\partial \epsilon} \tag{5.37}$$

where,

$$=\alpha_1\mathbf{E}+\alpha_2\mathbf{b}+\alpha_3(\mathbf{E}\times\mathbf{b})+\alpha_4\nabla\mathbf{T}+\alpha_5(\nabla\mathbf{T}\times\mathbf{b})+\alpha_6(\nabla\mathbf{T}\times\mathbf{E}). \tag{5.38}$$

Using Eqs.(5.38) and (5.37) in Eq.(5.36), we get,

$$\begin{aligned}\beta^2(\epsilon-\mu)f_0(1-f_0)L_3(p,\xi)\ \mathbf{v}.\nabla\mathbf{T}-\beta q f_0(1-f_0)L_1(p,\xi)\ \mathbf{v}.\mathbf{E}-\beta q f_0(1-f_0)L_2(p,\xi)\\ \{-\alpha_1|B|\mathbf{v}.(\mathbf{E}\times\mathbf{b})+\alpha_3|B|\ \mathbf{v}.\mathbf{E}-\alpha_4|B|\mathbf{v}.(\nabla\mathbf{T}\times\mathbf{b})+\alpha_5|B|\ \mathbf{v}.\nabla\mathbf{T}\}=\\ f^0(1-f^0)L_2(p,\xi)\frac{\epsilon}{\tau}\left\{\alpha_1\mathbf{v}.\mathbf{E}+\alpha_2\mathbf{v}.\mathbf{b}+\alpha_3\mathbf{v}.(\mathbf{E}\times\mathbf{b})+\alpha_4\mathbf{v}.\nabla\mathbf{T}+\alpha_5\mathbf{v}.(\nabla\mathbf{T}\times\mathbf{b})\right\}\end{aligned} \tag{5.39}$$

Now comparing the coefficients of the same tensor structure of from both sides, we get

$$\frac{\epsilon}{\tau}\alpha_1L_2(p,\xi)=-\alpha_3q|B|L_2(p,\xi)-qL_1(p,\xi) \tag{5.40}$$
$$\frac{\epsilon}{\tau}\alpha_2L_2(p,\xi)=0 \tag{5.41}$$
$$\frac{\epsilon}{\tau}\alpha_3L_2(p,\xi)=\alpha_1q|B|L_2(p,\xi) \tag{5.42}$$
$$\frac{\epsilon}{\tau}\alpha_4L_2(p,\xi)=\beta(\epsilon-\mu)L_3(p,\xi)-\alpha_5q|B|L_2(p,\xi) \tag{5.43}$$
$$\frac{\epsilon}{\tau}\alpha_5L_2(p,\xi)=\alpha_4q|B|L_2(p,\xi) \tag{5.44}$$

We find the values of $\alpha_1$, $\alpha_2$, $\alpha_3$, $\alpha_4$ and $\alpha_5$ from the above equations, which come out to be

$$\alpha_1 = -\frac{\tau}{\epsilon}\frac{q}{(1+\omega_c^2\tau^2)}\frac{L_1(p,\xi)}{L_2(p,\xi)} \tag{5.45}$$

$$\alpha_2 = 0 \tag{5.46}$$

$$\alpha_3 = -\frac{\tau^2}{\epsilon}\frac{\omega_c q}{(1+\omega_c^2\tau^2)}\frac{L_1(p,\xi)}{L_2(p,\xi)} \tag{5.47}$$

$$\alpha_4 = \frac{\tau}{\epsilon}\frac{\beta(\epsilon-\mu)}{(1+\omega_c^2\tau^2)}\frac{L_3(p,\xi)}{L_2(p,\xi)} \tag{5.48}$$

$$\alpha_5 = \frac{\tau^2}{\epsilon}\frac{\omega_c\beta(\epsilon-\mu)}{(1+\omega_c^2\tau^2)}\frac{L_3(p,\xi)}{L_2(p,\xi)} \tag{5.49}$$

$$\delta f = \frac{\tau}{(1+\omega_c^2\tau^2)}\mathbf{p}.\Bigg\{\left(-\frac{q}{\epsilon}\mathbf{E} - \frac{q\tau\omega_c}{\epsilon}(\mathbf{E}\times\mathbf{b})\right)\frac{L_1(p,\xi)}{L_2(p,\xi)} \tag{5.50}$$

$$+\left(\frac{\beta(\epsilon-\mu)}{\epsilon}\nabla\mathbf{T} + \frac{\beta\tau\omega_c(\epsilon-\mu)}{\epsilon}(\nabla\mathbf{T}\times\mathbf{b})\right)\frac{L_3(p,\xi)}{L_2(p,\xi)}\Bigg\}\frac{\partial f^0}{\partial\epsilon}L_2(p,\xi) \tag{5.51}$$

Similarly,we can compute the deviation in the anti quarks distribution function as

$$\delta\bar{f} = \frac{\tau}{(1+\omega_c^2\tau^2)}\mathbf{p}.\Bigg\{\left(\frac{q}{\epsilon}\mathbf{E} - \frac{q\tau\omega_c}{\epsilon}(\mathbf{E}\times\mathbf{b})\right)\frac{\bar{L}_1(p,\xi)}{\bar{L}_2(p,\xi)} \tag{5.52}$$

$$+\left(\frac{\beta(\epsilon+\mu)}{\epsilon}\nabla\mathbf{T} - \frac{\beta\tau\omega_c(\epsilon+\mu)}{\epsilon}(\nabla\mathbf{T}\times\mathbf{b})\right)\frac{\bar{L}_3(p,\xi)}{\bar{L}_2(p,\xi)}\Bigg\}\frac{\partial\bar{f}^0}{\partial\epsilon}L_2(p,\xi) \tag{5.53}$$

The $x$ and $y$ components of the induced current density can be written as

$$J_x = I_1E_x + I_2E_y + I_3\frac{\partial T}{\partial x} + I_4\frac{\partial T}{\partial y} \tag{5.54}$$

$$J_y = -I_2E_x + I_1E_y - I_4\frac{\partial T}{\partial x} + I_3\frac{\partial T}{\partial y} \tag{5.55}$$

The electric field components are related to the temperature gradients, Seebeck and Nernst coefficients via a matrix equation

$$\begin{pmatrix} E_x \\ E_y \end{pmatrix} = \begin{pmatrix} S & N|B| \\ -N|B| & S \end{pmatrix}\begin{pmatrix} \frac{\partial T}{\partial x} \\ \frac{\partial T}{\partial y} \end{pmatrix} \tag{5.56}$$

where

$$I_1 = \frac{q^2 g\beta}{3}\int \frac{d^3p}{(2\pi)^3}\,\frac{p^2}{\epsilon^2}\frac{\tau}{(1+\omega_c^2\tau^2)}\bigg\{f_0(1-f_0)L_1(p,\xi)+\bar{f}_0(1-\bar{f}_0)L_1(p,\xi)\bigg\} \tag{5.57}$$

$$I_2 = \frac{q^2 g\beta}{3}\int \frac{d^3p}{(2\pi)^3}\,\frac{p^2}{\epsilon^2}\frac{\omega_c\tau^2}{(1+\omega_c^2\tau^2)}\bigg\{f_0(1-f_0)L_1(p,\xi)-\bar{f}_0(1-\bar{f}_0)L_1(p,\xi)\bigg\} \tag{5.58}$$

$$I_3 = \frac{qg\beta^2}{3}\int \frac{d^3p}{(2\pi)^3}\,\frac{p^2}{\epsilon^2}\frac{\tau}{(1+\omega_c^2\tau^2)}\bigg\{(\epsilon+\mu)\bar{f}_0(1-\bar{f}_0)L_3(p,\xi) \tag{5.59}$$

$$-(\epsilon-\mu)f_0(1-f_0)L_3(p,\xi)\bigg\} \tag{5.60}$$

$$I_4 = \frac{qg\beta^2}{3}\int \frac{d^3p}{(2\pi)^3}\,\frac{p^2}{\epsilon^2}\frac{\omega_c\tau^2}{(1+\omega_c^2\tau^2)}\bigg\{-(\epsilon-\mu)f_0(1-f_0)L_3(p,\xi) \tag{5.61}$$

$$-(\epsilon+\mu)\bar{f}_0(1-\bar{f}_0)L_3(p,\xi)\bigg\} \tag{5.62}$$

In the state of equilibrium, the components of the induced current density along $x$ and $y$ direction vanish *i.e.* $J_x = J_y = 0$. We can write from Eqns. (5.54) and (5.55)

$$I_1E_x + I_2E_y + I_3\frac{\partial T}{\partial x} + I_4\frac{\partial T}{\partial y} = 0, \tag{5.63}$$

$$-I_2E_x + I_1E_y - I_4\frac{\partial T}{\partial x} + I_3\frac{\partial T}{\partial y} = 0, \tag{5.64}$$

We can further write Eqns (5.63) and (5.64) as

$$E_x = \left(-\frac{I_1I_3+I_2I_4}{I_1^2+I_2^2}\right)\frac{\partial T}{\partial x} + \left(-\frac{I_2I_3-1_1I_4}{I_1^2+I_2^2}\right)\frac{\partial T}{\partial y}, \tag{5.65}$$

$$E_y = \left(-\frac{I_1I_3+I_2I_4}{I_1^2+I_2^2}\right)\frac{\partial T}{\partial y} - \left(-\frac{I_2I_3-I_1I_4}{I_1^2+I_2^2}\right)\frac{\partial T}{\partial x}, \tag{5.66}$$

where,

$$S = -\frac{(I_1I_3+I_2I_4)}{I_1^2+I_2^2}, \tag{5.67}$$

$$N|B| = \frac{(I_2I_3-I_1I_4)}{I_1^2+I_2^2}. \tag{5.68}$$

The integrals $I_2$ and $I_4$ vanishes in the absence of the magnetic field. As a result, the Nernst coefficient also vanishes.

In what follows, we will compute the Seebeck and Nernst coefficients for the medium composed of the $u$ and $d$ quarks. In the medium The $x$ and $y$ components of the induced current in the medium can then be written as the sum of the individual flavour contributions as

$$J_x = \sum_{i=u,d} \left[ (I_1)_i E_x + (I_2)_i E_y + (I_3)_i \frac{\partial T}{\partial x} + (I_4)_i \frac{\partial T}{\partial y} \right], \tag{5.69}$$

$$J_y = \sum_{a=u,d} \left[ -(I_2)_i E_x + (I_1)_i E_y - (I_4)_i \frac{\partial T}{\partial x} + (I_3)_i \frac{\partial T}{\partial y} \right]. \tag{5.70}$$

We extract the Seebeck and Nernst coefficients for the QCD medium composed of u and d quarks by imposing the equilibrium condition ($i.e.$ putting $J_x = J_y = 0$) as

$$S_{tot}^{B'} = -\frac{(K_1 K_3 + K_2 K_4)}{K_1^2 + K_2^2}, \tag{5.71}$$

$$N|B| = \frac{(K_2 K_3 - K_1 K_4)}{K_1^2 + K_2^2}. \tag{5.72}$$

where,

$$K_1 = \sum_{i=u,d} (I_1)_i, \qquad K_2 = \sum_{i=u,d} (I_2)_i,$$
$$K_3 = \sum_{i=u,d} (I_3)_i, \qquad K_4 = \sum_{i=u,d} (I_4)_i. \tag{5.73}$$

## 5.4 Results and discussion

We begin with the results obtained at $B = 0$, followed by those obtained at finite $B$. Figs. (5.1a) and (5.1b) show the temperature variation of individual seebeck coefficients. That is to say, if the medium were composed exclusively of a single species of quark ($u$ or $d$), the Seebeck coefficients would vary with temperature as shown in the aforementioned figures. The positively charged $u$ quark gives rise to a positive Seebeck coefficient, while the same is negative in case of the negatively charged $d$ quark, which concurs with previous results [86, 90]. Positivity of the Seebeck coefficient indicates that the induced electric field is along the direction of the temperature gradient, whereas a negative value indicates that the induced electric field is in the direction opposite to the direction of the temperature gradient. It should be noted that we have used the convention where the direction of increasing temperature is considered positive. The magnitudes of both the individual Seebeck coefficients ($S_u$, $S_d$) decrease with temperature. Importantly, the magnitudes also decrease with the strength of

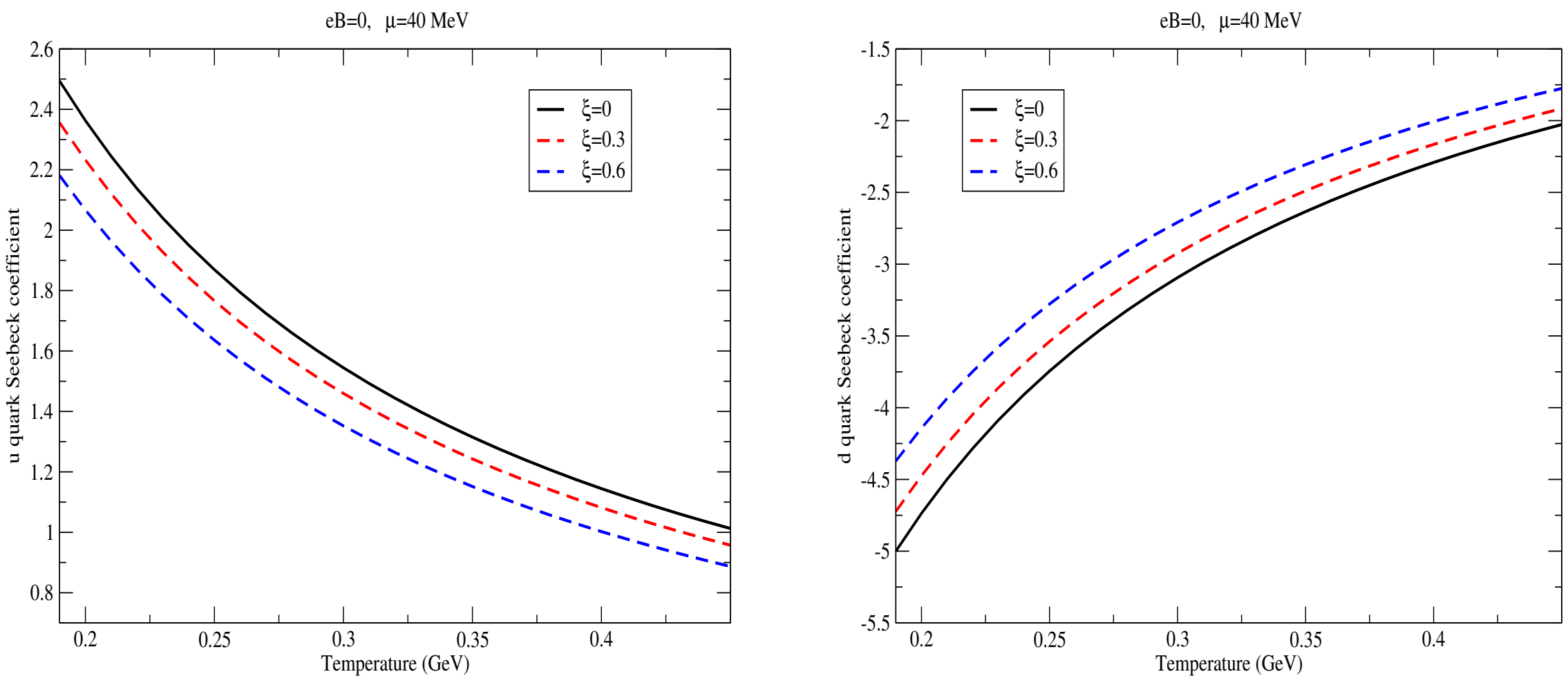

**(a)** **(b)**

**Figure 5.1:** (a) Temperature dependence of $u$ quark Seebeck coefficient in the absence of $B$, at a fixed value of $\mu$. (b) Temperature dependence of $d$ quark Seebeck coefficient in the absence of $B$, at a fixed value of $\mu$. The different curves correspond to different values of $\xi$.

anisotropy parameter $\xi$. $S_u$ decreases by 5.32% while going from $\xi = 0$ to $\xi = 0.3$, averaged over the entire temperature range. From $\xi = 0.3$ to $\xi = 0.6$, the decrease is 7.11%. Interestingly, the corresponding values for $S_d$ are identical upto two decimal places.

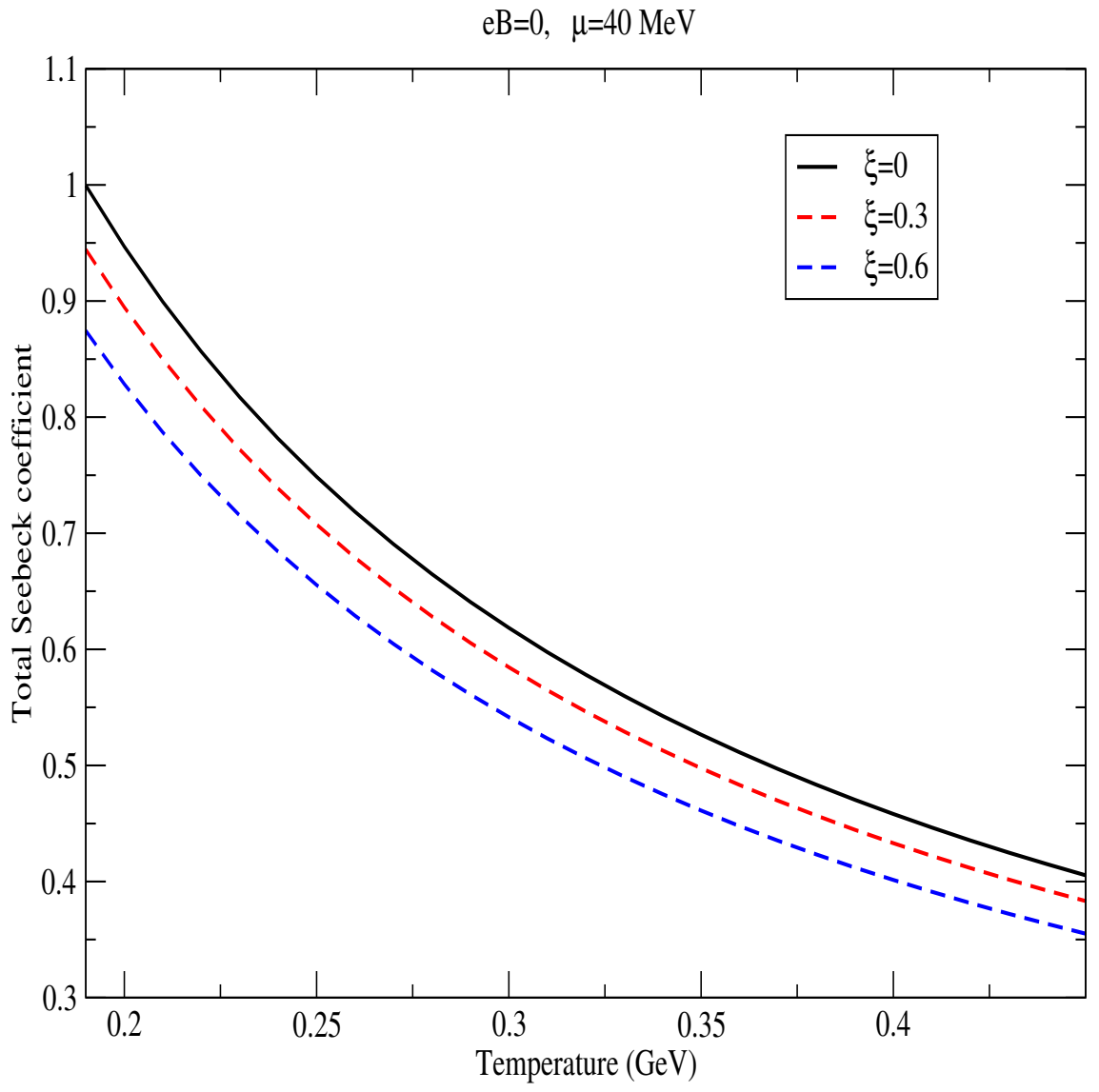

**Figure 5.2:** Seebeck coefficient of the composite medium in the absence of $B$ as a function of temperature.

Fig.(5.2) shows the Seebeck coefficient of the composite medium composed of $u$ and $d$ quarks-the total Seebeck coefficient ($S_{\text{tot}}$), as a function of temperature. $S_{\text{tot}}$ is

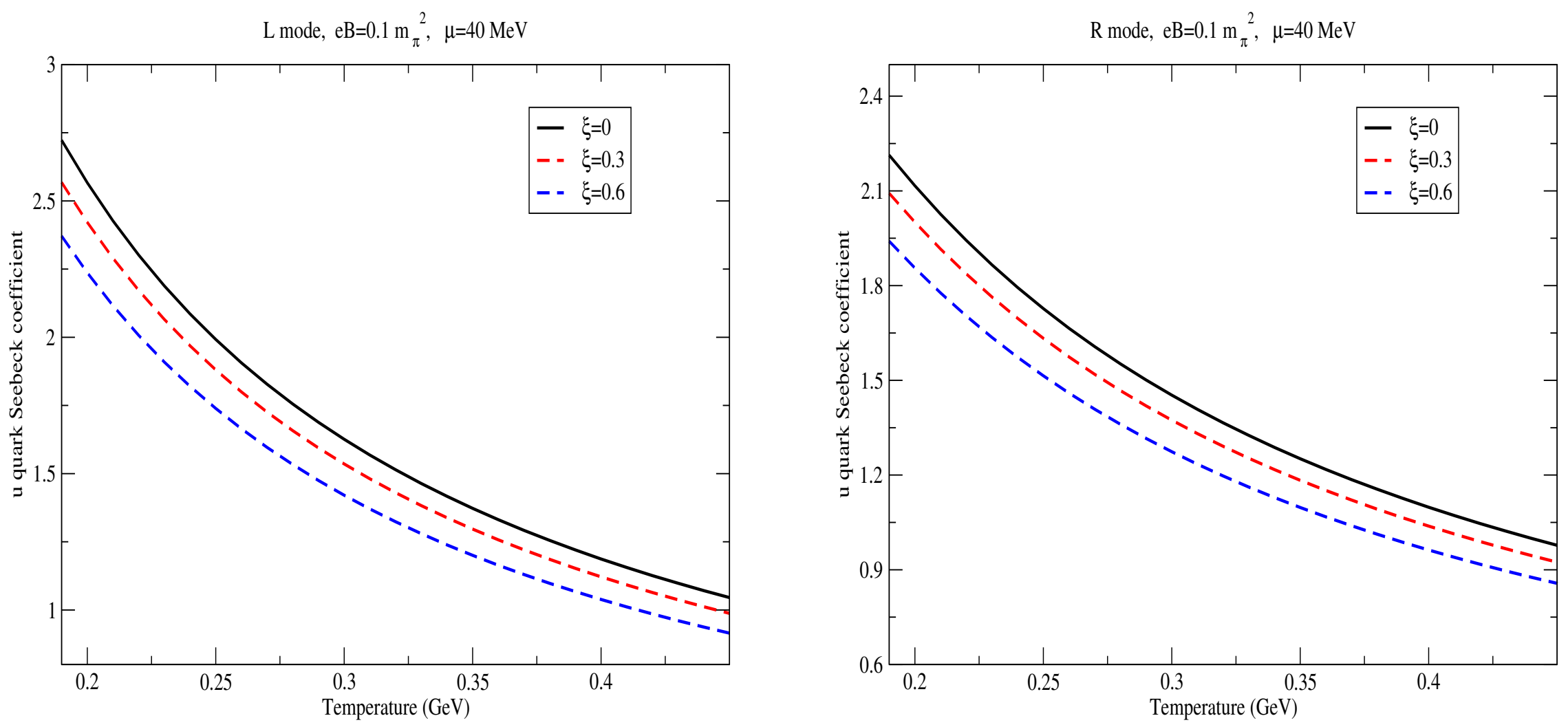

(a) (b)

**Figure 5.3:** (a) Temperature dependence of $u$ quark $L$ mode Seebeck coefficient at fixed values of $eB$ and $\mu$. (b) Temperature dependence of $u$ quark $R$ mode Seebeck coefficient at fixed values of $eB$ and $\mu$. The different curves correspond to different values of $\xi$

positive and decreases with temperature. So, the induced electric field points in along the temperature gradient. Also, it is to be noted that even though the magnitude of $d$ quark Seebeck coefficient is larger than that of the $u$ quark at a given temperature, the total coefficient is positive. This reflects the fact that a $u$ quark carries double the electric charge than a $d$ quark. Again, a finite anisotropy decreases the magnitude of $S_{\text{tot}}$. It decreases by 5.31% in going from $\xi = 0$ to $\xi = 0.3$, and by 7.11% when $\xi$ changes from 0.3 to 0.6. The decrease in the magnitude of transport coefficients with anisotropy has been observed earlier [58].

Figures (5.3) and (5.4) show the variation with temperature of $u$ quark and $d$ quark Seebeck coefficients, respectively. The coefficient is positive for positively charged $u$ quark and negative for the negatively charged $d$ quark, which is along expected lines. As can be seen from the figures, the coefficient magnitudes are decreasing functions of temperature. One can also see the effect of anisotropy on the coefficient magnitudes. Compared to the isotropic ($\xi = 0$) result, the anisotropic medium leads to a lesser value of the Seebeck coefficient magnitude for a particular value of temperature and magnetic field. Also, with the increase in the extent of anisotropy, the coefficient magnitude decreases. Comparing the graphs for the $L$ and $R$ modes shows that the Seebeck coefficient magnitudes for the $R$ mode are slightly smaller than its $L$ mode counterpart. This is due to the smaller effective mass of the $R$ mode compared to the $L$ mode at the same values of temperature and magnetic field. Specifically, the difference

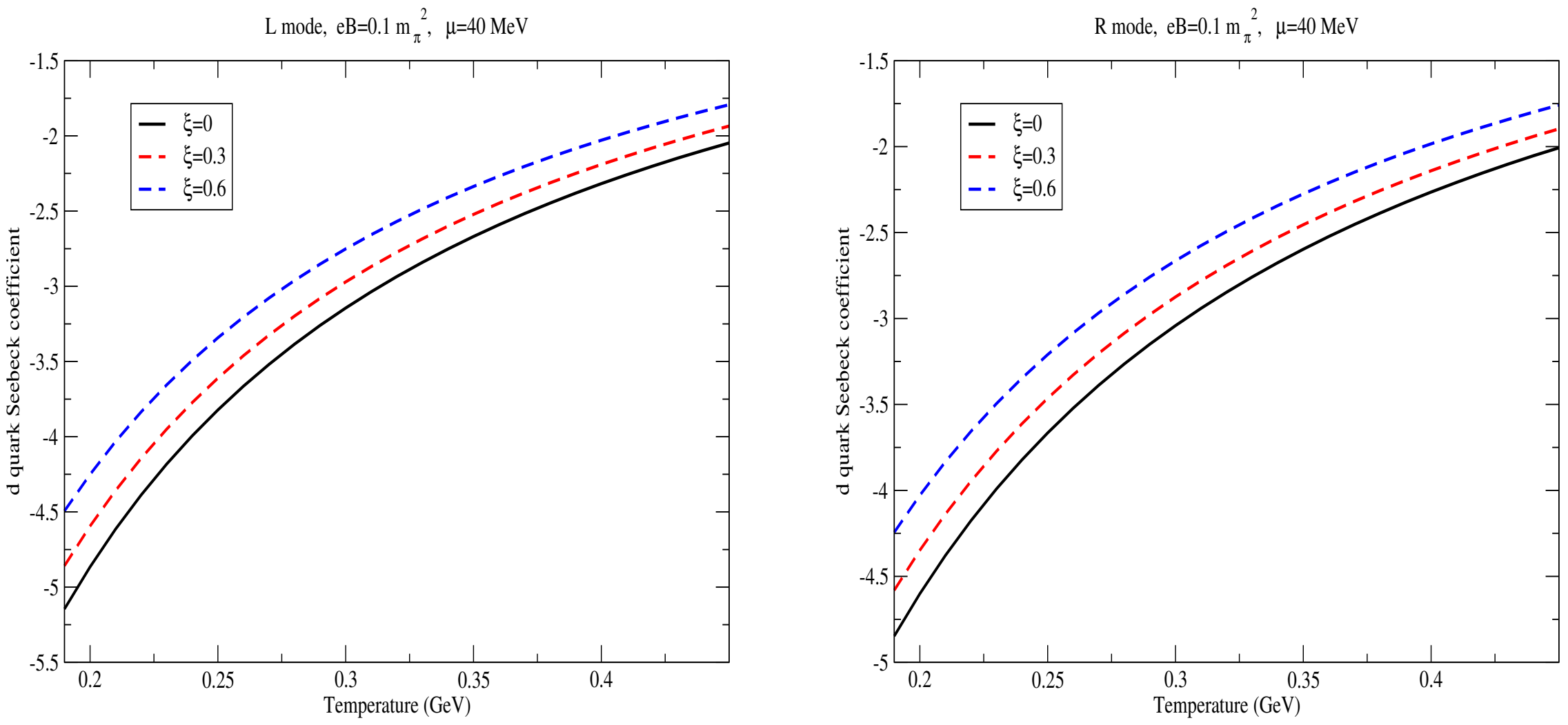

**(a)** **(b)**

**Figure 5.4:** (a) Temperature dependence of $d$ quark $L$ mode Seebeck coefficient at fixed values of $eB$ and $\mu$. (b) Temperature dependence of $d$ quark $R$ mode Seebeck coefficient at fixed values of $eB$ and $\mu$. The different curves correspond to different values of $\xi$.

between the magnitudes of the $L$ and $R$ modes is greatest at lower temperatures and decreases as the temperature rises. We can take the average in the entire temperature range and define an average percentage change corresponding to each value of $\xi$. We found that the percentage decrease in the $u$ quark coefficient magnitude as one goes from the $L$ mode to the $R$ mode is ∼9.95% for $\xi = 0$, ∼9.86% for $\xi = 0.3$ and ∼9.68% for $\xi = 0.6$. This shows that as the strength of anisotropy increases, the difference between the $L$ and $R$ magnitudes decreases. Taking the mean of the three values corresponding to the three $\xi$ values, we arrive at a mean percentage decrease value of $\sim$ 9.83%. Interestingly, for the $d$ quark Seebeck coefficient, these values are $\sim$ 3.06% for $\xi = 0$, $\sim$ 3.03% for $\xi = 0.3$ and $\sim$ 2.97% for $\xi = 0.6$. So, the effect of the difference in $L$ and $R$ mode quasiparticle masses is suppressed for the $d$ quark Seebeck coefficient compared to the $u$ quark result.

Figures (5.5a) and (5.5b) show the temperature dependence of Seebeck coefficient of the composite medium composed of $u$ and $d$ quarks. The total Seebeck coefficient is positive, which means that the induced electric field of the medium points in the direction of increasing temperature. The coefficient magnitude decreases with temperature and also decreases with increase in the anisotropy parameter $\xi$. As in the case of the individual coefficients, the $R$ mode Seebeck coefficient magnitude is slightly smaller than that of the $L$ mode, which again can be attributed to the smaller effective mass

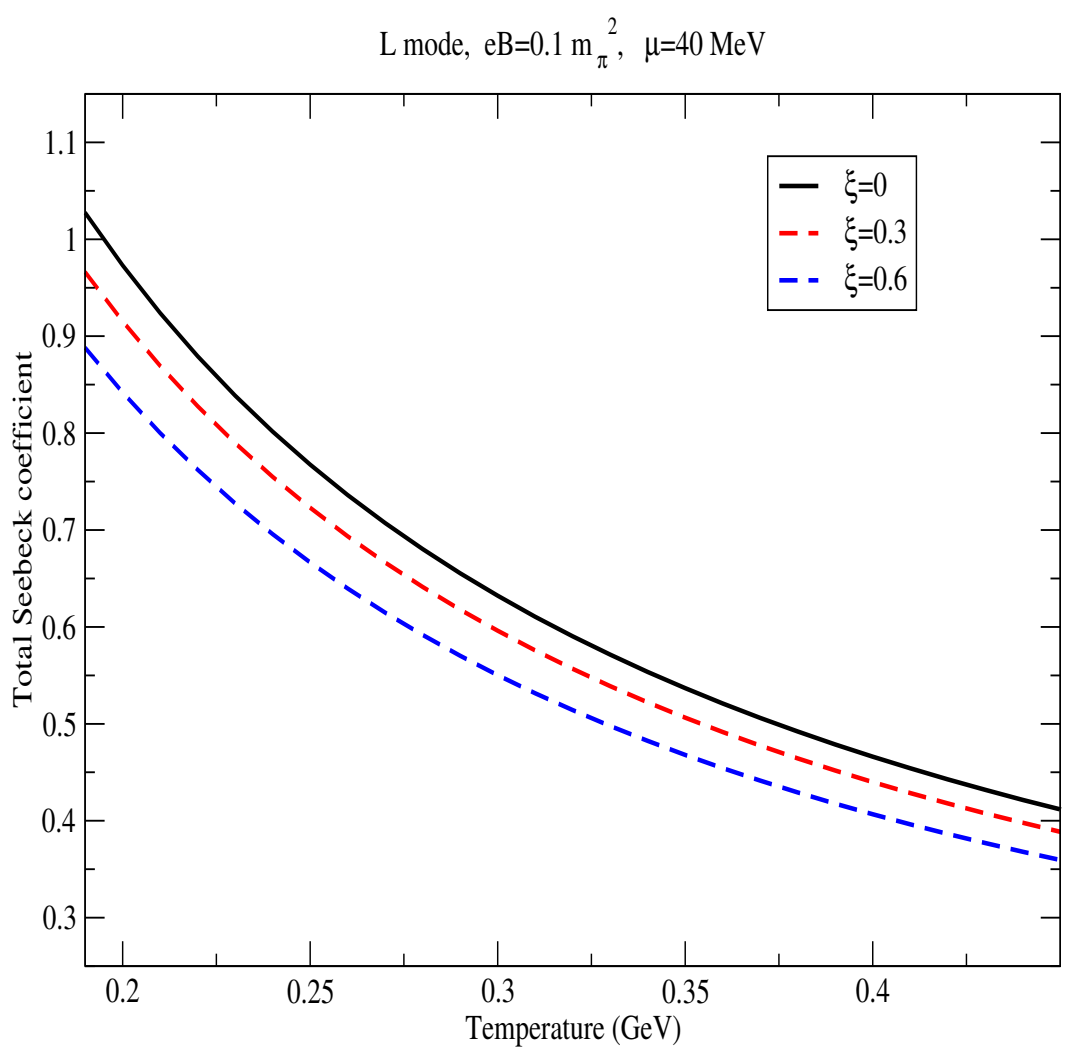

**(a)**

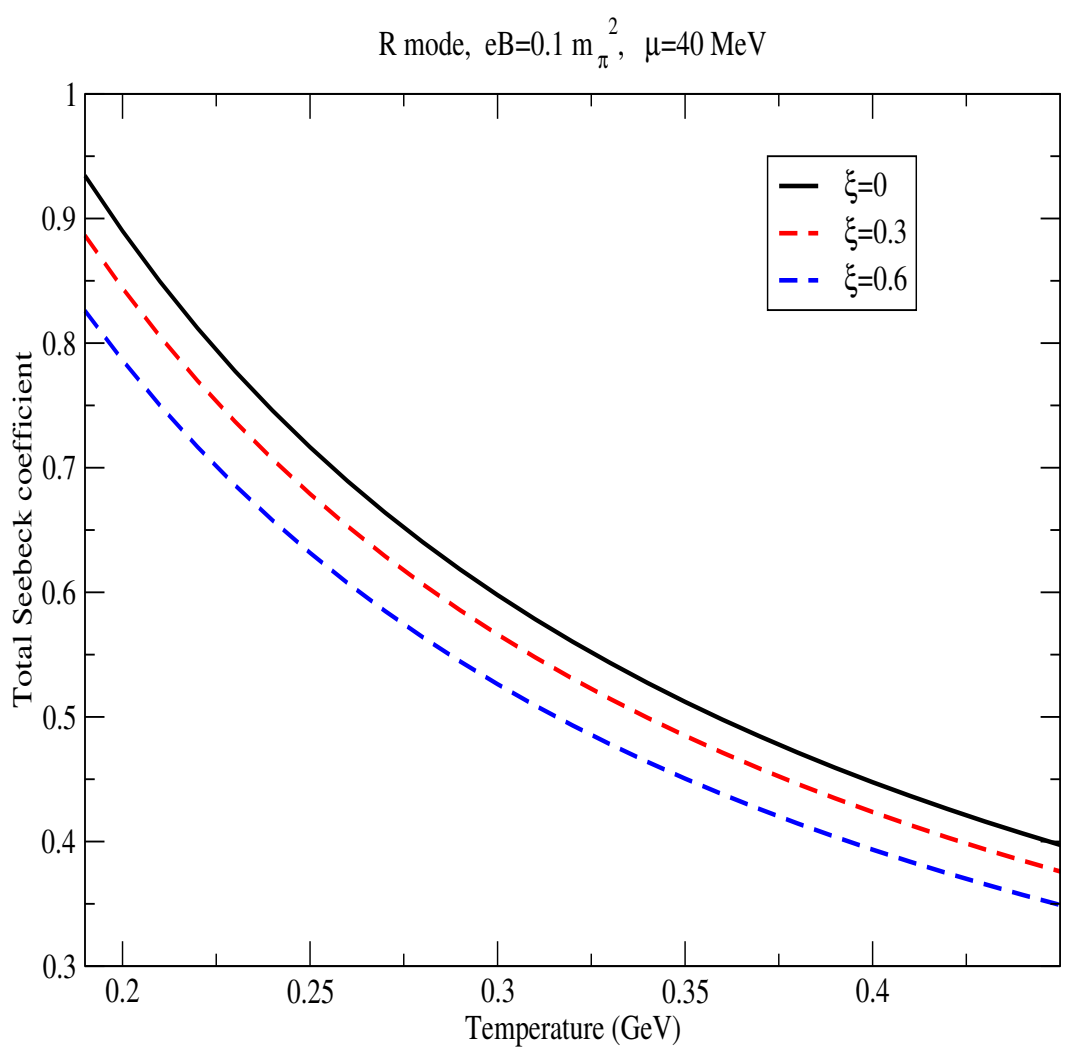

**(b)**

**Figure 5.5:** (a) Variation of total Seebeck coefficient of the medium ($L$ mode) with temperature at fixed values of $eB$ and $\mu$. (b) Variation of total Seebeck coefficient of the medium ($R$ mode) with temperature at fixed values of $eB$ and $\mu$. The different curves correspond to different values of $\xi$.

of the $R$ mode quasiquark. The average (over $T$ and $\xi$ both) percentage decrease as one goes from the $L$ mode to the $R$ mode composite Seebeck coefficient is $\sim 4.61\%$. Here also, the percentage difference between the two modes decreases with increasing anisotropy strength. Thus, anisotropic expansion of the medium hinders the ability of a thermal QCD medium to convert a temperature gradient to electric field.

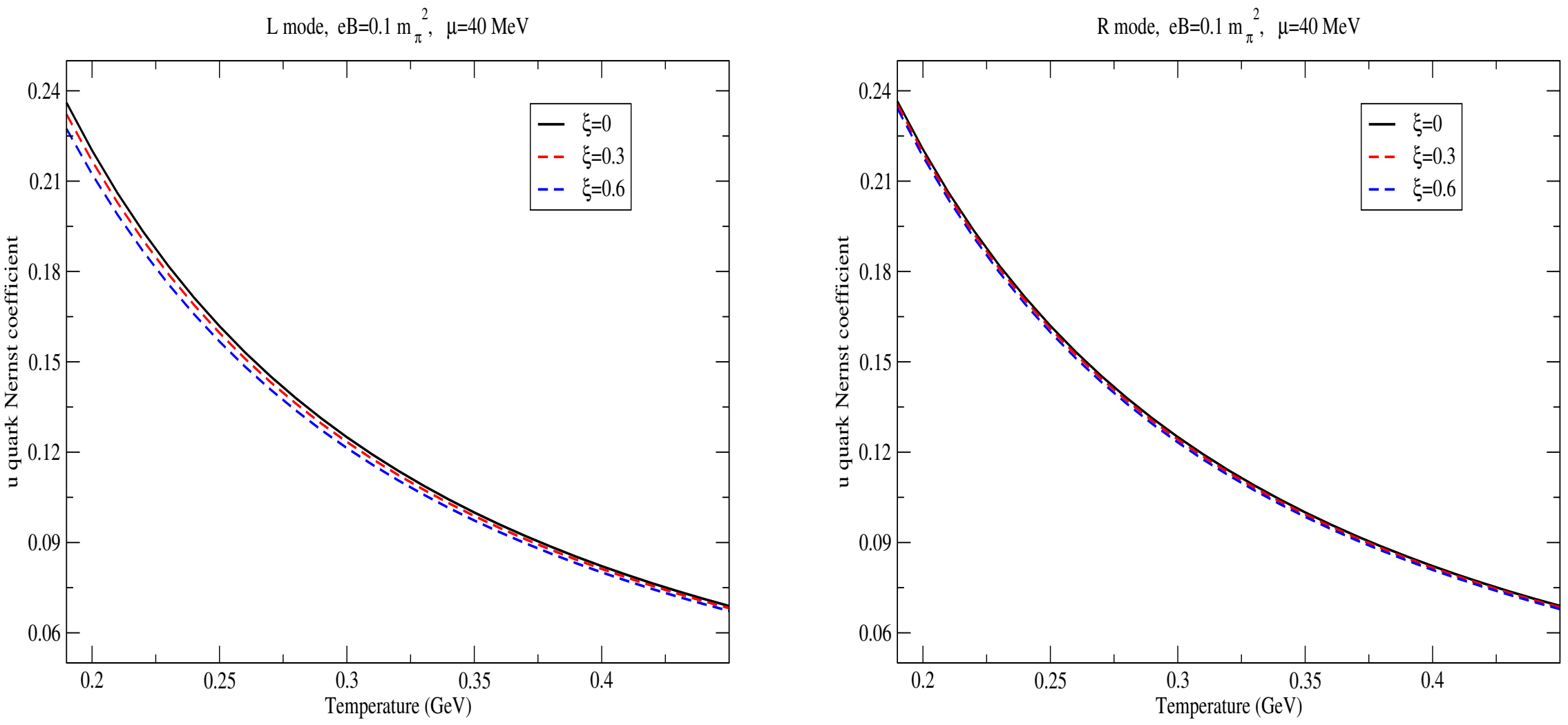

**(a)** **(b)**

**Figure 5.6:** (a) Variation of $u$ quark $L$ mode Nernst coefficient with temperature at fixed values of $eB$ and $\mu$. (b) Variation of $u$ quark $R$ mode Nernst coefficient with temperature at fixed values of $eB$ and $\mu$. The different curves correspond to different values of $\xi$.

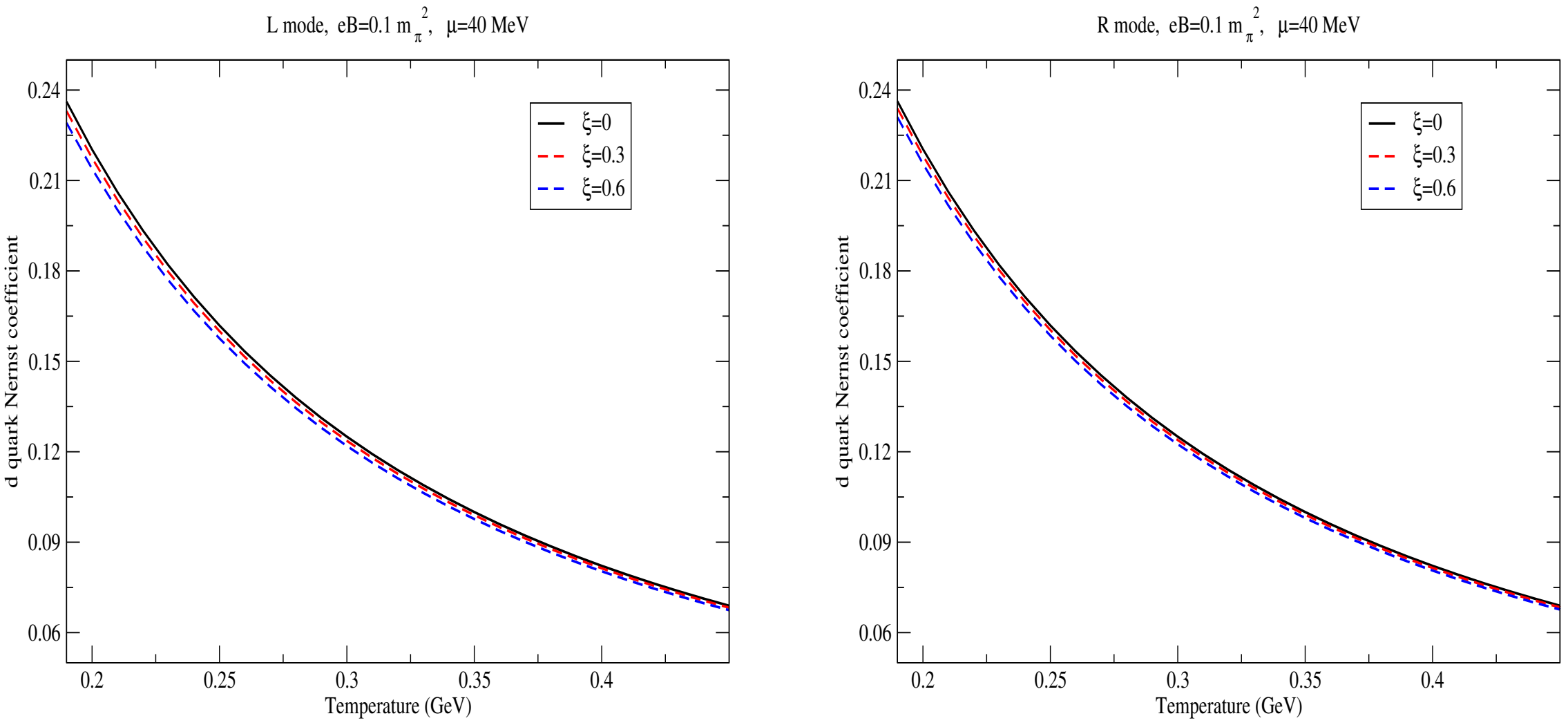

**(a)** **(b)**

**Figure 5.7:** (a) Variation of $d$ quark $L$ mode Nernst coefficient with temperature at fixed values of $eB$ and $\mu$. (b) Variation of $d$ quark $R$ mode Nernst coefficient with temperature at fixed values of $eB$ and $\mu$. The different curves correspond to different values of $\xi$..

Figures (5.6a) and (5.6b) show the variation with temperature, of the Nernst coefficient corresponding to the $L$ and $R$ modes, respectively, of a medium composed exclusively of $u$ quarks. Similar to the Seebeck coefficient, the Nernst coefficient magnitude decreases with temperature and also decreases with the value of the anisotropy parameter $\xi$. Comparison between the graphs corresponding to the $L$ and $R$ modes

reveals that $R$ mode magnitudes are slightly more than that of the $L$ mode. This trend is opposite to the individual Seebeck coefficient case where $R$ mode magnitudes were less than their $L$ mode counterparts. Also, the extent of difference between the two modes is much smaller compared to the Seebeck coefficient case. Specifically, for the $u$ quark nernst coefficient, the temperature averaged percentage decrease as one goes from $R$ mode to $L$ mode is 0.051% for $\xi = 0$, $\sim 0.65\%$ for $\xi = 0.3$ and $\sim 1.43\%$ for $\xi = 0.6$. Taking the mean of the values corresponding to the different $\xi$ values gives us an average value of 0.71%. Compared to the $u$ quark Seebeck coefficient, these values are almost an order of magnitude smaller. Also, unlike in the case of the Seebeck coefficients, the percentage change between the $L$ and $R$ mode increases sharply with increase in the strength of anisotropy,

The $d$ quark Nernst coefficients corresponding to the $L$ and $R$ modes shown in Figs.(5.7a) and (5.7b) respectively show that the magnitudes of Nernst coefficients corresponding to both the modes decreases with temperature. Also, the magnitudes decrease with increase in the degree of anisotropy, parameterized by the value of $\xi$. Similar to the $u$ quark Nernst coefficient, the $L$ mode absolute values of $d$ quark Nernst coefficients are smaller than their $R$ mode counterparts; the averaged (over $T$) percentage decrease being 0.015% for $\xi = 0$, $\sim 0.192\%$ for $\xi = 0.3$ and $\sim 0.422\%$ for $\xi = 0.6$. Averaging also over the different $\xi$ values yields a mean percentage decrease value of 0.21%. This trend is similar to the individual Seebeck coefficient case where the percentage changes for $d$ quark Seebeck coefficient was much smaller than that for the $u$ quark. Also from the magnitudes of the individual Seebeck and Nernst coefficients (both $u$ and $d$ quarks), it can be seen that magnitudes of the Nernst coefficients are $\sim 1$ order of magnitude smaller.

The major point of difference with the Seebeck coefficient, however, is that the Nernst coefficient for both $u$ and $d$ quarks is positive. To understand this, let us consider positively charged quarks moving in the $+\hat{x}$ direction under the influence of a temperature gradient. On application of a magnetic field in the $\hat{z}$ direction, the Lorentz force will cause them to drift in the $-\hat{y}$ direction. This will result in an induced electric field in the $+\hat{y}$ direction. If the electric charge of the quarks were negative instead, they would drift towards the $+\hat{y}$ direction and pile up there. This would again lead to an induced electric field in the $+\hat{y}$ direction. Thus, the direction of the induced field does not depend on the sign of the electric charge of the quark, unlike in the case of Seebeck coefficient, resulting in positive Nernst coefficients for both $u$ and $d$ quarks.

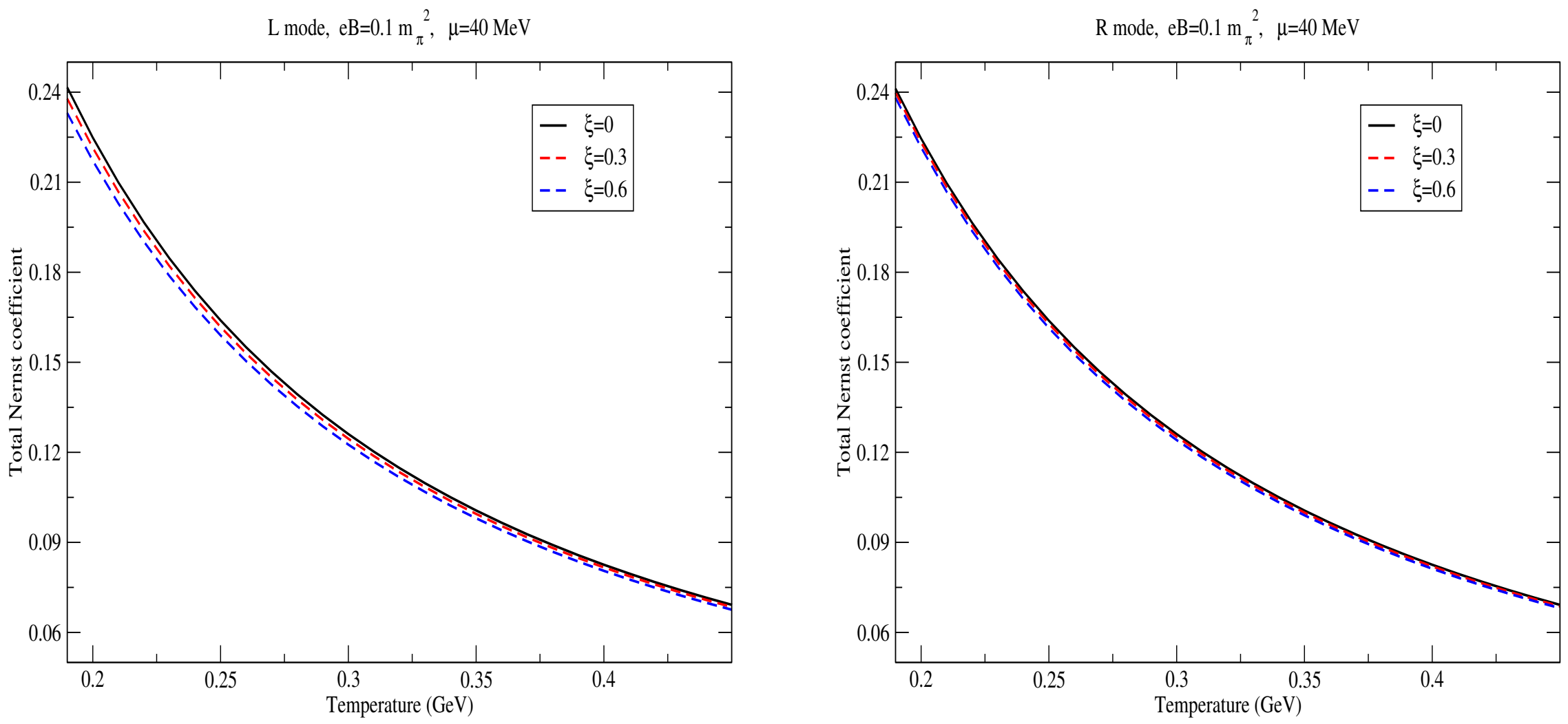

**(a)** **(b)**

**Figure 5.8:** (a) Variation of total Nernst coefficient of the medium ($L$ mode) with temperature at fixed values of $eB$ and $\mu$. (b) Variation of total Nernst coefficient of the medium ($R$ mode) with temperature at fixed values of $eB$ and $\mu$. The different curves correspond to different values of $\xi$.

Figures (5.8a) and (5.8b) show the temperature variation of the Nernst coefficients of the composite medium corresponding to $L$ mode quasiparticles and $R$ mode quasiparticles, respectively. The total Nernst coefficient of the medium is positive, and is a decreasing function of temperature. Like its Seebeck counterpart, its magnitude decreases with the strength of anisotropy. Compared to the individual Nernst coefficient magnitudes, the total Nernst coefficient magnitudes are an order of magnitude smaller. Also, compared to the total Seebeck coefficient, its values are an order of magnitude smaller. This suggests that the Nernst effect in a weakly magnetized thermal QCD medium is a weaker effect compared to the Seebeck effect. Just as in the case of the individual Nernst coefficients, the $L$ mode values of the total Nernst coefficient are slightly smaller in magnitude than their $R$ mode counterparts; the averaged (over $T$) percentage decrease being 0.04% for $\xi = 0$, $\sim$ 0.47% for $\xi = 0.3$ and $\sim$ 1.14% for $\xi = 0.6$. The mean of the three values corresponding to different $\xi$ is 0.55%. Interestingly, these numbers are $\sim$ an order of magnitude bigger than corresponding numbers for the individual Nernst coefficients, and an order of magnitude smaller than the corresponding values for the total Seebeck coefficient. Such a drastic change in going from the individual to the total coefficients was not observed for the Seebeck coefficient. Thus, the sensitivity to the mass difference between the $L$ and $R$

modes of quarks is much amplified for the composite medium, compared to a single flavor medium.

## 5.5 Conclusion

We have estimated the thermoelectric response of a deconfined hot QCD medium in the presence of a weak external magnetic field, taking into account the anisotropic expansion of the QGP fireball. The strength of the response, *i.e.* the ability to convert a temperature gradient into an electric field, is quantified by two coefficients, *viz.* Seebeck coefficient and Nernst coefficient. We have calculated the individual as well as the total response coefficients of the medium and checked their variation with temperature and anisotropy strength. We have presented results both in the absence and presence of a background magnetic field $B$. Our plots for finite $B$ have been generated for a constant background magnetic field of strength $eB = 0.1\, m_\pi^2$, and a constant chemical potential $\mu = 40 MeV$. Because of the lifting of mass degeneracy between the left-handed ($L$) and right-handed ($R$) chiral quark modes, we have shown each result for the two modes separately. We have found that the magnitudes of both the individual as well as the total coefficients-both Seebeck and Nernst, are decreasing functions of temperature, and decreasing functions of anisotropy strength, characterized by the anisotropy parameter $\xi$. It is important to note that the Seebeck coefficient vanishes for $\mu = 0$ irrespective of the magnetic field strength, and the Nernst coefficient vanishes for $|\boldsymbol{B}| = 0$, irrespective of the value of $\mu$. We have analysed the sensitivity to the $L$ and $R$ modes, *i.e.* to the difference in quasiparticle effective masses of the coefficient magnitudes; both for individual and total. To that end we have calculated the average percentage change in the coefficient magnitude as one goes from the $L$ mode to the $R$ mode or vice versa. For the Seebeck coefficient, the average (over $T$ and $\xi$) percentage change for the $u$ quark is $\sim 9.83\%$, whereas for the $d$ quark, it is $\sim 3.02\%$. This shows that the $d$ quark Seebeck coefficient is comparatively less sensitive to differences in quasiparticle masses. For the total Seebeck coefficient, the average percentage change in going from the $L$ mode to the $R$ mode is $\sim 4.61\%$, which is in between the values for the individual coefficients.

Certain differences arise in the case of Nernst coefficient. Firstly, for both the individual and total Nernst coefficients, the absolute values of the $R$ mode coefficients are greater than that of the $L$ mode. This is opposite to the case of Seebeck coefficients.

The average percentage change in going from the $R$ mode to the $L$ mode in case of $u$ quark Nernst coefficient is $\sim 0.71\%$; for the $d$ quark Nernst coefficient, this value is $\sim 0.21\%$. Compared to their Seebeck counterparts, these numbers are an order of magnitudes smaller, indicating that the individual Nernst coefficients are comparatively much less sensitive to the change in quasiparticle modes. A major difference between the individual Nernst and Seebeck coefficients is that while positively (negatively) charged quarks lead to positive (negative) Seebeck coefficients, the individual Nernst coefficients are independent of the electric charge of the quark. For the case of the total Nernst coefficient, the average percentage decrease from $R$ to $L$ modes is $\sim 0.55\%$. The absolute values of both the individual and total Nernst coefficients are about 1 order of magnitude smaller than their Seebeck counterparts, which shows that the Nernst effect is a weaker response than the Seebeck effect.

Chapter 6

# Heavy flavor dynamics in a weakly magnetized thermal QCD medium

ग्रहणां चक्रगमनं गुरुत्वाकर्षणाद्भवति।

*The circular motion of the planets is due to an attractive force.*

– **Sūrya Siddhānta (∼300 CE-400 CE)**

The second part of this thesis is contained in this chapter and deals with the propagation of heavy quarks (Charm and Bottom) in the QGP medium in the presence of a weak background magnetic field, using perturbative techniques. The medium is considered to be isotropic. This is based on the work *Dynamics of open heavy flavour in a weakly magnetized thermal QCD medium* by Debarshi Dey and Binoy Krishna Patra, **arXiv:2307.00420 [hep-ph] (2023).**

## 6.1 Introduction

Heavy-ion collisions at experimental facilities such as the Relativistic Heavy Ion Collider (RHIC) and the Large Hadron Collider (LHC) have presented strong evidence of formation of a deconfined thermal QCD medium, called the Quark Gluon Plasma (QGP) [12, 208]. When the heavy nuclei collide non-centrally, the spatial asymmetry of the initial overlap zone is carried over to the momenta of emitted particles, and can be seen experimentally in the final hadron $p_T$ spectra [209–212]. This asymmetric expansion of the QGP fireball is referred to as elliptic flow. A remarkable property of this medium is the very small value of the ratio of shear viscosity to entropy density $\eta/s$, making QGP one of the most perfect fluids known [213]. A microscopic explana-tion of these interesting transport properties is still a subject of intense investigation. To that end, heavy quarks (Charm, Bottom) are considered to be excellent probes of the QGP medium. The large mass of the heavy quark (HQ), compared to the temperature of the medium means that heavy quarks are necessarily formed with a hard initial momentum spectra during primordial $N$-$N$ collisions, before the formation of the thermal medium. Their thermalization times are much larger than that of the light quarks of the medium- again a consequence of their large masses, parametrically by a factor $\sim M_Q/T$, where, $T$ is the temperature of the QGP, and $M_Q$ the heavy quark mass. With an estimate of the QGP lifetime, $\tau_{\rm QGP} \sim 5$ fm$/c$, the HQ is expected not to thermalize and hence retain a "memory" of their interactions in the medium. Consequently, the spectra of heavy flavour hadrons can shed quantitative light on the interaction strength, and hence, the transport properties of the medium. Extensive reviews of HQ phenomenology can be found in Refs. [214, 215].

Apart from causing an anisotropic expansion of the created matter, non-central heavy ion collisions also lead to creation of large magnetic fields [191]. The decay rate of the magnetic field depends strongly on the electrical conductivity of the medium which is exposed to the field [192–201]. The thermalization of light quarks in the QGP could lead to a finite electric conductivity of the medium which would in turn affect the decay rate of the external magnetic field, in accordance with Lenz's law [216–218]. Assuming a large background magnetic field, several phenomena have been studied such as chiral magnetic effect (CME) [219],chiral magnetic wave [203, 220] leading to charge dependent elliptic flow [221, 222] magnetic catalysis (MC) [223–225], inverse magnetic catalysis (IMC) [137, 226–231], etc. For small conductivities however, the magnetic field would decay would be very fast and this has motivated the study of transport properties of the QGP in the presence of weak magnetic fields such as electric and Hall conductivities [58, 196, 232–234], shear viscosity [235, 236], thermoelectric coefficients [87, 91, 92], heavy quarkonia dissociation [237]. Further, the contribution of magnetic field in the thermalization of the QGP medium could be sub-dominant. This is because gluons play the major role in thermalization of the medium as opposed to quarks/antiquarks, and they are not directly affected by the magnetic field. In the context of heavy quarks, the diffusion of HQ in a thermal medium has been studied using perturbation theory [238–246], lattice QCD [247–250], AdS/CFT correspondence [251], in a Polyakov-loop plasma [252]. HQ diffusion in the early stages of heavy-ion collisions (glasma) has been studied in [253]. Next-to-leading order (NLO) calculation of the HQ diffusion has also been carried out [254]. Effect of momentum anisotropy on the dynamics of HQ has also been studied recently [190,255]. Recently, a non-perturbative study of HQ diffusion in strong magnetic fields has been carried out [256] Although HQ transport has been studied in the presence of a strong background magnetic field, using both imaginary and real time formalism [257–260], the literature using weak background magnetic field is rather scant. This motivates us to investigate the dynamics of open heavy flavour in the QGP medium, in the limit of a weak background magnetic field.

The HQ mass is the hardest scale in the problem, which is true even if the magnetic field is strong. The scale hierarchy considered in this problem is $M_Q \gg T \gg eB/T$. We calculate the energy loss $dE/dx$, momentum diffusion coefficient $\kappa$ and the drag coefficient $\eta_D$ of the HQ propagating in the QGP by evaluating the scattering rate of the HQ with the light thermal quarks. Cutting rules allow for determination of this scattering rate from the imaginary part of the HQ self energy [119]. This method was employed to study HQ dynamics for the first time in [261]. In what follows, the HQ self energy is evaluated using an effective gluon propagator, which, in turn, is calculated in the presence of a weak magnetic field, up to second order in $qB$. HTL perturbation theory is made use of throughout the calculations. Owing to its large mass, the problem of HQ immersed in a thermal bath of light particles is amenable to a non-relativistic treatment, in general, and a diffusion treatment, in particular, as will be justified in the next section. We go beyond the static limit and evaluate the aforementioned quantities for finite HQ momentum.

This chapter is organized as follows: In section (6.2) the description of heavy quarks in a thermal medium is presented. In section (6.3), the calculation of the scattering rate $\Gamma$ in the presence of a weak magnetic field is outlined. In section (6.4), the energy loss $dE/dx$ and momentum diffusion coefficients $\kappa$ are evaluated. In section (6.5), the results obtained are discussed. In section (6.6), phenomenological applications of the calculations are described. Finally, we conclude in section (6.7).

## 6.2 Description of heavy quarks in a thermal medium: static case and beyond

We consider a heavy quark of mass $M_Q$ propagating through a plasma of light quarks and gluons. The HQ thermal momentum $p \sim \sqrt{M_Q T} \gg T$ translates to a thermal velocity $v \sim \sqrt{T/M_Q} \ll 1$. Even if one considers hard scatterings of the HQ with the light medium particles (characterised by a momentum transfer of $\mathcal{O}(T)$), it takes a large number of collisions ($\sim M_Q/T$) to change the HQ momentum by $\mathcal{O}(1)$, since $p \gg T$. This implies that the momentum changes accumulate over time from uncorrelated "kicks", and the HQ momentum therefore evolves according to Langevin dynamics:

$$\frac{dp_i}{dt} = \xi_i(t) - \eta_D p_i\,, \qquad \langle \xi_i(t)\xi_j(t')\rangle = \kappa\, \delta_{ij}\delta(t-t'), \tag{6.1}$$

where, $(i,j) = (x,y,z)$. These are the macroscopic Langevin equations with $\eta_D$ being the momentum drag coefficient and $\kappa$ the momentum diffusion coefficient. The random forces $\xi(t)$ representing the uncorrelated momentum kicks are assumed to be white noises. The solution of Eq.(6.1) under the assumption $\eta_D^{-1} \ll t$ is given as

$$p_i(t) = \int_{-\infty}^{t} dt'\, e^{\eta_D(t'-t)}\xi_i(t'). \tag{6.2}$$

$\kappa$ can be determined by calculating the mean squared momentum transfer per unit time from the underlying microscopic theory:

$$\langle p^2\rangle = \int dt_1 dt_2 e^{\eta_D(t_1+t_2)}\langle \xi_i(t_1)\xi_j(t_2)\rangle = \frac{3\kappa}{2\eta_D}. \tag{6.3}$$

Equivalently, $\kappa$ can be defined as

$$3\kappa(\boldsymbol{p}) = \lim_{\Delta t\to 0}\frac{\left\langle(\Delta p)^2\right\rangle}{\Delta t}, \tag{6.4}$$

where, $p(t+\Delta t) - p(t)$. This leads to the following equations of motion for the heavy quark.

$$\frac{d}{dt}\langle p\rangle \equiv -\eta_D(p)p \tag{6.5}$$

$$\frac{1}{3}\frac{d}{dt}\left\langle(\Delta p)^2\right\rangle \equiv \kappa(p) \tag{6.6}$$

In a thermal medium of light quarks and gluons the random momentum kicks originate from the scattering processes $qH \to qH$ and $gH \to gH$ ($q \to$ quark, $g \to$ gluon). The former occurs only via $t$ channel Coulomb scattering. The latter, effectively also occurs via the same mechanism since its Compton amplitude is suppressed by $v^2 \sim T/M_Q$, in the rest frame of the plasma. This is especially true for the bottom quark (M=4.18

GeV), compared to the charm quark (M=1.28 GeV). In the general description where the directions of magnetic field and HQ velocity do not align, the momentum diffusion coefficient $\kappa$ gets decomposed into components such as $\kappa_L$, $\kappa_T$, $\kappa_\parallel$, $\kappa_\perp$ [238, 259], thus making the diffusion coefficient anisotropic. The drag coefficient or the relaxation rate $\eta_D$ is related to $\kappa$ via the fluctuation-dissipation relation,

$$\eta_D = \frac{\kappa}{2M_Q T}, \tag{6.7}$$

which follows from general thermodynamical arguments. Apart from $\eta_D$ and $\kappa$, we also have the spatial diffusion coefficient $D_s$ and the heavy quark energy loss $dE/dx$. The problem of HQ motion and its subsequent diffusion in a thermal medium can be characterised fully by these 4 quantities, which are related to each other. In particular,

$$D_s = \frac{T}{M_Q\,\eta_D} = 2T^2/\kappa, \tag{6.8}$$

as derived in [262]. We assume that the dominant mechanism for the HQ energy loss is coulomb scattering of the HQ with the light medium particles and ignore radiative energy loss (gluon bremsstrahlung), which is suppressed by an additional power in the strong coupling $\alpha_s$, as explained in [238]. The central quantity from which all the above mentioned dynamical quantities can be obtained is the scattering rate $\Gamma$, whose computation will be outlined in the next section. The energy loss and the momentum diffusion coefficient are given as

$$\frac{dE}{dx} = \frac{1}{v}\int d^3q \frac{d\Gamma(q)}{d^3q} q_0 \tag{6.9}$$

$$3\kappa = \int d^3q \frac{d\Gamma(q)}{d^3q} q^2. \tag{6.10}$$

$\frac{d\Gamma(q)}{d^3q}$ is the differential probability per unit time for the heavy quark momentum to change by $\boldsymbol{q}$. It can also be interpreted as the scattering rate of heavy quark via one-gluon exchange with thermal partons per unit volume of momentum transfer $\boldsymbol{q}$. $v$ and $q_0$ are the heavy quark velocity and energy respectively. The factor of 3 comes from assuming isotropicity of the momentum diffusion coefficient, which is valid if the heavy quark under consideration is assumed to be non relativistic and the background magnetic field is weak.

Beyond the static limit, the direction of motion of the HQ defines an anisotropy di-rection and the momentum diffusion coefficient breaks into longitudinal and transverse components as $3\kappa \rightarrow \kappa_L + 2\kappa_T$. This reflects the fact that there are two equivalent transverse directions.

$$\kappa_L = \int d^3q \frac{d\Gamma(q)}{d^3q} q_L^2 \tag{6.11}$$

$$\kappa_T = \frac{1}{2} \int d^3q \frac{d\Gamma(q)}{d^3q} q_T^2 \tag{6.12}$$

The HQ momentum can be diffused via collisions, in the direction of HQ momentum and also transverse to it, of which, $\kappa_L$ and $\kappa_T$ respectively are quantitative measures. In general, the HQ velocity and the direction of magnetic field need not align. However, for the purpose of simplicity, we fix the direction of HQ velocity to be along the z-axis, thereby aligning it with the magnetic field direction. As a result, the direction of magnetic field does not lead to an additional anisotropy direction.

## 6.3 Perturbative determination of scattering rate $\Gamma$

As mentioned earlier, we consider coulomb scattering of the propagating heavy quark with the thermal quarks and gluons. To leading order, these $2 \rightarrow 2$ processes are represented by the following tree level Feynman diagrams.

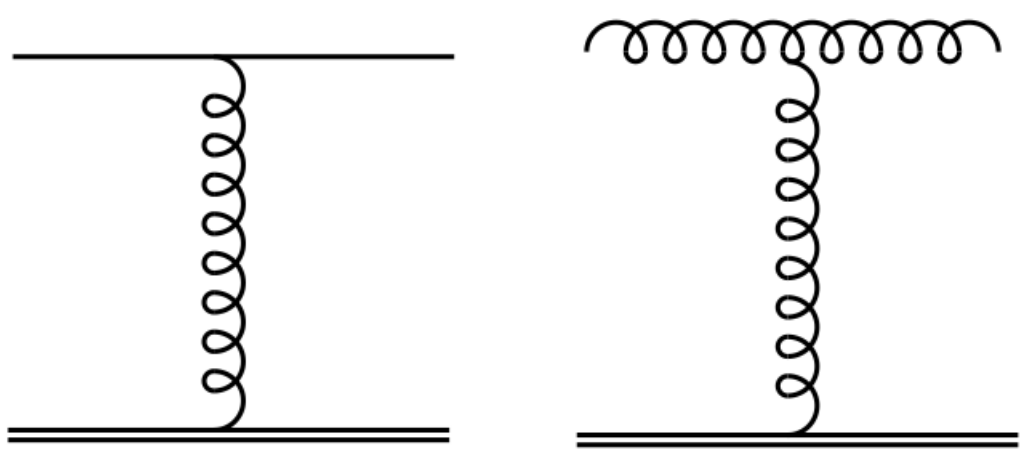

**Figure 6.1:** Feynman diagrams of processes contributing to heavy quark diffusion at leading order.

The double line represents the heavy quark, whereas the thermal light quark is represented by the single line. $\Gamma$ calculated using the tree level diagrams in Fig.(6.1) turns out to be quadratically infrared divergent which corroborates with the well known fact that the total rate of coulomb scattering in a plasma is quadratically infrared divergent[Landau]. Using a resummed gluon propagator in Fig.(6.1) instead of a bare one softens the divergence to a logarithmic one [261]. This arises because the dynamical screening of the magnetic interaction provided by the transverse effective propagator is not sufficient to completely screen the divergence from the long-range static magnetic interaction. However, the two additional powers of $q$ in Eq.(6.10) render $\kappa$ infrared finite. The presence of the logarithm reflects that $\Gamma$ receives contribution from both the soft and hard momentum transfers. Soft processes involve $\boldsymbol{q} \sim gT$ and occur at a rate $\Gamma_{\text{soft}} \sim g^2 T$, whereas the relatively scarce hard processes correspond to $\boldsymbol{q} \sim T$ and occur at a rate $\Gamma_{\text{hard}} \sim g^4 T$. In this article, we shall be evaluating the soft contribution to $\Gamma$, and therefore, to the heavy quark diffusion coefficient, since it dominates over hard processes.

An efficient method of calculating the scattering rate was put forward by Weldon [119] wherein, $\Gamma$ is evaluated from the imaginary part of the heavy quark self energy:

$$\Gamma(P \equiv E, \mathbf{v}) = -\frac{1}{2E}[1 - n_F(E)]\,\mathrm{Tr}\left[(\not{P} + M_Q)\,\mathrm{Im}\,\Sigma\,(p_0 + i\epsilon, \boldsymbol{p})\right]. \tag{6.13}$$

The imaginary part of the heavy quark self energy is related to the squared amplitude for coulomb scattering processes via the cutting rules, for the 2 loop self energy diagrams shown in Fig.(6.2). This procedure automatically rules out using one-loop self energy diagrams, since the cut (imaginary) parts of those diagrams correspond to processes which do not conserve energy-momentum and thus are unphysical [16].

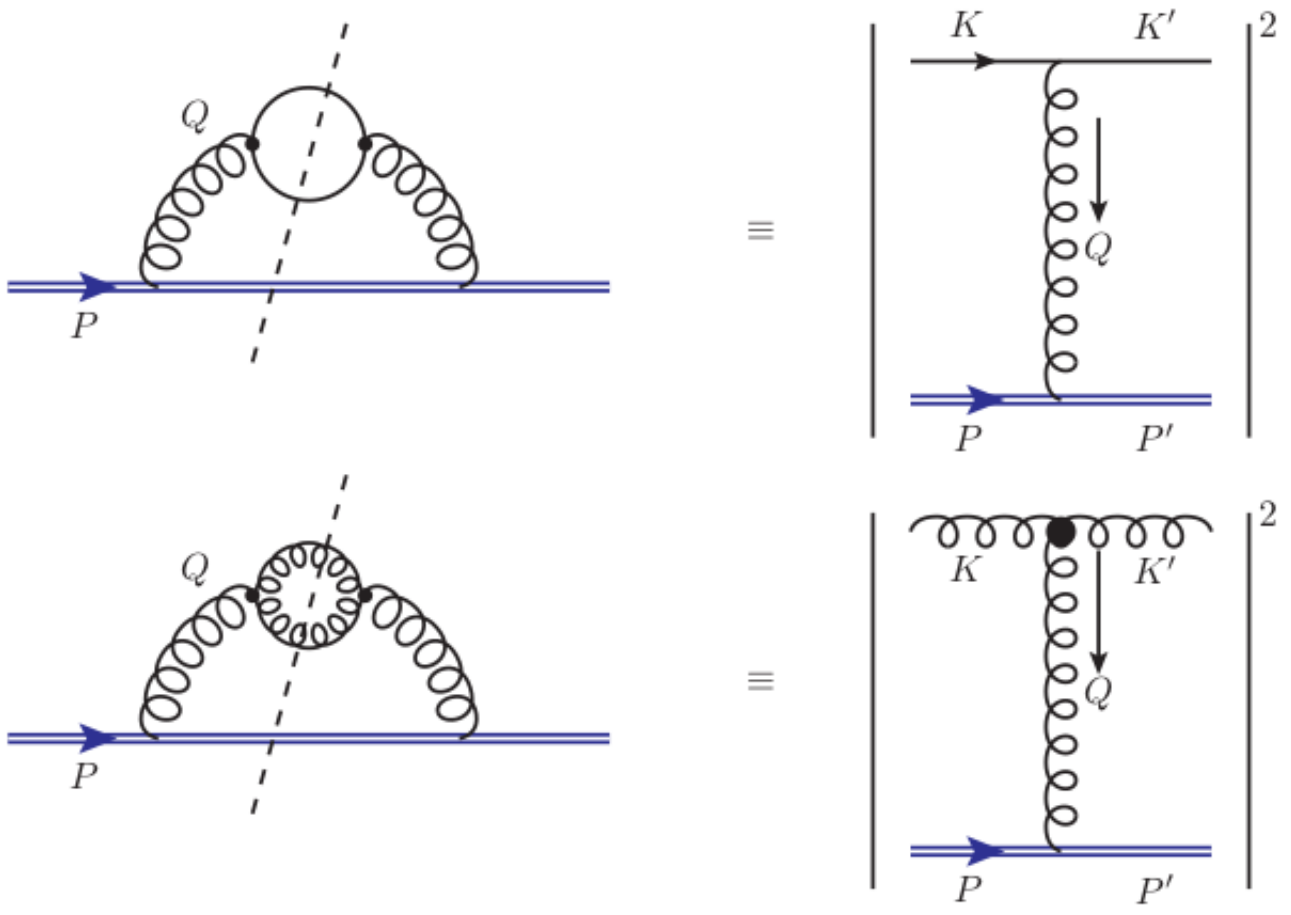

**Figure 6.2:** Cut (imaginary) part of heavy quark self energy diagrams yield the amplitude squared of $t$ channel scattering processes $qH \to qH$ and $gH \to gH$

The hard contribution to $\Gamma$ comes from the two loop self energy diagrams of Fig.(6.2). However, when the photon momentum is soft, hard thermal loop correc-tions to the photon propagator contribute at leading order in $g$ and therefore must be resummed. The diagram thus to be evaluated is shown in Fig.(6.3), where the blob on the gluon line represents a resummed/effective gluon propagator, and as is shown, is obtained by summing the geometric series of one-loop self energy corrections proportional to $g^2T^2$.

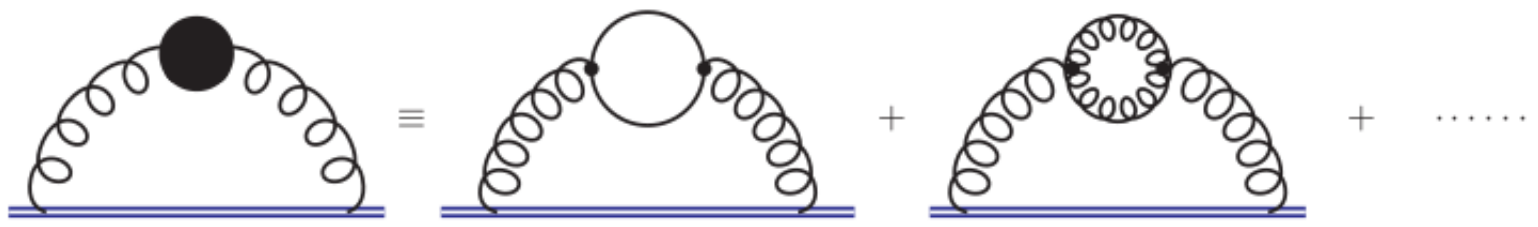

**Figure 6.3:** Heavy quark self energy with resummed gluon propagator. In addition to the leading order diagrams, resummation takes into account all higher order diagrams that contribute to leading order in $g$

We use the imaginary time formalism to compute the heavy quark self energy $\Sigma(P)$. Using Feynman rules, $\Sigma(P)$ in a weak background magnetic field is given by:

$$\Sigma(P) = ig^2 \int \frac{d^4Q}{(2\pi)^4} \mathcal{D}^{\mu\nu}(Q)\gamma_\mu S(P-Q)\gamma_\nu. \tag{6.14}$$

Since we work in the regime $\frac{qB}{M_Q} \ll 1$, we can ignore Landau quantization of the heavy quark energy levels, as has been done in [258], and write the heavy quark propagator as

$$iS(P-Q \equiv K) = i\frac{\not{K} + M_Q}{K^2 - M_Q^2}. \tag{6.15}$$

The effective gluon propagator in the presence of a weak magnetic field is expressed as [144]

$$\begin{aligned}\mathcal{D}^{\mu\nu}(Q) =& \frac{\xi Q^\mu Q^\nu}{Q^4} + \frac{\left(Q^2-d_3\right)\Delta_1^{\mu\nu}}{\left(Q^2-d_1\right)\left(Q^2-d_3\right)-d_4^2} + \frac{\Delta_2^{\mu\nu}}{Q^2-d_2} \\ &+ \frac{\left(Q^2-d_1\right)\Delta_3^{\mu\nu}}{\left(Q^2-d_1\right)\left(Q^2-d_3\right)-d_4^2} + \frac{d_4\Delta_4^{\mu\nu}}{\left(Q^2-d_1\right)\left(Q^2-d_3\right)-d_4^2},\end{aligned} \tag{6.16}$$

where,

$$d_1(Q) = \Delta_1^{\mu\nu}\Pi_{\mu\nu} \tag{6.17a}$$

$$d_2(Q) = \Delta_2^{\mu\nu}\Pi_{\mu\nu} \tag{6.17b}$$

$$d_3(Q) = \Delta_3^{\mu\nu}\Pi_{\mu\nu} \tag{6.17c}$$

$$d_4(Q) = \frac{1}{2}\Delta_4^{\mu\nu}\Pi_{\mu\nu} \tag{6.17d}$$

$\Pi_{\mu\nu}(Q)$ is the gluon self energy computed within the HTL approximation. $\Delta_i^{\mu\nu}$ are the projection tensors along which the gluon self energy and the effective gluon propagator are expressed in the presence of a finite magnetic field, details of which can be found in the Appendix. They are expressed as

$$\Delta_1^{\mu\nu} = \frac{1}{\bar{u}^2}\bar{u}^\mu \bar{u}^\nu, \tag{6.18a}$$

$$\Delta_2^{\mu\nu} = g_\perp^{\mu\nu} - \frac{Q_\perp^\mu Q_\perp^\nu}{Q_\perp^2}, \tag{6.18b}$$

$$\Delta_3^{\mu\nu} = \frac{\bar{n}^\mu \bar{n}^\nu}{\bar{n}^2}, \tag{6.18c}$$

$$\Delta_4^{\mu\nu} = \frac{\bar{u}^\mu \bar{n}^\nu + \bar{u}^\nu \bar{n}^\mu}{\sqrt{\bar{u}^2}\sqrt{\bar{n}^2}}. \tag{6.18d}$$

$u^\mu$ is the velocity of the heat bath and $n^\mu$ can be considered to define the direction of the background magnetic field. Evaluating the form factors $d_i(Q)$ of Eqs.(6.17) is akin to evaluating the effective gluon propagator. The calculation of these form factors under HTL approximation, along with the other tensors in Eq.(6.18) is detailed in appendix A.

Following [259], we next evaluate the trace in Eq.(6.13)

$$\begin{aligned} \mathrm{Tr}[(\not{P} + M_Q)\Sigma(P)] =& ig^2 \int \frac{d^4Q}{(2\pi)^4}\frac{1}{K^2 - M_Q^2} \\ &\times \sum_{i=1}^{4} \chi_i \,\mathrm{Tr}\left[(\not{P} + M_Q)\Delta_i^{\mu\nu}\gamma_\mu \left(\not{K} + M_Q\right)\gamma_\nu\right] \end{aligned} \tag{6.19}$$

Taking the gauge parameter $\xi$ in Eq.(6.16) to be 0, the coefficients $\chi_i$'s are given by:

$$\chi_1 = \frac{\left(Q^2 - d_3\right)}{\left(Q^2 - d_1\right)\left(Q^2 - d_3\right) - d_4^2}, \tag{6.20a}$$

$$\chi_2 = \frac{1}{\left(Q^2 - d_2\right)}, \tag{6.20b}$$

$$\chi_3 = \frac{\left(Q^2 - d_1\right)}{\left(Q^2 - d_1\right)\left(Q^2 - d_3\right) - d_4^2}, \tag{6.20c}$$

$$\chi_4 = \frac{d_4}{\left(Q^2 - d_1\right)\left(Q^2 - d_3\right) - d_4^2}. \tag{6.20d}$$

We evaluate the individual traces in Eq.(6.19).

$$\mathrm{Tr}\left[(\not{P}+M_Q)\Delta_1^{\mu\nu}\gamma_\mu\left(\not{K}+M_Q\right)\gamma_\nu\right]=\frac{4}{\bar{u}^2}\left[2(P.\bar{u})(K.\bar{u})+\bar{u}^2(M^2-P.K)\right] \tag{6.21a}$$

$$=A_1+B_1 \tag{6.21b}$$

$$\mathrm{Tr}\left[(\not{P}+M_Q)\Delta_2^{\mu\nu}\gamma_\mu\left(\not{K}+M_Q\right)\gamma_\nu\right]=4\left[2(P.K)_\perp-\frac{2(P.Q)_\perp(K.Q)_\perp}{Q_\perp^2}+(M^2-P.K)\right] \tag{6.21c}$$

$$=A_2+B_2 \tag{6.21d}$$

$$\mathrm{Tr}\left[(\not{P}+M_Q)\Delta_3^{\mu\nu}\gamma_\mu\left(\not{K}+M_Q\right)\gamma_\nu\right]=\frac{4}{\bar{n}^2}\left[2(P.\bar{n})(K.\bar{n})+\bar{n}^2(M^2-P.K)\right] \tag{6.21e}$$

$$=A_3+B_3 \tag{6.21f}$$

$$\mathrm{Tr}\left[(\not{P}+M_Q)\Delta_4^{\mu\nu}\gamma_\mu\left(\not{K}+M_Q\right)\gamma_\nu\right]=\frac{8}{\sqrt{\bar{n}^2}\sqrt{\bar{u}^2}}\left[(P.\bar{u})(K.\bar{u})+(P.\bar{n})+(\bar{u}\cdot\bar{n})(M^2-P.K)\right] \tag{6.21g}$$

$$=A_4+B_4 \tag{6.21h}$$

The traces have been separated into $q_0$ independent and $q_0$ dependent terms denoted by $A_i$ and $B_i$ respectively; $i=1,2,3,4$. This is done to facilitate the frequency sum over $q_0$, as will be seen later. The condition required for such a separation to be executed is that the transfer momentum four-vector $Q^\mu$ be spacelike, which is indeed the case for $t$ channel scattering processes. The $A_i'$s and $B_i'$s come out to be

$$A_1=4(2p_0^2-\boldsymbol{p}\cdot\boldsymbol{q}),\quad B_1=-8p_0^2\frac{q_0^2}{q^2}-8\frac{q_0^2(P.Q)^2}{Q^2q^2}+16p_0q_0\frac{P\cdot Q}{q^2} \tag{6.22}$$

$$A_2=4\left[2\{P_\perp^2-(P\cdot Q)_\perp\}-\boldsymbol{p}\cdot\boldsymbol{q}+\frac{2(P.Q)_\perp\{Q_\perp^2-(P.Q)_\perp\}}{Q_\perp^2}\right],\quad B_2=4p_0q_0 \tag{6.23}$$

$$A_3=\frac{8}{\bar{n}^2}\left[p_3^2-\frac{2p_3q_3}{q^2}(\boldsymbol{p}\cdot\boldsymbol{q})+\frac{q_3}{q^2}(\boldsymbol{p}\cdot\boldsymbol{q})^2-\frac{\bar{n}^2}{2}(\boldsymbol{p}\cdot\boldsymbol{q})\right],\quad B_3=4p_0q_0 \tag{6.24}$$

$$A_4=\frac{16}{\sqrt{\bar{n}^2}}\left[-p_0p_3+\frac{p_0p_3}{q^2}(\boldsymbol{p}\cdot\boldsymbol{q})\right]=\frac{16p_0p_3}{\sqrt{\bar{n}^2}}\left[\frac{\boldsymbol{p}\cdot\boldsymbol{q}}{q^2}-1\right] \tag{6.25}$$

$$B_4 = \frac{16}{\sqrt{\bar{n}^2}}\left[\frac{p_3 q_0^2 p_0}{q^2} - \frac{p_3 q_0}{q^2}(\boldsymbol{p}\cdot\boldsymbol{q}) - \frac{q_0^2 q_3 p_0}{Q^2 q^2}(\boldsymbol{p}\cdot\boldsymbol{q}) + \frac{q_0 q_3}{Q^2 q^2}(p\cdot\boldsymbol{q})^2\right] \tag{6.26}$$

$$\times\left\{\left(-\frac{q_0^2}{q^2}\right) + \text{ higher powers of } \frac{q_0^2}{q^2}\right\} \tag{6.27}$$

Next, we perform the frequency sum over $q_0$. To that end, a convenient method is to introduce spectral function representations for the propagators [96]. The fermion propagator is spectrally represented as

$$\frac{1}{K^2 - M_Q^2} = -\frac{1}{2E'}\int_0^\beta d\tau' e^{k_0\tau'}\left[(1-n_F(E'))e^{-E'\tau'} - n_F(E')e^{E'\tau'}\right], \tag{6.28}$$

where, $E' = \sqrt{k^2 + M_Q^2}$, $\beta = 1/T$ Similarly, pieces of the effective gluon propagator $\chi_i$ can be expressed as

$$\chi_i = -\int_0^\beta d\tau\, e^{q_0\tau}\int_{-\infty}^{\infty} d\omega\, \rho_i(\omega, q)\left[1 + n_B(\omega)\right] e^{-\omega\tau}. \tag{6.29}$$

$\rho_i$ are the spectral functions associated with $\chi_i$, and are odd functions of $\omega$. Each spectral function contains contributions from both spacelike and timelike frequencies, and is expressed as

$$\rho_i(\omega, q) = \rho_i^{\text{pole}}(\omega, q) + \rho_i^{\text{cut}}(\omega, q), \tag{6.30}$$

with

$$\rho_i^{\text{pole}}(\omega, q) = \rho_i^{\text{res}}\delta(\omega - \omega_i(q)) \tag{6.31}$$

$$\rho_i^{\text{cut}}(\omega, q) = \rho_i^{\text{dis}}\theta(q^2 - \omega^2) \tag{6.32}$$

Thus, the spectral functions have delta function contributions at the timelike points (poles) $\omega = \omega_i(q)$, where, $\omega_i(q)$ are the dispersion relations, and $\rho_i^{\text{res}}$ are the residues at those points. For spacelike frequencies $|\omega| < q$, $\rho_i'$s receive a discontinuous contribution from the imaginary part of the resummed propagator (Landau damping)

$$\rho_i^{\text{dis}}(\omega, q) = -\frac{1}{\pi}\text{Im}\left(\chi_i\Big|_{q_0=\omega+i\epsilon}\right). \tag{6.33}$$

Since we work in the regime $|\omega| < q$, only the cut part in Eq.(6.30) contributes and is denoted simply as $\rho$ hereafter. The calculation and final expressions of the $\rho_i$'s

are given in Appendix 2. The advantage of the spectral function representation is that it simplifies the evaluation of the frequency sums due to the appearance of delta functions in the integral, coming from

$$T\sum_{q_0} e^{q_0(\tau-\tau')} = \delta(\tau-\tau') \tag{6.34a}$$

$$T\sum_{q_0} q_0\, e^{q_0(\tau-\tau')} = \delta'(\tau-\tau') \tag{6.34b}$$

Using this, Eq.(6.19) becomes

$$\mathrm{Tr}[(\not{P}+M)\Sigma(P)] \tag{6.35}$$

$$= -ig^2\int\frac{d^4Q}{(2\pi)^4}\frac{1}{K^2-M_Q^2}\sum_{i=1}^{4}\chi_i[A_i+B_i] \tag{6.36}$$

$$= -g^2T\sum_{i=1}^{4}\int\frac{d^3q}{(2\pi)^3}\int_{-\infty}^{+\infty}d\omega\left[1+n_B(\omega)\right]\int_0^\beta d\tau'\int_0^\beta d\tau e^{p_0\tau'}e^{-\omega\tau} \tag{6.37}$$

$$\times\sum_{q_0}e^{q_0(\tau-\tau')}\left[A_i+B_i\right]\frac{\rho_i(\omega,q)}{2E'}\left[\left\{1-n_F(E')\right\}e^{-E'\tau'}-n_F\left(E'\right)e^{E'}\tau'\right] \tag{6.38}$$

$$= -g^2T\sum_{i=1}^{4}\int\frac{d^3q}{(2\pi)^3}\int_{-\infty}^{+\infty}d\omega\left[1+n_B(\omega)\right](I_1+I_2)\,, \tag{6.39}$$

where, using Eq.(6.34a),

$$I_1 = \int_0^\beta d\tau'\int_0^\beta d\tau\, e^{p_0\tau'}e^{-\omega\tau}A_i\,\delta(\tau-\tau')\frac{\rho_i(\omega,q)}{2E'} \tag{6.40}$$

$$\times\left[\left\{1-n_F(E')\right\}e^{-E'\tau'}-n_F\left(E'\right)e^{E'}\tau'\right]. \tag{6.41}$$

We use the $\delta$ function to integrate over $\tau'$ to obtain

$$I_1 = \int_0^\beta d\tau e^{(p_0-\omega)\tau}\frac{\rho_i(\omega,q)}{2E'}\left[\left\{1-n_F(E')\right\}e^{-E'\tau}-n_F\left(E'\right)e^{E'\tau}\right]A_i. \tag{6.42}$$

The $\tau$ integration ultimately yields

$$I_1 = -\sum_{j=\pm1}\frac{j\,n_F(jE')}{p_0-\omega+jE'}\left[e^{(p_0-\omega+jE')\beta}-1\right]A_i. \tag{6.43}$$

Since, $B_i$ is $q_0$ dependent, a sample term can be written as $B_i = q_0C_i$. Then, using

Eq.(6.34b) yields

$$I_2 = \int_0^\beta d\tau' \int_0^\beta d\tau\, e^{p_0\tau'} e^{-\omega\tau} \sum_i C_i\, \delta'(\tau-\tau') \frac{\rho_i(\omega,q)}{2E'} \left[\left\{1-n_F(E')\right\} e^{-E'\tau'} - n_F\left(E'\right) e^{E'\tau'}\right] \tag{6.44}$$

$$= -\int_0^\beta d\tau \frac{d}{d\tau} e^{(p_0-\omega)\tau} \left[\left\{1-n_F(E')\right\} e^{-E'\tau} - n_F\left(E'\right) e^{E'\tau}\right] C_i \tag{6.45}$$

$$= \sum_{j=\pm1} j\, n_F(jE') \left[e^{(p_o-\omega+jE')\beta} - 1\right] C_i. \tag{6.46}$$

$p_0$ is discrete since we are working in the imaginary time formalism. Specifically, $p_0 = i(2n+1)\pi/\beta$. At these discrete energies, $e^{p_0\beta} = -1$ and $p_0$ thus gets eliminated from the exponent in Eqs.(6.43) and (6.46). Thereafter, we analytically continue $p_0$ to real values via $p_0 \to E + i\omega$. The imaginary part is then extracted, which comes from energy denominator terms of the form

$$\mathrm{Im}\left(\frac{1}{p_0+E'-\omega}\right)\Big|_{p_0\to E+i\omega} = -i\pi\delta(p_0+E'-\omega). \tag{6.47}$$

Since there is no energy denominator in Eq.(6.46), $I_2$ does not have any imaginary part, and the contribution to the imaginary part of the self energy thus comes solely from $I_1$. Using Eq.(6.39), (6.43) and (6.47), we can write

$$\mathrm{Tr}\left[(\not{P}+M_Q)\,\mathrm{Im}\,\Sigma\left(p_0+i\epsilon,\boldsymbol{p}\right)\right] \tag{6.48}$$

$$= \pi g^2 \sum_{i=1}^{4} \int \frac{d^3q}{(2\pi)^3} \int_{-\infty}^{\infty} d\omega \left[1+n_B(\omega)\right] \frac{\rho_i(\omega,q)A_i}{2E'} \tag{6.49}$$

$$\times \sum_{j=\pm1} j n_F\left(\sigma E'\right) \left(e^{(\sigma E'-\omega)\beta}+1\right) \delta\left(E+jE'-\omega\right) \tag{6.50}$$

$$= \pi g^2 \left(e^{-E\beta}+1\right) \sum_{i=1}^{4} \int \frac{d^3q}{(2\pi)^3} \int_{-\infty}^{+\infty} d\omega \left[1+n_B(\omega)\right] \frac{\rho_i(\omega,q)A_i}{2E'} \tag{6.51}$$

$$\times \sum_{j=\pm1} j n_F\left(jE'\right) \delta\left(E+jE'-\omega\right). \tag{6.52}$$

Thus, using all the results, $\Gamma$ in Eq.(6.13) is given by

$$\Gamma(E,\boldsymbol{v}) = -\frac{\pi g^2}{2E}\sum_{i=1}^{4}\int\frac{d^3q}{(2\pi)^3}\int_{-\infty}^{+\infty}d\omega\left[1+n_B(\omega)\right]\frac{\rho_i(\omega,q)A_i}{2E'} \tag{6.53}$$

$$\times\sum_{j=\pm1} jn_F\left(jE'\right)\delta\left(E+jE'-\omega\right). \tag{6.54}$$

Eq.(6.54) is the scattering rate result without any approximations. We now simplify the expression further by recalling that $M_Q$, $p \gg T$. The delta function corresponding to $j=1$ does not contribute for $\omega \leq T$, and so can be dropped. For $E' \gg T$, the fermi distribution function is exponentially suppressed, so that $n_F(E') \approx 1$. Employing these approximations, we have

$$\Gamma(E,\boldsymbol{v}) = \frac{\pi g^2}{2E}\sum_{i=1}^{4}\int\frac{d^3q}{(2\pi)^3}\int_{-\infty}^{+\infty}d\omega\left[1+n_B(\omega)\right]\frac{\rho_i(\omega,q)A_i}{2E'}\delta(E-E'-\omega). \tag{6.55}$$

## 6.4 Energy loss and momentum diffusion coefficient

After having computed $\Gamma$, we use it to evaluate dynamic quantities such as the heavy quark energy loss and the momentum diffusion coefficient. We have

$$E' = \sqrt{(\boldsymbol{p}-\boldsymbol{q})^2+M_Q^2} \tag{6.56}$$

$$\simeq E\left(1-\frac{2\boldsymbol{p}\cdot\boldsymbol{q}}{E^2}\right)^{1/2} \tag{6.57}$$

$$\simeq E-\boldsymbol{v}\cdot\boldsymbol{q}. \tag{6.58}$$

Although $E-E' \sim \mathcal{O}(v)$, $\frac{1}{E}-\frac{1}{E'} \sim \mathcal{O}(v^2)$, which we neglect. Thus, $\frac{1}{E} \approx \frac{1}{E'}$. The energy loss of the heavy quark propagating through the high temperature QCD plasma is given by Eq.(6.9). Using Eq.(6.55) and the approximations mentioned above, we get, for the energy loss:

$$\frac{dE}{dx} = \frac{\pi g^2}{2Ev}\sum_{i=1}^{4}\int\frac{d^3q}{(2\pi)^3}\int_{-\infty}^{+\infty}d\omega\left[1+n_B(\omega)\right]\omega\,\frac{\rho_i(\omega,q)A_i}{2E}\delta(\omega-\boldsymbol{v}\cdot\boldsymbol{q}). \tag{6.59}$$

For $\omega \ll T$, the bose distribution function can be written as an expansion in $\omega/T$ so that

$$1 + n_B(\omega) \simeq \frac{T}{\omega} + \frac{1}{2} - \mathcal{O}\left(\frac{\omega}{T}\right) + \mathcal{O}\left(\frac{\omega}{T}\right)^2 - \cdots \tag{6.60}$$

Now, the $\rho_i^\prime$s in Eq.(6.59) are odd functions of $\omega$ [263]. Hence, only the even part of $1 + n_B(\omega)$ will contribute to the integral, since the integration over $\omega$ is symmetric. Thus, we have

$$\frac{dE}{dx} = \frac{\pi g^2}{8E^2 v} \sum_{i=1}^{4} \int \frac{d^3q}{(2\pi)^3} \int_{-\infty}^{+\infty} d\omega\, \omega\, \rho_i(\omega, q) A_i\, \delta(\omega - \boldsymbol{v} \cdot \boldsymbol{q}). \tag{6.61}$$

The momentum diffusion coefficients are given by

$$\kappa_L = \frac{\pi g^2}{2E} \sum_{i=1}^{4} \int \frac{d^3q}{(2\pi)^3} q_L^2 \int_{-\infty}^{+\infty} d\omega \left[1 + n_B(\omega)\right] \frac{\rho_i(\omega, q) A_i}{2E} \delta(\omega - \boldsymbol{v} \cdot \boldsymbol{q}). \tag{6.62}$$

$$\kappa_T = \frac{\pi g^2}{2E} \sum_{i=1}^{4} \int \frac{d^3q}{(2\pi)^3} q_T^2 \int_{-\infty}^{+\infty} d\omega \left[1 + n_B(\omega)\right] \frac{\rho_i(\omega, q) A_i}{2E} \delta(\omega - \boldsymbol{v} \cdot \boldsymbol{q}). \tag{6.63}$$

This time, only the odd part of $1 + n_B(\omega)$ will contribute to the integral. Thus, we have

$$\kappa_L = \frac{\pi g^2 T}{4E^2} \sum_{i=1}^{4} \int \frac{d^3q}{(2\pi)^3} q_L^2 \int_{-\infty}^{+\infty} d\omega \frac{\rho_i(\omega, q) A_i}{\omega} \delta(\omega - \boldsymbol{v} \cdot \boldsymbol{q}). \tag{6.64}$$

$$\kappa_T = \frac{\pi g^2 T}{4E^2} \sum_{i=1}^{4} \int \frac{d^3q}{(2\pi)^3} q_T^2 \int_{-\infty}^{+\infty} d\omega \frac{\rho_i(\omega, q) A_i}{\omega} \delta(\omega - \boldsymbol{v} \cdot \boldsymbol{q}). \tag{6.65}$$

For purposes of simplification, let us consider the HQ velocity to be along the $z$-axis. Then, $\boldsymbol{v} \cdot \boldsymbol{q} = vq\cos\theta \equiv vq\eta$, where, $\theta$ is the angle between $\boldsymbol{q}$ and the $z$-axis. The delta function is used to integrate over $\eta$ with $d^3q = 2\pi q^2 dq d\eta$, which sets $\omega = vq\eta$.

Since $-1 \leq \eta \leq 1$, $-vq \leq vq$. This finally leads to

$$\frac{dE}{dx} = \frac{\pi g^2}{8E^2\, v^2(2\pi)^2}\int dq\, q \int_{-vq}^{vq} d\omega\, \omega \sum_{i=1}^{4} \rho_i(\omega, q, \frac{\omega}{vq})A_i. \tag{6.66}$$

$$\kappa_L = \frac{\pi g^2 T}{4E^2\, v}\int dq\, q^3 \int_{-vq}^{vq} d\omega \sum_{i=1}^{4} \frac{\rho_i(\omega, q, \frac{\omega}{vq})A_i}{\omega}\left(\frac{\omega^2}{v^2q^2}\right). \tag{6.67}$$

$$\kappa_T = \frac{\pi g^2 T}{4E^2\, v}\int dq\, q^3 \int_{-vq}^{vq} d\omega \sum_{i=1}^{4} \frac{\rho_i(\omega, q, \frac{\omega}{vq})A_i}{\omega}\left(1 - \frac{\omega^2}{v^2q^2}\right). \tag{6.68}$$

## 6.5 Results and Discussions

In this section, we present the results of the heavy quark (Charm and Bottom) momentum diffusion coefficients and the heavy quark energy loss. The running coupling constant is taken up to one-loop:

$$g(\Lambda) = \left[\frac{48\pi^2}{(33 - 2N_f)\ln\left(\frac{\Lambda^2}{\Lambda^2_{\overline{MS}}}\right)}\right]^{1/2} \tag{6.69}$$

The renormalisation scale $\Lambda$ can be taken to be $2\pi T$ to introduce temperature dependence in the coupling. The $\overline{\text{MS}}$ scale is taken to be 176 MeV [264]. The use of this form of the coupling is justified since $q_f B \ll T^2$. In the strong field limit, use of momentum-dependent couplings might be more appropriate [128]. The bottom and charm quark masses are taken to be 4.18 GeV and 1.28 GeV, respectively. Heavy quark dynamics with temperature dependent couplings was studied first in [241].

An important point to note is that the integrals in Eqs.(6.66), (6.67), (6.68) are logarithmically U-V divergent and hence, require a U-V cut-off. Following the prescription of [242], we take the U-V cut-off to be $3.1g^{1/3}$. The reason for this divergence is that our calculations are confined to the region of soft gauge boson momentum transfer. In the $B = 0$ case, it is shown explicitly that the dependence on this cut-off vanishes once the full range of momentum transfers is taken into account. We expect the same to be true in the case of weak magnetic fields too. However, the full calculation, including hard scatterings, is left for a future work. As mentioned earlier, the soft scatterings contribute to $\mathcal{O}(g(T)^2)$ in $\Gamma$ whereas the hard contribution to $\Gamma$ will be of $\mathcal{O}(g(T)^4)$. As such, it can be inferred that the major contribution to the momentum diffusion of the HQ via elastic scatterings comes from soft gluon exchange with the thermal quarks and gluons of the heat bath. It is also worth mentioning that this U-V cut-off is not necessary if one uses the Lowest Landau Level (LLL) approximation for the HQ propagator in the presence of a strong ($q_f B \gg T^2$) magnetic field. This is because of the presence of the exponential factor $e^{-k_\perp^2 / |q_f B|}$ in the propagator.

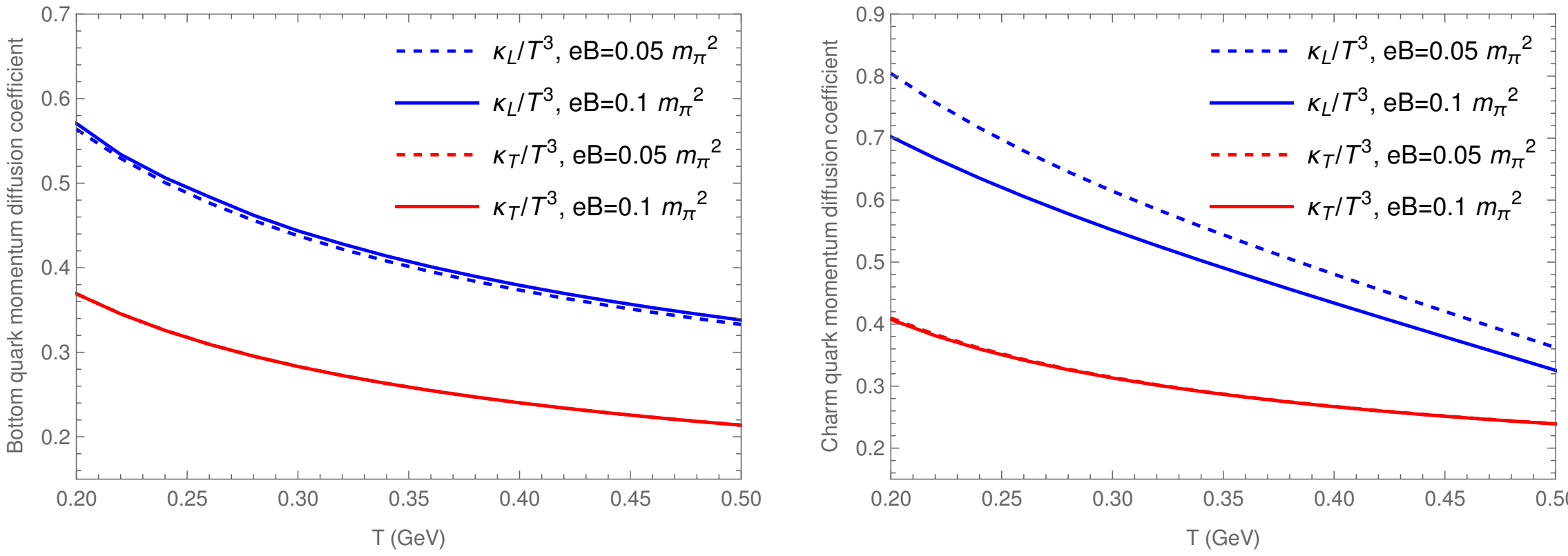

**(a)** **(b)**

**Figure 6.4:** Normalised momentum diffusion coefficients for Bottom(Left) and Charm (R) quarks as a function of temperature at different fixed values of background magnetic field.

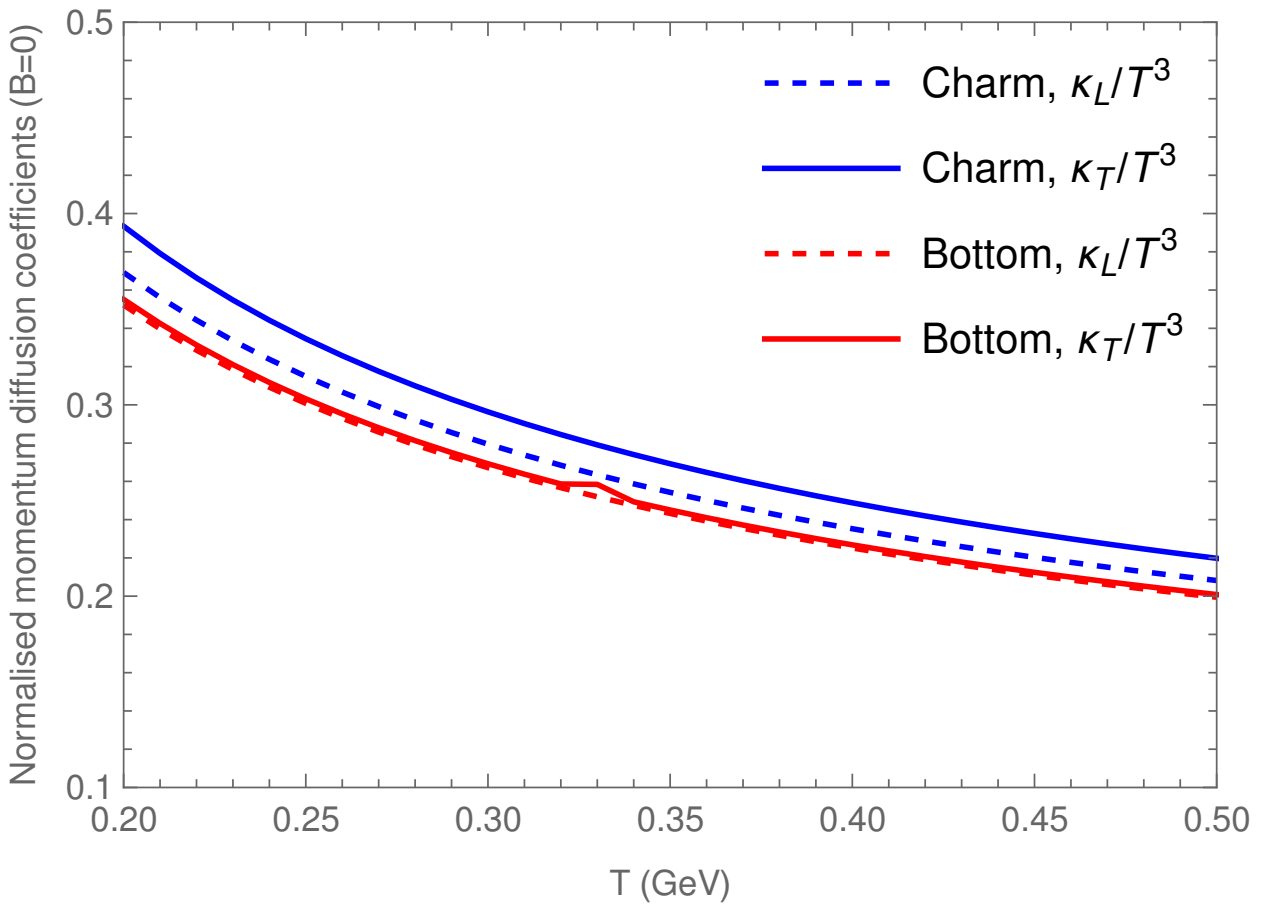

**Figure 6.5:** Normalised transverse and longitudinal momentum diffusion coefficients for both Charm and Bottom quarks as a function of temperature in the absence of a background magnetic field.

The figures show the temperature variation of the momentum diffusion coefficients for heavy quarks with a finite fixed momentum (p=0.3 GeV). For comparison, the $B = 0$ results are also shown. As can be seen from Figures (6.4a) and (6.4b), both the longitudinal and transverse momentum diffusion coefficients show a monotonous decrease with temperature. For bottom quark, we see that the effect of change in $B$ field strength is small; it being more discernible for the longitudinal component. With increasing $B$, the values of both $\kappa_L$ and $\kappa_T$ increase. For the charm quark, however, we notice a different trend, as can be seen in Fig.(6.4b), wherein, the values at $eB = 0.05m_{\pi^2}$ are greater than that at $eB = 0.1m_{\pi^2}$. This is something which is counter-intuitive and difficult to explain, since the $B = 0$ values are smaller than that at $eB = 0.1m_{\pi^2}$ [Fig.(6.5)]. Additionally, the anisotropy is larger at lower temperatures and $\kappa_L$ is significantly more sensitive to $B$ field strength. Fig.(6.5) shows the $B = 0$ result for the momentum diffusion coefficients. The degree of anisotropy for the charm quark is again seen to be greater than that for the bottom quarks. The momentum diffusion of Charm quarks is faster than that of the Bottom quark, owing to the smaller mass of the former. $\kappa_T$ is larger than $\kappa_L$ in the entire temperature range, for both the heavy flavours. A large value of the momentum diffusion coefficient would work towards decreasing the yield of bound states such as the $J/\psi$ (charmonium), bottomonium, etc.; the Brownian motion of the heavy-quarks overwhelming the screened potential holding the $q$-$\bar{q}$ pair together. On the other hand, a stronger energy loss, $dE/dx$, of the propagating HQ would result in the stopping of a $q$-$\bar{q}$ pair (not so much for $b$-$\bar{b}$), leading to the increase in yield of mesonic bound states involving heavy quarks. The fate of a $q$-$\bar{q}$ pair produced in the initial stages of a heavy-ion collision thus depends on these competing factors. This phenomenon has been elucidated in detail in [265].

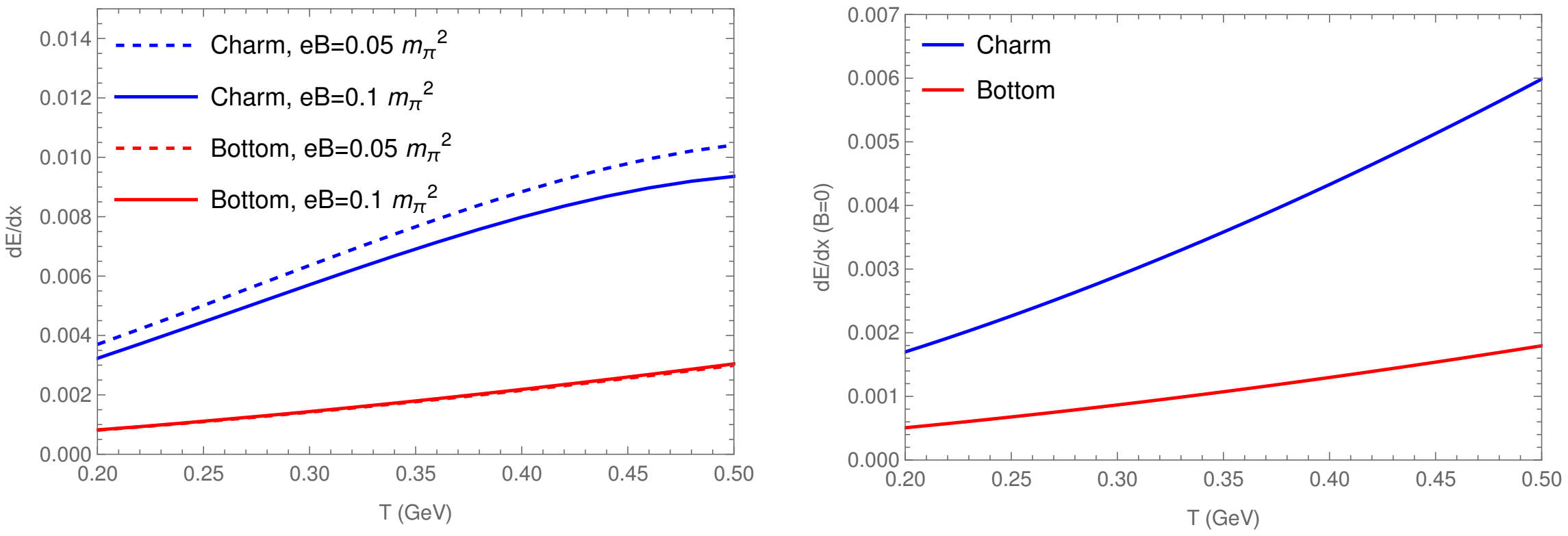

**(a)** **(b)**

**Figure 6.6:** (a) Energy loss of heavy quark as a function of temperature in the presence of fixed values of background magnetic field of strengths . (b) Energy loss of heavy quarks in the absence of background magnetic field.

Fig.(6.6a) shows the temperature variation of the HQ (Bottom and Charm) energy loss in the presence of a weak constant background field. In contrast to the momentum diffusion coefficients, the energy loss records an increasing trend with temperature.

Again, the variation with $B$ field strength is opposite for Charm and Bottom quarks. Whereas the bottom quark $dE/dx$ decreases with decreasing $B$, the charm $dE/dx$ increases. Also, the sensitivity to magnetic field is greater for the Charm quark compared to the bottom quark, which is reflected in Fig.(6.6a) by the discernible curves at $eB = 0.05m_\pi^2$ and $eB = 0.1m_\pi^2$. Fig.(6.6b) shows the variation of HQ energy loss with temperature in the absence of a background magnetic field. The energy loss for both the Charm and Bottom quarks increase with temperature, with both the magnitude as well as the rate of increase being greater for the Charm quark. Again, this can be attributed to the lighter mass of the Charm quark compared to Bottom.

## 6.6 Applications

### 6.6.1 Spatial diffusion coefficient

As mentioned in Section II, the drag coefficient $\eta$ can be obtained from the momentum diffusion coefficient via the fluctuation dissipation relation.

$$\eta_D = \frac{\kappa}{2M_Q T}. \tag{6.70}$$

The zero momentum value of the drag coefficient is then obtained from the zero momentum value of the momentum diffusion coefficient

$$\eta_D(p=0) = \frac{\kappa(p=0)}{2M_Q T}. \tag{6.71}$$

The spatial diffusion coefficient $D_s$ can be defined via $\eta_D(p=0)$ as [215]

$$D_s = \frac{T}{\eta_D(p=0)M_q}. \tag{6.72}$$

To evaluate $\kappa(p=0)$, we execute $p=0$ in the delta functions of Eqs.(6.64) and(6.65) which leads to $\omega = 0$. So, we have to calculate the momentum diffusion coefficients in the $\omega \to 0$ limit.

$$\sum_{i=1}^{4} \frac{\rho_i(\omega,q)A_i(\omega,q)}{\omega}\bigg|_{\omega\to 0} = \frac{A_1\rho_1}{\omega}\bigg|_{\omega\to 0} \tag{6.73}$$

All other terms vanish due to either the $A_i$'s or $\rho_i$'s vanishing in the $\omega \to 0$ limit. Finally, we are left with

$$\kappa_L = \frac{1}{2}\frac{\pi g^2 T}{4M_Q^2(2\pi)^2}\int_0^{q_{\max}} dq \int_0^{\pi} d\theta \frac{A_1(\omega=0)q^4}{\pi[q^4+q^2\,\Re_b(\omega=0)]^2}\left(q^4\sin^3\theta\right)\frac{\Im_b}{\omega}\bigg|_{\omega\to 0} \tag{6.74}$$

$$\kappa_T = \frac{\pi g^2 T}{4M_Q^2(2\pi)^2}\int_0^{q_{\max}} dq \int_0^{\pi} d\theta \frac{A_1(\omega=0)q^4}{\pi[q^4+q^2\,\Re_b(\omega=0)]^2}\left(q^4\sin\theta\cos^2\theta\right)\frac{\Im_b}{\omega}\bigg|_{\omega\to 0}. \tag{6.75}$$

Important thing to note is that $A_1(\omega=0)=M_Q^2$, and hence, the HQ mass dependence vanishes. Thus, we expect the momentum diffusion coefficient values for the charm and bottom quarks to be identical.

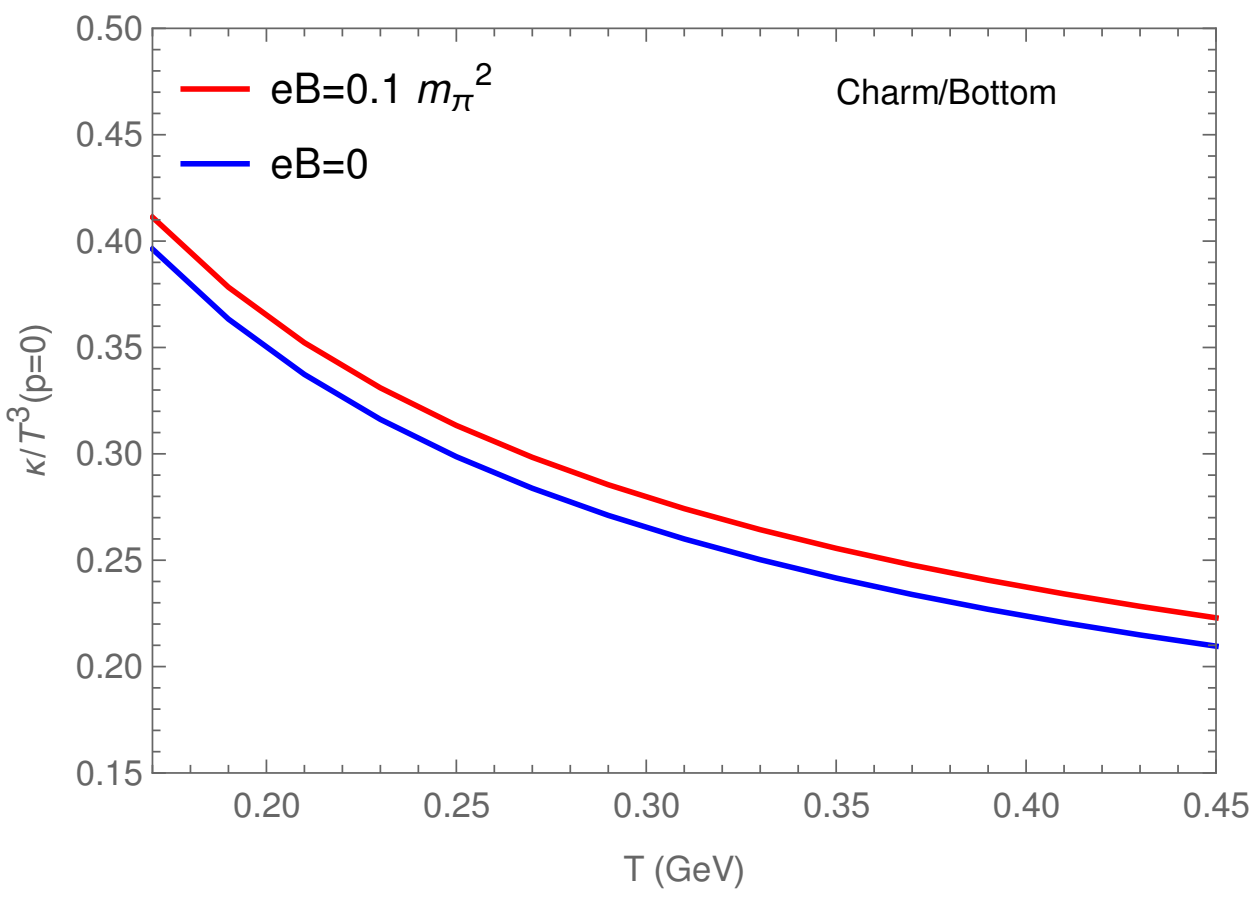

**Figure 6.7:** Normalised transverse and longitudinal momentum diffusion coefficients in the static limit ($\omega \to 0$) for both Charm and Bottom quarks as a function of temperature in the presence of a background magnetic field.

The momentum diffusion coefficient curves for the charm and bottom quarks overlap, as expected. However, the longitudinal and transverse components are also indistinguishable. Hence, in the static limit, the momentum diffusion of HQ is isotropic, even in the presence of a background magnetic field. As can be seen from Fig.(6.7), the static limit result of $\kappa/T^3$ for $B=0$ lies just below that of $B \neq 0$ with both the curves following a similar trend with respect to the temperature of the medium.

The case is similar for the spatial diffusion coefficient $D_s$ as well. From Eqs. (6.71) and (6.72), it can be seen that the HQ mass dependence cancels in $D_s$. In fact, this is one of the reasons why HQ diffusion is believed to carry generic information about the QCD medium.

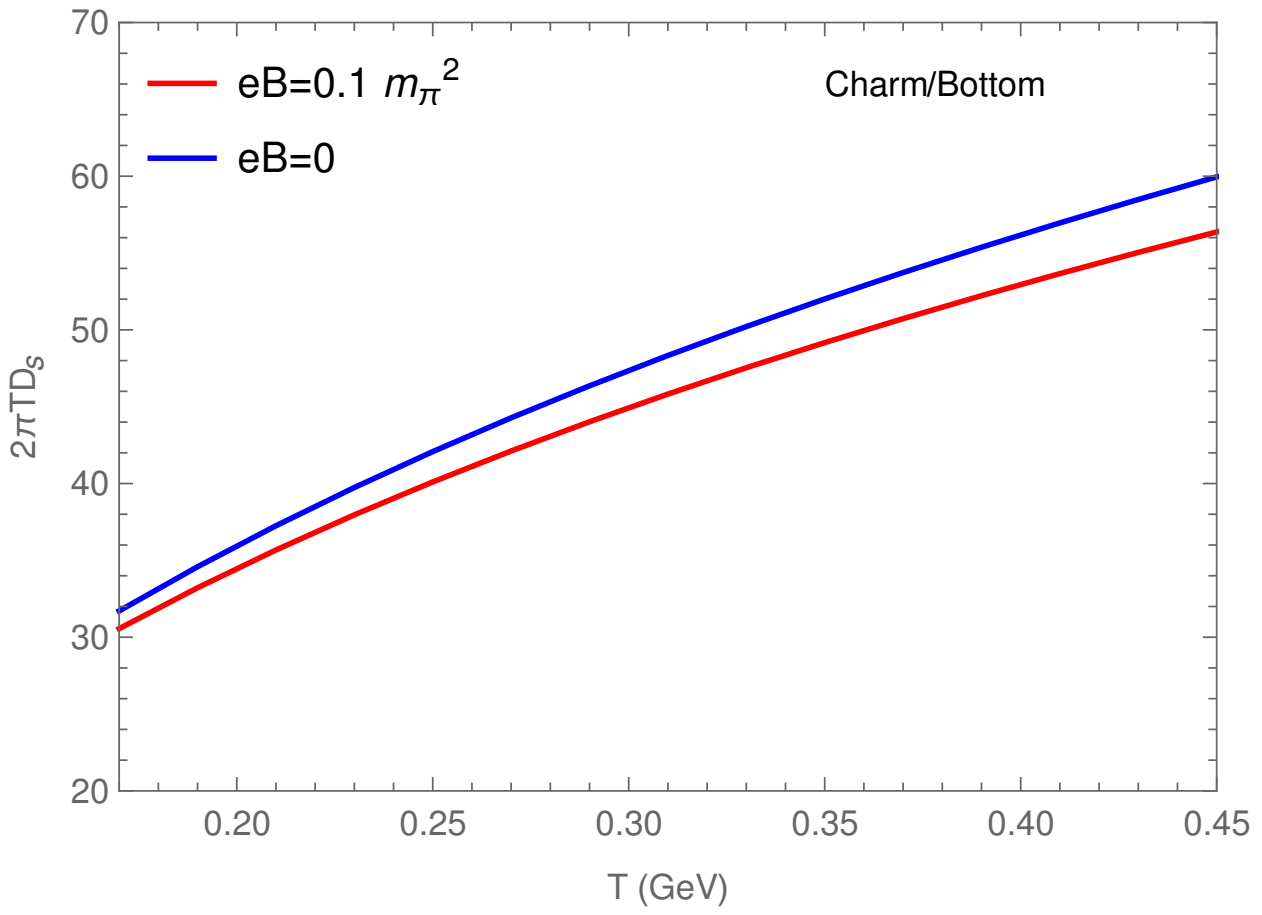

**Figure 6.8:** Spatial diffusion coefficient $D_s$ (multiplied with $2\pi T$) as a function of temperature in the presence of a background weak magnetic field.

Fig.(6.8) shows the variation of the spatial diffusion coefficient $(2\pi TD_s)$ with temperature in the presence of a background magnetic field of strength $eB = 0.1m_\pi^2$. For the sake of comparison, the $B = 0$ curve is also plotted. The curves corresponding to the charm and bottom quarks overlap, as expected. The increasing trend with temperature is similar to what has been observed in several pQCD leading order (LO) studies in the past, both with $T$-dependent and $T$-independent couplings [214, 238, 265].

## 6.7 Summary and Conclusions

In this work, we have investigated the dynamics of heavy quarks, *viz.* Charm and Bottom in the presence of a weak background magnetic field. In particular, we have calculated the momentum diffusion coefficients and the energy loss perturbatively up to first order in the strong coupling $\alpha_s$. The interaction rate is calculated by considering $2 \to 2$ elastic collisions of the form $Qq \to Qq$ and $Qg \to Qg$, by calculating the imaginary part of the heavy-quark self energy which is related to the squared matrix elements of the aforementioned collisional processes via the cutting rules. Gluon bremsstrahlung and Compton scattering processes are neglected since the former contributes only at higher order in $\alpha_s$ and the contribution of the latter is suppressed by powers of $M_Q/T$. There is a logarithmic U-V divergence present in the results of both the momentum diffusion coefficients and the energy loss. This is due to the fact that in this work, we have calculated the contribution to the dynamical quantities arising out of only soft gluon exchange. This can be justified since the interaction rate is dominated by processes involving soft gluon exchanges and hard scatterings contribute only at higher orders. Thus, it becomes necessary to put a U-V cutoff in the momentum integration, and hence, all our results are dependent on this cut-off.

We have investigated the temperature dependence of the momentum diffusion coefficients and energy loss for both the heavy flavours. The longitudinal and the transverse momentum diffusion coefficients decrease with increase in temperature for both the flavours For the values of magnetic fields considered, the sensitivity of Charm quark diffusion coefficients to the magnetic field is found to be greater than that of the Bottom quark, which can be attributed to the Charm quark being lighter. Further, the effect of increasing the magnetic field strength seems to have the opposite effects on the magnitudes of the Charm and Bottom diffusion coefficients; while the former decreases, the latter records an increase. This discrepancy in the Charm quark results could be an indication that processes such as Compton scattering and gluon bremsstrahlung are not really negligible for the charm quark because of its relatively lighter mass (hence higher velocity), and need to be taken into account. For comparison, the $B = 0$ results of the momentum diffusion coefficient ($\kappa$) have also been shown. It can be seen that the degree of anisotropy is larger for the Charm quark than that of Bottom quark, which suggests that the mass of the heavy quark under consideration plays a strong role in determining the isotropicity of $\kappa$. However, the extent of anisotropy at $B = 0$ is much less than that at $B \neq 0$. One of the key takeaways is that the presence of a background magnetic field increases the anisotropy in the momentum diffusion coefficients of the heavy quarks.

The heavy flavour energy loss is an increasing function of the temperature, both in the presence and absence of background magnetic field. Again, owing to its lighter mass, the sensitivity to both temperature and magnetic field is greater for the charmed quark, whereas for the Bottom quark, the curves corresponding to different magnetic fields almost overlap. The behaviour with $B$ is again different for the Charm quark compared to Bottom. The $B = 0$ results are similar with the rate of increase of the charm quark energy loss being steeper.

We have also looked at the $p = 0$ (static limit) results of momentum diffusion coefficients of the heavy quarks and found that the anisotropy in $\kappa$ completely vanishes in the said limit. Using this, we have looked at the spatial diffusion ($D_s$) of HQ. Interestingly, the mass HQ mass dependence cancels out both in $\kappa(p = 0)$ and $D_s$. The temperature variation of both $D_s$ and $\eta/s$ are similar to what has already been reported in the literature.

# Chapter 7

# Thesis summary and future prospects

प्रकाशस्य परिणामो द्विधा भवति, प्रतिबिम्बनं च विवर्तनं च ॥

*The change of light is two-fold: reflection and refraction.*

**– Bṛhat Saṃhitā by Varāhamihira (6$^{\text{th}}$ century CE)**

In this thesis, we have studied the thermoelectric response of the QGP medium via evaluation of the corresponding transport coefficients, *viz.* Seebeck and Nernst coefficients, in the presence of a background magnetic field. Specifically, both the limits of strong ($eB \gg T^2$) and weak ($eB \ll T^2$) magnetic field have been investigated in addition to the case of $B = 0$. We have studied the variation of the aforementioned coefficients with temperature, chemical potential ($\mu$), magnetic field strength ($B$), and also with current quark mass. A kinetic theory approach has been undertaken for the calculations wherein the relativistic Boltzmann transport equation is solved within the relaxation time approximation (RTA). In addition to the thermoelctric transport coefficients, we also study the transport coefficients associated with the propagation of heavy quarks (HQ) with a finite momentum in the QGP, *viz.* the HQ momentum and spatial diffusion coefficients and the HQ energy loss, up to leading order in the strong coupling $\alpha$.

In chapter 1, we provide a general introduction of the hot QCD matter produced in ultra-relativistic heavy-ion collisions at the RHIC and LHC, the possible production mechanisms, its signatures, and techniques of theoretical investigation.

In chapter 2, we discuss the theoretical tools and formalisms required in the subsequent chapters which includes the imaginary time formalism (ITF), HTL resummation, Boltzmann equation, and Langevin dynamics.

In Chapter 3, we study the thermoelectric response in a QGP in the absence of magnetic field as well as in the presence of a strong magnetic field. The response in both the situations is quantified by the Seebeck coefficient. We use the relativistic Boltzmann transport equation within the relaxation time approximation to carry out our calculations. We first calculate the Seebeck coefficients of hypothetical media consisting of a single quark species (*individual* Seebeck coefficient), followed by the evaluation in the composite medium (*Total* Seebeck coefficient). We study the Seebeck coefficient both using current quark masses **and** quasiparticle masses to explore how the interactions among partons described in perturbative thermal QCD in the quasiparticle framework, affect thermoelectric response. Thus, effectively, four different scenarios have been analysed:

1. Current mass description with $B = 0$.
2. Current mass description with $B \neq 0$.
3. Quasiparticle description with $B = 0$.
4. Quasiparticle description with $B \neq 0$.

Comparison between the cases 1 and 3 is highlights the effect of intereactions among the partons in the absence of strong magnetic field. We find that the magnitudes of individual as well as total Seebeck coefficients get sightly enhanced in the quasiparticle description in comparison to the current mass description. Comparison between cases 2 and 4 brings out the effect of the quasiparticle description in the presence of a strong magnetic field, where, it is seen that the magnitude of the coefficients are amplified in the presence of a strong $B$. Lastly, the comparison between cases 3 and 4 brings forth the effect of strong constant magnetic field on the Seebeck effect within the quasiparticle description. The variation of individual and total Seebeck coefficients with temperature and chemical potential are found to show similar trends in both the cases but with enhanced magnitudes in the latter case.

The sign of the individual Seebeck coefficients is positive for positively charged particles ($u$ quark) and negative for negatively charged particles ($d$ quark) for all the cases. The total Seebeck coefficient is positive for all cases. The trend of overall decrease (increase) of the magnitude of Seebeck coefficient with the increase in temperature (chemical potential) is also observed for all cases. Thus, for an interacting QGP, our study reveals that the thermoelectric response is much stronger in the presence of a strong magnetic field compared to the $B = 0$ scenario.

In Chapter 4 we study the thermoelectric response in the presence of a weak background magnetic field ($eB \ll T^2$), first by using the same quasiparticle masses as in the previous section, where, the magnetic field dependence in the quasiparticle mass enters through the coupling $g(T,B)$. In the presence of weak $B$, the thermoelectric response becomes a $2 \times 2$ matrix, instead of a scalar, with the diagonal elements representing the Seebeck coefficient and the off-diagonal elements representing the Nernst coefficient. The major observation is that the sign of the individual Nernst coefficients do not depend on the electric charge of the quarks. This is because the induced electric field in this case points in the same direction regardless of the charge of the quarks. We find that the Seebeck coefficient magnitude of the composite medium is less than that of the strong $B$ case, but more than that in $B = 0$. Thus, the Seebeck coefficient is maximum at strong $B$, followed by that at weak $B$, and is weakest at $B = 0$. Thereafter, we use the quasiparticle masses derived explicitly from the one-loop quark self-energy in weak magnetic field. The Left ($L$) and Right ($R$) chiral modes of these quasiparticle masses come out to be different, unlike in the case of strong $B$, where they are degenerate. Using these masses, we evaluate the Seebeck and Nernst coefficients, and compare them to results obtained in the $B = 0$ and strong $B$ cases. The main observation is that these quasiparticle masses can be used to define the range of values for $T$ and $B$ that can be used simultaneously in the weak field limit, by imposing the condition $m_R^2 > 0$.

In chapter 5, we evaluate the same coefficients by taking into account the anisotropic expansion of the QGP. We use the spheroidally asymmetric distribution function introduced by Romatschke-Strickland and use it to evaluate the Seebeck and nernst coefficients in the presence of a weak magnetic field. We have found that the magnitudes of both the individual as well as the total coefficients-both Seebeck and Nernst, are decreasing functions of temperature, and decreasing functions of anisotropy strength (characterized by the anisotropy parameter $\xi$). It is important to note that the Seebeck coefficient vanishes for $\mu = 0$ irrespective of the magnetic field strength, and the Nernst coefficient vanishes for $|\boldsymbol{B}| = 0$, irrespective of the value of $\mu$. We have analysed the coefficient sensitivity to the $L$ and $R$ modes, *i.e.* to the difference in quasiparticle effective masses, both for individual and total coefficients. To that end we have calculated the average percentage change in the coefficient magnitude as one goes from the $L$ mode to the $R$ mode or vice versa. For the Seebeck coefficient, the average (over $T$ and $\xi$) percentage change for the $u$ quark is $\sim 9.83\%$, whereas for the $d$ quark, it is $\sim 3.02\%$. This shows that the $d$ quark Seebeck coefficient is comparatively less sensitive to differences in quasiparticle masses. For the total Seebeck coefficient, the average percentage change in going from the $L$ mode to the $R$ mode is $\sim 4.61\%$, which is in between the values for the individual coefficients.

Certain differences arise in the case of Nernst coefficient. Firstly, for both the individual and total Nernst coefficients, the absolute values of the $R$ mode coefficients are greater than that of the $L$ mode. This is opposite to the case of Seebeck coefficients. The average percentage change in going from the $R$ mode to the $L$ mode in case of $u$ quark Nernst coefficient is $\sim 0.71\%$; for the $d$ quark Nernst coefficient, this value is $\sim 0.21\%$. Compared to their Seebeck counterparts, these numbers are an order of magnitudes smaller, indicating that the individual Nernst coefficients are comparatively much less sensitive to the change in quasiparticle modes. For the case of the total Nernst coefficient, the average percentage decrease from $R$ to $L$ modes is $\sim 0.55\%$. The absolute values of both the individual and total Nernst coefficients are about 1 order of magnitude smaller than their Seebeck counterparts, which shows that the Nernst effect is a weaker response than the Seebeck effect.

In chapter 6, we have investigated the dynamics of heavy quarks, *viz.* Charm and Bottom in the presence of a weak background magnetic field. In particular, we have calculated the momentum diffusion coefficients and the energy loss perturbatively upto first order in the strong coupling $\alpha_s$. We have investigated the temperature dependence of the momentum diffusion coefficients and energy loss for both the heavy flavours. The longitudinal and the transverse momentum diffusion coefficients decrease with increase in temperature for both the flavours; the rates being similar for the transverse coefficient, and higher for the charm quark with regard to the longitudinal coefficient. For the values of magnetic fields considered, the sensitivity of Charm quark diffusion coefficients to the magnetic field is found to be greater than that of the Bottom quark, which can be attributed to the Charm quark being lighter. Further, the effect of increasing the magnetic field strength seems to have the opposite effects on the magnitudes of the Charm and Bottom diffusion coefficients; while the former decreases, the latter records an increase. For comparison, the $B = 0$ results of the momentum diffusion coefficient ($\kappa$) have also been shown. It can be seen that the degree of anisotropy is much larger for the Charm quark than that of Bottom quark, which suggests that the mass of the heavy quark under consideration plays a strong role in determining the isotropicity of $\kappa$. The heavy flavour energy loss is an increasing function of the temperature, both in the presence and absence of background magnetic field. Again, owing to its lighter mass, the sensitivity to both temperature and magnetic field is greater for the charmed quark, whereas for the Bottom quark, the curves corresponding to different magnetic fields almost overlap. The $B = 0$ results are similar with the rate of increase of the charm quark energy loss being steeper. We have also looked at the $p = 0$ (static limit) results of momentum diffusion coefficients of the heavy quarks and found that the anisotropy in $\kappa$ completely vanishes in the said limit. Using this, we have looked at the spatial diffusion ($D_s$) of HQ. Interestingly, the mass HQ mass dependence cancels out both in $\kappa(p = 0)$ and $D_s$.

As we have seen, the weak $B$ leading order perturbative results of HQ observables are U-V cut-off dependent. This work can be extended to remove the cut-off dependence if one includes hard scatterings of HQs with the thermal partons. A fast developing area of research is investigation of the effects of a rotating QGP medium on various of its properties. It would be interesting to study transport coefficients of QGP in the presence of magnetic field in a rotating QGP.

# Appendix A

## A.1 Tensor structure of resummed gluon propagator in the presence of magnetic field

We begin by discussing the 4-vectors that characterize the system under consideration. The fluid 4-velocity in local rest frame (LRF), and the metric tensor is given by

$$u^\mu = (1, 0, 0, 0), \quad g^{\mu\nu} = \text{diag}(1, -1, -1, -1). \tag{A.1}$$

The direction of the external magnetic field is specified by the projection of the EM field tensor $F^{\mu\nu}$ along $u^\mu$:

$$n_\mu = \frac{1}{2B}\epsilon_{\mu\nu\rho\lambda} u^\nu F^{\rho\lambda} = (0, 0, 0, 1) \tag{A.2}$$

Introduction of these 4-vectors allows one to define a Lorentz invariant energy and momentum component as

$$q^0 = q_0 = Q \cdot u, \quad q^3 = -q_3 = Q \cdot n. \tag{A.3}$$

We define parallel and perpendicular components of vectors and the metric tensor in the LRF as

$$Q^\mu_\parallel = (Q.u)u^\mu + (Q\cdot n)n^\mu = \left(q^0,0,0,q^3\right); \tag{A.4}$$

$$Q^\mu_\perp = Q^\mu - Q^\mu_\parallel = \left(0,q^1,q^2,0\right) \tag{A.5}$$

$$Q^2_\parallel = q_0^2 - q_3^2\,, \quad Q^2_\perp = -(q_1^2+q_3^2) = -q_\perp^2 \tag{A.6}$$

$$g^{\mu\nu}_\parallel = u^\mu u^\nu - n^\mu n^\nu = \mathrm{diag}(1,0,0-1), \tag{A.7}$$

$$g^{\mu\nu}_\perp = g^{\mu\nu} - g^{\mu\nu}_\parallel = \mathrm{diag}(0,-1,-1,0), \tag{A.8}$$

One can further redefine $u^\mu$ and $n^\mu$ as

$$\bar{u}^\mu = u^\mu - \frac{(Q\cdot u)Q^\mu}{Q^2} = u^\mu - \frac{q_0 Q^\mu}{Q^2} \tag{A.9}$$

$$\bar{n}^\mu = n^\mu - \frac{(\tilde{Q}\cdot n)\tilde{Q}^\mu}{\tilde{Q}^2} = n^\mu - \frac{q_3 Q^\mu}{q^2} + \frac{q_0 q_3 u^\mu}{q^2}, \tag{A.10}$$

where, $\tilde{Q}^\mu = Q^\mu - (Q\cdot u)$. $\bar{u}^\mu$ and $\bar{n}^\mu$ so defined are orthogonal to $Q^\mu$ and $\tilde{Q}^\mu$, respectively. In the presence of a magnetic field, a set of basis tensors that are mutually orthogonal, can be constructed out of the 4-vectors mentioned above:

$$\Delta_1^{\mu\nu} = \frac{\bar{u}^\mu\bar{u}^\nu}{\bar{u}^2}, \tag{A.11}$$

$$\Delta_2^{\mu\nu} = g^{\mu\nu}_\perp - \frac{Q^\mu_\perp Q^\nu_\perp}{Q^2_\perp}, \tag{A.12}$$

$$\Delta_3^{\mu\nu} = \frac{\bar{n}^\mu\bar{n}^\nu}{\bar{n}^2}, \tag{A.13}$$

$$\Delta_4^{\mu\nu} = \frac{\bar{u}^\mu\bar{n}^\nu + \bar{u}^\nu\bar{n}^\mu}{\sqrt{\bar{u}^2}\sqrt{\bar{n}^2}}. \tag{A.14}$$

These tensors satisfy the following properties:

$$(\Delta_4)^{\mu\rho}(\Delta_4)_{\rho\nu} = (\Delta_1)^\mu_\nu + (\Delta_3)^\mu_\nu\,, \tag{A.15}$$

$$(\Delta_k)^{\mu\rho}(\Delta_4)_{\rho\nu} + (\Delta_4)^{\mu\rho}(\Delta_k)_{\rho\nu} = (\Delta_4)^\mu_\nu\,, \tag{A.16}$$

$$(\Delta_2)^{\mu\rho}(\Delta_4)_{\rho\nu} = (\Delta_4)^{\mu\rho}(\Delta_2)_{\rho\nu} = 0, \tag{A.17}$$

Any second rank tensor can be expanded in terms of these basis tensors. As such, the gluon self energy can be written as

$$\Pi^{\mu\nu}(q_0,\boldsymbol{q}) = b(q_0,\boldsymbol{q})\Delta_1^{\mu\nu} + c(q_0,\boldsymbol{q})\Delta_2^{\mu\nu} + d(q_0,\boldsymbol{q})\Delta_3^{\mu\nu} + a(q_0,\boldsymbol{q})\Delta_4^{\mu\nu}, \tag{A.18}$$

where, $b$, $c$, $d$, $a$ are Lorentz invariant form factors. The Schwinger-Dyson equation relates the bare propagator, resummed propagator and the self energy of the particle under consideration. For the gluon propagator, we have

$$\mathcal{D}^{-1}_{\mu\nu} = \mathcal{D}^0_{\mu\nu} - \Pi_{\mu\nu}, \tag{A.19}$$

where, $\mathcal{D}^0_{\mu\nu}$ is the bare gluon propagator. We recall that any rank-2 tensor (and it's inverse) can be written in terms of the basis tensors $\Delta_i's$. Then, using Eq.(A.18), (A.19), and the fact that $\mathcal{D}^{-1}_{\mu\rho}\mathcal{D}^{\rho\nu} = g^\nu_\mu$, we can derive the structure of the resummed gluon propagator as mentioned in Eq.(6.16)

## A.2 Form factors in weak magnetic field

The fermion propagator in a weak background magnetic field is written as a series expansion in powers of $qB$ as (upto $\mathcal{O}(qB)^2$)

$$\begin{aligned} iS(K) &= i\frac{(\not K + m_f)}{K^2 - m_f^2} - q_f B\frac{\gamma_1\gamma_2\left(\not K_\parallel + m_f\right)}{\left(K^2 - m_f^2\right)^2} - 2i\left(q_f B\right)^2 \frac{\left[K_\perp^2\left(\not K_\parallel + m_f\right) + \not K_\perp\left(m_f^2 - K_\parallel^2\right)\right]}{\left(K^2 - m_f^2\right)^4} \\ &\equiv S_0(K) + S_1(K) + S_2(K) \end{aligned}$$

The quark loop (fermion) contribution to the gluon self energy is then given by

$$\begin{aligned} \Pi_f^{\mu\nu}(Q) =& -\sum_f \frac{ig^2}{2}\int\frac{d^4K}{(2\pi)^4}\,\mathrm{Tr}\left[\gamma^\nu\left\{S_0(K)+S_1(K)+S_2(K)\right\}\times\gamma^\mu\left\{S_0(P)+S_1(P)+S_2(P)\right\}\right] \\ =&\Pi^{\mu\nu}_{(0,0)}(Q) + \Pi^{\mu\nu}_{(1,1)}(Q) + 2\Pi^{\mu\nu}_{(2,0)}(Q) + O\left[\left(q_f B\right)^3\right], \end{aligned}$$

where,

the first term is of $\mathcal{O}(qB)$ and the remaining are of $\mathcal{O}((qB)^2)$. The $\mathcal{O}(qB)$ term vanishes owing to Furry's theorem. They are given as:

$$\Pi^{\mu\nu}_{(0,0)}(Q) = \sum_f i2g^2 \int \frac{d^4K}{(2\pi)^4} \frac{\left[P^\mu K^\nu + K^\mu P^\nu - g^{\mu\nu}\left(K\cdot P - m_f^2\right)\right]}{\left(K^2 - m_f^2\right)\left(P^2 - m_f^2\right)} \tag{A.20}$$

$$\Pi^{\mu\nu}_{(1,1)}(Q) = \sum_f 2ig^2\left(q_f B\right)^2 \int \frac{d^4K}{(2\pi)^4} \frac{\left[P^\mu_\parallel K^\nu_\parallel + K^\mu_\parallel P^\nu_\parallel + \left(g^{\mu\nu}_\parallel - g^{\mu\nu}_\perp\right)\left(m_f^2 - K_\parallel\cdot P_\parallel\right)\right]}{\left(K^2 - m_f^2\right)^2\left(P^2 - m_f^2\right)^2} \tag{A.21}$$

$$\Pi^{\mu\nu}_{(2,0)}(Q) = -\sum_f 4ig^2\left(q_f B\right)^2 \int \frac{d^4K}{(2\pi)^4} \left[\frac{M^{\mu\nu}}{\left(K^2 - m_f^2\right)^4\left(P^2 - m_f^2\right)}\right], \tag{A.22}$$

where,

$$M^{\mu\nu} = K_\perp^2\left[P^\mu K^\nu_\parallel + K^\mu_\parallel P^\nu - g^{\mu\nu}\left(K_\parallel\cdot P - m_f^2\right)\right] + (m_f^2 - K_\parallel^2)\left[P^\mu K^\nu_\perp + K^\mu_\perp P^\nu - g^{\mu\nu}\left(K_\perp\cdot P\right)\right]$$

The complete gluon self energy is expressed as:

$$\Pi^{\mu\nu}(Q) = \Pi^{\mu\nu}_{\rm YM}(Q) + \Pi^{\mu\nu}_f(Q),$$

where, $\Pi^{\mu\nu}_{\rm YM}$ refers to the Yang-Mills contribution to the gluon self energy coming from the ghost and gluon loops, which is unaffected by the magnetic field. It is given by

$$\Pi^{\mu\nu}_{\rm YM}(Q) = -\frac{N_c\, g^2T^2}{3}\int\frac{d\Omega}{2\pi}\left(\frac{q_0\hat{K}^\mu\hat{K}^\nu}{\hat{K}\cdot Q} - g^{\mu 0}g^{\nu 0}\right).$$

Using the properties of the tensors $\Delta_i$, the form factors can be expressed as:

$$b(Q) = \Delta_1^{\mu\nu}(Q)\Pi_{\mu\nu}(Q) = \Delta_1^{\mu\nu}(\Pi^{\rm YM}_{\mu\nu} + \Pi^f_{\mu\nu}) = b_{\rm YM}(Q) + b_{0f}(Q) + b_{2f}(Q) \tag{A.23}$$

$$c(Q) = \Delta_2^{\mu\nu}(Q)\Pi_{\mu\nu}(Q) = \Delta_2^{\mu\nu}(\Pi^{\rm YM}_{\mu\nu} + \Pi^f_{\mu\nu}) = c_{\rm YM}(Q) + c_{0f}(Q) + c_{2f}(Q) \tag{A.24}$$

$$d(Q) = \Delta_3^{\mu\nu}(Q)\Pi_{\mu\nu}(Q) = \Delta_3^{\mu\nu}(\Pi^{\rm YM}_{\mu\nu} + \Pi^f_{\mu\nu}) = d_{\rm YM}(Q) + d_{0f}(Q) + d_{2f}(Q) \tag{A.25}$$

$$a(Q) = \frac{1}{2}\Delta_4^{\mu\nu}(Q)\Pi_{\mu\nu}(Q) = \frac{1}{2}\Delta_4^{\mu\nu}(\Pi^{\rm YM}_{\mu\nu} + \Pi^f_{\mu\nu}) = a_{\rm YM}(Q) + a_{0f}(Q) + a_{2f}(Q). \tag{A.26}$$

In terms of powers of $qB$, the form factors can be expressed as

$$F(Q) = F_0(Q) + F_2(Q) = \left[F_{\rm YM}(Q) + F_{0f}(Q)\right] + F_{2f}(Q), \quad F = b, c, d, a \tag{A.27}$$

## B.1 $\mathcal{O}(qB)^0$ terms of form factors

Using Eqs.(A.11-A.14) and Eq.(A.18), we can write

$$\Delta_1^{00} = \bar{u}^2, \quad \Delta_2^{00} = \Delta_3^{00} = \Delta_4^{00} = 0, \quad \Pi^{00} = b\bar{u}^2$$

Thus,

$$b_0(Q) = \frac{1}{\bar{u}^2}\left[\Pi_{00}^{\rm YM}(Q) + \Pi_{00}^{(0,0)}(Q)\right] \tag{A.28}$$

In the HTL approximation ($K \sim T$, $Q \sim gT$)

$$\Pi_{00}^{(0,0)}(Q) = \frac{N_f\, g^2T^2}{6}\left(1 - \frac{q_0}{2q}\log\frac{q_0+q}{q_0-q}\right), \quad \Pi_{00}^{\rm YM} = \frac{N_c\, g^2T^2}{3}\left(1 - \frac{q_0}{2q}\log\frac{q_0+q}{q_0-q}\right) \tag{A.29}$$

Thus,

$$b_0(Q) = \frac{m_D^2}{\bar{u}^2}\left(1 - \frac{q_0}{2q}\log\frac{q_0+q}{q_0-q}\right), \tag{A.30}$$

where, $m_D^2 = \left(\Pi_{00}^{(0,0)} + \Pi_{00}^{\rm YM}\right)\Big|_{\substack{p_0=0\\ \mathbf{p}\to 0}} = \frac{g^2T^2}{3}\left(N_c + \frac{N_f}{2}\right)$ is the QCD Debye screening mass in the absence of magnetic field.

An alternate way of evaluating the form factors is to calculate the self energy diagrammatically. As an example, the quark loop contribution to $c_0$ will be evaluated this way.

$$c_0^f(Q) = \left(g_\perp^{\mu\nu} - \frac{Q_\perp^\mu Q_\perp^\nu}{Q_\perp^2}\right)\Pi_{\mu\nu}^{(0,0)} = T1 - T2, \tag{A.31}$$

where,

$$T1 = g_\perp^{\mu\nu}\Pi_{\mu\nu}^{(0,0)}, \quad T_2 = \frac{Q_\perp^\mu Q_\perp^\nu}{Q_\perp^2}\Pi_{\mu\nu}^{(0,0)}$$

Using the expression of $\Pi_{\mu\nu}^{(0,0)}$ from Eq.(A.20) under the HTL approximation, we get,

$$T1 = -\sum_f 4g^2(I_1 - I_2), \tag{A.32}$$

where,

$$I_1 = \int \frac{d^3k}{(2\pi)^3} T \sum_n \frac{K_\perp^2}{(K^2 - m_f^2)(P^2 - m_f^2)} \tag{A.33}$$

$$I_2 = \int \frac{d^3k}{(2\pi)^3} T \sum_n \frac{K^2}{(K^2 - m_f^2)(P^2 - m_f^2)} \tag{A.34}$$

$I_2$ is a well known integral which, under the HTL approximation, and in the limit of $m_f \to 0$ is $T^2/24$. The $I_1$ integral after summing over matsubara frequencies simplifies to

$$I_1 = -\frac{1}{2} \int \frac{d^3k}{(2\pi)^3} \left[ \frac{n_F(E_1)}{E_1} - \left(1 - \frac{q_0}{q_0 - q\cos\theta}\right) \frac{dn_F(E_1)}{dk} \right] (\cos^2\theta - 1) \tag{A.35}$$

Here, $E_1 \approx k$ is the energy of the fermion propagator having 4-momentum $K$, $\theta$ is the polar angle made by $\boldsymbol{k}$, and $n_F$ is the Fermi-Dirac distribution. First term of Eq.(A.35) evaluates to $T^2/72$. The second term can be expanded as

$$\frac{1}{8\pi^2} \int dk\, d(\cos\theta)\, k^2 \frac{dn_F(k)}{dk} \left[ \cos^2\theta - 1 - \frac{q_0 \cos^2\theta}{q_0 - q\cos\theta} + \frac{q_0}{q_0 - q\cos\theta} \right].$$

This evaluates term by term to

$$-T^2/72 + T^2/24 - \frac{q_0 T^2}{48q} \left[ 2\frac{q_0}{q} + \frac{q_0^2}{q^2} \log\frac{q_0 + q}{q_0 - q} - \log\frac{q_0 + q}{q_0 - q} \right]$$

. Thus,

$$I_1 - I_2 = -\frac{T^2}{48q^2} \left[ 2q_0^2 - (q_0^2 - q^2)\frac{q_0}{q} \log\frac{q_0 + q}{q_0 - q} \right] \tag{A.36}$$

Similarly, it can be shown that

$$T2 = -\sum_f 2g^2(-I_1 + I_2 + I_3), \tag{A.37}$$

where,

$$I_3(Q) = \int \frac{d^3k}{(2\pi)^3} T \sum_n \frac{2k_1 k_2}{(K^2 - m_f^2)(P^2 - m_f^2)} = 0$$

Thus,

$$c_0^f(Q) = T_1 + T_2 = -\sum_f 2g^2[I_1 - I - 2] = \frac{N_f\, g^2T^2}{6}\frac{1}{2q^2}\left[q_0^2 - (q_0^2 - q^2)\frac{q_0}{2q}\log\frac{q_0+q}{q_0-q}\right] \tag{A.38}$$

The Yang-Mills contribution is given by

$$\frac{N_c\, g^2T^2}{3}\frac{1}{2q^2}\left[q_0^2 - (q_0^2 - q^2)\frac{q_0}{2q}\log\frac{q_0+q}{q_0-q}\right]$$

Hence, finally,

$$c_0(q_0, q) = \frac{m_d^2}{2q^2}\left[q_0^2 - (q_0^2 - q^2)\frac{q_0}{2q}\log\frac{q_0+q}{q_0-q}\right] \tag{A.39}$$

Next, we have

$$d_0(q_0, q) = \frac{\bar{n}^\mu\bar{n}^\nu}{\bar{n}^2}\left(\Pi^{\rm YM}_{\mu\nu} + \Pi^{(0,0)}_{\mu\nu}\right) \tag{A.40}$$

It turns out that

$$d_0(q_0, q) = c_0(q_0, q) = \frac{m_d^2}{2q^2}\left[q_0^2 - (q_0^2 - q^2)\frac{q_0}{2q}\log\frac{q_0+q}{q_0-q}\right] \tag{A.41}$$

The form factor $a_0$ is given by

$$\begin{aligned} a_0(q_0, q) &= \frac{1}{2}\Delta_4^{\mu\nu}(\Pi^{\rm YM}_{\mu\nu} + \Pi^{(0,0)}_{\mu\nu}) && \text{(A.42)}\\ &= \frac{1}{2\sqrt{\bar{u}^2}\sqrt{\bar{n}^2}}\left[-2\frac{\bar{u}\cdot n}{\bar{u}^2}\left[\Pi^{\rm YM}_{00} + \Pi^{(0,0)}_{00}\right] + 2\left[\Pi^{\rm YM}_{03} + \Pi^{(0,0)}_{03}\right]\right] && \text{(A.43)}\\ &= 0 && \text{(A.44)} \end{aligned}$$

## B.2 $\mathcal{O}(qB)^2$ terms of form factors

$$b_2(q_0, q) = \frac{u^\mu u^\nu}{\bar{u}^2}\left[\Pi^{(1,1)}_{\mu\nu} + 2\Pi^{(2,0)}_{\mu\nu}\right] \tag{A.45}$$

Using Eqs. (A.21) and (A.22) in the above equation, we get

$$b_2(q_0, q) = -\sum_f \frac{2g^2\left(q_fB\right)^2}{\bar{u}^2}\int\frac{d^3k}{(2\pi)^3}T\sum_n\left\{\frac{K^2 + k^2\left(1+\cos^2\theta\right) + m_f^2\Big)}{\left(K^2 - m_f^2\right)^2\left(P^2 - m_f^2\right)^2} + \frac{8\left(k^4 + k^2K^2\right)\left(1 - \text{co}\right.}{\left(K^2 - m_f^2\right)^4\left(P^2 - \right.}\right.$$

We make use of the HTL simplifications mentioned in Appendix C of [bithika] to further simplify $b_2$ to obtain

$$b_2 = -\sum_f \frac{2g^2\left(q_f B\right)^2}{\bar{u}^2}\int\frac{d^3k}{(2\pi)^3}T\sum_n\left\{\frac{1}{\left(K^2-m_f^2\right)^2\left(P^2-m_f^2\right)}+\frac{(-7+9c^2)k^2+2m_f^2}{\left(K^2-m_f^2\right)^3\left(P^2-m_f^2\right)}\right.$$
$$\left.-\frac{8(1-c^2)(k^4+m_f^2k^2)}{\left(K^2-m_f^2\right)^4\left(P^2-m_f^2\right)}\right\},$$

where, $c = \cos\theta$. Next, we perform the frequency sum using

$$T\sum_n \frac{1}{\left(\omega_n^2+E_k^2\right)\left[\left(\omega_n-\omega\right)^2+E_{k-q}^2\right]} = \frac{\left[1-n_F\left(E_k\right)-n_F\left(E_{k-q}\right)\right]}{4E_kE_{k-q}}\left\{\frac{1}{i\omega+E_k+E_{k-q}}-\frac{1}{i\omega-E_k-E_{k-q}}\right\}$$
$$+\frac{\left[n_F\left(E_k\right)-n_F\left(E_{k-q}\right)\right]}{4E_kE_{k-q}}\left\{\frac{1}{i\omega+E_k-E_{k-q}}-\frac{1}{i\omega-E_k+E_{k-q}}\right\},$$

where, $E_k = \sqrt{k^2+m_f^2}$, $E_{k-q} = \sqrt{(k-q)^2+m_f^2}$. We write the expression in terms of mass derivatives to finally obtain

$$b_2\left(q_0,q\right) = \sum_f \frac{2g^2q_f^2B^2}{\bar{u}^2}\left\{\left(\frac{\partial^2}{\partial^2\left(m_f^2\right)}+\frac{5}{6}m_f^2\frac{\partial^3}{\partial^3\left(m_f^2\right)}\right)\int\frac{d^3k}{(2\pi)^3}\frac{n_F\left(E_k\right)}{E_k}\left(\frac{q_0}{q_0-q\cos\theta}-1\right)\right.$$
$$+\left(\frac{\partial}{\partial\left(m_f^2\right)}+\frac{5}{6}m_f^2\frac{\partial^2}{\partial^2\left(m_f^2\right)}\right)\int\frac{d^3k}{(2\pi)^3}\frac{n_F\left(E_k\right)}{2E_k^3}\left(\frac{q_0}{q_0-q\cos\theta}\right)$$
$$-\left(\frac{\partial^2}{\partial^2\left(m_f^2\right)}+\frac{m_f^2}{2}\frac{\partial^3}{\partial^3\left(m_f^2\right)}\right)\int\frac{d^3k}{(2\pi)^3}\frac{n_F\left(E_k\right)}{E_k}\cos^2\theta\left(\frac{q_0}{q_0-q\cos\theta}-1\right)$$
$$\left.-\left(\frac{\partial}{\partial\left(m_f^2\right)}+\frac{m_f^2}{2}\frac{\partial^2}{\partial^2\left(m_f^2\right)}\right)\int\frac{d^3k}{(2\pi)^3}\frac{n_F\left(E_k\right)}{2E_k^3}\cos^2\theta\left(\frac{q_0}{q_0-q\cos\theta}\right)\right\}$$

After simplification, we finally obtain

$$b_2 = \frac{\delta m_D^2}{\bar{u}^2} + \sum_f \frac{g^2\left(q_f B\right)^2}{\bar{u}^2\pi^2} \times \left[\left(g_k + \frac{\pi m_f - 4T}{32 m_f^2 T}\right)(A_0 - A_2)\right. \tag{A.46}$$

$$\left. + \left(f_k + \frac{8T - \pi m_f}{128 m_f^2 T}\right)\left(\frac{5A_0}{3} - A_2\right)\right]. \tag{A.47}$$

Here, $\delta m_D^2$ is the correction to the debye mass due to weak magnetic field given by:

$$\delta m_D^2 = \left[\Pi_{\mu\nu}^{(1,1)} + 2\Pi_{\mu\nu}^{(2,0)}\right]_{q_0=0,\, q\to 0} = \sum_f \frac{g^2}{12\pi^2 T^2}(q_f B)^2 \sum_{l=1}^{\infty}(-1)^{l+1} l^2 K_0\left(\frac{m_f l}{T}\right) \tag{A.48}$$

$$f_k = -\sum_{l=1}^{\infty}(-1)^{l+1}\frac{l^2}{16T^2}K_2\left(\frac{m_f l}{T}\right) \tag{A.49}$$

$$g_k = \sum_{l=1}^{\infty}(-1)^{l+1}\frac{l}{4m_f T}K_1\left(\frac{m_f l}{T}\right). \tag{A.50}$$

$$A_0 = \int \frac{d\Omega}{4\pi}\frac{q_0 c^0}{Q\cdot\hat{K}} = \frac{q_0}{2q}\log\left(\frac{q_0+q}{q_0-q}\right) \tag{A.51}$$

$$A_2 = \int \frac{d\Omega}{4\pi}\frac{q_0 c^2}{Q\cdot\hat{K}} = \frac{q_0^2}{2q^2}\left(1 - \frac{3q_3^2}{q^2}\right)\left(1 - \frac{q_0}{2q}\log\frac{q_0+q}{q_0-q}\right) + \frac{1}{2}\left(1 - \frac{q_3^2}{q^2}\right)\frac{q_0}{2q}\log\frac{q_0+q}{q_0-q} \tag{A.52}$$

$K_0$, $K_1$, $K_2$ are the modified Bessel functions of the second kind. Similarly, the form factor $c_2$ is given by

$$c_2(q_0, q) = \left(g_\perp^{\mu\nu} - \frac{Q_\perp^\mu Q_\perp^\nu}{Q_\perp^2}\right)\left[\Pi_{\mu\nu}^{(1,1)} + 2\Pi_{\mu\nu}^{(2,0)}\right] \tag{A.53}$$

Using Eqs.(A.21) and (A.22), we get

$$c_2(q_0,q) = -\sum_f \frac{g^2\left(q_fB\right)^2}{2}\int\frac{d^3k}{(2\pi)^3}T\sum_n\left[\frac{4k_0^2-4k_3^2-4m_f^2}{\left(K^2-m_f^2\right)^2\left(P^2-m_f^2\right)^2}+\frac{4\left(4k_3^2-4k_0^2+4m_f^2\right)}{\left(K^2-m_f^2\right)^3\left(P^2-m_f^2\right)}\right.$$
$$\left.-\frac{4\left(k_0^2-k_3^2-m_f^2\right)\left(8k_\perp^2-4K^2+4m_f^2+8(\boldsymbol{k}\cdot\boldsymbol{q})_\perp^2/q_\perp^2\right)}{\left(K^2-m_f^2\right)^4\left(P^2-m_f^2\right)}\right]$$

Using HTL approximations to simplify as earlier, we write $c_2$ in terms of mass derivatives as earlier

$$c_2 = -\sum_f 2g^2\left(q_fB\right)^2\int\frac{d^3k}{(2\pi)^3}T\sum_n\left[\frac{1}{2}+\frac{1}{4}\left(1-\cos^2\theta\right)\cos^2\phi+\frac{7}{4}\sin^2\theta\left(1+\cos^2\phi\right)\right.$$
$$\left.-\frac{5}{4}\sin^4\theta\left(1+\cos^2\phi\right)\right]\times\frac{\partial}{\partial\left(m_f^2\right)}\frac{1}{\left(K^2-m_f^2\right)\left(P^2-m_f^2\right)}$$

After performing the frequency sum followed by the integral, we finally obtain

$$c_2(q_0,q) = -\sum_f\frac{4g^2\left(q_fB\right)^2}{3\pi^2}g_k+\frac{g^2\left(q_fB\right)^2}{2\pi^2}\left(g_k+\frac{\pi m_f-4T}{32m_f^2T}\right)\times\left[-\frac{7}{3}\frac{q_0^2}{q_\perp^2}+\left(2+\frac{3}{2}\frac{q_0^2}{q_\perp^2}\right)A_0\right. \tag{A.54}$$

$$\left.+\left(\frac{3}{2}+\frac{5}{2}\frac{q_0^2}{q_\perp^2}+\frac{3}{2}\frac{q_3^2}{q_\perp^2}\right)A_2-\frac{3q_0q_3}{q_\perp^2}A_1-\frac{5}{2}\left(1-\frac{q_3^2}{q_\perp^2}\right)A_4-\frac{5q_0q_3}{q_\perp^2}A_3\right], \tag{A.55}$$

The remaining $A$ integrals are

$$A_1 = -\frac{q_0 q_3}{q^2}\left[1-\frac{q_0}{2q}\log\left(\frac{q_0+q}{q_0-q}\right)\right] \tag{A.56}$$

$$A_3 = \frac{q_0}{2q}\frac{q_3}{q}\left(1-\frac{5}{3}\frac{q_3^2}{q^2}\right)-\frac{3}{2}\frac{q_0}{q}\frac{q_3}{q}\left(1-\frac{q_0^2}{q^2}-\frac{q_3^2}{q^2}+\frac{5}{3}\frac{q_0^2}{q^2}\frac{q_3^2}{q^2}\right)\times\left(1-\frac{q_0}{2q}\log\frac{q_0+q}{q_0-q}\right) \tag{A.57}$$

$$A_4 = \frac{3}{8}\left(1-\frac{q_3^2}{q^2}\right)^2-\frac{q_0^2}{8q^2}\left(1-\frac{5q_3^2}{q^2}\right)^2+\frac{5}{3}\frac{q_0^2}{q^2}\frac{q_3^4}{q^4}-\frac{3}{8}\left\{\left(1-\frac{q_0^2}{q^2}\right)^2-\frac{2q_3^2}{q^2}\left(1-\frac{3q_0^2}{q^2}\right)^2\right. \tag{A.58}$$

$$\left.+\frac{q_3^4}{q^4}\left(1-\frac{5q_0^2}{q^2}\right)^2+\frac{8q_0^4}{q^4}\frac{q_3^2}{q^2}\left(1-\frac{5q_3^2}{3q^2}\right)\right\}\times\left(1-\frac{q_0}{2q}\log\frac{q_0+q}{q_0-q}\right). \tag{A.59}$$

It should be noted that the imaginary parts of the form factors come from the imaginary parts of the $A_i'$s. We write down the final expressions. The detailed derivation can be found in [144]

$$d_2(q_0,q) = \frac{\bar{n}^\mu\bar{n}^\nu}{\bar{n}^2}\left[\Pi_{\mu\nu}^{(1,1)}+2\Pi_{\mu\nu}^{(2,0)}\right] \tag{A.60}$$

$$= F_1+F_2, \tag{A.61}$$

where,

$$F_1 = -\sum_f\frac{g^2\left(q_fB\right)^2q^2}{\pi^2q_\perp^2}\times\left[g_k\left\{\frac{q_0^2q_3^2}{3q^4}+\frac{A_0}{4}-\left(\frac{3}{2}+\frac{q_0^2q_3^2}{q^4}\right)A_2+\frac{5}{4}A_4\right\}+\left(\frac{\pi}{32m_fT}-\frac{1}{8m_f^2}\right)\right. \tag{A.62}$$

$$\times\left\{\frac{A_0}{4}-\left(\frac{3}{2}+\frac{q_0^2q_3^2}{p^4}\right)A_2+\frac{5}{4}A_4\right\}-f_k\frac{q_0^2q_3^2}{q^4}\left(\frac{14}{3}-5A_0+A_2\right)+\frac{q_0^2q_3^2}{q^4}\frac{8T-\pi m_f}{128Tm_f^2}\left(5A_0-A_2\right. \tag{A.63}$$

$$F_2 = -\sum_f\frac{g^2\left(q_fB\right)^2}{6\pi^2m_fT}\frac{q^0q^3}{q_\perp^2}\frac{1}{1+\cosh\frac{m_f}{T}}\times\left(\frac{3A_1}{2}-A_3\right). \tag{A.64}$$

Finally,

$$a_2(q_0,q) = \frac{1}{2}\left(\frac{\bar{u}^\mu\bar{n}^\nu + \bar{u}^\nu\bar{n}^\mu}{\sqrt{\bar{u}^2}\sqrt{\bar{n}^2}}\right)\left[\Pi^{(1,1)}_{\mu\nu} + 2\Pi^{(2,0)}_{\mu\nu}\right] \tag{A.65}$$

$$= G_1 + G_2, \tag{A.66}$$

$$G_1 = \sum_f \frac{4g^2\left(q_fB\right)^2}{2\pi^2\sqrt{\bar{u}^2}\sqrt{\bar{n}^2}} \times \left[\frac{q_0q_3}{q^2}\left\{\left(\frac{2}{3} - A_0 + A_2\right)g_k + \left(\frac{4}{3} - \frac{5A_0}{3} + A_2\right)f_k\right\}\right. \tag{A.67}$$

$$\left. + \left\{(-A_0 + A_2)\frac{\pi m_f - 4T}{32Tm_f^2} \; -\frac{1}{6}\left(5A_0 - 3A_2\right)\frac{8T - \pi m_f}{64Tm_f^2}\right\}\right]. \tag{A.68}$$

$$G_2 = \sum_f \frac{g^2\left(q_fB\right)^2}{\sqrt{\bar{u}^2}\sqrt{\bar{n}^2}6\pi^2 m_f T\left(1 + \cosh\frac{m_f}{T}\right)} \times \left(-5A_1 + 4A_3\right). \tag{A.69}$$

## A.3 Calculation of spectral functions, $\rho_i$

The cut part of the spectral functions are evaluated from the discontinuity in the pieces of the gluon propagators, which in turn is given by their imaginary parts analytically continued to real values of energy

$$\rho_1(\omega,q) = -\frac{1}{\pi}\,\mathrm{Im}\left(\chi_1|_{q_0=\omega+i\epsilon}\right) \tag{A.70}$$

$$= -\frac{1}{\pi}\,\mathrm{Im}\left(\left.\frac{(Q^2-d)}{(Q^2-b)(Q^2-d)-a^2}\right|_{q_0=\omega+i\epsilon}\right) \tag{A.71}$$

$$= -\frac{1}{\pi D}\left[\Im_b\left(\Im_d^2 + \Re_d^2 + Q^4 - 2Q^2\Re_d\right) + 2\Im_a\Re_a\left(Q^2 - \Re_d\right) + \Im_d\left(\Re_a^2 - \Im_a^2\right)\right]. \tag{A.72}$$

Here $\Im$ and $\Re$ respectively depict the imaginary and real parts of the form factors.

$$\rho_2(\omega,q) = -\frac{1}{\pi}\,\mathrm{Im}\left(\chi_2|_{q_0=\omega+i\epsilon}\right) \tag{A.73}$$

$$= -\frac{1}{\pi}\,\mathrm{Im}\left(\left.\frac{1}{(Q^2-c)}\right|_{q_0=\omega+i\epsilon}\right) = -\frac{1}{\pi}\left[\frac{\Im_c}{\Im_c^2 - (Q^2-\Re_c)^2}\right]. \tag{A.74}$$

$$\rho_3(\omega,q) = -\frac{1}{\pi}\,\mathrm{Im}\left(\chi_3|_{q_0=\omega+i\epsilon}\right) \tag{A.75}$$

$$= -\frac{1}{\pi}\,\mathrm{Im}\left(\left.\frac{\left(Q^2-b\right)}{\left(Q^2-b\right)\left(Q^2-d\right)-a^2}\right|_{q_0=\omega+i\epsilon}\right) \tag{A.76}$$

$$= -\frac{1}{\pi D}\left[\Im_d\left(\Im_b^2+\Re_b^2+Q^4-2Q^2\Re_b\right)\;+2\Im_a\Re_a\left(Q^2-\Re_b\right)+\Im_b\left(\Re_a^2-\Im_a^2\right)\right]. \tag{A.77}$$

$$\rho_4(\omega,q) = -\frac{1}{\pi}\,\mathrm{Im}\left(\chi_4|_{q_0=\omega+i\epsilon}\right) \tag{A.78}$$

$$= -\frac{1}{\pi}\,\mathrm{Im}\left(\left.\frac{a}{\left(Q^2-b\right)\left(Q^2-d\right)-a^2}\right|_{q_0=\omega+i\epsilon}\right) \tag{A.79}$$

$$= -\frac{1}{\pi D}\left[\Im_a\left\{-\Im_b\Im_d+\Re_b\Re_d+\Re_a^2+\Im_a^2+Q^4-Q^2\left(\Re_b+\Re_d\right)\right\}\right. \tag{A.80}$$

$$\left.+\Re_a\left(Q^2\left(\Im_b+\Im_d\right)\Im_d\Re_b-\Im_b\Re_d\right)\right]. \tag{A.81}$$

Here the denominator $D$ is expressed as

$$D = \left[\left(-\Im_b Q^2-\Im_d Q^2+\Im_d\Re_b+\Im_b\Re_d-2\Im_a\Re_a\right)^2+\left(-\Im_b\Im_d+\Im_a^2+\left(Q^2-\Re_b\right)\left(Q^2-\Re_d\right)-\Re_a^2\right)\right.$$